\documentclass[12pt,twoside,a4paper,BCOR12mm,footexclude]{scrreprt}
\usepackage[english]{babel}
\usepackage{amssymb}
\usepackage{color}
\usepackage{psfrag}
\usepackage{epsfig}
\usepackage{url}
\usepackage{physik}
\usepackage{feynmf}

\begin{document}

\newcommand{\bsf}{\sffamily\bfseries}

\newcommand{\pt}{p_\perp}
\newcommand{\kt}{k_\perp}
\newcommand{\mt}{m_\perp}
\newcommand{\df}{\Delta \phi}
\newcommand{\mean}[1]{\langle #1 \rangle}

\renewcommand{\topfraction}{0.9}
\renewcommand{\textfraction}{0.1}
\renewcommand{\floatpagefraction}{0.2}

\pagestyle{empty}

{\centering

\LARGE
\textsc{Faculty of Physics and Astronomy}

\vskip5mm

\Large
\textsc{University of Heidelberg}

\vfill

\normalsize

Diploma thesis\\[5pt]
in Physics\\
\vspace{.7cm}
submitted by\\[5pt]
\textbf{Korinna Zapp}\\[5pt]
born in Eckernf\"orde\\
\vspace{.7cm}
2005

\par}

\cleardoublepage

{\centering

\vspace{1cm}

{\bsf\large
The Soft Scattering Contribution
to Jet Quenching in a Quark-Gluon Plasma
and General Properties of \\Partonic Energy Loss
\par}

\vfill

This diploma thesis has been carried out by Korinna Zapp at the \\
Institute of Physics\\
under the supervision of\\
Prof.~Dr.~Johanna Stachel 

\par}

\cleardoublepage

% abstract -------------------------------------------------------------------- 

\subsubsection*{Der Beitrag von weicher Streuung zum Jet Quenching in einem
Quark-Gluon Plasma und allgemeine Eigenschaften des Energieverlusts von 
Partonen}

Ausgehend von der Beobachtung, dass weiche Streuungen eines hart gestreuten
Partons am Farbladung tragenden Proton\"uberrest in Proton-Proton Kollisionen
eine wichtige Rolle spielen, wurde ein \"ahnliches Modell f\"ur weiche Streuung
in einem Quark-Gluon Plasma konstruiert und als Monte Carlo Generator
implementiert. Die haupts\"achliche Frage war, inwieweit diese weichen
Streuungen zum Energieverlust beitragen k\"onnen, den ein  hochenergetisches
Parton erleidet, wenn es ein Quark-Gluon Plasma durchquert. Das Ergebnis war,
dass sie bis zu 50\% des beobachteten Energieverlusts erkl\"aren k\"onnen.

Au\ss erdem wurde die Frage untersucht, was man von heute verf\"ugbaren Daten
\"uber die allgemeinen Eigenschaften des Energieverlusts von Partonen erfahren 
kann. Dazu wurde ein Monte Carlo Modell benutzt, das zwar die volle Simulation
des Quark-Gluon Plasmas beinhaltet aber f\"ur den Energieverlust einen 
allgemeineren Ansatz verwendet. Dadurch wird es m\"oglich verschiedene
Szenarien wie koh\"arente Gluonbremsstrahlung, Streuung etc.\ nachzubilden.
Leider stellte sich heraus, dass es sehr schwierig ist die verschiedenen 
M\"oglichkeiten zu unterscheiden, sodass zum jetzigen Zeitpunkt keine
eindeutige Entscheidung zugunsten der einen oder anderen m\"oglich ist.

\vspace{2cm}

\subsubsection*{The Soft Scattering Contribution to Jet Quenching in a
Quark-Gluon Plasma and General Properties of Partonic Energy Loss}

Starting from the observation that soft rescattering of hard scattered partons
from the colour charged remnant plays an important role in  proton-proton
collisions a similar model for soft scatterings in a quark-gluon plasma was
constructed and implemented as a Monte Carlo event generator. The main emphasis
was put on the question to what extend these soft scatterings can contribute to
the energy loss that an energetic parton suffers, when it traverses a
quark-gluon plasma. It was found that the soft scattering can account for up to
50\% of the observed energy loss.

Furthermore, it was investigated what information present data reveal about the
general features of partonic energy loss. This study was carried out with a
Monte Carlo model that includes the full simulation of the quark-gluon plasma
but uses a more general ansatz for the energy loss which makes it possible to
emulate different scenarios like coherent gluon bremsstrahlung, scattering etc.
Unfortunately, it turned out that it is difficult to differentiate between them
so that at present no clear decision in favour of one or the other is possible.

\cleardoublepage

% thesis ----------------------------------------------------------------------

\pagestyle{headings}
\pagenumbering{roman}
\setcounter{page}{1}

\tableofcontents
\cleardoublepage

\pagenumbering{arabic}
\setcounter{page}{1}

\chapter{Introduction}
The quark-gluon plasma (QGP) is a state of deconfined quarks and gluons and
restored chiral symmetry predicted by Quantum Chromodynamics (QCD), the theory
of the strong interactions. It is expected to be formed at very high energy
densities. It presumably existed in the early universe and can possibly still
today be found in some kinds of  neutron stars or other exotic objects. Great
efforts are being made to produce the QGP in the laboratory by means of heavy
ion collisions. This is the only possibility to reach high enough energy
densities in man-made experiments. The price that has to be paid is that the
QGP -- if it is formed at all -- has a very short lifetime because of rapid
expansion and thus cooling of the system.

The detection and study of the QGP is a highly non-trivial task that has kept
both experimentalists and theorists busy for many years and is still far from
completed. The observables can be grouped into soft and hard probes. The hard
probes are sensitive to the very early stage of the collision whereas the soft
ones probe the latest stage. Hard observables such as jet quenching
(suppression of particles with high transverse momentum) have only recently
become accessible with the Relativistic Heavy Ion Collider (RHIC) in
Brookhaven, where data taking started in 2000. They are considered an adequate
tool for investigating the properties of the medium created in the collisions
wherefore sometimes the term "jet tomography" is used. For the proof of QGP
formation on the other hand other observables like thermal and chemical
equilibration are more suited.

Most of the phenomena connected to heavy ion collisions and QGP formation and
its evolution belong to the non-perturbative regime of QCD where it is very
difficult to derive a quantitative description from first principles. Thus
models of different scenarios have to be constructed and the results compared
to data in order to learn something about the physics. This is often a
laborious business and it is in the majority of cases not possible to obtain
clear cut conclusions.

\enlargethispage{-\baselineskip}

In this thesis a new model for the jet quenching is presented and compared to
data. It is built on partonic energy loss due to soft colour interactions. The
starting point was the Soft Colour Interaction (SCI) model \cite{sci} that
successfully describes a wealth of data mainly on deep inelastic scattering and
diffraction. The SCI model has been extended in order to be applicable for the
jet quenching phenomena. 

There is no perfect model for the jet quenching, so this new model can
hopefully help to gain a better understanding of the effects in QGP and soft
physics in general. The SCI jet quenching model is implemented as a Monte Carlo
event generator which has the advantage that the processes can be simulated and
studied in great detail.

A second issue of this thesis are the general properties of the energy loss
mechanism. Different theoretical approaches lead to quite different behaviours.
The idea is to see what presently available data reveal about the main features
of the sought energy loss mechanism without having a detailed model. This
information may then help to get to the right description in the end.

\smallskip

In the first part of this report a short overview of QCD, the search for the
QGP, jet quenching as a possible signature of the QGP and a few basics of heavy
ion physics is given. There is also a discussion of experimental results with
the main focus on jet quenching. After a brief introduction to Monte Carlo
methods and the SCI model the new jet quenching model is described and compared
to data in more detail. In the following section an investigation of general
features of energy loss in the QGP  is carried out with the help of an adjusted
Monte Carlo model.

\chapter{QCD for Beginners}
\label{chap_qcd}
This section gives an introduction to the basics of QCD and some topics that
are relevant for this study. A more systematic and detailed discussion can be
found in \cite{nachtmann,qcd} or other standard books.

\section{Introduction} %------------------------------------------------------- 
\label{sec_qcdintro} 
 
Quantum Chromodynamics (QCD) is the theory of the strong interaction. It 
describes the interaction of quarks via exchange of gluons. There are the six 
quarks called down ($d$), up ($u$), strange ($s$), charm ($c$), bottom ($b$) 
and top ($t$) which are fermions and the massless gluons which are bosons. The
relevant charge, which is the source of the field, is called colour and the
three states are labelled red ($r$), blue ($b$) and green ($g$). Quarks come as
colour triplets, which means that a particular quark can carry any of the three
colours. The antiquarks form the corresponding antitriplet (i.e.\ they carry
the anticolours $\bar r$, $\bar b$ and $\bar  g$). Quarks cannot be observed as
free particles but are always confined into hadrons, which are colour neutral
objects. A colour neutral combination is also called a singlet. There are at
least two possible neutral colour combinations: a colour and its matching
anticolour (giving rise to mesons that are composed of one quark and one
antiquark), or the three (anti)colours (leading to (anti)baryons that consist
of three (anti)quarks). There is also a possibility to combine two quarks, but
it is not clear to what extend these diquarks are bound states. Nevertheless,
they can be very useful, e.g.\ for the treatment of remnants as will be
discussed later. The combination of two triplet charges gives an antitriplet
(or a sextet, but this is not interesting in this context), two antitriplets
form a triplet. 

The gluons are also colour charged, but unlike the quarks they carry octet 
charges which can be viewed as the combination of a colour and an anticolour 
(the combination of colour and anticolour gives a singlet and an octet). The
name octet already indicates that it contains eight states. The naively expected
nineth gluon does not occur because as a colour singlet combination it does
not correspond to an interaction. The  fact that the gluons carry colour charge
gives rise to the gluon self-interaction, which means that gluons can directly
couple to each other. 

\smallskip 
 
The coupling strength of the strong interaction $\alpha_s$ is not constant but
depends on  the momentum transfer such that it increases with decreasing energy
scale.  Therefore processes involving large momentum transfers can be treated 
analytically since it is possible to make a perturbative expansion in powers 
of $\alpha_s$. In regimes with small energy scales, where the  coupling is
large, pertubation theory is not applicable and other ways to treat these
problems have to be found. One possibility is to solve the equations 
numerically, which has its own problems. This task is highly non-trivial and to
a great extent limited by  the available computing power. In many cases one has
to rely on phenomenological models that describe the observed phenomena without
having a firm theoretical basis. 

\section{Structure of the proton and parton showers} %------------------------- 
 
The proton is a baryon built of three valence quarks (two $u$- and one 
$d$-quark) which carry its electric charge and baryon number. In addition, 
there are also gluons and sea quarks that are fluctuations of gluons into 
quark-antiquark pairs. The quarks carry approximately 50\% of the proton's 
energy-momentum, the other half is carried by the gluons. 
 
\bildtex{plots/dis.pictex}{fig_dis}{Inelastic lepton-proton scattering via 
exchange of a virtual photon} 
 
The best way to investigate the structure of the proton is in lepton-proton 
scattering via photon exchange (Fig.~\ref{fig_dis}). The standard variables
are 

\begin{eqnarray} 
 Q^2 &=& -q^2 \\ 
 x   &=& \frac{-q^2}{2p\cdot q} \\ 
 y   &=& \frac{q\cdot p}{k\cdot p}  
\end{eqnarray} 
where $p$, $q$ and $k$ are the proton's, the photon's and the lepton's
four-momenta, respectively ($k$ refers to the initial and $k'$ to the final
vector).   

At small momentum transfers $Q^2$ the photon interacts with the proton as a 
whole, but at higher $Q^2$ the structure of the proton can be resolved. The 
momentum transfer can be viewed as the resolution: With a bigger $Q^2$ smaller 
structures can be resolved (the resolution is $d \sim \unit[0.2]{fm}/\sqrt{Q^2} 
[\unit{GeV}]$). The cross section can be written as 
 
\begin{equation} 
\frac{\tn{d}^2\sigma}{\tn{d}Q^2 \tn{d}x} = \frac{4\pi}{xQ^4} ((1-y)F_2 + 
     xy^2 F_1)
\end{equation} 
where the structure functions $F_i(x,Q^2)$ parametrise the proton structure.
They are approximately independent of $Q^2$ (Bjorken scaling) which implies
that the photon interacts with quasi-free pointlike constituents (partons) of
the proton. In  the frame where the proton has infinite momentum, $x$ can be
interpreted as the limit of the fraction of the proton's momentum carried by
the parton. The structure function $F_2$ can to lowest order be written as 

\begin{equation} 
  F_2(x,Q^2) = \sum_a e_a^2xf_a(x,Q^2) 
\end{equation}
where $a$ runs over all flavours and antiflavours, $e_a$ is the electric charge
of the respective quark or antiquark and $f_a(x,Q^2)$ is the probability of
finding a quark of flavour $a$ carrying the momentum fraction $x$ inside the
proton when probing with momentum transfer $Q^2$. The $f_a$ are called parton
distribution functions (pdf's), they include the contributions from both the
valence and the sea quarks. Scattering of the lepton from gluons is not
possible since the gluons do not carry electric charge. 

$F_1$ is connected to $F_2$ via the Callan-Gross relation

\begin{equation} 
F_2 = 2xF_1 
\end{equation}

The $Q^2$ dependence of the parton distribution arises from the fact that the 
(anti)quark may have radiated gluons (initial state radiation)  before it was
actually struck by the photon. The $Q^2$ dependence can be  calculated in
pertubation theory and is governed by the DGLAP\footnote{\textbf{D}okshitzer
\textbf{G}ribov \textbf{L}ipatov \textbf{A}ltarelli \textbf{P}arisi} evolution 
equations  

\begin{equation}
   \del[f_q(x,Q^2)]{\ln Q^2} = \frac{\alpha_s(Q^2)}{2\pi} \int \limits_x^1 
      \frac{\tn{d}y}{y}\ \left[ P_{qq}\left( \frac{x}{y} \right) f_q(y,Q^2) +  
	P_{qg}\left( \frac{x}{y} \right) f_g(y,Q^2) \right] 
\end{equation}
\begin{equation} 
 \del[f_g(x,Q^2)]{\ln Q^2} = \frac{\alpha_s(Q^2)}{2\pi} \int \limits_x^1 
      \frac{\tn{d}y}{y}\ \left[ P_{gq}\left( \frac{x}{y} \right) f_q(y,Q^2) +  
	P_{gg}\left( \frac{x}{y} \right) f_g(y,Q^2) \right] 
\end{equation}
where $f_q$ and $f_g$ are the quark and gluon pdf's, respectively. The
$P_{ij}$ are called splitting functions and describe the evolution of a
parton $j$ into a parton $i$ carrying the energy fraction $z=x/y$ of the
original parton. $P_{qq}$ for instance describes the radiation of a gluon
from a quark where the quark keeps the energy fraction $z$; $P_{gq}$
describes the same situation when the gluon gets the fraction $z$ and the
quark obtaines $1-z$. 

There are similar equations for gluon radiation after the photon-parton
scattering (final  state radiation). The gluons tend to be emitted with small 
angles relative to the (anti)quark. The branching processes can explicitly be
simulated using parton showers.

In contrast to the $Q^2$ dependence, the $x$ dependence of the pdf's has to
be  parametrised and fitted to data. There are of course parton densities for
all hadrons, it is common to write  $f_{a/h}$ for the pdf of the parton $a$ in
hadron $h$.    

For (perturbatively hard) hadron-hadron processes any inclusive cross section
can be factorised into the form 

\begin{equation} 
  \sigma(p_1,p_2) = \sum_{a,b}\int\!\tn{d}x_1\,\tn{d}x_2\ f_a(x_1,\mu^2)  
      f_b(x_2,\mu^2) \hat \sigma_{a,b}(x_1p_1,x_2p_2;Q^2/\mu^2)
\end{equation}
where $\hat \sigma_{a,b}$ is the parton level cross section for the parton 
species $a,b$ and $\mu^2$ is an arbitrary factorisation scale separating the
hard from the soft regime. (It is a  factorisation of the process in pdf's and
the parton-parton process described by $\hat \sigma_{a,b}$.)

\section{Fragmentation} %------------------------------------------------------ 
 
Quarks and gluons cannot be observed as free particles but are confined in 
colour neutral hadrons. The process of hadron formation from partons (quarks 
and gluons) falls into the domain of non-perturbative QCD and it has not been 
possible to derive it from first principles (it might become possible with 
lattice QCD). Nevertheless, a profound understanding of these mechanisms is of 
great importance for all kinds of activities in high energy physics, where the
dynamics on the parton level has to be deduced from the observed hadronic final
state. 

There are basically three models: string fragmentation, independent 
fragmentation and cluster fragmentation. The main ideas of the Lund string 
model \cite{lund} and the independent fragmentation (\cite{indep}, see also
\cite{pythia} for an overview) will be reviewed briefly here. 

\subsection{String Fragmentation}
\label{sec_lund} 

\enlargethispage{\baselineskip}

\bildtex{plots/lund1.pictex}{fig_lund1}{Quark-quark scattering via gluon 
exchange as an example of inelastic pp scattering with the resulting colour 
field topology represented by strings (dotted lines)} 
 
The jet production from a hard quark-quark scattering in a proton-proton 
collision can serve as a simple example for the fragmentation of a quark. In
this interaction the protons are broken apart: The two quarks are scattered to
large angles and the remnants (in this case diquarks) continue in the beam
directions. The incoming protons are colour singlets, after the scattering the
(colour-)charged constituents can again be grouped into two colour neutral
systems (Fig.~\ref{fig_lund1}). A field stretches between a colour charge
(triplet charge) and its matching anticolour (antitriplet, in this case the
diquark). Due to the gluon self-interaction the field does not extend
transversly in space but can be viewed as a colour flux tube with transverse
size $\sim \unit[1]{fm}$. The potential rises linearly with the distance
between the charges and at some point it becomes possible to form a new
quark-antiquark pair from the field energy. This process is sketched in
Figure~\ref{fig_sbreak}. 

\bildtex{plots/stringbreak.pictex}{fig_sbreak}{Creation of a $q\bar q$ pair 
  from the field energy during the separation of a quark ($q$) and a diquark 
  ($qq$)} 
 
In the Lund model the colour flux tube is idealised to a one-dimensional 
massless relativistic string with energy density $\kappa \simeq 
\unit[1]{GeV/fm}$. A newly produced $q\bar q$ pair breaks the string into two 
independent parts. The creation of a $q\bar q$ pair is described as a quantum 
mechanical tunneling process so that the probability becomes \cite{lund}
 
\begin{equation} 
  P \sim \exp \left( -\frac{\pi m_\perp^2}{\kappa} \right) 
       = \exp \left( -\frac{\pi m^2}{\kappa} \right) \exp \left( -\frac{\pi 
	   \pt^2}{\kappa} \right) 
\label{eq_prob} 
\end{equation} 
 
This leads to a Gaussian distribution of the two transverse components of the 
produced (anti)quark's momentum ($p_x$ and $p_y$ if the string is stretched 
along the $z$-axis). The transverse momentum is compensated when the string
breaks between the created quark  and the antiquark. Equation~\ref{eq_prob}
shows also that the  creation of heavy flavours is suppressed so that charm,
bottom and top quarks  are not expected to be produced. Since the quark masses
are not well defined in  this context the suppression of strangeness relative
to $u$ (or $d$) quarks is  left as a parameter with $\gamma_s= \frac{s\bar
s}{u\bar u} \sim 0.3$. 

\enlargethispage{\baselineskip}

It is assumed that a meson is formed when the invariant mass of a string piece 
becomes small. Its $\pt$ is given by the sum of the transverse momenta of the 
string endpoints, which also fix the flavour composition. The spin of the 
meson is chosen to be either 0 or 1 according to a suitable probability, 
orbital excitations are expected to be rare and are therefore excluded in the
default Lund model.  
 
The different break-ups of the string are assumed to be independent of each 
other. It is therefore possible to start at one end of the string, make a 
break-up and be left with a meson and a shorter string. The procedure can be 
iterated until the available energy is used up (the termination requires some 
extra treatment that will not be discussed here). Let us consider the string in 
the c.m.\ frame of the quark and the diquark with the quark moving in the 
$+z$-direction and start with the fragmentation from the quark end. The first 
step is to choose the flavour, spin and transverse momentum of the created 
$q\bar q$ pair as described above. The identity and transverse mass of the 
meson are then fixed and only the longitudinal momentum (or the energy) 
remains to be determined. This is done by assigning the meson a fraction $z$ 
of the available $E+p_z$. What is left to the new string is \cite{lund} 
 
\begin{eqnarray} 
  (E + p_z)_\tn{new} &=& (1-z)(E + p_z)_\tn{old} \\ 
  (E - p_z)_\tn{new} &=& (E-p_z)_\tn{old} - \frac{m_\perp^2}{z(E+p_z)_\tn{old}} 
\end{eqnarray} 
 
The values of $z$ are distributed according to a probability distribution 
$f(z)$. The constraint that the result should be independent of the choice 
from which end to start leads to the 'Lund symmetric fragmentation function' 
 
\begin{equation} 
  f(z) \propto z^{-1}(1-z)^a \exp\left(-\frac{bm_\perp^2}{z} \right) 
\end{equation} 
 
One important point is that the hadron that contains the end-(di)quark
of  the string will typically have the highest momentum in the ensemble. It 
should further be noted that in the lab system it is the slowest hadrons that
are  formed first, because all particles have a formation time of $\sim 
\unit[1]{fm}$ in their rest frame. The difference in formation time in the lab 
frame is a boost effect. This ordering is frame dependent and therefore  not
important for the fragmentation algorithm, which is Lorentz invariant. 

\bildtex{plots/lund2.pictex}{fig_lund2}{Inelastic pp scattering with
  additional  gluon radiation resulting according to the Lund model in a
  colour string field topology (dashed lines) where radiated gluons are
  connected in the same string as the scattered quarks} 

The formation of baryons is somewhat more complicated but works in principle 
like the meson production. In the 'popcorn model' baryons arise from cases 
where the produced $q\bar q$ pair does not match the colour of the string ends.
It  is then possible that another pair with the third colour is produced and a 
baryon and an antibaryon is formed. Baryon production is suppressed since
pairs  with the "wrong" charge can only exist as fluctuations and also because
two  $q\bar q$ pairs must be produced. Another source of baryon production in
the Lund model are the remnant diquarks, which are turned into a baryons via
normal string break-ups with formation of a $q\bar q$ pair. 

Gluons carry colour octet charges which means that they cannot serve as 
endpoint for a triplet string. But they can be situated in the middle of a 
string with the colour connected to the anticolour-end and the anticolour with 
the colour-end of the string. If a quark for example radiates gluons in a 
parton shower \`a la DGLAP the gluons will be aligned in the same string as the 
quark (Fig.~\ref{fig_lund2}). 
 
\subsection{Independent Fragmentation}
\label{sec_indep}

\enlargethispage{\baselineskip}

In the framework of independent fragmentation it is assumed that each parton
hadronises on its own, i.e.\ the fragmentation of a jet system  is an
incoherent sum of the fragmentations of each parton. The procedure has to be
carried out in the overall c.m. system.

\smallskip

There is an iterative approach somewhat similar to the Lund model: A jet arising
from the fragmentation of a quark is split
into a meson and a remainder jet with lower energy and momentum. The sharing of
energy-momentum is governed by a fragmentation function, actually the same
functions can be used for string and independent fragmentation. Flavour and
transverse momentum are conserved in each break-up. It is assumed that the
formation of
a meson does not depend on the energy of the remainder jet so that the step can
be iterated resulting in a sequence of hadrons. There is, however, one problem
with very small values of $z$: They lead to backward moving hadrons (i.e.\
$p_\parallel < 0$) that have to be rejected. 

Although flavour and transverse momentum are conserved locally this is not the
case in the global balance. In the end there will always be an unpaired
(anti)quark left behind. There may also be hadrons with $p_\parallel < 0$, that
were removed their energy and momentum being lost for the jet. Thus overall
energy, momentum, charge and flavour are not conserved. This also applies to jet
systems since each parton hadronises separately.

\smallskip

There are several possibilities for the treatment of gluon jets. Since a gluon
is expected to result in a softer jet it is sensible to split the gluon
perturbatively in a $q \bar q$ pair, then the same fragmentation function as
for the quark jets can be used.

\smallskip

Apart from the non-conservation of energy-momentum and flavour there are two
more conceptual weaknesses: the issues of Lorentz invariance and collinear
divergences. The result of independent fragmentation depends on the coordinate
frame and is thus not Lorentz invariant. This problem is circumvented by
requiring the fragmentation to take place in the overall c.m. frame. The
colliniear divergence, on the other hand, leads to problems in connection with
parton showers. A system of collinear partons leads to a much higher hadron
multiplicity than a single parton with the same energy.

\chapter{Hunting the Quark Gluon Plasma}
\section{QCD Predictions}

At high temperatures and densities the long range interactions between quarks
are dynamically screened (similar to Debye screening). Only the very short
range interactions remain but here the coupling is weak so that the quarks and
gluons are quasi-free and thus deconfined. Furthermore chiral symmetry is
restored at apparantly the same critical temperature $T_c$. This phase of
deconfined quarks and gluons and restored chiral symmetry is called the
quark-gluon plasma (QGP) \cite{bass}. This is not unexpected since
spontaneously broken symmetries (such as chiral symmetry) are often restored
at high temperatures through a phase transition \cite{harris}. An important
question is now whether there is a phase transition from hadronic to
deconfined matter. Recent results of simulations of QCD on the lattice
indicate a phase transition at a critical temperature $T_c \simeq
\unit[170]{MeV}$ which corresponds to an energy density of $\epsilon_c \simeq
\unit[1]{GeV\, fm^{-3}}$ \cite{snellings}. This is too low for pertubation
theory to be applicable and one has to rely on lattice QCD. Although great
progress has been made in this field there are still major problems. It has,
for instance, so far not been possible to determine the order of the phase
transition and many calculations are done for vanishing baryon chemical
potential (i.e.\ vanishing baryon number) which is a good approximation for
the early universe and the LHC but not for AGS, SPS and RHIC
\cite{bass,harris}.

\section{Models}

\enlargethispage{\baselineskip}

There is a large variety of different models for heavy-ion collisions that
are  based on largely different ideas and assumptions. Only a short overview
over  the main classes can be given here \cite{bass,harris}.

\begin{description}
\item[Statistical models] (e.g.\ \cite{stachel}) assume that local thermal and
  chemical equilibrium is achieved during the collision. The starting point is
  a hadron gas that can either be created directly in the collision or the
  product of the hadronisation of a QGP. It is described as an ideal hadron
  gas using a canonical or grand canonical formalism. The gas expands until
  inelastic interactions cease, then  the composition of the system is fixed
  (chemical freeze-out). At some point the mean free path becomes so long that
  also the elastic interactions stop (thermal freeze-out). When resonance
  decays are included statistical models can yield the relative abundancies of
  hadron species as they are measured by experiments.\\    
  Comparison of the model results with data can help to clarify if, or to what
  extent, equilibrium is achieved. The freeze-out temperature and (hadron)
  chemical potential can be determined by fitting the model to data.  
\item[Parton Cascades] (e.g.\ \cite{part-cas}) are detailed microscopic models
  that describe the collision of two nuclei in a perturbative QCD framework.
  The first step is the decomposition of the nuclei in partons according to
  measured structure functions. The interactions during the collisions are
  treated as perturbative scatterings with initial and final state radiation.
  The last stage is the hadronisation of the partons using the Lund string
  model. \\  Parton cascades predict a rapid thermalisation (proper time scale 
  \unit[0.3-0.5]{fm}) and a chemical equilibration that takes somewhat longer
  (several \unit{fm}). The plasma is initially gluon rich due to the larger
  cross sections for gluons. 
\item[Hadronic Transport models] (e.g.\ \cite{had-trans}) are formulated in a
  hadron basis although some also include non-hadronic elements such as quarks
  and strings. The heavy-ion  collision is described as a sequence of
  collisions of constituents (mesons,  baryons, quarks,\dots). Partonic degrees
  of freedom are not treated  explicitly so that no phase transition can
  occur.\\ 
  Hadronic models provide a very useful background for other models because it
  is important to understand which phenomena can be described in terms of
  hadronic  physics. 
\item[Hydrodynamic models] (e.g.\ \cite{bjorken,hydro}) are macroscopic kinetic
  models that are based on the assumption of local equilibrium and
  energy-momentum conservation. The nuclei  are described as fluids and in some
  models a third fluid can be created in the collisions. Starting from the
  colliding nuclei, an equilibrated QGP or hadronic matter the time evolution
  of the system can be studied until hadronic freeze-out.\\ 
  These are the only dynamical models in which a phase transition can be
  incorporated explicitly via the equations of state. 
\end{description}

\section{Observables}

The number of observables that have been suggested as signatures of QGP
formation is so large that it is impossible to discuss all of them here.
Instead three of the most popular will be presented without going into details.

\subsubsection*{Strangeness Enhancement}

The production of strange hadrons is suppressed in pp collisions and the
suppression increases with the strangeness content of the respective hadron.
This has been argued to be due to the higher strange quark mass
\cite{bass,harris}. In a QGP strangeness saturation via $s \bar s$ pair
production is expected which would significantly increase the yields of
strange particles \cite{harris}. The estimated time scale of several \unit{fm}
for strangeness equilibration is maybe too long for a complete saturation, but
an increased strangeness content leads to increased strange particle yields in
any case. If it is the statistical hadronisation of a strangeness-enhanced
deconfined phase that is observed a stronger enhancement of multistrange
hadrons is expected. The enhancement factor for a hadron containing $N$
strange quarks is $E_s^N$ where $E_s$ is the global enhancement factor.
Hadron rescattering scenarios lead to the opposite behaviour \cite{scomparin}.

A QGP formed at AGS or SPS would have nonzero chemical potentials for $u$ and
$d$ leading to an ordering in the quark abundances: The densities of $u$ and
$d$ is higher than that of $s$ and $\bar s$, which is higher than the $\bar u$
and $\bar d$ density. This means that at freeze-out the combination of a $\bar
s$ with a $u$ or $d$ to $K^+$ or $K^0$ is more likely than the $s \bar u$ and
$s \bar d$ combination ($K^-$ and $\bar K^0$). Thus the $K^+/\pi^+$ ratio
should be different from $K^-/\pi^-$ in case of QGP formation. Unfortunately,
the argumentation has several drawbacks. One is that the strange particle
abundancies after freeze-out of a QGP are very close to those in an
equilibrated hadron gas with the same entropy content \cite{bass} so that it is
difficult to unambiguously relate different ratios to QGP formation.

\subsubsection*{Chemical Equilibrium and Freeze-Out}

If the observed hadrons are produced in the hadronisation of an equilibrated QGP
they should inherit the property of chemical equilibrium. This also includes the
disappearance of the strangeness suppression. From a statistical model fit to
data the freeze-out temperature and chemical potential can be found. These
values can then be compared to lattice QCD calculations for the phase boundary.
A chemical freeze-out close to or at the phase boundary suggests that the hadrons
originate from a deconfined medium and that the chemical composition is
established during the phase transition.

\subsubsection*{Charmonium Suppression} 

In a QGP colour screening reduces the range of the attractive force between
quarks and antiquarks and thereby prevents $c \bar c$ pairs from binding. It is
therefore expected that the more loosely bound $\psi'$ and $\chi_c$ states
start to be suppressed at a lower temperature than $J/\psi$ \cite{scomparin}.
There is evidence from lattice calculations that the $1S$ states $J/\psi$ and
$\eta_c$ survive up to $1.5\, T_c$ whereas the $\chi_c$ states are dissolved
already at $1.1\, T_c$ \cite{karsch}.

$c \bar c$ pairs are produced as small configurations and the evolution to the
larger charmonium state takes approximately \unit[1]{fm}. Thus the $J/\psi$
should survive if it escapes fast enough, i.e.\ if it has a high transverse
momentum or if the QGP expands very rapidly \cite{harris}.

There is, however, a problem with charmonium suppression as a clear signal of
a QGP and that is the charmonium suppression observed in pA collisions.
Interactions with comoving particles can break the charmonium apart, the
broadened intrinsic transverse momentum distribution (Cronin effect) and the
absorption on nucleons also contribute to the suppression. The hard
contribution to the charmonium-nucleon cross section can be calculated using
perturbative QCD and is in good agreement with pA data when the formation
length and feeding of the $J/\psi$ from $\psi'$ and $\chi_c$ are taken into
account \cite{zschiesche,scherer}. A QGP would thus manifest itself in an
anomalously high charmonium suppression.  

Hadronic cascade models attempt to explain the charmonium suppression in AA
collisions by interactions with comoving hadrons. They also predict a stronger
suppression of $\psi'$ than of $J/\psi$ \cite{harris}.

The statistical hadronisation model \cite{stachelc} assumes that all charm
quarks are produced in hard interactions in the early stage of the collision
and are then equilibrated in a QGP (thermal but not chemical equilibrium). The
charmed hadrons are formed at freeze-out according to statistical laws.

A similar behaviour is expected for the bottonium states.

\bigskip

Finding signatures that can unambiguously be related to the presence of a QGP
is very difficult and it is often possible to describe the effects, that where
believed to be a clear signal, with hadronic scenarios. It is thus likely that
the proof for QGP formation will be based on several effects, each of which
alone cannot provide a convincing proof.

The search at AGS and SPS was based on soft physics such as the strangeness
enhancement. They are mostly sensitive to the latest stage of the collision
after hadronisation. At RHIC also hard probes like jet quenching become
accessible, which provide information on the very early phase of the collision
since hard scatterings occur even before the equilibration of the QGP.

\chapter{Jet Quenching}
\label{chap_jet-quenching}
When a hard (i.e.\ large momentum transfer) scattering takes place in the
interaction of two protons the two scattered partons leave the protons with
high transverse momentum. In the centre-of-momentum frame they are emitted
back-to-back in azimuthal angle due to momentum conservation. The fragmentation
of each of these energetic partons gives rise to a jet, i.e.\ a spray of
hadrons with small angles relative to the momentum of the parton. The energy
distribution inside the jet is determined by the fragmentation function
(Sec.~\ref{sec_lund}). The typical jet signature is a large energy deposition
localised in a small solid angle. In two-jet events the two jets are opposite
in azimuth and have nearly the same energy (small deviations arise from
differences in the momentum fraction $x$ carried by the partons). Partons with
$\pt \gtrsim \unit[3-4]{GeV}$ are expected to give rise to jets. 

In a soft interaction the cross section for producing a given total transverse
momentum rises only slowly with the collision energy $\sqrt{s}$ since the
particle multiplicity rises logarithmically with $s$ and the mean $\pt$ also
depends only little on $s$. In hard interactions on the other hand the cross
section for producing a given total transverse momentum rises rapidly with $s$
because the required $x \sim \sum |\pt|/\sqrt{s}$ gets smaller and the parton
distribution functions rise rapidly towards smaller $x$. Therefore the
hard parton-parton scattering dominates the cross section for sufficiently high
$\sum |\pt|$ and $s$ \cite{barger}.

\smallskip

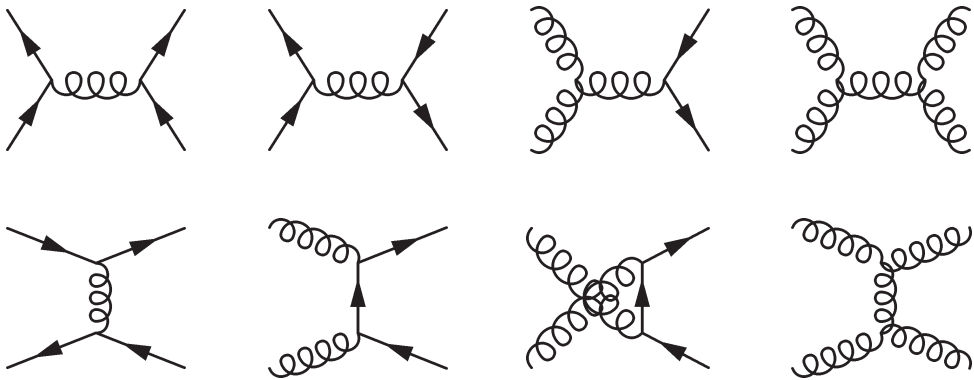
\begin{figure}[ht]
\begin{minipage}{.225\textwidth}
\begin{fmffile}{fmf1pic}
\begin{fmfchar*}(84,53)
  \fmfleft{em,ep} 
  \fmf{fermion}{em,Zee,ep}
  \fmf{gluon}{Zee,Zff}
  \fmf{fermion}{fb,Zff,f}
  \fmfright{fb,f}
\end{fmfchar*}
\end{fmffile} 
\end{minipage}
\begin{minipage}{.225\textwidth}
\begin{fmffile}{fmf2pic}
\begin{fmfchar*}(84,53)
  \fmfleft{em,ep} 
  \fmf{fermion}{em,Zee,ep}
  \fmf{gluon}{Zee,Zff}
  \fmf{fermion}{f,Zff,fb}
  \fmfright{fb,f}
\end{fmfchar*}
\end{fmffile} 
\end{minipage}
\begin{minipage}{.225\textwidth}
\begin{fmffile}{fmf3pic}
\begin{fmfchar*}(84,53)
  \fmfleft{em,ep} 
  \fmf{gluon}{em,Zee,ep}
  \fmf{gluon}{Zee,Zff}
  \fmf{fermion}{f,Zff,fb}
  \fmfright{fb,f}
\end{fmfchar*}
\end{fmffile} 
\end{minipage}
\begin{minipage}{.225\textwidth}
\begin{fmffile}{fmf4pic}
\begin{fmfchar*}(84,53)
  \fmfleft{em,ep} 
  \fmf{gluon}{em,Zee,ep}
  \fmf{gluon}{Zee,Zff}
  \fmf{gluon}{f,Zff,fb}
  \fmfright{fb,f}
\end{fmfchar*}
\end{fmffile} 
\end{minipage}

\vspace{1cm} 

\begin{minipage}{.225\textwidth}
\begin{fmffile}{fmf5pic}
\begin{fmfchar*}(84,53)
  \fmfleft{em,ep} 
  \fmf{fermion}{fb,Zee,em}
  \fmf{gluon}{Zee,Zff}
  \fmf{fermion}{ep,Zff,f}
  \fmfright{fb,f}
\end{fmfchar*}
\end{fmffile}
\end{minipage}
\begin{minipage}{.225\textwidth}
\begin{fmffile}{fmf6pic}
\begin{fmfchar*}(84,53)
  \fmfleft{em,ep} 
  \fmf{gluon}{em,Zee}
  \fmf{gluon}{Zff,ep}
  \fmf{fermion}{fb,Zee,Zff,f}
  \fmfright{fb,f}
\end{fmfchar*}
\end{fmffile}
\end{minipage}
\begin{minipage}{.225\textwidth}
\begin{fmffile}{fmf7pic}
\begin{fmfchar*}(84,53)
  \fmfleft{em,ep} 
  \fmf{fermion,tension=2}{fb,Zee,Zff,f}
  \fmf{gluon}{ep,Zee}
  \fmf{gluon}{Zff,em}
  \fmfright{fb,f}
  \fmfforce{(.6w,.25h)}{Zee}
  \fmfforce{(.6w,.75h)}{Zff}
\end{fmfchar*}
\end{fmffile}
\end{minipage}
\begin{minipage}{.225\textwidth}
\begin{fmffile}{fmf8pic}
\begin{fmfchar*}(84,53)
  \fmfleft{em,ep} 
  \fmf{gluon}{em,Zee}
  \fmf{gluon}{Zff,ep}
  \fmf{gluon}{fb,Zee,Zff,f}
  \fmfright{fb,f}
\end{fmfchar*}
\end{fmffile}
\end{minipage}
\label{fig_feyn}
\caption{Examples for Feynman diagrams for parton-parton scattering to lowest order}
\end{figure}

The lowest order parton-parton scattering ($a+b \to c+d$) cross section is of
the form \cite{barger}

\begin{equation}
 \der[\hat \sigma]{\hat t} (ab\to cd) = \frac{1}{16 \pi^2 \hat s^2}
 |\mathcal{M}|^2
\end{equation}
The quantities with hats refer to the parton level and $|\mathcal{M}|^2$ is the
matrix element squared. Spins and colours are averaged in the initial and
summed in the final state. $\hat s$, $\hat t$, $\hat u$ are the Mandelstam
variables

\begin{eqnarray}
\hat s\!\!\!&=&\!\!\!(p_a+p_b)^2 = (p_c+p_d)^2 \\
\hat t\!\!\!&=&\!\!\!(p_a-p_c)^2 = (p_b-p_d)^2 \\
\hat u\!\!\!&=&\!\!\!(p_a-p_d)^2 = (p_b-p_c)^2 
\end{eqnarray}
Figure~\ref{fig_feyn} shows a few examples of processes that contribute to
lowest order to the parton-parton scattering. 

With the help of the relation

\begin{equation}
 \pt^2 = \frac{\hat u \hat t}{\hat s} = \frac{\hat s}{4} \sin^2 \hat \theta 
\end{equation}
and neglecting the intrinsic transverse momentum of the partons the cross
section can be written as 

\begin{equation}
 \der[\hat \sigma]{(\pt^2)}
  = \frac{1}{\cos \hat \theta}\der[\hat \sigma]{\hat t}
  = \frac{\hat s}{\hat t - \hat u}\der[\hat \sigma]{\hat t}
\end{equation}
Then the lowest order QCD cross section for two-jet production in a collision of
two hadrons $A$ and $B$ is given by \cite{barger}

\begin{eqnarray}
 && \!\!\!\!\!\!\!\!\der[\sigma]{(\pt^2)}(AB\to 2\ \tn{jets})  \\ 
 && \!\!\! = \sum_{abcd} \int\limits_0^1\!\tn{d} x_a\!\int\limits_0^1\!\tn{d} x_b\,
 \Theta\left(x_ax_b-\frac{4 \pt^2}{s}\right) f_{a/A}(x_a,Q^2) f_{b/B}(x_b,Q^2) 
 \der[\hat \sigma]{(\pt^2)}(ab\to cd) \nonumber
\end{eqnarray} 

\smallskip

There are many ways to express the differential cross section, a very useful
relation is

\begin{equation}
 E \frac{\tn{d}^3 \sigma}{\tn{d}^3 p}
  = \frac{\tn{d}^3 \sigma}{\pt \tn{d} \phi\, \tn{d} y\, \tn{d}\pt}
  = 2\frac{\tn{d}^3 \sigma}{\tn{d} \phi\, \tn{d} y\, \tn{d}(\pt^2)}
\end{equation}  

\begin{figure}[ht]
\centering
\begin{minipage}[b]{.45\textwidth}
 \centering
 \includegraphics[scale=1]{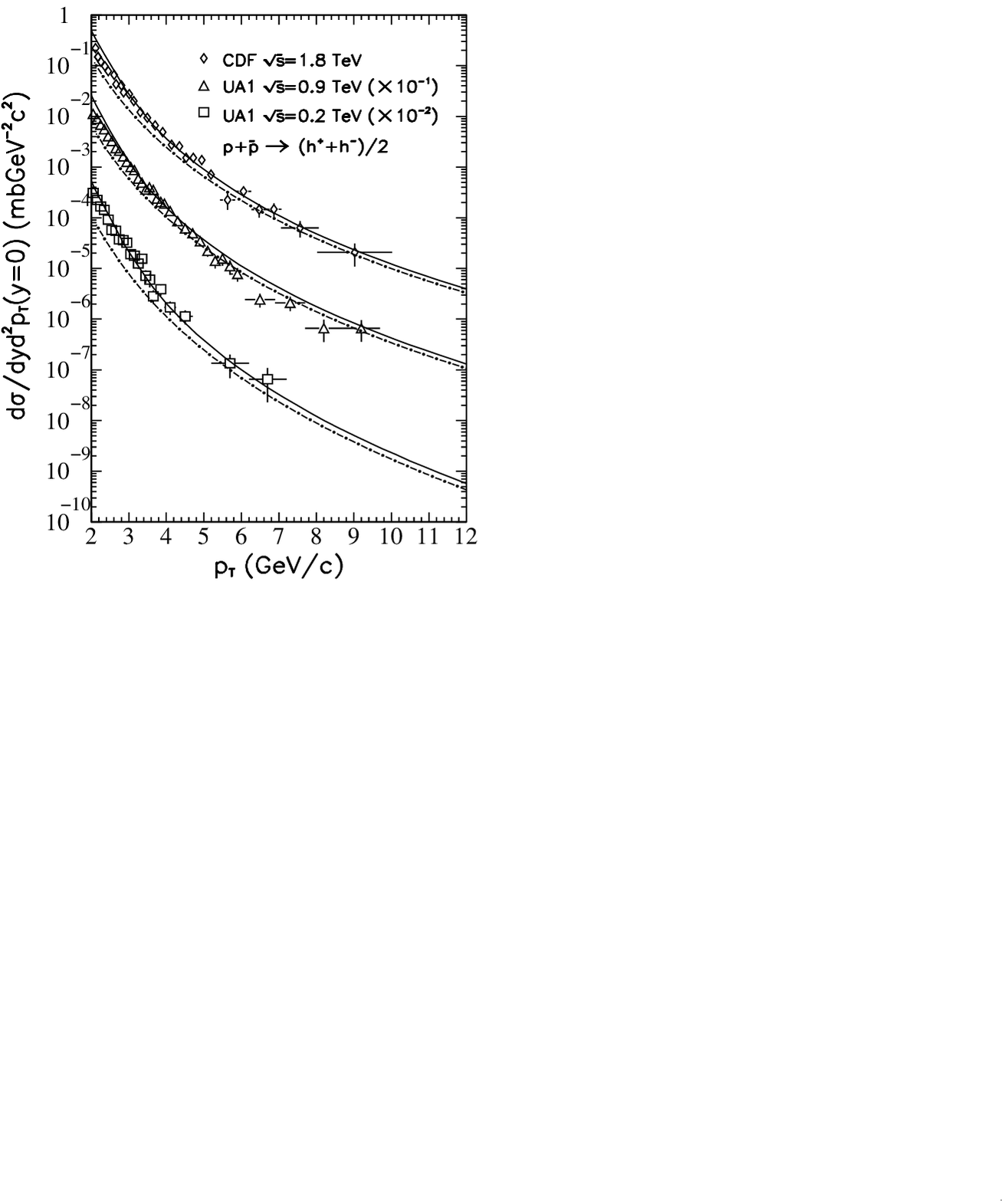}
\end{minipage}%
\begin{minipage}[b]{.55\textwidth}
 \centering \includegraphics[scale=0.5]{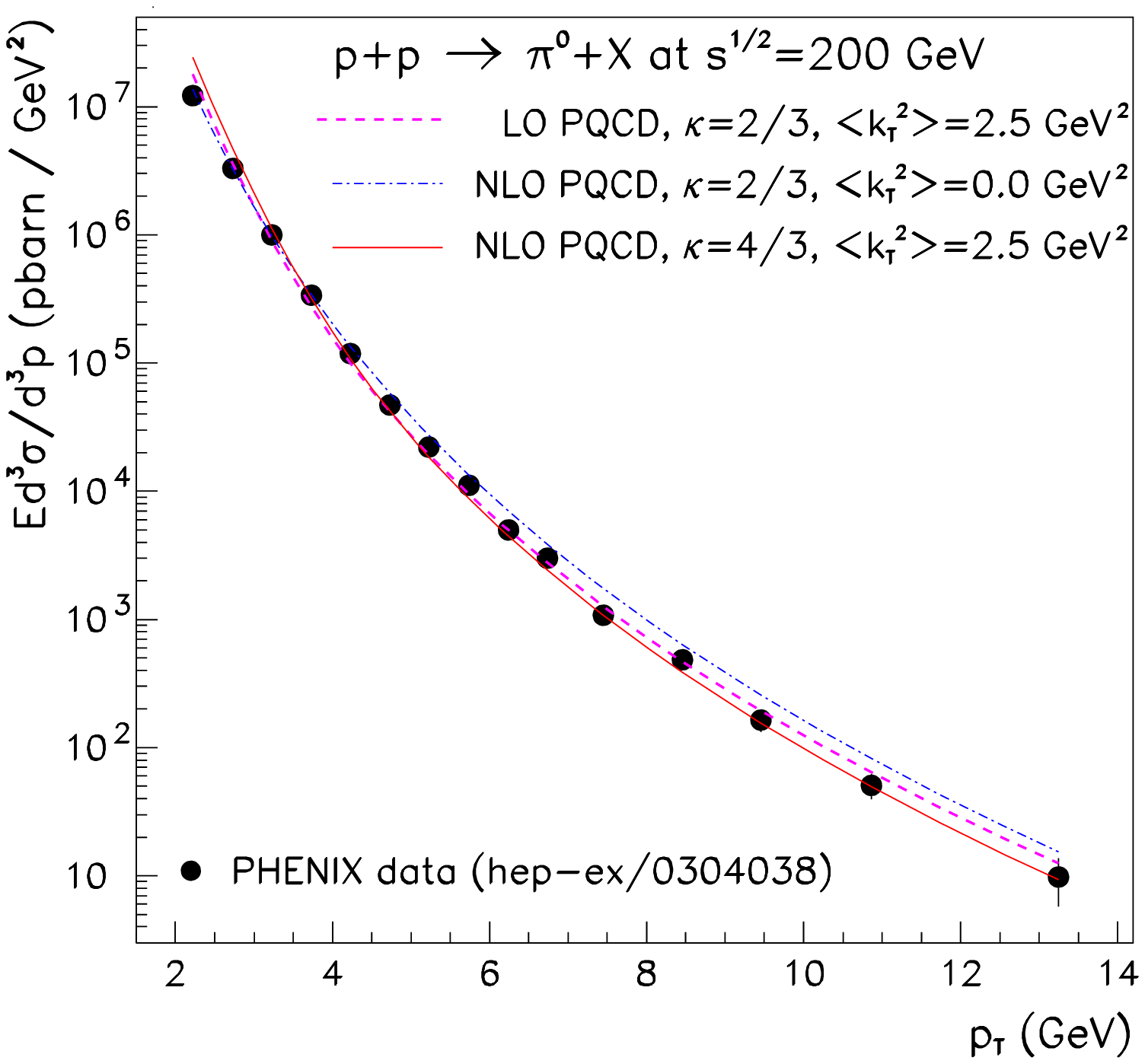}
\end{minipage}\\[-10pt]
\begin{minipage}[t]{\textwidth}
\caption{Left: Comparison of pQCD calculations for spectra of charged hadrons
   to data, solid lines are calculations with intrinsic $\kt$, dash-dotted lines
   are without \cite{wang2}. 
   Right: Invariant cross section for neutral pion production, lines are QCD
   calculations with different parameters \cite{levai}.}
 \label{fig_wang+levai}
\end{minipage}
\end{figure}

The agreement between the parton model calculations and data is impressive, a
few examples are shown in Figure~\ref{fig_wang+levai}. 

\smallskip

Insofar as Bjorken scaling holds in deep inelastic scattering this means that
the dependence on any dimensionful parameter has disappeared. A similar property
in hadron-hadron collisions leads to factorised cross sections such that
single-particle inclusive cross sections are for high $\pt$ of the form

\begin{equation}
 E \frac{\tn{d}^3 \sigma}{\tn{d}^3 p} \longrightarrow  \pt^{-n} 
 F(x_\perp, \theta) \qquad x_\perp = \frac{2 \pt}{\sqrt{s}}
\end{equation}
where $\theta$ is the centre-of-mass production angle \cite{jacob}. This 
factorisation in a power of $\pt$ and a dimensionless function $F$ is called 
\emph{power law scaling}.
 
The hadron-hadron cross sections are well described by a parametrisation of the
form \cite{albrecht}

\begin{equation}
 E \frac{\tn{d}^3 \sigma}{\tn{d}^3 p} = C \left(\frac{p_0}{\pt+p_0}\right)^n
\end{equation} 

where $C$, $p_0$ and $n$ are free parameters.  Although it yields a very good
description of the shape of the cross sections this parametrisation is not
suited for extrapolations since it doesn't include the dependence on energy
and rapidity. 

\medskip

\enlargethispage{\baselineskip}

The hard part of the $\pt$-spectra in pp collisions is well understood in terms
of pQCD. Collisions of nuclei are more complicated, nevertheless, because of
the hard scale involved, high-$\pt$ hadrons are especially suited to
investigate nuclear modifications. The perturbative processes happen on very
short distance and time scales so that it is likely that the hard scattering in
nucleus-nucleus collisions will look exactly like in proton-proton because it
doesn't feel the surrounding. However, the pdf's might be different from the
ordinary nucleon pdf's.

In the following considerations it is assumed that a quark-gluon plasma is
formed in central nucleus-nucleus collisions. Hard scatterings can thus take
place anywhere in the overlap region of the colliding nuclei, i.e.\ inside the
plasma. Consequently the scattered partons have to traverse it. Since the
quark-gluon plasma is a phase with very high colour charge density the partons
should interact strongly with it. The parton, that has a much higher energy
than its environment, is expected to lose a considerable fraction of it in the
plasma. In the subsequent hadronisation this leads to a softer jet with lower
mean momentum and number of hadrons. This expected suppression of high-$\pt$
hadrons as compared to scaled pp yields is commonly referred to as \textit{jet
quenching}.

The amount of energy loss depends on the length of the parton's path trough the
plasma and is thus given by the geometry of that particular event (i.e.\
centrality of the collision, position of the hard interaction in the plasma,
scattering angle of the partons). If the energy loss in the plasma is large it
should be possible that a parton with a long path loses essentially all its
energy and does not produce a jet. This jet-disappearance would be a clear
signal for the QGP, but it is very difficult to measure the non-existence of a
jet experimentally. A possible way out is offered by two-jet correlations.
In fact, if the hard scattering took place off the centre of the plasma and one
parton has a shorter way out so that it loses less energy. Consequently the other one
has to travel a longer distance through the plasma and may get thermalised
(i.e.\ lose essentially all its energy).  The parton with the shorter path
length in the plasma may still have enough energy to produce a jet containing
high-$\pt$ hadrons. This can then be seen as an event with only one clear jet
that can be identified experimentally.

The observation of jet quenching, i.e.\ suppression of high-$\pt$ hadrons and
disappearance of two-jet events would be a signal for
the formation of a QGP. This search is currently going on at RHIC
(Ch.~\ref{sec_RHIC}).

\smallskip

\enlargethispage{\baselineskip}

A currently accepted scenario for the energy loss mechanism is induced gluon
radiation. One possible approach to this problem is to consider the scattering
off static scattering centres \cite{g-radiation}. In the approximation that
successive scatterings are independent and the incident parton has very high
energy the radiation spectrum induced by multiple scattering can be derived
(Sec.~\ref{sec_gluonrad}).

Nevertheless, this may not be the only explanation. It is the aim of this study
to investigate if soft gluon exchanges can also contribute significantly to the
jet quenching. A model for this will be dicussed in detail in
Chapter~\ref{chap_scijetmodel}. Furthermore the general properties of partonic
energy loss will be studied in Chapter~\ref{chap_genprop}. The main focus
will be on the question if the experimental data clearly exhibit signs that
enable to discriminate between a bremsstrahlung and a scattering dominated
parton energy loss in the plasma phase.

It is in fact probable that both effects -- scattering and radiation --
contribute to the overall jet quenching. The parton scatters off a QGP
constituent and is deflected (and loses energy) which induces gluon
bremsstrahlung (and thereby additional energy loss). There might also be
quantum mechanical interference between the two processes. In the gluon
bremsstrahlung models the effect of the scattering is neglected. The model
studied here concentrates in a complementary fashion on the scattering and does
not take into account gluon radiation. When both effects have been studied
separately it will hopefully be possible to combine the two into a more
complete model.

\chapter{Basics of Heavy Ion Physics}
\label{chap_heavy-ion}
Collisions of relativistic heavy nuclei are very complex processes with many
interesting effects that are not well understood today. This chapter cannot
cover all of these exciting phenomena, instead only the effects that are
relevant for this study will be presented briefly.

\section{Glauber models} %-----------------------------------------------------
\label{sec_glauber}

In Glauber models it is assumed that the interactions of two nuclei can be 
viewed as an incoherent superposition of nucleon-nucleon interactions. A
collision of two nuclei is then  characterised by the mean number
of binary nucleon-nucleon collisions $\mean{N_\tn{coll}}$ and the mean number
of participating nucleons $\mean{N_\tn{part}}$. Both numbers depend on the
impact parameter $b$. Participating nucleons are those that encountered at
least one binary collision. 

There exist many variations on the theme, the model introduced in \cite{eskola} 
will be dicussed briefly as a representative. 

The nuclei $A$ and $B$ are characterised by the spherical nuclear density 
$n_{A,B}(r)$ satisfying 

\begin{equation} 
  \int\!\tn{d}^3r\, n_A(r) = A 
\end{equation} 
The nuclear density is described rather well by a Woods-Saxon distribution 
\begin{equation} 
  n_A^\tn{WS}(r) = \frac{n_0}{e^{(r-R_A)/d} + 1} \quad \tn{with} \quad 
      n_0 = \frac{3A}{4\pi R_A^3 (1 + \pi^2d^2/R_A^2)} 
\end{equation} 
A simpler alternative that unfortunately doesn't yield as good results is the 
sharp sphere
\begin{equation}      
  n_A^{ss}(r) = \frac{3A}{4\pi R_A^2} \Theta(R_A^2 - r^2) 
\end{equation} 
  
The $z$-direction is chosen to be the beam axis and the impact parameter $\vec 
b$ lies in the plane perpendicular to $z$ and points from beam to target. The
thickness function $T_A(b)$ is  given by 

\begin{equation} 
  T_A(b) = \int\limits_{-\infty}^\infty\!\tn{d} z\, n_A(\sqrt{b^2+z^2})  
\end{equation} 
with 
 
\begin{equation} 
  \int\!\tn{d}^2 b\, T_A(b)= A  
\end{equation}   
$T_A$ can be interpreted as the part of the nucleus $A$ "seen" by a nucleon 
passing through it with impact parameter $b$. For central collisions (i.e.\
small values of $b$) $T_A$ is proportional to $A^{1/3}$. When going to
nucleus-nucleus collisions the product of the two thickness functions has to be
integrated in the overlap region in order to get the nuclear overlap
$T_{AB}(b)$: 

\begin{equation} 
  T_{AB}(b) = \int\!\tn{d}^2 b_1\,\tn{d}^2 b_2\, \delta^2(\vec b-\vec b_1-\vec b_2) T_A(b_1) T_B(b_2)\, 
\end{equation}   
with
 
\begin{equation}  
  \int\!\tn{d}^2 b\, T_{AB}(b)= AB 
\end{equation}   
 
The product $\sigma^\tn{NN}_\tn{inel} T_{AB}(b)$, where
$\sigma^\tn{NN}_\tn{inel}$ is the  total inelastic nucleon-nucleon cross
section at the respective collision energy, can be interpreted as the mean 
number of binary collisions $\mean{N_\tn{coll}}$ at impact parameter  $b$. The
total nucleus-nucleus cross section is obtained by integrating 
$\sigma^\tn{NN}_\tn{inel} T_{AB}(b)$ over $b$: 

\begin{equation} 
  \sigma^{AB}_\tn{inel} = \int\!\tn{d}^2 b\, \sigma^\tn{NN}_\tn{inel} T_{AB}(b)
    = AB \sigma^\tn{NN}_\tn{inel} 
\end{equation}
Similarly $T_{AB}(b)$ can be multiplied with any cross section to get the mean 
number of events of a particular kind per $AB$ collision at impact parameter 
$b$. 
 
\smallskip 
 
The mean number of participants is given by 
 
\begin{equation} 
  \mean{N_\tn{part}(b)} = \int\!\tn{d}^2 b_1\, \tn{d}^2 b_2\,  
    \delta(\vec b-\vec b_1-\vec b_2) \left \{ 
    T_A(b_1) p_{\ge 1}^{(B)}(b_2) + T_B(b_2) p_{\ge 1}^{(A)}(b_1) \right\}\,  
\end{equation}     
where $p_{\ge 1}^{(A)}(b_1)$ is the probability for a nucleon passing through 
the nucleus $A$ with impact parameter $b_1$ to take part in at least one 
interaction. It can be obtained from the Binomial distribution: 
 
\begin{eqnarray} 
 p_{\ge 1}^{(A)}(b_1) &=& 1 - {A\choose 0}\left( \frac{\sigma^\tn{NN}_\tn{inel} 
  T_A(b_1)}{A} \right)^0 \left( 1-\frac{\sigma^\tn{NN}_\tn{inel} T_A(b_1)}{A} 
  \right)^{A-0} \nonumber \\ 
  &=& 1-\left( 1-\frac{\sigma^\tn{NN}_\tn{inel} T_A(b_1)}{A} \right)^A \\
  &\approx & 1 - e^{-\sigma^\tn{NN}_\tn{inel} T_A(b_1)} 
        \qquad \tn{for\ large\ } A   
\end{eqnarray}      
 
\smallskip 
 
For the most central and symmetric (i.e.\ $A=B$) collisions
$\mean{N_\tn{part}}$ is linear in $A$ whereas  $\mean{N_\tn{coll}}$ is
proportional to $A^{4/3}$. It turns out that soft  particle production in
heavy ion collisions scales with the number of  participants while hard
processes scale with the number of binary collisions. 

\section{Quantum Statistics of an Ideal Quark-Gluon Gas} %----------------------
\label{sec_idgas}

As a consequence of the energy dependence of the strong coupling constant
$\alpha_s$ the interactions among quarks and gluons become weak at high energy
densities. Thus a gas of non-interacting ultra-relativistic quarks and
gluons can be used as an approximation. 

The particle numbers are not fixed so that the grand canonical formalism has to
be used (see e.g.\ \cite{landau}). Let $\vec \nu$ be a set of quantum numbers
characterising a one-particle state, $n(\vec\nu)$ the occupation number of
that state and $E(\vec\nu)$ the corresponding one-particle energy.
$\{n(\vec\nu)\}$ is the complete set of occupation numbers. Then the grand
canonical partition function for bosons is 

\begin{eqnarray}
\Xi^{(B)}(T,V,\mu) &=& \sum_\tn{all\ states} \exp(-\beta(E_\tn{tot} - \mu N))
				\nonumber \\
	&=& \sum_{\{n(\vec\nu)\}} \exp\left(-\beta \sum_{\vec\nu} n(\vec\nu)
		(E(\vec\nu) - \mu)\right) \nonumber \\
	&=& \sum_{\{n(\vec\nu)\}} \prod_{\vec\nu} \{\exp(-\beta(E(\vec\nu) -
		\mu)) \}^{n(\vec\nu)} \nonumber \\
	&=& \prod_{\vec\nu} \sum_{n(\vec\nu)=0}^\infty \{\exp(-\beta(E(\vec\nu) -
		\mu)) \}^{n(\vec\nu)} \nonumber \\
	&=&  \prod_{\vec\nu} \frac{1}{1-\exp(-\beta(E(\vec\nu) - \mu))}
\end{eqnarray}
where $\beta = 1/T$, $\mu$ is the chemical potential and $E_\tn{tot}$ and $N$
denote the total energy and number of particles in the ensemble. The partition
function for fermions can be derived in a similar fashion.
\begin{eqnarray}
\Xi^{(F)}(T,V,\mu) &=& \prod_{\vec\nu} \sum_{n(\vec\nu)=0}^1 \{\exp(-\beta(E(\vec\nu) -
		\mu)) \}^{n(\vec\nu)} \nonumber \\
		&=&  \prod_{\vec\nu} \frac{1}{1+\exp(-\beta(E(\vec\nu) - \mu))}
\end{eqnarray}

\begin{equation}
\Rightarrow \ln \Xi(T,V,\mu) = \pm \sum_{\vec\nu} \ln\left(1 \pm e^{-\beta(E(\vec\nu) - \mu)} \right) 
\end{equation}
where the plus sign applies to fermions and the minus sign to bosons. If the 
energy levels are not discrete but continuous the sum can be transformed into an
integral yielding
\begin{equation}
\ln \Xi(T,V,\mu) = \pm \int\!\frac{\tn{d}^3q\,\tn{d}^3p}{(2\pi)^3}\,
	\ln\left(1 \pm e^{-\beta(E(\vec\nu) - \mu)} \right) 
\end{equation}
where $p$ and $q$ are momentum and spacial coordinates.

The mean occupation number of a state is given by

\begin{equation}
 \bar n(\vec\nu) = - \del[\ln\Xi]{(\beta E(\vec \nu))} = \frac{1}		
		{e^{-\beta(E(\vec\nu) - \mu)} \pm 1}
\end{equation}

For the quark gluon gas it is helpful to assume that the QGP is net baryon-free
as will be the case in ultra-relativistic ion-ion collisions. Then the gluon as
well as the quark chemical potential vanishes and the numbers of quarks and
antiquarks are equal. Furthermore, it has to be taken into account that the
energy levels are degenerate so that the mean occupation number is

\begin{equation}
\bar n_p = \frac{g}{e^{E/T} \pm 1} \quad \tn{with} \quad  
  E = \sqrt{\vec p\,^2+m^2}
\end{equation}
where again the plus (minus) sign applies to fermions (bosons) and $g$ is the
degeneracy. It is given by the number of flavours times the number of colours
times the number of polarisations:

\begin{eqnarray}
 & & g_g =  2 (\tn{polarisation}) \cdot 8 (\tn{colour})= 16 \\
 & & g_q = g_{\bar q} = N_f (\tn{flavour}) \cdot  2 (\tn{polarisation}) \cdot 3
 (\tn{colour})= 6N_f
\end{eqnarray}

If the masses are neglected the particle and energy densities can easily be
calculated:
 
\begin{eqnarray}
 n_g \!\!\!&=&\!\!\! \frac{1}{V}\frac{1}{(2\pi)^3} \int\!\tn{d}^3q\,\tn{d}^3p\, 
 									\bar n_p 											 
       = \frac{4\pi g_g}{(2\pi)^3} \int\limits_0^\infty\!\tn{d}p\, 
	 				\frac{p^2}{e^{p/T}-1} \nonumber \\
    \!\!\!&=&\!\!\!\frac{g_g}{2\pi^2} T^3 \int\limits_0^\infty\!\tn{d}\xi\,
     		\frac{\xi^2}{e^\xi-1} 
       = \frac{g_g}{2\pi^2} T^3 2\,\zeta(3) 
       \simeq  1.2\,\frac{g_g}{\pi^2} T^3 \\
 n_q \!\!\!&=&\!\!\! n_{\bar q} = \frac{1}{V}\frac{1}{(2\pi)^3}
 		\int\!\tn{d}^3q\,\tn{d}^3p\, \bar n_p\ 
	 =  \frac{4\pi g_q}{(2\pi)^3} \int\limits_0^\infty\!\tn{d}p\,
	 				\frac{p^2}{e^{p/T}+1} \nonumber \\
     \!\!\!&=&\!\!\! \frac{g_q}{\pi^2}\,T^3 d(3)
       \simeq 0.9\,\frac{g_q}{\pi^2} T^3 	\\
 \epsilon_g \!\!\!&=&\!\!\! \frac{1}{V}\frac{1}{(2\pi)^3}
 		\int\!\tn{d}^3q\,\tn{d}^3p\, p\,\bar n_p\ 
	 = \frac{3g_g}{\pi^2}\,T^4\,\zeta(4)
	 = \frac{\pi^2 g_g}{30}\,T^4 \label{eq_eps-T} \\
 \epsilon_q \!\!\!&=&\!\!\!\epsilon_{\bar q} = \frac{3g_q}{\pi^2}\,T^4\,d(4)
       = \frac{7\pi^2 g_q}{240}\,T^4 	 
\end{eqnarray}
with

\begin{equation}
\zeta(\xi) = \frac{1}{\Gamma(\xi)} \int \limits_0^\infty\!\tn{d}\alpha\, 
 \frac{\alpha^{\xi-1}}{e^\alpha - 1} \qquad \tn{and} \qquad
d(\xi) = \frac{1}{\Gamma(\xi)} \int \limits_0^\infty\!\tn{d}\alpha\, 
 \frac{\alpha^{\xi-1}}{e^\alpha + 1}
\end{equation}       
Thus the energy per particle is given by

\begin{eqnarray}
 \frac{\epsilon_g}{n_g} = 3T\,\frac{\zeta(4)}{\zeta(3)} \simeq 2.7\,T \\
 \frac{\epsilon_q}{n_q} = 3T\,\frac{d(4)}{d(3)} \simeq 3.2\,T 
\end{eqnarray} 
and the total pressure can be obtained from the equation of state:

\begin{equation}
 p_{qg} = \frac{1}{3}\,\epsilon_{qg} = \frac{1}{3} (\epsilon_g + \epsilon_q +
   \epsilon_{\bar q}) = \frac{\pi^2}{90}\left( 16 + \frac{21}{2}N_f \right) T^4
\end{equation}

\section{The Bjorken Model} %---------------------------------------------------
\label{sec_bjorken}

The Bjorken model \cite{bjorken} is applicable to energetic ion-ion collisions
with $\sqrt{s} \ge \unit[100]{A\, GeV}$. It is based on the assumption that the
inclusive particle multiplicity as a function of rapidity shows a "central
plateau", i.e.\ particle production looks the same in all centre-of-mass-like
frames. The second assumption is that the nuclei are essentially transparent to
each other. The net baryon number is thus contained in the receding remnants of
the nuclei and the central rapidity region is net baryon-free (i.e.\ the baryon
chemical potential vanishes). 

After the collision the two Lorentz contracted nuclei recede from the collision
point with nearly speed of light. Between them is a dense and hot system that
expands longitudinally. The expansion velocity is $z/t$ where $z=0$ and $t=0$
refer to the collision point and time. From the charged particle 
multiplicity observed in nucleon-nucleus collisions the initial energy density
is estimated:
\begin{equation}
 \epsilon_0 = \epsilon(\tau_0) \approx \unit[1-10]{\frac{GeV}{fm^3}} 
\end{equation}
for an initial proper time
\begin{equation}
 \tau_0 = \unit[1]{fm/c} \qquad \tau = \sqrt{t^2-z^2}  
\end{equation}
This relatively high number suggests that the system rapidly comes into local
thermal equilibrium and afterwards follows a hydrodynamic evolution. It is thus
possible to define a local energy density $\epsilon(\tau,y)$, pressure
$p(\tau,y)$, temperature $T(\tau,y)$ and four-velocity $u_\mu(\tau,y)$.

The first assumption implies that the initial condition possesses a symmetry
under Lorentz transformations which is preserved during the subsequent
evolution. This means that the energy density, pressure etc. cannot depend on
the rapidity.

\bildtex{plots/space-time.pstex_t}{pic_space-time}{Space-time diagram of
longitudinal evolution of the quark-gluon plasma \cite{bjorken}}

Neglecting the viscosity and heat conduction the energy-momentum tensor becomes
\begin{equation}
 T_{\mu\nu} = (\epsilon + p)u_\mu u_\nu -g_{\mu\nu} p 
\end{equation}
It is conserved, i.e.
\begin{equation}
 \del[T_{\mu\nu}]{x_\mu} = 0\ . 
\end{equation}
 
Making use of the fact that the occuring quantities depend only on the proper
time this expression simplifies to
\begin{equation}
\label{eq_1}
 \der[\epsilon]{\tau} = -\frac{\epsilon+p}{\tau} 
\end{equation} 

This can also be written as
\begin{equation}
 \der[\epsilon]{\tau} = \der[\epsilon]{p} \der[p]{T} \der[T]{\tau} = 
   -\frac{\epsilon+p}{\tau} = -\frac{Ts}{\tau}
\end{equation}
where $s$ is the entropy density. With the help of
\begin{equation}
\label{eq_2}
 \der[p]{T}=\frac{S}{V}=s \quad \tn{and} \quad
 \der[\epsilon]{p}=\frac{1}{v_s^2} 
\end{equation}
($v_s$ is the velocity of sound) the differential equation for the
time dependence of the temperature is obtained:
\begin{equation}
\label{eq_3}
 \frac{1}{T} \der[T]{\tau} = -\frac{v_s^2}{\tau}
\end{equation}

\medskip 
 
In the case of an ideal ultra-relativistic gas the equation of state is
$\epsilon=3p$. Inserting this into Equations~\ref{eq_1}, \ref{eq_2} and
\ref{eq_3} leads to the result that the time dependence of the energy
density  and temperature is given by
\begin{equation}
 \epsilon(\tau) \propto \tau^{-4/3} 
\label{eq_eps-time} 
\end{equation}
and
\begin{equation}
 T(\tau) \propto \tau^{-1/3} 
\end{equation}
respectively.

\bigskip
 
The simple picture of purely longitudinal expansion has to be modified when the
distance between the nuclei becomes comparable to their diameter.
There is a rarefaction front moving inward from the periphery at the velocity of
sound. Under the assumption of a time independent velocity of sound the equation
for the rarefaction front is 
\begin{equation}
 \rho(t)=R-\int\limits_0^{\sqrt{t^2-z^2}}\!\tn{d} t'\, v_s  = R - v_s \sqrt{t^2-z^2}
\end{equation}
At transverse distances smaller than the rarefaction front the system continues
to expand longitudinally because the information that there is a boundary has
not yet reached this region. Outside the front the gas expands radially outward
and cooles faster than during longitudinal expansion. 

\bildtex{plots/rarefaction.pstex_t}{pic_rarefaction}{Geometry of fluid expansion near
the edge of the nuclei \cite{bjorken}}

\section{Azimuthal Anisotropy} %------------------------------------------------
\label{sec_flow}

\bildtex{plots/overlap.pictex}{fig_overlap}{Overlap region in a non-central
  collision}

The beam axis ($z$ axis) and the impact parameter $b$ define the reaction
plane. In non-central collisions the overlap region has an almond like shape
(Fig.~\ref{fig_overlap}) so that the pressure gradient depends on the
direction. This leads to an asymmetric particle emission that can be described
by

\begin{equation}
 E \frac{\tn{d}^3N}{\tn{d}^3p} = \frac{1}{2\pi} \frac{\tn{d}^2N}{\pt \tn{d}
  \pt \tn{d}y} \left[ 1 + \sum_{n=1}^\infty 2v_n \cos(n\phi) \right] 
\label{eq_flow}
\end{equation}  
where $\phi$ is the azimuthal angle with respect to the reaction plane
\cite{snellings}.  

An important feature is that the odd Fourier coefficients change sign at
mid-rapidity whereas the even are symmetric. It is common to discuss collective
flow in terms of the lowest order coefficients in the Fourier expansion in
Equation~\ref{eq_flow}. They can be determined experimentally in the following
way:
$v_1$ is called directed flow and is associated with the mean transverse
momentum in the reaction plane:

\begin{equation}
 v_1 = \frac{\mean{p_x}}{\mean{\pt}} 
\end{equation}
$v_2$ is called elliptic flow and can be written as

\begin{equation}
 v_2 = \mean{\cos 2\phi} 
\end{equation}
For negative $v_2$ the emission is dominantly perpendicular to the reaction
plane whereas a positive $v_2$ favours emissions in the reaction plane. $v_2$
depends on the beam energy: It is negative at low energies (BEVALAC/SIS) and
becomes positive at AGS and SPS with a larger value at SPS. This behaviour is
predicted by theory. The pressure gradient leads to an emission that is
predominantly in the reaction plane, but shadowing by spectator nucleons
is expected to turn it into an out-of-plane emission. This effect becomes
unimportant at higher beam energies because the spectators are then too far
away from mid-rapidity to affect the distributions there
\cite{stachel2,stachelt}.

\section{Cronin effect} %------------------------------------------------------
\label{sec_cronin} 
 
Already in 1975 it was discovered that the production of hadrons with high 
transverse momentum is enhanced in proton-nucleus collisions. This effect is 
called the Cronin effect \cite{cronin} after its discoverer and is commonly
quantified by the Cronin ratio $R$ of the $\pt$-spectra obtained in
proton-nucleus collisions using two  different nuclei $A$ and $B$ ($A > B$): 

\begin{equation} 
  R_{AB}(\pt) = \frac{B}{A} \frac{\tn{d}\sigma^{pA} / 
       \tn{d}\pt}{\tn{d}\sigma^{pB} / \tn{d}\pt} 
\label{eq_cronin} 
\end{equation}
It has roughly the shape shown in Figure~\ref{fig_cronin} with a maximum at 
medium $\pt$. The details depend on $A$, $B$ and the collision energy. 

\bildtex{plots/croninR.pictex}{fig_cronin}{Approximate behaviour of the
  Cronin  ratio $R$ (Eq.~\ref{eq_cronin}), $p_1$ is usually of the order
  \unit[1]{GeV} and $p_\tn{max}$ a few \unit{GeV} \cite{accardi}}   

The enhancement can be accounted for in terms of multiple scattering of the 
proton as it traverses the nucleus. There are different models \cite{accardi} 
to describe the rescattering, but they all lead in one way or another to a 
broadening of the $\pt$-distribution. They differ in the object that 
experiences rescattering (the proton or its partons) and the hardness, i.e.\ 
the momentum transfer, of the scattering processes. 
 
\smallskip 
 
A short review of the model introduced in \cite{wang} and \cite{zhang} will be 
given here. In this case it is the nucleons that undergo rescattering with a 
predominantly low momentum transfer.  
 
Already in pp collisions it has to be taken into account that the partons have 
an intrinsic transverse momentum $\kt$ that reflects the size of the proton
via  the uncertainty principle and the Fermi motion. In initial state parton
showers  further transverse momentum can be built up. These two  effects are
described together in a phenomenological approach in \cite{wang,zhang}. The
parton  distributions are assumed to factorise in two parts that depend on the 
longitudinal and transverse momentum respectively: 

\begin{equation} 
  g_N(\kt,Q^2)f_{a/N}(x,Q^2)\tn{d}x\,\tn{d}^2\kt 
\end{equation} 
 
$f_{a/N}(x,Q^2)$ are the normal parton distribution functions ($x$ is the 
longitudinal momentum fraction carried by the parton and $Q$ is the momentum 
transfer in the hard scattering). The transverse momentum distribution 
$g_N(\kt,Q^2)$ is assumed to have a Gaussian form: 
 
\begin{equation} 
  g_N(\kt,Q^2) = \frac{1}{\pi \mean{\kt^2}_N}e^{-\kt^2/\mean{\kt^2}_N} 
\end{equation}   
 
The variance $\mean{\kt^2}_N$ depends on $Q$, because both the intrinsic $\kt$ 
and the effect of parton showers are included here. It is taken to be 
 
\begin{equation} 
  \mean{\kt^2}_N(Q^2) = \unit[1.2]{GeV^2} + 0.2 \alpha_s(Q^2)Q^2 
\end{equation} 
and the parameters are chosen such that the experimental data are reproduced. 
 
In proton-nucleus collisions the proton may experience several soft scatterings 
before the hard process. It is assumed in this model that the $\kt$-
distribution is still given by a Gaussian. 
 
\begin{equation} 
  g_A(\kt,Q^2) = \frac{1}{\pi \mean{\kt^2}_A}e^{-\kt^2/\mean{\kt^2}_A} 
\end{equation}   
But now the variance is larger due to the soft scatterings: 
 
\begin{equation} 
  \mean{\kt^2}_A(Q^2) = \mean{\kt^2}_N(Q^2) +\delta^2(Q^2)  
        (\nu_a(b)-1) 
\label{eq_ktbr}	   
\end{equation}	   
 
In \cite{wang} $\nu_a(b)$ is taken to be $\mean{N_\tn{coll}}(b)$ while it is 
argued in \cite{zhang} that $\nu_a(b)$ cannot become larger than 4 because of 
the dissociation of the proton. But a highly excited proton that does not
interact as a whole any more will not cease to interact. The condition $\nu_a
< 4$ is therefore not sensible. The values of $\delta^2$ needed to describe
the data are considerably larger with the restricted $\nu_a$, which is a
strong indication that also the interactions of excited or even dissociated
protons have to be taken into account. 

The best fit to experimental data is obtained in \cite{wang} with 
 
\begin{equation} 
  \delta^2(Q^2) = 0.225\frac{\ln^2(Q/\unit{GeV})}{1 + \ln(Q/\unit{GeV})}  
   \unit{GeV^2} 
\end{equation} 
 
In nucleus-nucleus collisions the $\kt$-distributions of both partons are
broadened  accoring to Equation~\ref{eq_ktbr}, so this part is a
straightforward  generalisation of the above results. However, there are more
effects that  influence the production of high-$\pt$ particles and lead to
large  uncertainties. An example is the modification of the parton distributions
in a  nuclear environment. 

\smallskip 
 
The Cronin effect is expected to be relatively small in pAu collisions at RHIC 
energies. The different models predict $R_\tn{max} \simeq 1.1\dots1.6$ and 
$p_\tn{max} \simeq 2.5\dots3.5$ \cite{accardi}. Predictions for AuAu are much 
more difficult due to other nuclear effects as mentioned above. 

\section{Radiative Energy Loss} %-----------------------------------------------
\label{sec_gluonrad}

A high energy parton traversing a colour charged medium is expected to radiate
gluons. This medium induced gluon radiation is the QCD analogue to the QED
bremsstrahlung. There is no unique way to treat this highly involved problem,
here only a review of the main features of the model introduced in
\cite{g-radiation} will be given without detailed calculations.

\smallskip

The matter is  represented by static scattering centres each of which creates a
screened Coulomb potential 

\begin{equation}
 \mathcal{V}_i (\vec x) = \frac{g}{4 \pi}\frac{e^{-\mu |\vec x - \vec x_i|}}
   {|\vec x - \vec x_i|}
\end{equation}
where $g$ is the QCD coupling constant and $\mu$ is the Debye mass, i.e.\ the
inverse of the range of the screened potential. It is assumed that the range of
the potential is small compared with the mean free path $\lambda$ of the
parton, i.e.\ the successive scatterings are independent. The calculation of
the radiation amplitudes is performed in time-ordered pertubation theory. The
scattering centres are very heavy so that the energy loss in the scattering
vanishes.

Furthermore all partons are assumed to be quarks of very high energy so that the
soft gluon approximation 

\begin{equation}
\omega \ll E
\end{equation}
can be made.
 
\bildtex{plots/gluonrad.pstex_t}{fig_gluonrad1}{Radiation amplitude induced by a
single scattering; $A$, $A'$, $B$, $B'$, $a$ and $b$ denote the colour indices
\cite{g-radiation}}

The three diagrams that contribute to the gluon emission amplitude induced by
one scattering are depicted in Figure~\ref{fig_gluonrad1}. The gluon energy
spectrum, which is given by the ratio between the radiation and the elastic
cross section, is found to be

\begin{equation}
  \omega \frac{\tn{d}I}{\tn{d}\omega \tn{d}^2 \vec \kt} = 
  N_c \frac{\alpha_s}{\pi^2} \mean{\vec J(k,q)^2}
\end{equation}
where $\vec J$ is the emission current
 
\begin{equation}
 \vec J(k,q) = \frac{\vec \kt}{\kt^2} - \frac{\vec \kt - \vec q_\perp}{(\vec \kt
 - \vec q_\perp )^2}  
\end{equation}
and 

\begin{equation}
 \mean{\vec J(k,q)^2} \equiv \int\!\tn{d}^2 \vec q_\perp\, 
 \frac{\mu^2}{\pi(q_\perp^2 + \mu^2)^2} \vec J(k,q)^2
\end{equation} 

This result can be generalised to an arbitrary number of scatterings.
Figure~\ref{fig_gluonrad2} shows as an example the case of two scatterings.

\bildtex{plots/gluonrad2.pstex_t}{fig_gluonrad2}{Gluon emission amplitude
induced by two scatterings; $a$, $b$ and $c$ denote the colour indices
\cite{g-radiation}}

The end result becomes quite complicated for a finite number of scatterings and
the heuristic discussion is more instructive. Here it is assumed that the number
of scatterings is large

\begin{equation}
  \frac{1}{N} \ll \kappa \ll 1 \qquad \kappa = \frac{\lambda \mu^2}{2\omega}
\label{eq_cond}
\end{equation}
There are three different regimes: the incoherent Bethe-Heitler (BH) regime of
small gluon energies, the coherent regime for intermediate $\omega$ and the
factorisation limit corresponding to the highest energies. In the BH regime the
radiation is due to $N$ incoherent scatterings, in the coherent regime
$1/\sqrt{\kappa}$ scatterings act together as a single one
(LPM\footnote{\textbf{L}andau \textbf{P}omeranchuk \textbf{M}igdal} effect) and
in the factorisation limit finally all $N$ centres behave as a single
scattering centre. This has to do with the formation length of the radiated
gluons which increases with the gluon energy. The longer the formation length
is the more scatterings occur during the radiation of the respective gluon. The
condition \ref{eq_cond} can also be expressed in terms of the gluon energy:

\begin{equation}
 \omega_\tn{BH} \sim \lambda \mu^2 \ll \omega \ll \omega_\tn{fact} \sim
 \frac{\mu^2L^2}{\lambda} \le E
\end{equation}
The last part $\omega_\tn{fact}\le E$ is satisfied when the length $L$ of the
traversed medium is smaller than the critical length

\begin{equation}
 L \le L_\tn{cr} = \sqrt{\frac{\lambda E}{\mu^2}}
\end{equation}

\pagebreak

Then the radiation spectrum per unit length is 

\begin{equation}
 \omega \frac{\tn{d}I}{\tn{d}\omega\, \tn{d}z} \simeq \left\{
  \begin{array}{ll}
  \frac{\alpha_s}{\lambda} & \omega < \omega_\tn{BH} \\
  \frac{\alpha_s}{\lambda}\sqrt{\frac{\lambda\mu^2}{\omega}} & \omega_\tn{BH} < \omega < \omega_\tn{fact} \\
  \frac{\alpha_s}{L}       & \omega_\tn{fact} < \omega \\
  \end{array} \right.
\end{equation}  
The spectrum can be integrated in the range $0 \le \omega \le E$ and $0 \le z
\le L$ in order to get the total energy loss. It is found that the energy loss

\begin{equation}
 \Delta E (L) \sim \alpha_s \frac{\mu^2 L^2}{\lambda} \left( 1 +
 O\left(\frac{1}{N}\right)\right)
\label{eq_gluonrad1}
\end{equation}
is proportional to $L^2$ and independent of $E$ in the high energy limit. In
addition there is a factorisation contribution which is proportional to
$\alpha_s E$. For the case that $L > L_\tn{cr}$ or equivalently $E \le
E_\tn{cr} = L^2\mu^2/\lambda$ the energy loss becomes

\begin{equation}
  \Delta E \simeq \alpha_s \sqrt{\frac{\mu^2 E}{\lambda}} L
\label{eq_gluonrad2}
\end{equation}
Now $\Delta E (L)$ grows only linearly with $L$ but depends on $E$. The
different scenarios are reflected in different $E$- and $L$-dependencies of the
overall energy loss.

\chapter{Experimental Results}
\section{AGS and SPS} %========================================================

\subsubsection*{Strangeness Enhancement} %--------------------------------------

A clear enhancement of strangeness is observed in various systems at
AGS and SPS. This does not automatically mean that a QGP is formed since
the strangeness enhancement can partly result from hadronic interactions
\cite{harris}.

At AGS an increase of the kaon to pion ratio $K/\pi$ is seen and
$K^+/\pi^+ > K^-/\pi^-$. The data are described by purely hadronic
models and the strange particle ratios can be fitted by a statistical
model with an equilibrated hadronic fireball. This does not mean that
the system always was in a hadronic phase \cite{bass,harris,stachel}.

At SPS a global enhancement factor 2 for kaons in central PbPb collisions is
observed and again $K^+/\pi^+ > K^-/\pi^-$. The enhancement increases with the
strangeness content of the respective hadron species \cite{bass,scomparin}.
The antihyperon ratio $\bar \Xi/\bar \Lambda$ is found to smoothly increase
from pp over pA to AA collisions. Furthermore a strong enhancement of
multistrange hadrons from light to heavy nuclei is observed, which can be
counted as evidence for a non-hadronic enhancement \cite{bass}. Hadronic
transport models fail to describe the SPS data unless they invoke
non-hadronic scenarios while microscopic transport models are in good
agreement with data \cite{bass,harris}. The strange particle ratios can be
fitted with a hadronic hadron gas and a QGP \cite{bass,harris}. 

This may be seen as evidence for QGP formation, but it has not been 
possible to unambiguously relate the data on strangeness enhancement to 
the presence of a QGP. A more detailed discussion of practical and 
conceptual problems can be found in \cite{bass}. 

\subsubsection*{Chemical Equilibrium and Freeze-Out} %--------------------------

At top AGS and higher energies mid-rapidity particle production can be well
described in the framework of a statistical model in full equilibrium while in
the integrated results strangeness is undersaturated. The freeze-out points are
below the calculated phase boundary for all except the top SPS energy
\cite{stachel}. 

\subsubsection*{Charmonium Suppression} %---------------------------------------

$J/\psi$ suppression can be quantified using the ratio of the  $J/\psi$ cross
section to the cross section for the Drell-Yan process.  The Drell-Yan process
is the creation of a lepton pair from $q \bar q$  annihilation ($q \bar q \to
l^+ l^-$). It is a hard process that scales  with the number of binary
nucleon-nucleon collisions in a  nucleus-nucleus collision. The ratio
$\sigma_{J/\psi}/\sigma_{DY}$ can be interpreted as the probability to produce
a $J/\psi$ per binary collision. A big advantage is that many experimental
uncertainties drop out in the ratio. It is often given as a function of $L$,
the mean length of nuclear  matter traversed by a $c \bar c$ pair
\cite{scomparin}. It is, however, advisable to be careful since this is not a
measured quantity and model dependent \cite{bass}. 

$\sigma_{J/\psi}/\sigma_{DY} (L)$ has been measured at CERN for  different
systems. The results for pA, SU and peripheral PbPb ($L <  \unit[8]{fm}$) are
consistent with the exponential absorption in nuclear  environment while in
central PbPb ($L > \unit[8]{fm}$) a significantly  larger suppression is seen
\cite{scomparin,bass,na50-j/psi}. Hadronic transport  models reproduce the data
but depend heavily on details in the models  and need a very high comover
density \cite{bass,stachel2} which is not typical for hadronic matter.

The cross section ratio $\sigma_{J/\psi}/\sigma_{DY} (E_\perp)$ for  PbPb
collisions as a function of transverse energy shows a clear suppression in
addition to the extrapolated absoption in nuclear matter. For peripheral events
the data are consistent with the suppression expected from normal absorption,
but for mid-central collisions the measured cross section ratio drops
significantly below the normal suppression. It continues to decrease for more
central events \cite{na50-j/psi}. 

\smallskip

Additional observations are:
\begin{itemize}
\item The loosely bound $\psi'$ is similarly absorbed in SU and PbPb
   \cite{bass,scomparin}.	
\item The statistical hadronisation model describes the centrality dependence
	of $J/\psi$ at SPS, but only with an increased charm production cross
	section \cite{stachelc}.
\item The anomalous $J/\psi$ suppression observed at SPS can be counted as a 
	signal for QGP formation, but there are still many uncertainties and
	the theoretical debate is far from settled \cite{bass,scherer}. 
\end{itemize}

\section{RHIC} %=============================================================== 
\label{sec_RHIC} 
 
At RHIC gold nuclei collide with  energies up to \unit[200]{GeV} per nucleon
pair (which corresponds to a total  c.m.s.\ energy of $\sqrt{s} \sim
\unit[40]{TeV}$). The energy density reached is  expected to be at least a few
$\unit{GeV/fm^3}$ and thus high enough for the  formation of a quark-gluon
plasma. These calculations are, however, heavily  model dependent. 

At these centre-of-mass energies, hadrons with high transverse momentum are 
produced in the fragmentation of partons that underwent a hard scattering. The 
study of these high-$\pt$ hadrons is one of the main goals of the four 
experiments PHENIX, STAR, BRAHMS and PHOBOS. 
 
The particle multiplicities are so high in ion-ion collisions that jets cannot
be identified with the normal algorithms which search for a large energy
deposition in neighboring calorimeter cells. Instead only  single high-$\pt$
hadrons can be identified and from the relative angle it can be seen whether
they belong to the same jet. But since there is a huge background of low-$\pt$
particles it is impossible to identify also the softer part of the jet and
therefore also the total jet energy as well as the jet cone is unknown. 

There have also been p+p and d+Au runs at the same energy ($\sqrt{s_\tn{NN}} = 
\unit[200]{GeV}$) and a Au+Au run at $\sqrt{s_\tn{NN}} = \unit[130]{GeV}$ to 
which the results can be compared. 
 
\subsection{Suppression of high-$\pt$ Hadrons} %------------------------------- 
 
In pp collisions hadrons with $\pt \gtrsim \unit[2]{GeV}$ are produced in hard 
scattering processes. If this is also the case in AA collisions the yield of 
hadrons with large $\pt$ is expected to scale with the number of binary 
collisions (Ch.~\ref{sec_glauber}). It is therefore often characterised by 
the ratio $R_\tn{AB}$ of the yield per nucleon-nucleon collision in AB 
collisions to the yield in pp collisions: 
 
\begin{equation} 
 R_\tn{AB}(\pt,\eta) = \left(\frac{1}{N_\tn{evt}} \frac{\tn{d}^2 N^\tn{AB}} 
   {\tn{d}\pt\tn{d}\eta}\right) \cdot \left( \frac{\langle N_\tn{coll} 
   \rangle}{\sigma_\tn{inel}^\tn{pp}} \frac{\tn{d}^2 \sigma^\tn{pp}} 
   {\tn{d}\pt\tn{d}\eta}\right)^{-1} 
\label{eq_RAA}   
\end{equation}   
 
The inelastic pp cross section is from measurements known to be 
$\sigma_\tn{inel}^\tn{pp} \simeq \unit[42]{mb}$ at $\sqrt{s} = \unit[200]{GeV}$. 
$R_\tn{AB}$ is called the nuclear modification factor since all nuclear 
effects will lead to a deviation from unity. $\langle N_\tn{coll}\rangle$ has 
to be calculated for each centrality class (i.e.\ range in impact parameter) 
separately. It should be noted that in the region below $\pt \simeq 
\unit[2]{GeV}$ the $\pt$-spectrum is dominated by hadrons that were produced 
in soft interactions. This contribution is naively expected to scale with the 
number of participants rather than the number of binary collisions. This means 
that even without any nuclear effects $R_\tn{AB}$ will drop below unity for 
small $\pt$. In summary, the nuclear modification factor is a measure of all 
kinds of nuclear effects and deviations from scaling with $\langle 
N_\tn{coll}\rangle$ provided hard scattering processes are the by far dominant 
source of high-$\pt$ hadrons. 
 
\medskip 
 
The nuclear modification factor is measured by all four experiments 
\cite{phenix_AuAu,star_AuAu,brahms_AuAu+dAu,phobos_AuAu}. There is no need  to
discuss all of them since the results are similar and therefore only  the PHENIX
data will be presented here. PHENIX measures charged hadrons and  neutral pions
covering the pseudorapidity range $|\eta| < 0.35$, but for the  charged
particles a cut of $|\eta| < 0.18$ was applied. The events are divided  into
nine centrality classes that are listed in Table~\ref{tab_centr-cl}  together
with the corresponding values of $\langle N_\tn{coll}\rangle$ and  $\langle
N_\tn{part}\rangle$. 

\begin{tabelle}{|ccc|}{tab_centr-cl}{Centrality classes with the number of 
 participants and binary collisions as used by PHENIX, the percent values
 characterising the centrality classes give the fraction of the total cross
 section $\sigma^\tn{AuAu} = \unit[6.9]{b}$ \cite{phenix_AuAu}} 
\hline 
centrality & $\langle N_\tn{coll}\rangle$ & $\langle N_\tn{part}\rangle$ \\ 
\hline 
$0-10$\%  & $955.4 \pm 93.6$       & $325.2 \pm 3.3$       \\ 
$10-20$\% & $602.6 \pm 59.3$       & $234.6 \pm 4.7$       \\ 
$20-30$\% & $373.8 \pm 39.6$       & $166.6 \pm 5.4$       \\ 
$30-40$\% & $219.8 \pm 22.6$       & $114.2 \pm 4.4$       \\ 
$40-50$\% & $120.3 \pm 13.7$       & $74.4 \pm 3.8$       \\ 
$50-60$\% & $61.0 \pm 9.9$       & $45.5 \pm 3.3$       \\ 
$60-70$\% & $28.5 \pm 7.6$       & $25.7 \pm 3.8$       \\ 
$70-80$\% & $12.4 \pm 4.2$       & $13.4 \pm 3.0$       \\ 
$80-92$\% & $4.9  \pm 1.2$       & $6.3  \pm 1.2$       \\
\hline 
min.\ bias & $257.8 \pm 25.4$       & $109.1 \pm 4.1$       \\ 
\hline 
\end{tabelle} 
 
The charged particle spectrum measured by PHENIX in the pp run does not reach 
far enough in $\pt$, so the pp reference spectrum was constructed from the 
$\pi^0$ spectrum measured by PHENIX and the observed charged hadron to neutral
pion ratio (PHENIX and other experiments \cite{phenix_AuAu}). 

\smallskip 
 
\begin{figure}[ht!] 
 \centering 
 \includegraphics[scale=0.75]{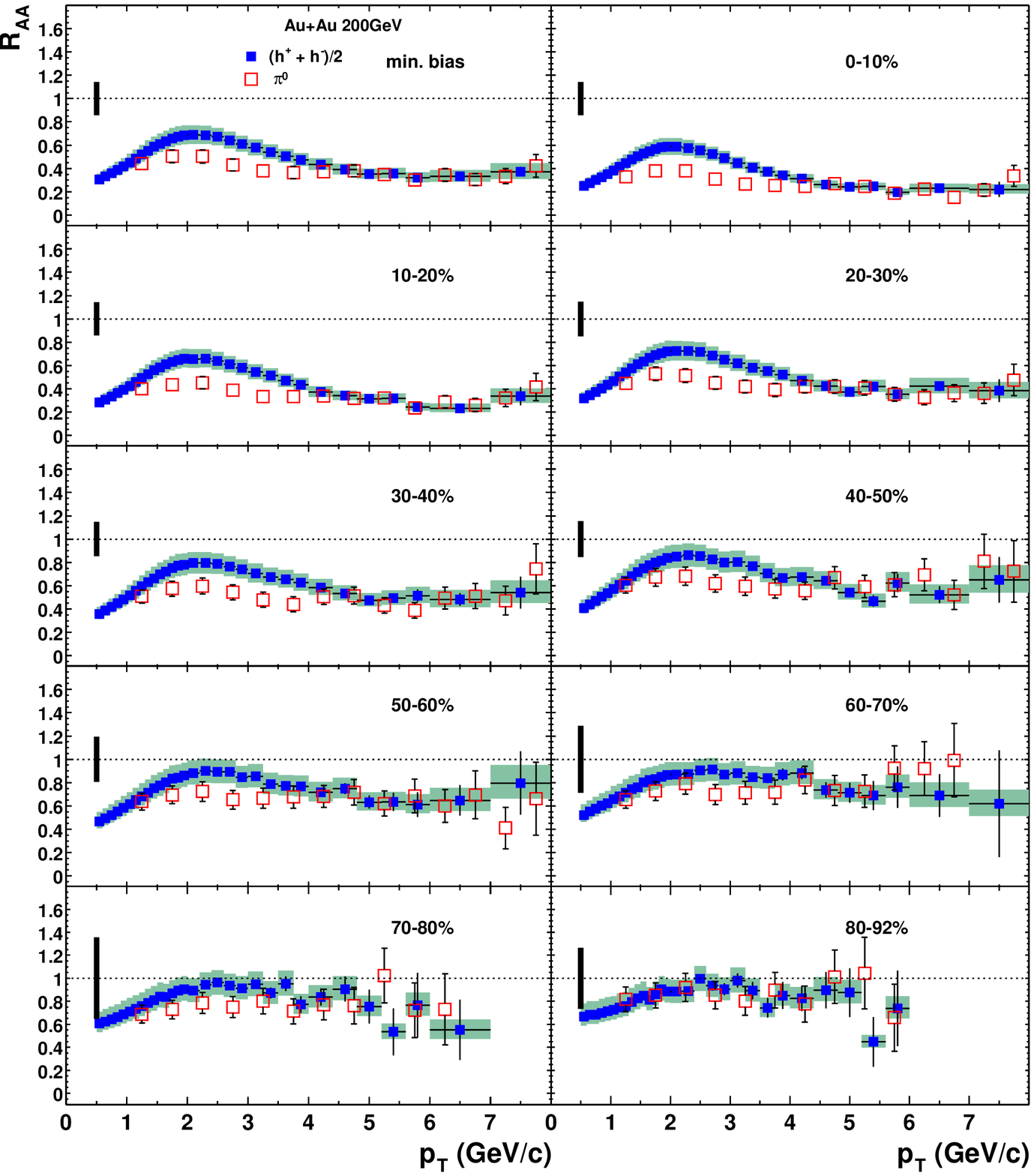} 
 \caption{$R_\tn{AuAu}(\pt)$ for $\pi^0$ and 
  charged hadrons at mid-rapidity for the different centrality classes 
  measured by PHENIX \cite{phenix_AuAu}} 
 \label{fig_raa} 
\end{figure} 

The results for $R_\tn{AuAu}$ at $\sqrt{s_\tn{NN}} = \unit[200]{GeV}$ are 
shown in Figure~\ref{fig_raa}. There is a clear suppression of high-$\pt$ 
hadrons that gradually increases with increasing centrality. The event
centrality is determined from the number of spectator neutrons and the number
of fast secondaries. The drop in the  region $\pt \lesssim \unit[2]{GeV}$ is
certainly caused at least partly by the $\mean{N_\tn{coll}}$ instead of
$\mean{N_\tn{part}}$ scaling in Equation~\ref{eq_RAA}. The ratio is flat for
$\pt \gtrsim \unit[4.5]{GeV}$ indicating that particle production is dominated
by hard scattering in this regime although the yield per binary collision is
not as high as in pp collisions. The scaling with $x_\perp$ ($x_\perp =
2\pt/\sqrt{s}$) gives further support to this hypothesis \cite{phenix_AuAu}.
The increase of the suppression with centrality is consistent with a plasma
scenario since the size and thereby the energy loss suffered by hard partons
are also expected to increase with centrality. From the Cronin effect an
enhancement of particles with intermediate $\pt$ (Ch.~\ref{sec_cronin}) is
expected so that the suppression is possibly even higher. 

\smallskip 
 
It has been argued in \cite{surface} that in case of plasma formation and high
energy loss in the deconfined medium only the partons produced near the surface
could escape since they would be thermalised if the way through the plasma was
too long. This would result in a scaling with $\mean{N_\tn{part}}$ instead of
$\mean{N_\tn{coll}}$ also for the high-$\pt$ hadrons \cite{surface}. There is,
however, no clear evidence for that. 

\smallskip

A surprising result is the centrality dependence of the nuclear modification
factor: It scales neither with $\mean{N_\tn{coll}}$ nor with
$\mean{N_\tn{part}}$ but is linear as a function of percent of the total cross
section. Figure~\ref{fig_cendep} shows the centrality dependence of the mean
$R_\tn{AuAu}$ for $\pt > \unit[3]{GeV}$ of the PHENIX $\pi^0$ data, the linear
fit is

\begin{equation}
 \mean{R_\tn{AuAu}(\pt > \unit[3]{GeV})} = (0.00835 \pm 0.00034)\cdot
 \tn{centrality\ [\%]} + (0.186 \pm 0.012)
\end{equation}
There is no convincing explanation for this behaviour so far.

\bildtex{plots/phenix-cendep-2.pstex_t}{fig_cendep}{Centrality dependence of the
  mean $R_\tn{AuAu}$ for $\pt > \unit[3]{GeV}$ for PHENIX $\pi^0$ data
  \cite{phenix_pi0}}

\smallskip 
 
\bild[0.4]{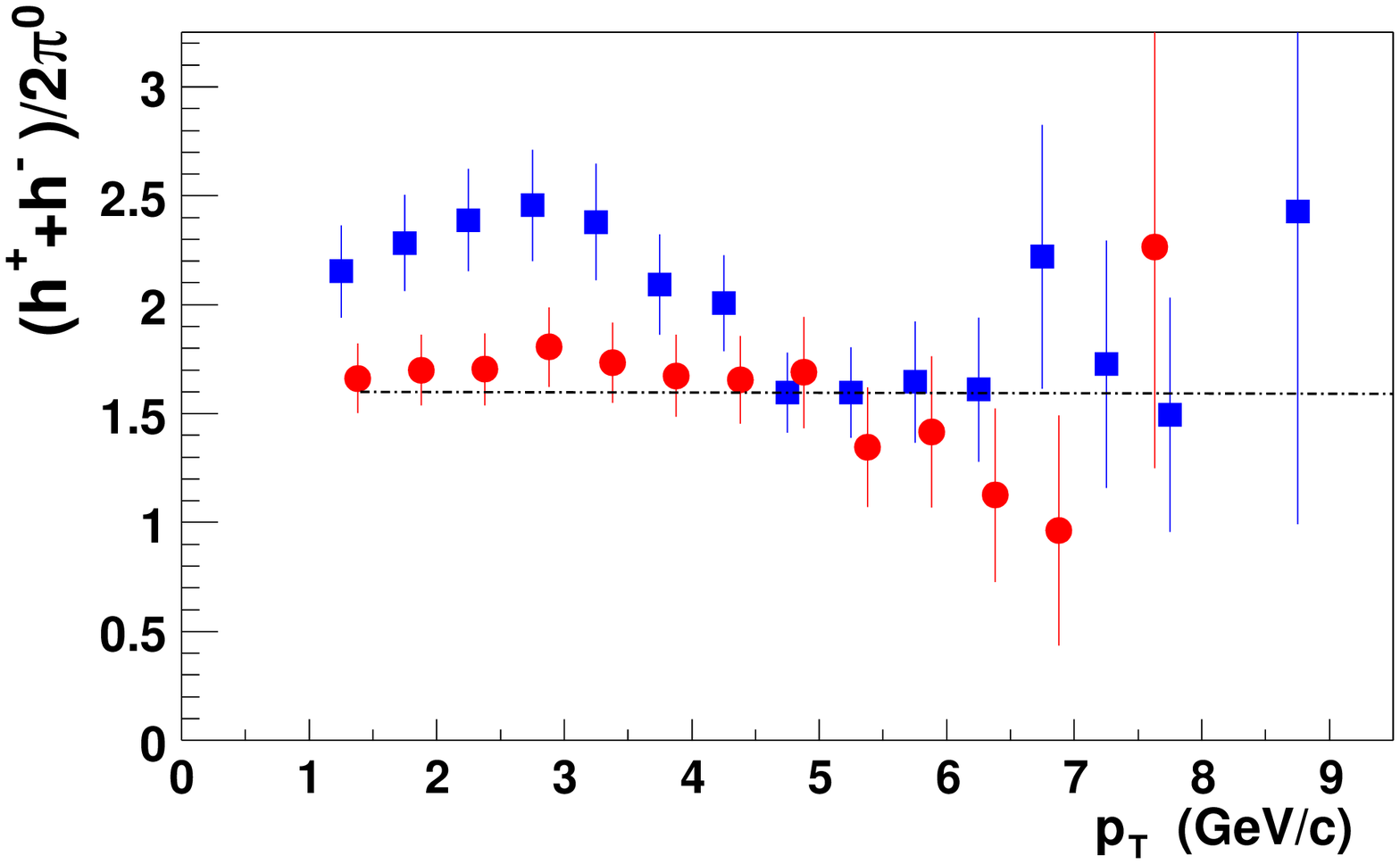}{fig_char/pi0}{Charged hadron to 
 neutral pion ratio in central ($0-10\%$ - squares) and peripheral ($80-92\%$ 
 - circles) AuAu collisions (the peripheral data points are offset by 
 \unit[130]{MeV}); PHENIX \cite{phenix_pinAuAu}} 
 
\bild[0.65]{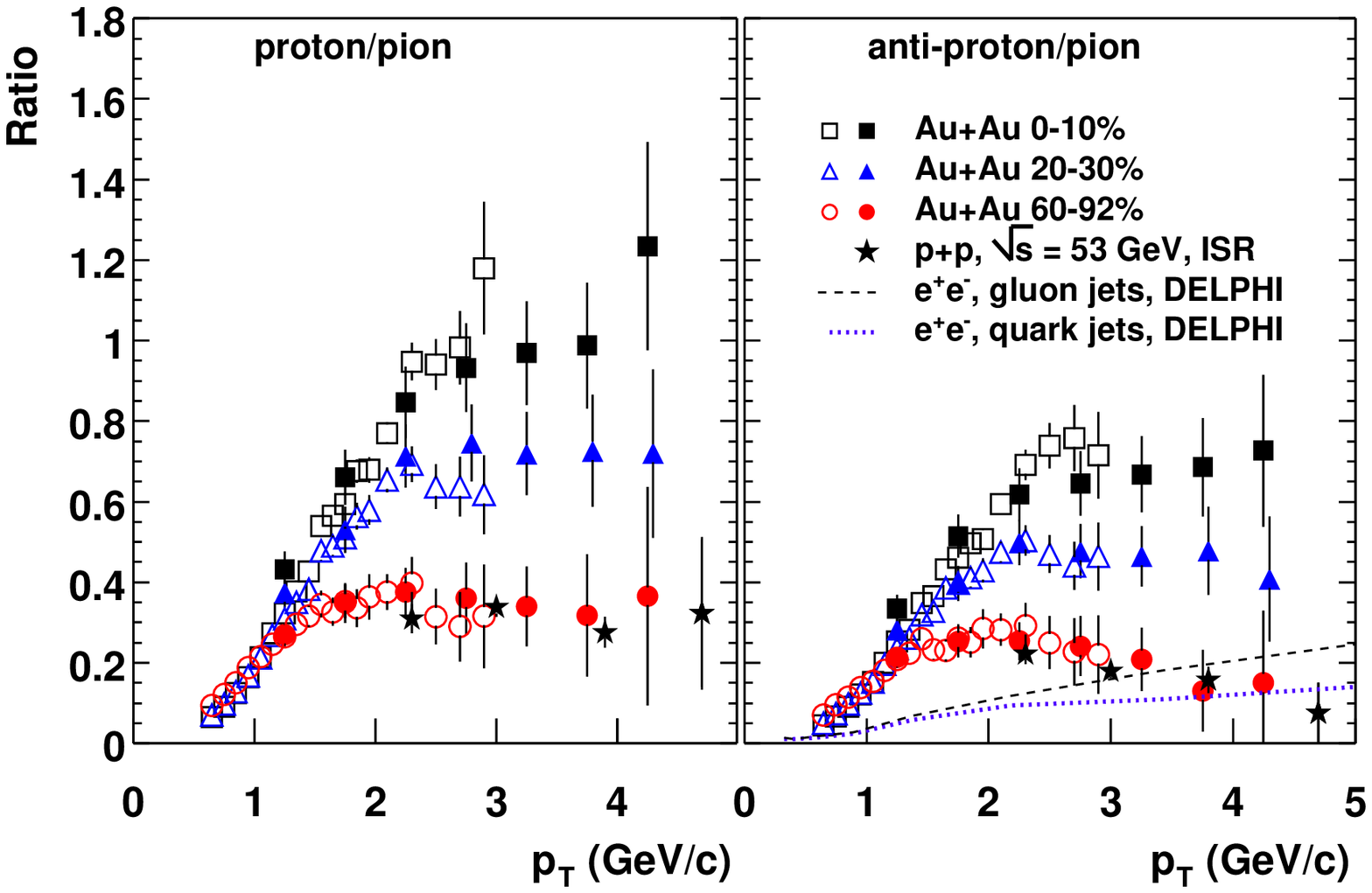}{fig_p/pions}{$p/\pi$ (left) and $\bar 
 p/\pi$ (right) ratios; open (filled) points are for charged (neutral) pions; 
 PHENIX \cite{phenix_pinAuAu}} 
 
An astonishing feature of Figure~\ref{fig_raa} is the stronger suppression of 
neutral pions at intermediate $\pt$. It turns out that this is not a stronger 
suppression of $\pi^0$ but a higher production rate of protons and
antiprotons. Figure~\ref{fig_char/pi0} shows the ratio of charged hadrons to
neutral pions in AuAu for central and peripheral collisions. The dotted line
at 1.6 indicates the ratio measured in $e^+e^-$ and pp collisions. This plot
nicely illustrates the excess of charged particles at intermediate $\pt$.
From Figure~\ref{fig_p/pions} it becomes clear that this is caused by an
enhanced (anti)proton production in more central AuAu collisions indicating a
deviation from the standard picture of particle production in the medium-$\pt$ 
range ($\unit[2]{GeV} \lesssim \pt \lesssim \unit[4.5]{GeV}$). 

\smallskip 
 
\bild[0.5]{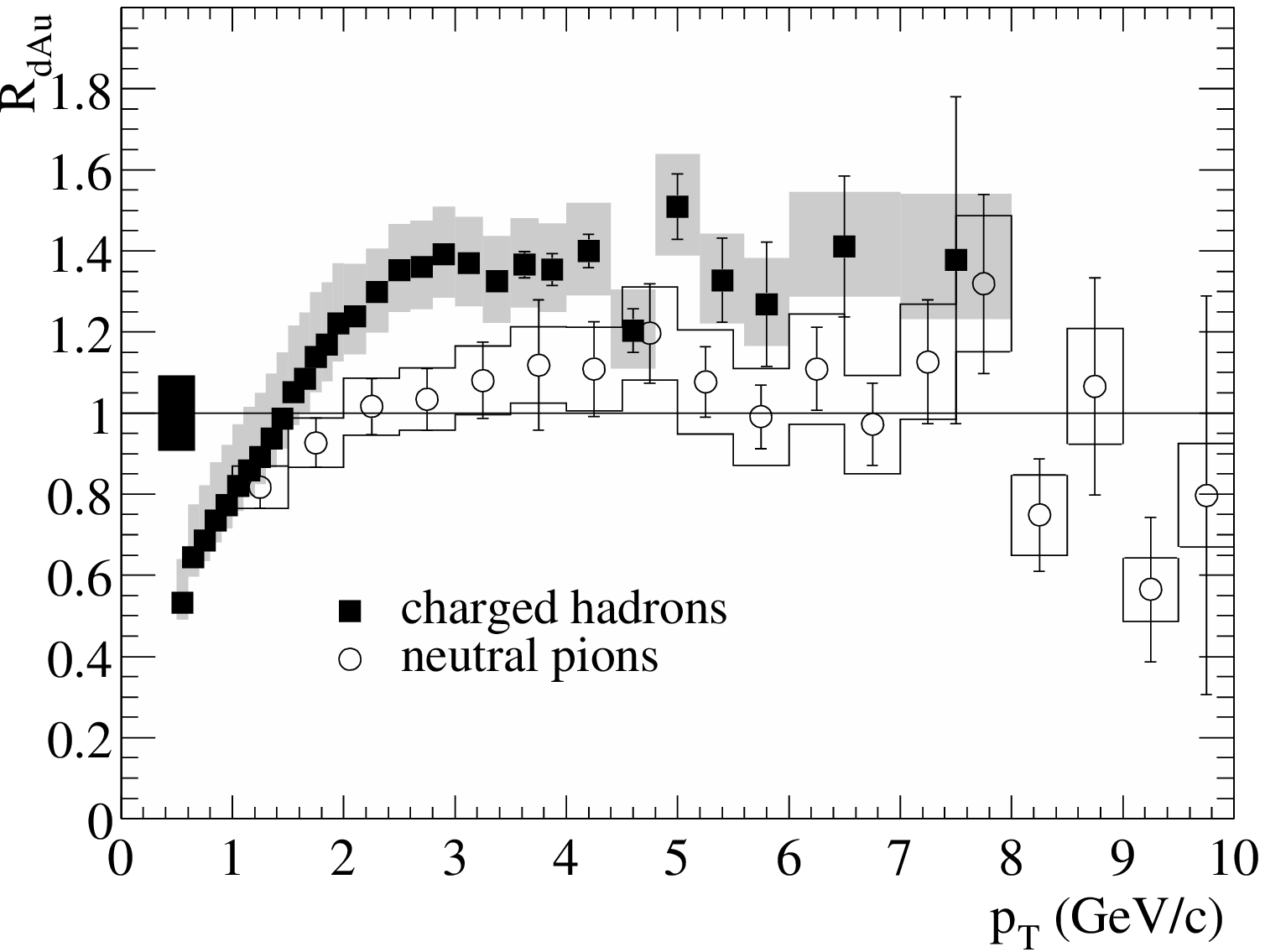}{fig_dAu}{$R_\tn{dAu}$ for charged hadrons and 
 neutral pions at $\sqrt{s_\tn{NN}} = \unit[200]{GeV}$; PHENIX 
 \cite{phenix_dAu}} 
 
In deuteron-gold collisions at RHIC no suppression of high-$\pt$ hadrons is 
observed. At mid-rapidity there is an enhancement 
\cite{brahms_AuAu+dAu,phenix_dAu,star_dAu} that is usually interpreted as the 
Cronin effect while the ratio is consistent with unity at larger rapidities 
\cite{phobos_dAu}. This further strengthens the case for a plasma scenario 
since the suppression in AuAu collisions is obviously related to final state 
effects (initial state effects should be present in the dAu data as well).
Figure~\ref{fig_dAu} shows the nuclear modification factor $R_\tn{dAu}$ in dAu
collisions measured by PHENIX, the values of $\langle N_\tn{coll} \rangle$ and
$\langle N_\tn{part} \rangle$ are 8.5 and 1.7 respectively. 

\smallskip 

\enlargethispage{\baselineskip}
 
In summary, there is a substantial suppression of high-$\pt$ hadrons in AuAu 
collisions that smoothly increases with increasing centrality. The spectral 
shape is unchanged as compared to pp for large $\pt$ indicating that 
fragmentation of hard partons is the dominant particle production mechanism in 
this regime. At medium $\pt$ a large relative proton and antiproton yield is 
observed, which is a hint to a deviation from the "normal" particle 
production. All observables approach their pp values in the most peripheral 
events. It should be kept in mind that the scaling behaviour is different for 
small $\pt$. No suppression is observed in dAu collisions indicating final
state effects as the most probable origin. These results are consistent with a
scenario where a quark-gluon plasma is formed in the overlap region of the
colliding nuclei and energetic partons lose a substantial amount of energy
when traversing the plasma. However, this is an indication and no proof. 

Surprisingly the nuclear modification factor increases linearly with centrality,
this feature is presently not understood.
 
\FloatBarrier 
 
\subsection{Disappearance of Back-to-Back Hadron Correlations} %--------------- 
\label{sec_2partcor}
 
STAR also measures the azimuthal correlation of charged particles at 
mid-rapidity ($|\eta| < 0.7$) \cite{star_dAu,star_phi}. The azimuthal 
distribution is defined as 
 
\begin{equation} 
\label{eq_d(deltaphi)}
 D(\Delta \phi) = \frac{1}{N_\tn{trig}}\der[N]{(\Delta \phi)} 
\end{equation} 
 
Particles with $4 < \pt < \unit[6]{GeV}$ are defined as trigger particles. For 
each trigger particle the azimuthal separation $\Delta \phi$ from all other 
particles satisfying $\unit[2]{GeV} < \pt < \pt (\tn{trig})$ is calculated 
yielding $N(\Delta \phi)$. In pp collisions $D(\Delta \phi)$ shows clear 
jet-like peaks at $\Delta \phi \sim 0$ and $\Delta \phi \sim \pi$
(Fig.~\ref{fig_DAuAu} and \ref{fig_Dpp+dAu}(a)). The distribution is
characterised by a Gaussian for each peak: 

\begin{equation} 
 D(\Delta \phi) = \frac{A_N}{\sqrt{2\pi}\sigma_N}e^{-(\Delta \phi)^2 / 
  2\sigma_N^2} + \frac{A_B}{\sqrt{2\pi}\sigma_B}e^{-(|\Delta \phi|-\pi)^2 / 
  2\sigma_N^2} + P 
\label{eq_fit} 
\end{equation} 
where the indices $N$ and $B$ stand for the near-side and back-to-back 
respectively and $P$ is a constant offset. The fit is shown as the solid line in 
Figure~\ref{fig_Dpp+dAu}(a), the parameters are given in Table~\ref{tab_fit}. 
 
\begin{tabelle}{|c|ccc|}{tab_fit}{Fit parameters from Equation~\ref{eq_fit}, 
 for the AuAu central sample only one Gaussian was fitted to the data after
 subtraction of the flow and pedestal contribution \cite{star_dAu}} 
\hline 
      & pp min.\ bias   & dAu min.\ bias  & AuAu central   \\ 
\hline 
$A_N$   & $0.081 \pm 0.005$ & $0.074 \pm 0.003$ & $0.093 \pm 0.008$ \\ 
$\sigma_N$ & $0.18 \pm 0.01$ & $0.20 \pm 0.01$ & $0.22 \pm 0.02$ \\ 
$A_B$   & $0.119 \pm 0.007$ & $0.097 \pm 0.004$ & $\diagup$     \\ 
$\sigma_B$ & $0.45 \pm 0.03$ & $0.48 \pm 0.02$ & $\diagup$     \\ 
$P$    & $0.008 \pm 0.001$ & $0.039 \pm 0.001$ & $0.004 \pm 0.003$ \\ 
\hline	 
\end{tabelle} 
 
\smallskip 
 
In nucleus-nucleus collisions there will also be a contribution from the 
elliptic flow anistropy (Ch.~\ref{sec_flow}): 
 
\begin{equation} 
 \der[N_\tn{ef}]{(\Delta \phi)} = B[1 + 2v_2^2 \cos(2\Delta \phi)] 
\label{eq_ef}  
\end{equation}   
where $v_2$ has been measured independently in the same set of events and is 
assumed to be constant in $2 < \pt < \unit[6]{GeV}$, whereas $B$ has to be 
fitted. The values for the different centrality classes are listed in
Table~\ref{tab_v2}. 

\begin{tabelle}{|cccc|}{tab_v2}{Values of $N_\tn{part}$, $v_2$ ($2 < \pt < 
 \unit[6]{GeV}$) and the normalisation constant $B$ for the different 
 centrality classes \cite{star_phi}} 
\hline  
centrality & $N_\tn{part}$ & $v_2$	   & $B$        \\  
\hline 
0-5\%   & $352 \pm 7$  & $0.07 \pm 0.01$ & $1.442 \pm 0.003$ \\ 
5-10\%   & $298 \pm 10$ & $0.10 \pm 0.01$ & $1.187 \pm 0.008$ \\ 
10-20\%  & $232 \pm 11$ & $0.15 \pm 0.01$ & $0.931 \pm 0.006$ \\ 
20-30\%  & $165 \pm 13$ & $0.19 \pm 0.01$ & $0.633 \pm 0.005$ \\ 
30-40\%  & $144 \pm 13$ & $0.21 \pm 0.01$ & $0.420 \pm 0.005$ \\ 
40-60\%  & $61 \pm 10$ & $0.22 \pm 0.01$ & $0.231 \pm 0.003$ \\ 
60-80\%  & $20 \pm 6$  & $0.24 \pm 0.04$ & $0.065 \pm 0.003$ \\ 
\hline 
\end{tabelle} 
 
{\psfrag{Delta phi}{$\Delta \phi\ [\tn{rad}]$} 
\psfrag{D(Delta phi)}{$D(\Delta \phi)$} 
\bild[0.8]{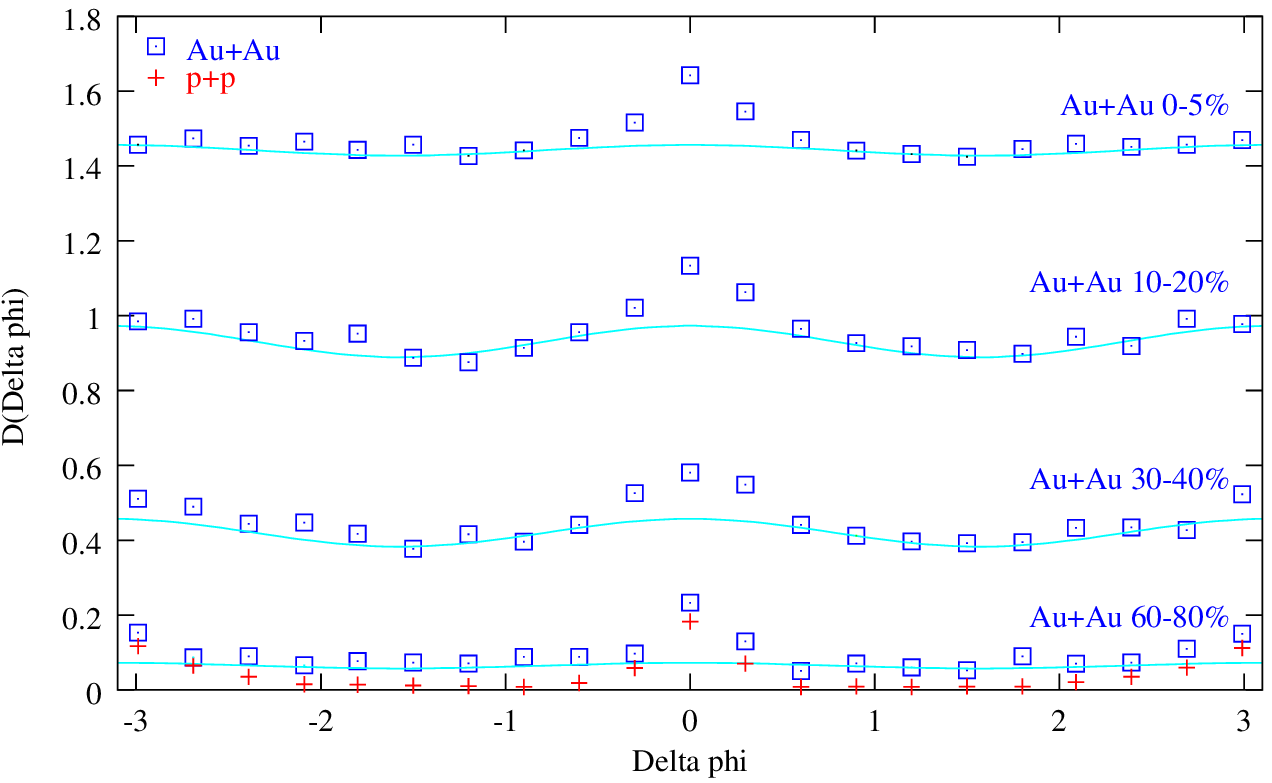}{fig_DAuAu}{Azimuthal distribution $D(\df)$ for pp and 
 different centralities in AuAu measured by STAR, the lines show the elliptic 
 flow and pedestal contribution (Eq.~\ref{eq_ef},Tab.~\ref{tab_v2}) 
 \cite{star_phi}}} 
 
\bild{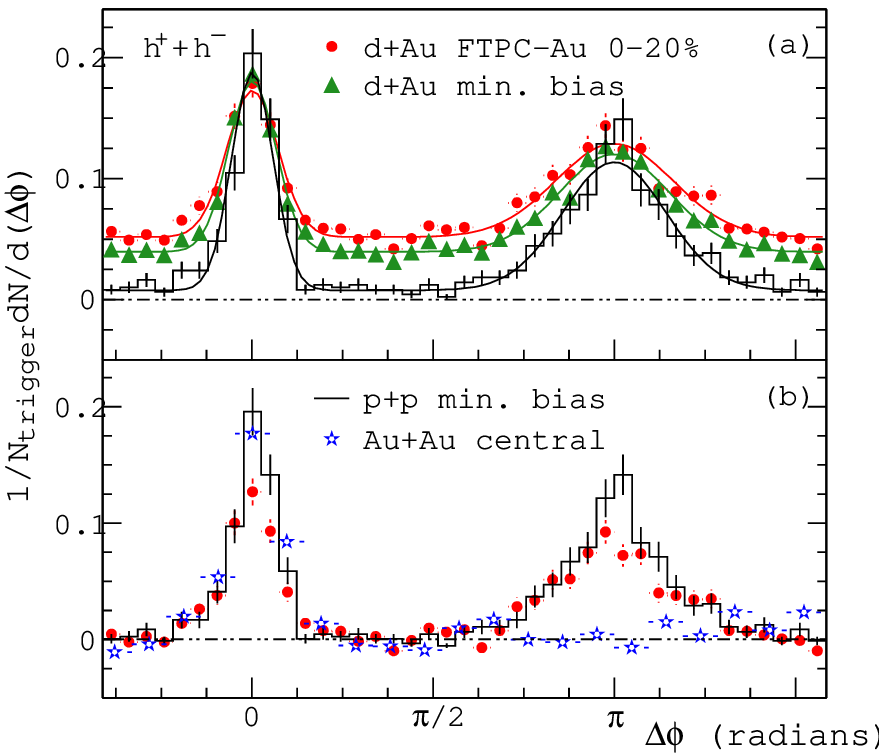}{fig_Dpp+dAu}{Azimuthal distribution $D(\df)$ for pp, dAu 
 and central AuAu (elliptic flow and pedestal distributions have been 
 subtracted in (b)), solid lines in (a) are Gaussian fits (Eq.~\ref{eq_fit}, 
 Tab.~\ref{tab_fit}), STAR \cite{star_dAu}} 
 
Figure~\ref{fig_DAuAu} shows the measured azimuthal distribution for different 
centralities in AuAu together with the elliptic flow contribution (solid
lines)  and the pp data. Jet-like correlations can be seen as an excess of the
measured  correlation over the elliptic flow contribution. The jet on the
trigger side  ($\Delta \phi \sim 0$) is present for all centralities but the
away-side jet  ($\Delta \phi \sim \pi$) vanishes in more central events. In
Figure~\ref{fig_Dpp+dAu}(b) the distribution in pp collisions is compared to
central  AuAu collisions after subtraction of the elliptic flow and pedestal 
contributions. It is not surprising that the correlation on the trigger side
is  nearly as strong as in pp collisions since the trigger condition introduces
a  bias. The important point is that the jet on the other side disappears 
completely in the most central collisions. The suppression increases again 
smoothly with centrality. 

\smallskip 
 
The results from the dAu measurements are shown in Figure~\ref{fig_Dpp+dAu}(a) 
and Table~\ref{tab_fit}. There is no substantial suppression of the away-side 
jet which again points towards final state effects being responsible for the 
disappearance of the second jet in central AuAu collisions. 
 
The results are commonly interpreted in terms of a plasma scenario in the 
following way. The data samples for the AuAu collisions are heavily biased 
towards events where the hard scattering took place near the surface of the 
overlap region due to the trigger condition. Particles with $\pt > 
\unit[4]{GeV}$ are rare already in pp collisions. If the energetic parton is 
produced inside the plasma and has to traverse a substantial part of it, it
will  lose a sizable fraction of its energy in interactions with the plasma.
The  leading hadron formed in the fragmentation will then have a transverse 
momentum which is too small to pass the trigger condition. If now the hard
scattering happened near the surface with one parton escaping undisturbed the
other one will typically have to traverse a substantial part of the plasma for
geometrical reasons. The mean path length through the plasma increases with
increasing centrality and one would therefore expect an increasing suppression
of the away-side jet. In dAu collisions, where no plasma is formed, the
azimuthal correlation should be similar to pp distribution. 

\enlargethispage{\baselineskip}

This argumentation sounds convincing, nevertheless one has to be careful.
Firstly the $\pt$-spectrum has a tail to very high energies and these partons
may produce trigger particles even after heaving lost some energy. Therefore
the trigger bias does not automatically mean that the hard scattering took
place near the surface. This point can only be clarified with the number of
trigger particles produced per binary collision which is much smaller in the
scenario where only scatterings that happen to be located near the surface can
give rise to trigger particles. Secondly the lifetime of the QGP which is
estimated to be of the order of a few \unit{fm} limits the path lengths. In the
surface emission scenario this would lead to saturation as soon as the
geometrically available path length becomes shorter than the lifetime. And
finally the matrix element is symmetric in azimuth which means that there is a
sizable probability that both partons escape nearly undisturbed when they are
emitted from the surface. Consequently the observed azimuthal correlation seems
to contradict the surface-emission scenario and the observed jets stem from
some greater depth.

The conclusion can be drawn that also the measured azimuthal correlation of
high-$\pt$  hadrons is qualitatively consistent with the hypothesis of
quark-gluon plasma  being formed in relativistic nucleus-nucleus collisions. 

%\bildtex{plots/integrale.pstex_t}{fig_integrals}{Integrals of the two peaks of
%the azimuthal correlation for different event topologies. Long arrows denote
%particles with $\unit[4]{GeV} < \pt < \unit[6]{GeV}$ (trigger particles) and
%short arrows stand for particles with $\unit[2]{GeV} < \pt < \unit[4]{GeV}$.}

\subsection{Correlation with the Reaction Plane} %------------------------------

\enlargethispage{\baselineskip}

The almond-like shape of the overlap region in mid-central collisions leads to
a dependence of the two-particle correlation (Eq.~\ref{eq_d(deltaphi)}) on the
orientation of the pair relative to the reaction plane. A parton that is
emitted in the reaction plane has on average to traverse a shorter distance in
the QGP than one that is emitted perpendicular to it. Consequently the
suppression of the away-side jet is stronger in the latter case.

\smallskip

The effects of different emission angles with respect to the reaction plane
have alse been studied by the STAR experiment \cite{star_reacplane}. The
procedure is analogous to the one described in Section~\ref{sec_2partcor} but
here all tracks with $|\eta|<1$ are considered. A pair is defined to be
in-plane if the azimuthal angle of the trigger particles satisfies
$|\phi^\tn{trig}-\Psi_2| < \pi/4$ or $|\phi^\tn{trig}-\Psi_2| > 3\pi/4$ where
$\Psi_2$ is the reaction plane angle of the event. A trigger is out-of-plane if
$\pi/4 < |\phi^\tn{trig}-\Psi_2| < 3\pi/4$.

The elliptic flow contribution is \cite{star_reacplane}

\begin{equation}
\label{eq_fit2}
\der[N_\tn{out}^\tn{in}]{\df} = B \left[ 1 + 2v_2^\tn{assoc} \left( 
  \frac{\pi v_2^\tn{trig} \pm 2\mean{\cos(2\Delta\Psi)}}{\pi \pm
  4v_2^\tn{trig}\mean{\cos(2\Delta\Psi)}} \right) \cos (2\df) \right]
\end{equation}  
where $v_2^\tn{assoc}$ and $v_2^\tn{trig}$ are the elliptic flow of the
associated and trigger particles, respectively, and $\mean{\cos(2\Delta\Psi)}$
is the reaction plane resolution. The values for 20\%-60\% centrality are
$v_2^\tn{assoc} = 0.20$, $v_2^\tn{trig} = 0.18$ and $\mean{\cos(2\Delta\Psi)} =
0.70$. 

\smallskip

\bild[0.5]{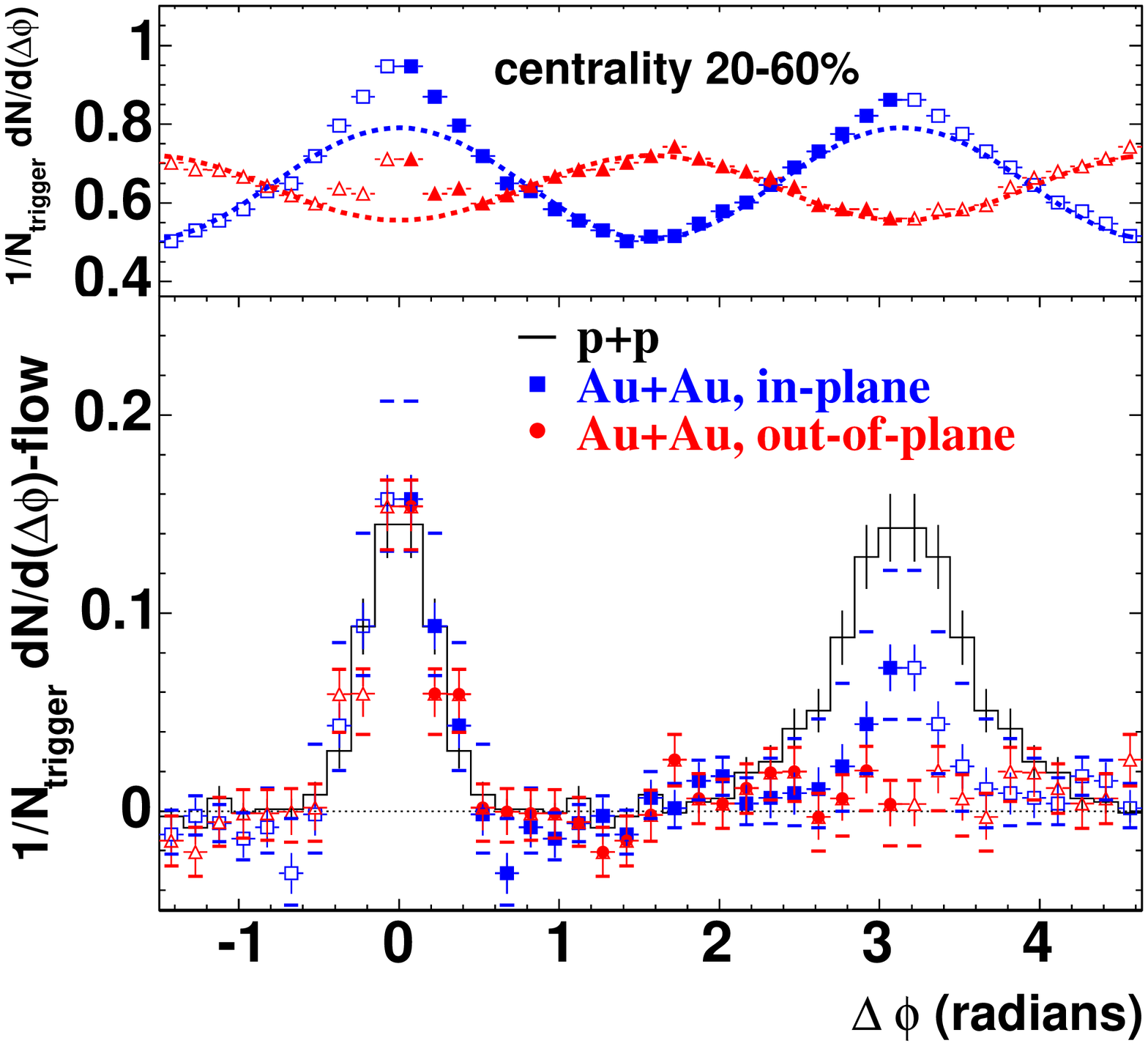}{fig_reacplane}{Upper panel: Two-particle
correlations for trigger particles in-plane (squares) and out-of-plane
(triangles) in mid-central AuAu. The contribution from elliptic flow is shown
by dashed lines. Lower panel: Azimuthal distributions after subtracting the
elliptic flow and pp data.\cite{star_reacplane}}

The azimuthal distributions are shown in the upper panel of
Figure~\ref{fig_reacplane}. The solid symbols are data points whereas the open
symbols are reflections around $\df=0$ and $\df=\pi$. The elliptic flow
contributions from Equation~\ref{eq_fit2} were fitted with $B$ as the only free
parameter. The results, shown as dashed lines, are
$B=0.649\pm0.004(stat.)\pm0.005(sys.)$ for the in-plane and
$B=0.638\pm0.004(stat.)\pm0.002(sys.)$ for the out-of-plane distribution.

The lower panel of  Figure~\ref{fig_reacplane} shows the distributions after
subtraction of the elliptic flow together with the pp measurement. The data
clearly confirm a stronger quenching of jets perpendicular to the reaction
plane where the original parton had to travel a longer distance inside the QGP
and thus experienced larger energy loss. The large uncertainty is mainly due to
the uncertainty in the elliptic flow contribution. 

The fact that a substantial part of the away-side jets survives in the reaction
plane also illustrates that the QGP cannot be completely opaque since the
smallest diameter of the overlap region in a 20\% centrality event is already
$\sim \unit[4]{fm}$. It can thus be concluded that a considerable part of the
jets survives a journey of a few \unit{fm} through the QGP.

\subsection{Other Observables} %------------------------------------------------

The yields of multi-strange particles are enhanced as compared to SPS. The data
indicate strangeness saturation for central collisions \cite{star_strange}.

The particle ratios are again beautifully described by a statistical model
assuming full chemical equilibrium. The freeze-out point is found to be
consistent with the phase boundary calculated with lattice QCD
\cite{stat-rhic}. Furthermore it has been shown that the chemical freeze-out
temperature coincides with the critical temperature since it is not possible
to maintain chemical equilibrium in a hadronic phase even close to the phase
transition \cite{wetterich}. This provides a strong argument for QGP formation.

The first measurements of $J/\psi$ production show a suppression but the
experimental uncertainty (mainly lack of statistics) is still too large to
make more detailed statements \cite{phenix_jpsi}.

\chapter{Comparison of Identified Particle Spectra from SPS and RHIC}
\enlargethispage{-\baselineskip}

While the hard part of the $\pt$-spectra in nucleon-nucleon collisions is
theoretically under control and shows the power-law behaviour typical for hard
parton-parton scattering this is not the case for the soft part. Here the main
contribution comes from non-perturbative processes that are not well understood
so that one has to rely on paramtrisations of measured cross sections.
However, the situation is not as bad in ion-ion collisions where the soft part
of the $\pt$-spectra is better understood and contains valuable information
about the properties of the created medium.

In collisions of relativistic nuclei the low-$\pt$ part of the hadron spectra
is determined by the hydrodynamic behaviour of the system, primarily the
freeze-out temperature and the pressure which drives the transverse expansion.
A still open question is whether the physics at SPS energies is qualitatively
different from the physics at RHIC. The hard part of the spectra is hardly
accessible at SPS, but the soft part can be used to quantify differences
between RHIC and SPS. In this chapter a detailed comparison of the shapes of
the $\pt$- spectra of identified particles is carried out.

\smallskip

A thermal source at rest emitting particles from the origin leads to a spectrum
of the form 

\begin{equation}
\label{equ_expon}
\frac{\tn{d}^2N}{\mt \tn{d}\mt \tn{d}y} \propto e^{-\frac{\mt-m_0}{T}} 
\end{equation}
which is completely determined by the temperature. $\mt = \sqrt{\pt^2 + m_0^2}$
is the transverse mass and $m_0$ is the respective particle's rest mass. The
transverse expansion leads to deviations from this form which are, however,
model dependent. Further distortions arise from resonance decays. 

The freeze-out temperature can be extracted from the ratios of different
particle yields  and is found to be nearly the same for SPS and RHIC, namely
$T=\unit[(168\pm 2.4)]{MeV}$ at SPS \cite{stat-sps} and $T=\unit[(174\pm
7)]{MeV}$ at RHIC \cite{stat-rhic}. Differences in the shape of the spectra
should thus stem from different expansion velocities.

\begin{figure}[ht]
\centering
\begin{turn}{-90}
\includegraphics[scale=0.4]{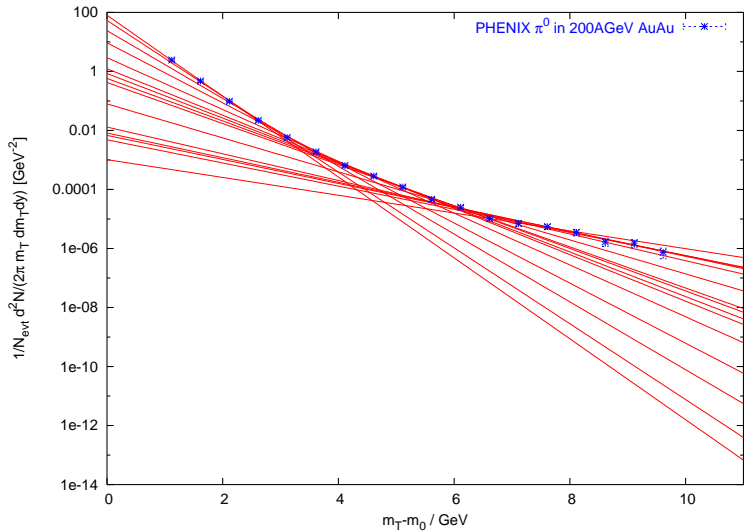}
\end{turn}
\caption{$\pi^0$ spectrum measured by the PHENIX experiment \cite{phenix_pi0}
 with locally fitted exponentials}
\label{fig_pi0}
\end{figure}

In order to avoid the problem with the model dependence of the transverse
expansion contribution an exponential $a\exp{\{-b(\mt-m_0)\}}$ is fitted
locally, i.e.\ in a small $\pt$-interval, to the $\pt$-spectrum. $b$ will then
depend on $\pt$ and cannot be interpreted as an inverse temperature, but it can
be used to quantify differences in the slopes of the spectra. Since this
comparison relies solely on the shape of the spectrum the parameter $a$ which
describes the yields is irrelevant. To get comparable results the spectra from
central collisions near mid-rapidity were used. A large number of different
identified particle spectra were analysed in order to improve the significance
and give an overview over the present status in terms of experimental data. All
analysed spectra are listed in Table~\ref{tab_spectra}. Concerning protons and
antiprotons only spectra that are corrected for feed-down from week decays were
considered.  Figure~\ref{fig_pi0} shows as an example a measured pion spectrum
with the fitted exponentials.

\enlargethispage{-\baselineskip}

\begin{figure}[ht]
\centering
\begin{turn}{-90}
\includegraphics[scale=0.4]{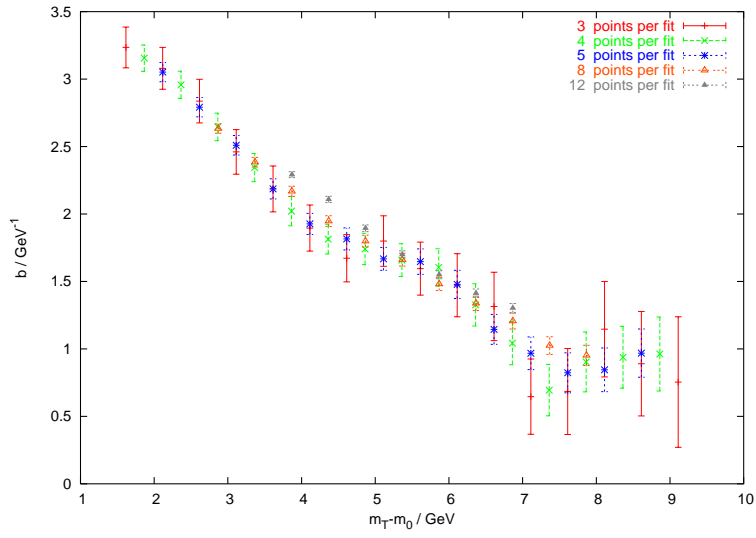}
\end{turn}
\caption{Fit parameters for the $\pi^0$ spectrum from Figure~\ref{fig_pi0} 
obtained with different interval lengths}
\label{fig_intlen}
\end{figure}

A potentially critical point is the possible dependence of the results on the
interval length. Fortunately it was found be small as is illustrated in
Figure~\ref{fig_intlen}. A compromise between a small interval, which is
desirable because of better sensitivity and resolution, and a small error has
to be made. The interval length was therefore chosen individually for each
spectrum depending mainly on the experimental uncertainties. 

\smallskip

The results for the kaons are shown in Figure~\ref{fig_kaon} as a representative,
the rest can be found in Appendix~\ref{app_spectra}. The fact that $b$ changes
with $\pt$ illustrates that transverse expansion does indeed play a role. The
differences between SPS and RHIC are small, the RHIC points are continuously
slightly higher. The discrepancies among the experiments at the same accelarator
are also sizable, in fact they are of the same magnitude as the differences
between RHIC and SPS. There is also no visible difference between the SPS data
at \unit[80]{A\,GeV} and \unit[158]{A\,GeV} beam energy, except maybe in the
$\Lambda$ spectra. The sum of the results clearly suggests that there is no
dramatic change in (soft) physics when going from SPS to RHIC. There is room for
a small quantitative difference.

\begin{figure}[ht]
\centering
\begin{turn}{-90}
\includegraphics[scale=0.4]{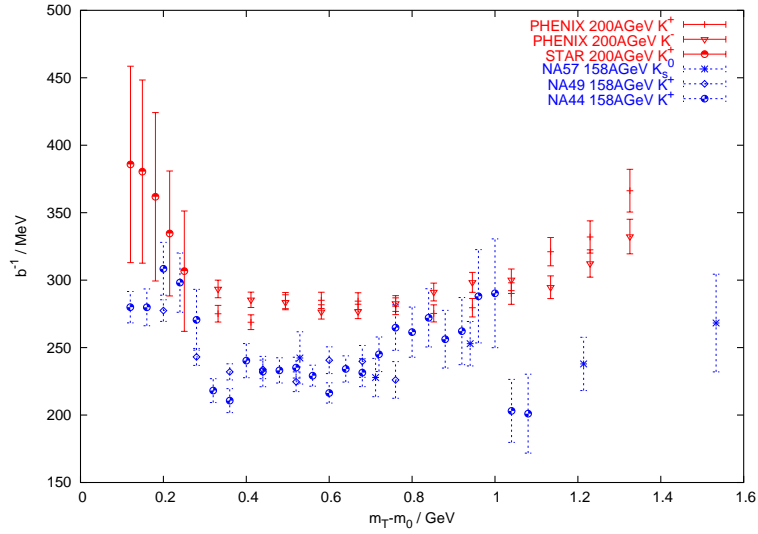}
\end{turn}
\caption{Fit parameters for the kaon spectra}
\label{fig_kaon}
\end{figure}

\smallskip

A comparison of the elliptic flow leads to the same conclusion, here the
differences between SPS and RHIC are also not very big. Indeed the two observables
are correlated because the pressure, that drives the transverse expansion, also
has an influence on the pressure gradient, which is responsible for the elliptic
flow. The transverse size of the system also turned out to be similar at SPS and
RHIC \cite{phenix-HBT}. Together these observation lend even stronger support to
the conclusion that at least the soft physics is similar at SPS and RHIC.

\begin{tabelle}{|l|l|c|c|c|c|c|}{tab_spectra}{List of all analysed spectra}
\hline
Particle &Experiment &Energy  &System &\# points&Centrality &\\ 	
         &           &[\unit[A]{GeV}]&&  per fit&  &\\ \hline
$K^+$		&PHENIX	 &200 	  &Au+Au	   &5 	    &5.0\%	& \cite{phenix1}\\
$K^-$		&PHENIX	 &200 	  &Au+Au	   &5 	    &5.0\%	& \cite{phenix1}\\  	    
$K^+$		&STAR 	 &200 	  &Au+Au	   &6 	    &5.0\%	& \cite{star1}\\
$K_s$		&NA57 	 &158 	  &Pb+Pb	   &4 	    &4.5\%	& \cite{na57}\\  	    
$K^+$		&NA44 	 &158 	  &Pb+Pb	   &6 	    &3.7\%	& \cite{na44}\\
$K^+$		&NA49 	 &158 	  &Pb+Pb	   &5 	    &5.0\%	& \cite{na49-1}\\  	    
$\Lambda$	&PHENIX	 &130 	  &Au+Au	   &4 	    &5.0\%	& \cite{phenix2}\\  	    
$\Lambda$	&STAR 	 &130 	  &Au+Au	   &4 	    &5.0\%	& \cite{star2}\\
$\Lambda$	&NA49 	 &80  	  &Pb+Pb	   &6 	    &7.2\%	& \cite{na49-2}\\  	    
$\Lambda$	&NA49 	 &158 	  &Pb+Pb	   &7 	    &10.0\%	& \cite{na49-2}\\
$\Lambda$	&NA57 	 &158 	  &Pb+Pb	   &4 	    &4.5\%	& \cite{na57}\\
$\bar \Lambda$ &PHENIX	 &130 	  &Au+Au	   &5 	    &5.0\%	& \cite{phenix2}\\  	    
$\bar \Lambda$ &STAR	 &130	    	  &Au+Au       &4		    &5.0\%  & \cite{star2}\\
$\bar \Lambda$ &NA49	 &80	        &Pb+Pb       &4		    &7.2\%  & \cite{na49-2}\\
$\bar \Lambda$ &NA49	 &158	        &Pb+Pb       &4		    &10.0\% & \cite{na49-2}\\
$\bar \Lambda$ &NA57	 &158	        &Pb+Pb       &4		    &4.5\%  & \cite{na57}\\
$p$		&PHENIX	 &200 	  &Au+Au	   &5 	    &5.0\%	& \cite{phenix1}\\  	    
$p$		&NA44 	 &158 	  &Pb+Pb	   &6 	    &3.7\%	& \cite{na44}\\
$p$		&NA49 	 &80  	  &Pb+Pb	   &5 	    &7.0\%	& \cite{na49-3}\\
$p$		&NA49 	 &158 	  &Pb+Pb	   &6 	    &10\%	& \cite{na49-3}\\
$\bar p$	&PHENIX	 &200 	  &Au+Au	   &5 	    &5.0\%	& \cite{phenix1}\\  	    
$\bar p$	&NA44 	 &158 	  &Pb+Pb	   &7 	    &3.7\%  & \cite{na44}\\
$\bar p$	&NA49 	 &80  	  &Pb+Pb	   &7 	    &7.2\%  & \cite{na49-3}\\
$\bar p$	&NA49 	 &158 	  &Pb+Pb	   &7 	    &10.0\%	& \cite{na49-3}\\
$\pi^0$	&PHENIX	 &200 	  &Au+Au	   &5 	    &10.0\%	& \cite{phenix_pi0}\\	    
$\pi^+$	&PHENIX	 &200 	  &Au+Au	   &5 	    &5.0\%	& \cite{phenix1}\\  	    
$\pi^+$	&STAR 	 &200 	  &Au+Au	   &5 	    &5.0\%	& \cite{star1}\\
$\pi^+$	&NA44 	 &158 	  &Pb+Pb	   &8 	    &3.7\%	& \cite{na44}\\
$\pi^\pm$	&NA45 	 &158 	  &Pb+Au	   &5 	    &6.1\%	& \cite{na45}\\
$\pi^-$	&NA49 	 &158 	  &Pb+Pb	   &5 	    &5.0\%	& \cite{na49-1}\\
$\pi^0$	&WA98 	 &158 	  &Pb+Pb	   &5 	    &10\% of min.bias& \cite{wa98}\\
\hline
\end{tabelle}

\chapter{Monte Carlo Techniques}
A very general definition would be that Monte Carlo is everything that makes
use of random numbers to solve a problem. Monte Carlo methods are mainly used
to simulate processes, estimate integrals or select random variables according
to some probability density (or a combination of these). The processes to which
Monte Carlo methods are applicable do not necessarily have to be of stochastic
nature, although this is normally the case in high energy physics. 

\medskip

The values that a random variable takes are unpredictable, so there have to
be at least two different possible values of that variable. The distribution
of a random variable is given by a probability density $g(u)\tn{d}u$, which is
the probability that the value lies in the range $[u,u+\tn{d}u]$. The
scattering angle in a $a+b \to a+b$ elastic scattering is an example for a
random variable. In this case the probability density is given by the
differential cross section $\tn{d}\sigma = f(\vartheta)\tn{d}\vartheta$. This
example also illustrates that probability densities in physical problems are
often not normalised to unity, because the integral may also have a physical
meaning (in this example the total cross section).

\smallskip

The integral of a function $f(x)$ in an interval $[a,b]$ can be estimated by
choosing $N$ values $x_i$ distributed randomly in  $[a,b]$, summing the
function values $f(x_i)$ in these points and dividing by $N$. According to
the law of large numbers the Monte Carlo estimate of an integral converges (in
the statistical sense) to its exact value as the number of points approaches
infinity.

\begin{equation} 
  \frac{1}{N}\sum_{i=1}^N f(x_i) \stackrel{N\to \infty}{\longrightarrow}
      \frac{1}{b-a} \int \limits_a^b\!\tn{d}x\, f(x)  
\label{eq_mc1}
\end{equation}	

Convergence in the statistical sense means that for every probability $p$ and
positive number $\epsilon$ there is a $k$ such that for all $N>k$ the
probability of the difference between the left- and the right-hand side of
Equation~\ref{eq_mc1} being smaller than $\epsilon$ is greater than $p$.

The Monte Carlo estimate is a function of the random numbers $\{x_i\}$. For
finite $N$ the value obtained with a different set of numbers $\{x'_i\}$ will
be different from the first one. The central limit theorem says that for large
$N$ the Monte Carlo estimate is a Gaussian distribution. Its expectation is
the true value of the integral. The expectation of a function $f(x)$ is
defined as

\begin{equation}
  E(f) = \int\!\tn{d}x\, f(x)g(x) 
\end{equation}  
and the variance (average of the squared deviation from the expectation) is

\begin{equation}
  V(f) = E[(f-E(f))^2] = \int\!\tn{d}x\, (f-E(f))^2 g(x) 
\end{equation}  

The variance of the Monte Carlo estimate is $V(f)/N$. The speed, with which the
Monte Carlo estimate converges, does not depend explicitly on the number of
dimensions (although other calculations in the algorithm might take more time
in higher dimensions). This means that for small number if dimensions numerical
quadrature will be much more efficient while the Monte Carlo clearly wins for
higher number of dimensions. Since the variance also depends on the variance of
the function $f$, the performance of the Monte Carlo can also be improved by
reducing $V(f)$. These methods will not be discussed explicitly here but can be
found for example in \cite{james}. The variance reducing techniques described
below for the selection from a given probability density can be adopted to the
estimation of integrals.

\medskip

Given a random number generator that produces numbers $R$ uniformly distributed
in $]0,1[$ these can be used to obtain numbers distributed according to any
other density function $f(x)$ defined on an interval $[a,b]$. This is
extremely useful in simulations of physical processes. 

\smallskip

The simplest way to do that is the hit-or-miss Monte Carlo. First an upper
limit $f_\tn{max}$ of the function has to be found, i.e.\ $f_\tn{max} \ge
f(x)\  \forall x\in [a,b]$. Then the procedure is to select an $x$ from a flat
distribution, i.e.\ $x = a + R(b - a)$ where $R$ is a random number in $]0,1[$.
Then another random number $R$ in the interval $]0,1[$ is selected and if $R <
f(x)/f_\tn{max}$ the $x$-value is accepted. Otherwise it is rejected and the
procedure continues with choosing a new $x$.

The advantage of this algorithm is that it works for practically all functions
without singularities. The efficiency is 

\begin{equation}
  \frac{ \int\!\tn{d}x\, f(x)}{f_\tn{max}(b-a)}
\end{equation}

If $f$ varies a lot the efficiency becomes very bad, so it might be useful to
divide the interval into suitable subintervals. Then a decision which
subinterval to choose has to be made first (the relative probability is given
by the integral in the subinterval).

\smallskip

Another method is the direct sampling, that can be used if $f$ has a primitive
function $F$ whose inverse $F^{-1}$ can be calculated. Then $x$-values can be
found using the relation

\begin{equation}
  \int \limits_a^x\!\tn{d}x'\, f(x') = R \int \limits_a^b\!\tn{d}x\, f(x)
\end{equation}

\begin{equation}
  \Rightarrow x = F^{-1}(F(a) + R(F(b)-F(a)))
\end{equation}
This is a very elegant and efficient method with the main drawback that the
number of functions to which it can be applied is very limited.

\smallskip

The efficiency can also be improved by generating  more points in regions
where the function values are large (importance sampling). This is
mathematically equivalent to a change of integration variable:

\begin{equation}
  f(x)\tn{d}x = \frac{f(x)}{g(x)} \tn{d}G(x) \quad \tn{where} 
     \quad \tn{d}G(x) = g(x) \tn{d}x 
\end{equation}

In practice it works like this: First a suitable function $g(x)$ has to be
chosen. Suitable means that it satisfies (i) $g(x) \ge f(x)\,\forall x\in
[a,b]$, (ii) has an invertible primitve function or a $g$-distributed random
number generator is available and (iii) the ratio $f(x)/g(x)$ should be as
constant as possible. Then select a $x$ according to $g(x)$. Then a random
number $R$ is selected, the corresponding $x\in [a,b]$ is accepted if $R <
f(x)/g(x)$ and rejected otherwise.  The importance sampling will improve the
efficiency if $V(f/g)<V(f)$, but it suffers from the small number of available
functions $g$.

\medskip

A crucial point in all Monte Carlo calculations or simulations is the random
number generator, since reliable random numbers are needed. Truly random
numbers are unpredictable and therefore also unreproducible. They can be
obtained from physical processes such as radioactive decay or thermal noise in
an electronic circuit. However, these truly random numbers have some
disadvantages when it comes to their application. One point is that it is
difficult to construct a device that generates random numbers by observing
suitable processes being at the same time accurate and very fast. In fact
speed is of great importance, since extended simulations consume huge amounts
of random numbers. 

An alternative to using truly random numbers is to generate pseudo-random
numbers from a mathematical formula. These are not random in the sense that
they are reproducible, but are just as good for Monte Carlo calculations as
long as they are uncorrelated (which is of course the critical point). There
are also advantages in the sequence of numbers generated being always the same
for one particular starting point, e.g.\ debugging of a program using random
numbers becomes much easier. When generated on a computer the sequence of
numbers has a finite length and at some point there will be a number that has
already occured before. From now on the sequence will repeat itself and
therefore be useless. So the period is an important characteristic of a
(pseudo-) random number generator. The generator used in this study has a
period of over $10^{43}$.

The ideal algorithm for producing pseudo-random numbers does not exist, so
testing is an issue. But even with a lot of tests made it is never possible to
be sure that the numbers are suitable for a certain problem that is to be 
investigated using Monte Carlo methods. It should therefore be kept in mind
that the generator one is using is not perfect and that problems may occur. The
lesson to learn is that random number generators have to be handled with care.

\bigskip

Monte Carlo event generators are powerful tools for the simulation of
complicated processes like particle collisions, where one has to deal with many
degrees of freedom. Instead of trying to solve the whole problem in one step
(which is practically impossible), the process is divided into a sequence of
subprocesses each of which is of reasonable complexity. For high energy
physics, for instance, the first part consists of the particles before the
interaction, where they are treated in terms of their parton density functions.
The next step is the hard interaction that can be calculated using pQCD matrix
elements. Then parton showers \`a la DGLAP are added, after that the event is
hadronised and as a last step follows the decay of unstable particles. In each
part the dynamical variables are chosen from their probability distributions.
The outcome of this whole machinery is a list of all particles in the final
states with their masses and momenta, but also the complete history of the
event is provided. When generating many events the cross sections can be
estimated at the same time. Ideally the mean and variance of a sample of
generated events should be the same as for the real physical process.

For this study the \textsc{Pythia} 6.2 package \cite{pythia} was used, but
with an additional phase, representing the interaction of the scattered
partons with the plasma, added between the generation of parton showers and
hadronisation.

\chapter{Soft Colour Interactions}
\label{chap_sci}
As was already discussed in Chapter~\ref{sec_qcdintro} only QCD processes
involving a large momentum transfer can be treated in pertubation theory,
since the coupling strength $\alpha_s$ increases with decreasing energy. This
is exactly the reason why it cannot be assumed that interactions below the
cutoff for perturbative calculations do not occur. They may be negligible in
many situations but there are observables related to the hadronic final state
that are sensitive to the soft interactions. 

\smallskip 
 
The Soft Colour Interaction (SCI) model \cite{sci} is an attempt to gain a 
better understanding of non-pQCD (non-perturbative QCD) processes. It
describes the interaction of  partons originating from a hard scattering in
$ep$ or $p\bar p$ collisions  with the colour background field inside the
hadron. The basic assumption is that the change in the dynamics caused by the
soft interactions is negligible. The important point is that the partons can
exchange colour leading to a  different string topology, which may then result
in a different final state after hadronisation of the string. 

The background field is represented by the remnants of the hadron(s), i.e.\ 
those consituents that did not take part in the hard scattering
(Ch.~\ref{chap_qcd}). In the case where a valence quark was kicked out the
remainder is a diquark in a colour antitriplet state. If the scattered parton
is a gluon the remnant (namely the three valence quarks in a colour octet
state) is split into a quark and a diquark. If a seaquark was scattered its
partner has to be taken into account to conserve quantum numbers. The remnant
then consists of the three valence quarks (again split into a quark and a
diquark) and the seaquark partner. Together they form a (anti)triplet,
depending on whether the seaquark or the sea-antiquark is left. 

The interactions consist of an exchange of a colour octet (consisting of 
colour and anticolour) between a pair of partons, i.e.\ this can be viewed as 
a soft gluon exchange. The model is implemented as add-on routines to 
\textsc{Pythia} at a stage between the hard scattering process including 
parton showers and the hadronization. Each perturbativly produced parton may 
interact with each of the remnant consituents with a probability $P_\tn{int}$
which  cannot be calculated from theory and is the only free parameter of the
model. The outcome of this procedure will vary from event to event. 

\bildzweiA{0.75}{0.75}{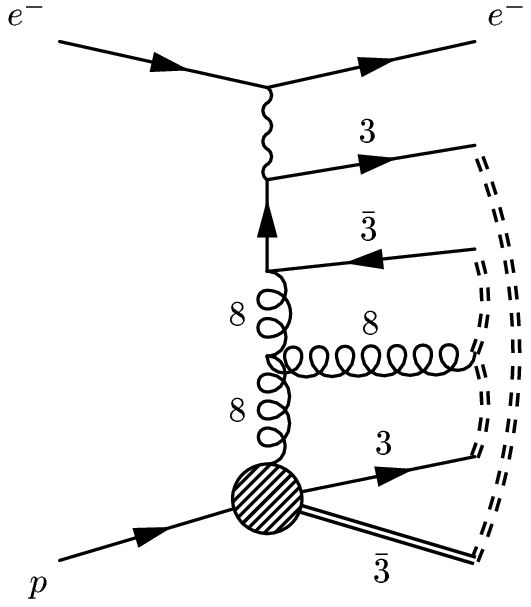}{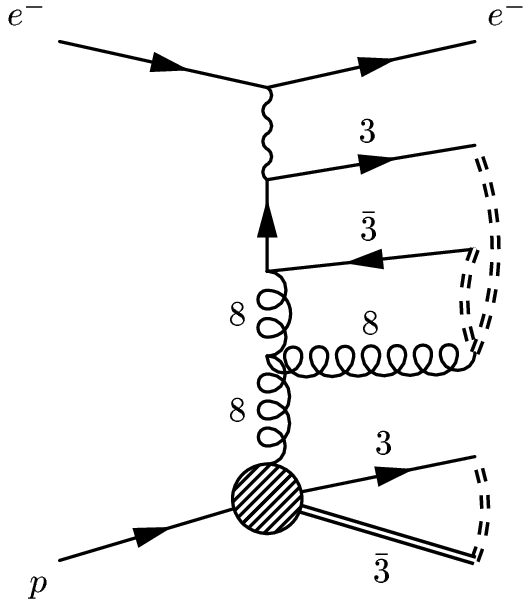}{fig_sci}{Photon-gluon 
 fusion in deep inelastic scattering with the strings drawn as double dashed 
 lines; left: configuration in the Lund model, right: configuration after the 
 exchange of a soft gluon (colour octet) between the quark from the $2\to 2$ 
 scattering and the quark in the remnant \cite{sci}.} 
 
One example for a deep inelastic scattering event is shown in
Figure~\ref{fig_sci}. In the conventional Lund model the remnant is in a colour
octet state with the colour and anticolour charges matching those of the gluon
so that gluon and remnant together form a singlet. In the case shown in
Figure~\ref{fig_sci} the radiated gluon carries the anticolour charge of the
original gluon (which belongs to the charge of the quark in the remnant) and
its colour is the anticolour of the original gluon after the radiation. When
this gluon splits into a $q\bar q$ pair the quark gets the colour and the
antiquark the anticolour charge resulting in the string configuration shown in
Figure~\ref{fig_sci}. The exchange of a soft gluon between the two quarks in the
diagram leads to an interchange of the two colour charges resulting in a
different string topology (right-hand side of Figure~\ref{fig_sci}). Now the
remnant system is in a singlet state. If instead the antiquark had interacted
with the diquark, the scattered quark-antiquark pair would form a colour
singlet system and the remnant, together with the radiated gluon, would form the 
other singlet. 

\smallskip 
 
The Generalized Area Law (GAL) model \cite{gal} is formulated in a string basis
instead of  a parton basis, since the soft gluons may not resolve single
partons. Here the  interactions are between strings and constructed such that
the area swept out  by the strings in energy-momentum space is minimised. The
basic idea,  however, is the same as in the SCI model. 

\smallskip 
 
The two models give very similar results, although there are differences when 
attention is paid to the details. In this study only the SCI model was used.
The parameter $P_\tn{int}$ was chosen to reproduce the rate of diffractive
rapidity gap events observed at HERA (a rapidity gap is an interval in rapidity
or polar angle where no particles are produced) \cite{sci}. Its value is
$P_\tn{int}=0.5$, but the observables studied so far do not depend strongly on
the exact value. Once $P_\tn{int}$ is fixed the models describe successfully
not only the diffractive data from HERA, but also production of $W$, jets and
beauty in diffractive events and double leading proton events at the Tevatron
\cite{sci}. Furhermore it provides a very general description in the sense that
it does not introduce substantial changes in non-diffractive events (events
without rapidity gaps). 

\smallskip

The SCI model has recently received a firmer theoretical basis in terms of
parton rescattering in QCD \cite{sci-theory}.

\chapter{The Soft Colour Interaction Jet Quenching Model}
\label{chap_scijetmodel}
\section{Description of the Model} %=============================================

The model is built on the same concept of non-perturbative QCD-interactions 
between (colour-) charged objects as the SCI model (Ch.~\ref{chap_sci}). But 
now a scattered parton has to traverse a part of the QGP instead of only the 
proton remnant and will thus experience much more interactions. It is therefore 
necessary to consider also the momentum-transfer since even a small energy-loss 
per interaction may sum up to give a sizeable amount in the end. 
 
\medskip 

The simulation is based on proton-proton collisions in \textsc{Pythia}, where 
the additional interactions with the plasma are added after the perturbative 
part but before hadronisation. This is justified since at $\sqrt{s} =
\unit[200]{A\,GeV}$ the probability to have more than one hard scattering per
ion-ion collision is very small. All soft processes like hadronisation of the
plasma, collective flow etc. are not included in the model which is only
concerned with phenomena that are connected to high-$\pt$ partons. The only
non-perturbative process in the model is the interaction of these partons with
the QGP. In the hard scattering a transverse momentum transfer of at least
\unit[3]{GeV} is required. It is necessary to have a $\pt$-cut since the matrix
element is applicable only in the perturbative regime and diverges at $\pt \to
0$.  Here a relatively high value is used because non-perturbative interactions
influence the particle production up to $\sim \unit[2]{GeV}$. Furthermore the
value of \unit[3]{GeV} allows to accumulate a good statistics in the high-$\pt$
regime up to $\sim \unit[10]{GeV}$ in reasonable run times.

\subsection{Geometry} %---------------------------------------------------------

The coordinate system is chosen such that the origin is in the centre of the
overlap region, the $z$ direction corresponds to the beam direction and the $x$
direction is defined by the impact parameter (Fig.~\ref{fig_geo}).

\bildtex{plots/geometry.pstex_t}{fig_geo}{Definition of the coordinate system}
The diameters of the overlap region in $x$ and $y$ direction are

\begin{eqnarray}
 d_x &=& 2R_\tn{Au} - b\\
 d_y &=& \sqrt{4 R_\tn{Au}^2 - b^2}
\end{eqnarray} 

The geometrical effects are modeled with the help of a simple Glauber model
(Sec.~\ref{sec_glauber}, \cite{eskola}) with a sharp sphere potential. The
centrality (defined as fraction of the total cross section) is proportional to
$b^2$ since $\tn{d}\sigma \propto b\,\tn{d}b$ for geometric reasons. From a fit
to a Glauber model calculation \cite{eskola}

\begin{equation}
  b = \left[ \unit[(1.579\pm 0.003)]{fm^2}\cdot \tn{centrality}\,[\%]
  \right]^{1/2}
\end{equation}   
is obtained.  In the model a centrality range has to be given which is then
translated into a range of the impact parameter and sampled according to
$\tn{d}\sigma \propto b\tn{d}b$.

For the Monte Carlo simualtion it is necessary to have a simple parametrisation
of the nuclear thickness function since it cannot be afforded to frequently
perform time-consuming calculations. The parametrisation used here is
(Fig.~\ref{fig_glaub1})

\begin{equation}
 T_\tn{Au}(r) = \left\{ \begin{array}{lcl} a_0 + a_2 r^2 + a_4 r^4 &;&
 								 r \le R_\tn{Au} \\
							0 &;& \tn{otherwise} \\
							\end{array} \right.
\label{eq_param}
\end{equation}								 
The radius of the nucleus $R_\tn{Au} \simeq \unit[6.86]{fm}$ is taken as the
zero of the polynomial. The parameters are

\begin{eqnarray*}
  a_0 &=& \unit[(2.193\pm 0.13)]{fm^{-2}} \\ 
  a_2 &=& \unit[(-0.0157\pm 0.0018)]{fm^{-4}} \\ 
  a_4 &=& \unit[(-0.000654\pm 0.000046)]{fm^{-6}}  
\end{eqnarray*}

\bildtex{plots/glauber1.pstex_t}{fig_glaub1}{Glauber calculation \cite{eskola}
and parametrisation \ref{eq_param} for the thickness function of a gold
nucleus}

\medskip

The hard scattering events have to be distributed over the overlap region.
Since the two nuclei are strongly Lorentz contracted their longitudinal
extension is neglected so that all hard scattering points lie in the $z=0$
plane. The instant of overlap between the two nuclei also defines the time
origin, i.e.\ all hard scatterings occur at $t=0$. It is reasonable to assume
that in the transverse plane the probability $u(x,y)$ for hard interactions is
largest in the region where the number of binary nucleon-nucleon collisions is
highest. This implies that $u(x,y)$ is coupled to the nuclear thickness
function:

\begin{equation}
  u(x,y) \propto T_\tn{Au}(x-b/2,y) \cdot T_\tn{Au}(x+b/2,y)
\end{equation}
Consequently the hard scatterings take place preferentially near the centre of
the overlap region.

There is a possibility to define a transverse area for the QGP which is smaller
than the overlap region. Consequently not all hard scatterings take place
inside the QGP volume and there is a finite propbability for these partons to
escape undisturbed.

\subsection{Cronin Effect} %----------------------------------------------------
\label{sec_des-cronin}

The Cronin effect is taken into account by a broadening of the intrinsic
$\kt$-distribution of the partons (Ch.~\ref{sec_cronin}). In this model the
protons experience rescattering before the hard scattering. The additional
transverse momentum built up by the protons translates into a bigger intrinsic
transverse momentum of the partons:

\begin{equation} 
 \sigma_{\kt}^2(x,y,b) = \sigma_{\kt,0}^2 + \alpha\cdot (N_\tn{coll}(x,y,b) -1) 
\end{equation}
where $\sigma_{\kt,0} = \unit[1]{GeV}$ is the width of the $\kt$-distribution 
used in proton-proton collisions and $\alpha$ is a constant. Here $\sigma_{\kt}$ 
does not depend on $Q$, since parton showers are done explicitly in the 
simulation. This includes the evolution down to a fixed scale $Q_0$ and the 
intrinsic $\kt$ only has to cover the range below that. 
 
\smallskip 
 
The number of collisions that a proton experiences before the hard scattering is
estimated again with the help of the Glauber model. It depends on the point in
the transverse plain where the proton is located (and where the hard scattering
will happen) and the impact parameter. The maximum number of these
rescatterings is given by 
\begin{equation}
  N_\tn{coll}^\tn{max}(x,y,b) = \sigma_\tn{inel}^{pp} T_\tn{Au}(x\pm b/2,y)
\end{equation}
with $\sigma_\tn{inel}^{pp}\approx \unit[42]{mb}$ at $\sqrt{s}=\unit[200]{GeV}$. The
sign in the argument of $T_\tn{Au}$ depends on the membership of the nucleon
under consideration in one or the other nucleus. 
 
For simplicity it is assumed that the probability for a hard interaction in a
binary nucleon-nucleon collision is $P_\tn{hs}$ and that there is no
possibility that the pair doesn't interact at all. Then the probability for the
nucleon to make a hard interaction with the $n$th nucleon from the other
nucleus is
\begin{equation}
  P_n = P_\tn{hs}(1 - P_\tn{hs})^{n-1}
\end{equation}   
This is used in the model to choose for each nucleon the actual number of
rescatterings. Then the mean of the two values is taken because in
\textsc{Pythia} the $\kt$-distribution is always the same for both participants.
For $P_\tn{hs}$ the ratio of the cross section for hard scattering to the total
cross section is used so that $P_\tn{hs}\approx 0.04$.

\smallskip 
   
The value of $\alpha$ can be determined from the dAu data. In that case only
the nucleon from the  deuteron has undergone rescattering and the one from the
gold-nucleus is  unchanged. This situation is equivalent to a symmetric
increase with $\alpha/2$  instead of $\alpha$ on both sides. For the comparison
to data standard proton-proton collisions were simulated with the
$\kt$-broadening as the only change. 

\subsection{Model of the QGP: Energy Density Distribution, EOS and Evolution} %-

Since it is not a priori clear what the energy density distribution should be, 
there are two options available in the model: One is a constant, i.e.\ the
energy density depends only on the proper time $\tau$, and in the other
$\epsilon$ is a function of the transverse coordinates $x$ and $y$ and $\tau$.
In the latter case it is assumed that the energy density is proportional
to the number of binary collisions, i.e.
\begin{equation}
  \epsilon(x,y,b,\tau=\unit[1]{fm/c}) = \mathcal{N} \cdot T_\tn{Au}(x-b/2,y)
  								\cdot T_\tn{Au}(x+b/2,y)
\end{equation}
It is plausible that the energy deposition and thus the resulting energy density
is larger in areas where more particles collide.
The normalisation is chosen such that the mean energy density for $b=0$ and at
$\tau=\unit[1]{fm/c}$ is equal to the initial energy density.  
\begin{equation}
  \mathcal{N} = \epsilon_0 \left( \frac{2\pi}{\pi R_\tn{Au}^2} 
    \int\limits_0^{R_\tn{Au}}\!r\,\tn{d}r\, T_\tn{Au}^2(r) \right)^{-1}
     = \frac{\epsilon_0}{\unit[2.21]{fm^{-4}}}
\end{equation}
As a consequence the energy density is higher in the centre than near the
periphery (Fig.~\ref{fig_endis}) and on average decreases with increasing $b$.
In this scenario it is natural to have the contour, where the energy density
falls below the critical value, as boundary of the QGP and not the overlap
region or some other artificial choice.

\bildtex{plots/energy.pstex_t}{fig_endis}{Energy density distribution in the
transverse plain at $\tau = \unit[1]{fm/c}$ for $b = \unit[4]{fm}$}

\medskip

The equation of state of an ideal ultra-relativistic gluon gas is used
(Sec.~\ref{sec_idgas}). The contribution from quarks is neglected, because the
plasma is gluon-rich in the beginning due to the large gluon content in the
proton at the typical $x$ of interest at RHIC. Besides the gluons come into
equilibrium very fast, because the gluon cross sections are large. Much more
time is needed for the quarks. The crucial quantity is the gluon number
density which is given by

\begin{equation}
  n = \frac{g}{\pi^2} T^3\,\zeta(3) = \frac{16\cdot 1.2}{\pi^2} T^3
\end{equation}
Another useful relation is the one between the temperature and the energy
density (Eq.~\ref{eq_eps-T}):

\begin{equation}
  T = \left( \frac{30\,\epsilon}{16\,\pi^2}\right)^{1/4}
\end{equation}

\medskip

For the evolution of the QGP a Bjorken-like scenario with only longitudinal
expansion (Sec.~\ref{sec_bjorken}) is used. With the initial condition
$\epsilon(\tau_0) = \epsilon_0$ the energy density follows the evolution
(Eq.~\ref{eq_eps-time})

\begin{equation}
  \epsilon(\tau) = \epsilon_0 \left(\frac{\tau}{\tau_0}\right)^{-4/3}
\end{equation}  
while the temperature changes according to

\begin{equation}
  T(\tau) = \left( \frac{30\,\epsilon_0}{16\,\pi^2}\right)^{1/4} 
  				\cdot \left( \frac{\tau}{\tau_0}\right)^{-1/3}
\end{equation}
This relation can be used to specify the lifetime of the QGP, defined as
the point when the temperature drops below the critical temperature $T_c$.

\begin{equation}
  \frac{\tau_f}{\tau_0} =\left( \frac{30\,\epsilon_0}{16\,\pi^2}\right)^{3/4}
  		\cdot T_c^{-3}
\end{equation}

The time evolution of the gluon density is correspondingly given by

\begin{equation}
  n(\tau) = \frac{16\cdot 1.2}{\pi^2} \cdot \left( \frac{30\,\epsilon_0}{16\,
		  \pi^2}\right)^{3/4} \cdot \left( \frac{\tau}{\tau_0}\right)^{-1}
\end{equation}

\smallskip

The plasma gluons have a mean thermal energy of $2.7\,T$ which is in practice
smeared out in the form of a Gaussian distribution with width $\sigma_g$. This
motion is undirected. In addition there is the directed expansion motion in
$\pm z$ direction with velocity $z/t$. Furthermore the gluons have a small
effective mass $m_g$ to account for interactions.

\bigskip

The rapidity density of QGP gluons can be estimated easily if the thermal
motion which is undirected and will cancel on average is neglected. Then there
is a simple relation between the rapidity $y$ and $x_3$ (here the spacial
coordinates are denoted with $x_1$, $x_2$ and $x_3$ to avoid name conflicts
with the rapidity):

\begin{equation}
  y = \frac{1}{2}\ln \left( \frac{p^0+p_\parallel}{p^0-p_\parallel} \right)
    = \frac{1}{2}\ln \left( \frac{p^0+p^0 x_3/t}{p^0-p^0 x_3/t} \right)
    = \frac{1}{2}\ln \left( \frac{1+x_3/\sqrt{\tau^2+x_3^2}}{1-x_3/\sqrt{\tau^2+x_3^2}}
    \right)
\end{equation}
which can be solved for $x_3$ leading to

\begin{equation}
 x_3 = \tau \sinh(y)
\end{equation}    
The rapidity density of plasma gluons at mid-rapidity is found to be 

\begin{eqnarray}
 \left. \der[N_g]{y}\right|_{y=0} 
 &=& \int\!\!\int\!\tn{d}x_1\tn{d}x_2 \, \frac{\tn{d}^3N}{\tn{d}x_1 \tn{d}x_2 \tn{d}x_3} \left. 
   \der[x_3]{y}\right|_{y=0}  \nonumber \\
 &=& \pi R_\tn{Au}^2 n(\tau) \tau \nonumber \\
 &\simeq& 993 
\end{eqnarray}
where it was assumed that the energy density is constant over the transverse
plane.  
This is consistent with the numbers derived in \cite{vitev}.

\subsection{Energy Loss} %------------------------------------------------------
 
The soft interactions (SI) of a traversing energetic parton with the QGP are 
treated as elastic scatterings from the plasma gluons. These are
non-perturbative interactions and their characteristics can therefore not be
derived from theory. So elastic scattering is the natural ansatz for such
interactions between pointlike partons. It is assumed that the individual
scatterings are independent. The parton interacts with each gluon on its way
out with a probability $P_\tn{int}$. $P_\tn{int}$ can be inherited from the SCI
model where it was found to be  roughly 0.5 (Ch.~\ref{chap_sci}). 

A suitable variable to describe the scatterings is the Mandelstam variable $t$,
taken to be Gaussian distributed since the matrix elements calculated in the
framework of pertubation theory are not applicable in this regime. The width
$\sigma_t$ and mean $\mu_t$ can not be larger than a few hundred \unit{MeV} for
the process to be non-perturbative. The $t$-range is limited by the available
energy: 

\begin{equation} 
t_1 = 4(m_p^2 - E_p^2) \le t \le 0 = t_2 
\label{eq_trange}
\end{equation}
where $m_p$ and $E_p$ are the parton's mass and energy in the c.m. system. The 
scattering angle in the c.m. system is given by 
 
\begin{equation} 
 \cos\vartheta_{cm} = \frac{t}{2(E_p^2 - m_p^2)} + 1 
\end{equation} 
 
In the lab system the parton is typically much more energetic than the plasma
gluons so that it loses energy due to the boost and suffers only a small
deflection. 

Gluons (as parton) should interact more strongly than quarks since they carry 
colour octet charge. In the model this is taken into account by giving the
gluon a second chance if it didn't interact with a certain background gluon in
the first try. Two interactions with the same gluon are excluded so that the
effective interaction probability for a gluon becomes $P_\tn{int}^g=0.75$. 
 
\smallskip 
 
The parton is sensitive to a cylindrical volume with radius $R_\tn{cyl}$ around
its track, i.e.\ it has the possibility to interact with all gluons within this
volume. The total number of encountered gluons is given by

\begin{equation}
  N_g = \pi R_\tn{cyl}^2 \int \limits_{\tau_\tn{in}}^{\tau_\tn{out}}\!
    \tn{d}\tau\, n(\tau)\, = 
    \pi R_\tn{cyl}^2\,\frac{16\cdot 1.2}{\pi^2}\, 
    \left( \frac{30\,\epsilon_0}{16\,\pi^2}\right)^{3/4} \frac{1}{\tau_0} 
    \ln \frac{\tau_\tn{out}}{\tau_\tn{in}}
\end{equation}		  
where $\tau_\tn{in}$ and $\tau_\tn{out}$ are the proper times when the parton
enters and leaves the QGP. In a scenario where the QGP fills nearly the entire 
overlap region $\tau_\tn{in}$ is simply the QGP formation time $\tau_i$,
otherwise the geometry has to be considered as well. Similarly $\tau_\tn{out}$
is the point when the parton geometrically leaves the QGP but it is limited by
the lifetime of the plasma:

\begin{eqnarray}
   \tau_\tn{in} \!\!\!&\ge&\!\!\! \tau_i \\
   \tau_\tn{out} \!\!\!&\le&\!\!\! \tau_f 
\end{eqnarray}

\smallskip

The interactions with the QGP are simulated explicitely in an iterative way:
Starting from the entrance point the position of the next QGP constituent is
estimated and from a Gaussian around that point a choice is made for the
actual position. Then the gluon is generated with appropriate thermal energy
and expansion velocity and the soft scattering is simulated. The procedure is
repeated going out from the interaction point until $\tau_\tn{out}$ is
overstepped or -- if the energy density distribution is inhomogeneous -- the
local energy density falls below the critical value.

\smallskip 
 
It is not clear, how the proton remnants should be treated. Luckily, the 
observables studied here are not sensitive to the treatment of the remnants,
since the analysis is performed only in the central rapidity region and the
remnants do not contribute to the high-$\pt$ particle yields at mid-rapidity.
Therefore, they are neglected and do not interact with the plasma at all. 

\subsection{Hadronisation} %----------------------------------------------------
 
The interactions with the background also lead to changes in the string 
topology. The standard Lund string topology in shown in Figure~\ref{fig_lund2}.
The interactions with the QGP change the strings and the partons presumably get
connected to the plasma instead. It is also important to note that gluons
radiated from a scattered parton in final state parton showers do not any more
belong to the same string as the original parton.

However, it is not clear how this system hadronises and what happens if the
formation time of \unit[1]{fm/c} in the particle's rest frame lies within the
QGP. The matter clearly calls for detailed investigations. For the study
presented here a simple interim solution is employed in the SCI jet quenching
model: The partons are hadronised with independent fragmentation
(Sec.~\ref{sec_indep}). This is sensible insofar as the parton will in any case
have lost the connection to the proton remnant in the interactions with the
background. The non-conservation of energy-momentum is not a problem in this
context because the QGP acts as a kind of 'energy reservoir'.

The pp reference and dAu runs are hadronised with Lund string fragmentation
since there are no soft colour interactions in these cases. Strictly speaking
parton systems that are created outside the QGP and do not interact at all
would also have to be hadronised with the Lund model, however, this is not done
for practical reasons. The effect would be very small anyway in central and
mid-central collisions, because the transverse overlap area outside the QGP is
small compared to the QGP area and scarcely populated due to the concentration
of hard scatterings towards the centre. The situation might be different in
peripheral collisions, depending on the size of the possible hadronic layer
around the plasma.

\subsection{Model Parameters} %-------------------------------------------------

The parameters of the model are given in Table~\ref{tab_param} together with
their default values. The size of a hadronic layer can only be given when the
energy density distribution is flat, otherwise the size of the QGP is defined
by the contour of the critical energy density. The values of parameters like
the QGP formation time have been constrained in independent investigations so
that they are not free, but can be varied in a certain range only.

The parameter for the Cronin effect is fixed with the help of dAu data. This is
independent of the rest of the model, since in dAu collisions no QGP is formed 
and the part that describes the plasma and the interactions with it is
irrelevant. The only difference to ordinary pp collisions is the
$\kt$-broadening.

The interaction probability is taken from the original SCI model and is not
available as a free parameter in the jet quenching extension.

The width of the $t$-distribution is taken to be constant but scaled down for 
parton-gluon pairs with small invariant mass. This is done to prevent the 
routine from becoming very inefficient when a lot of values have to be 
rejected because they lie outside the allowed range (see Eq.~\ref{eq_trange}). 

\begin{tabelle}{|p{1.2cm}|p{4cm}|l|l|l|}{tab_param}{Parameters of the SCI jet
quenching model, their default values and constraints}
\hline
& \multicolumn{2}{c|}{Parameter} & default value & sensible range \\ \hline
QGP & size of hadronic layer (only with homogeneous energy density) & $d$ 
			& \unit[0]{fm} & \unit[0\dots3]{fm} (?) \\
& QGP formation time & $\tau_i$ & \unit[0.2]{fm/c} & \unit[0.2\dots 1]{fm/c}
							\cite{huovinen} \\
& initial energy density  ($\tau_0 = \unit[1]{fm/c}$) & $\epsilon_0$ &
		$\unit[5.5]{GeV\,fm^{-3}}$ & 	\unit[4\dots 6.5]{GeV\,fm$^{-3}$}
		\cite{huovinen} \\
& critical temperature & $T_c$ & \unit[0.175]{GeV} & \unit[0.17\dots
		0.18]{GeV} \\		
& width of gluon energy distr. & $\sigma_g$ & \unit[0.1]{GeV} & ? \\
& gluon mass & $m_g$ & \unit[0.2]{GeV} & ? \\ \hline
SI & interaction probability & $P_\tn{int}$ & 0.5 & $\sim$ 0.5 \cite{sci} \\
& radius of parton cylinder & $R_\tn{cyl}$ & \unit[0.3]{fm} & $\sim$
			\unit[0.3]{fm} \\ 
& width of $t$ distr. & $\sigma_t$ & $\min(\unit[0.5]{GeV^2},|t_1|)$ 
			& \unit[0.1\dots 0.6]{GeV$^2$} (?) \\ \hline
Cronin effect& increase of $\sigma_{\kt}^2$ per scattering & $\alpha$ & 
								$\unit[0.25]{GeV^2}$ & ? \\
\hline
\end{tabelle}

\section{Results} %=============================================================

The event analysis concentrates on the following parts:

\begin{itemize}
\item Ratio $R_\tn{AuAu}$ between the $\pt$-spectra for charged particles and
 neutral pions with $|\eta| < 0.35$ and a pp reference 
\item Centrality dependence of $R_\tn{AuAu}$
\item Azimuthal correlation of charged particles with $|\eta| < 0.7$:
 Particles  with $4 < \pt < \unit[6]{GeV}$ are defined as trigger particles,
 for each  trigger particle the azimuthal separation $\df$ from all other
 particles with  $2 < \pt < \pt(\tn{trig})$ in the event is calculated. In the
 analysis considered here several trigger particles in one event are also
 possible, but this happens only in $\sim $2\% of all events containing trigger
 particles.  
\item Change of the azimuthal correlation with respect to the reaction plane
 orientation: trigger particles with $\phi^\tn{trig} < \pi/4$ and
 $\phi^\tn{trig} > 3\pi/4$ are defined to be in-plane and those with $\pi/4 <
 \phi^\tn{trig} < 3\pi/4$ are out-of-plane
\end{itemize} 
The pp reference was generated by running standard pp in the same event
generator. The Cronin effect is the only nuclear effect that is explicitly
taken into account. One difference between the measured and the model nuclear
modification factor is the effect of different $A$-scaling. In the data this
leads to a decrease at small $\pt$ even in the absence of all nuclear effects
which is not present in the model. However, the absence of this scaling-effect
in the simulation does not influence the results because the model is not
applicable to the low-$\pt$ regime anyway. 

The azimuthal correlation should be the same as the one measured by the STAR
collaboration after their subtraction of the elliptic flow and pedestal
contribution (collective flow is not included in the model). 

The particle production mechanism has not been modified which means that the 
relative yields of hadron species are unaltered. In particular there is no 
excess of (anti)protons over pions as observed in experiment
(Ch.~\ref{sec_RHIC}). The model results should thus in the first hand be
compared to the $\pi^0$ data when possible. In the model the effects are the
same for all particle species and the ratio is calculated for the sum of the
spectra for neutral pions and charged particles in order to get a better
statistics. 

\smallskip 
 
There are several reasons why the model cannot be considered trustworthy for 
$\pt \lesssim \unit[3]{GeV}$. On the parton level only hard processes with 
$\pt > \unit[3]{GeV}$ are generated, all softer (perturbative and 
non-pertubative) interactions as well as multiple interactions that affect the
low-$\pt$ region are not taken into account. In a strict sense the $\pt$-cut on
the matrix element influences the $\pt$-spectra of hadrons up to 
\unit[4-5]{GeV}: Because of the intrinsic $\kt$ too few particles are
produced in this regime. This effect will cancel only partly in the
$R_\tn{AuAu}$ since the effect of $\kt$-broadening is not present in the pp
reference.

The $\pt$-cut has practically no influence on the azimuthal correlation since a
trigger particle with $\pt > \unit[4]{GeV}$ is required. Then the hard
scattered parton also had a $\pt > \unit[4]{GeV}$ (larger than the hadron's
$\pt$) and thus stems from a region that is hardly affected by the $\pt$-cut.
In the framework of the SCI jet quenching model all final state particles
originate from the fragmentation of the high-$\pt$ parton, this process is
entirely independent from the hard scattering and cannot be affected by the
cut-off.

Another aspect that is not included in the model is the  hadronisation of the
plasma. This is also expected to contribute to the  low-energy particle
yields. 

\medskip

All results shown in this section were obtained with the default values of the
parameters given in Table~\ref{tab_param}. "SCI jet quenching model I (II)"
refers to the ansatz with constant (inhomogeneous) energy density distribution.

The statistical uncertainty on the model result for the nuclear modification
factor increases with $\pt$: It is $\lesssim 5$\%  below \unit[6]{GeV}, $ \sim
$10\% below \unit[8]{GeV} and $>20 $\% above  \unit[8]{GeV}. On the azimuthal
correlations it is smaller, namely of the order of a few percent.

\subsection{Cronin Effect} %----------------------------------------------------

The parameter $\alpha$ for the Cronin effect was fixed by a comparison to the
dAu  data. This was done by running the simulation with the $\kt$-broadening as
only modification to standard pp. In \textsc{Pythia} the $\kt$ is always the
same for both colliding partons, but in dAu only one nucleon (that from the
deuteron) has experienced rescattering. This situation is mathematically
equivalent to a symmetric increase of the variance with $\alpha/2$ instead of
$\alpha$ (the Gaussian distributions of the two protons are convoluted leading
to a Gaussian whose variance is given by the sum of the variances of the
initial distributions). The effect of the second nucleon in the deuteron has
been neglected. In fact is has been shown experimentally in \cite{phenix_dAu}
that $R_\tn{dAu}$ and $R_\tn{pAu}$ are very similar. 

\turnpic[0.4]{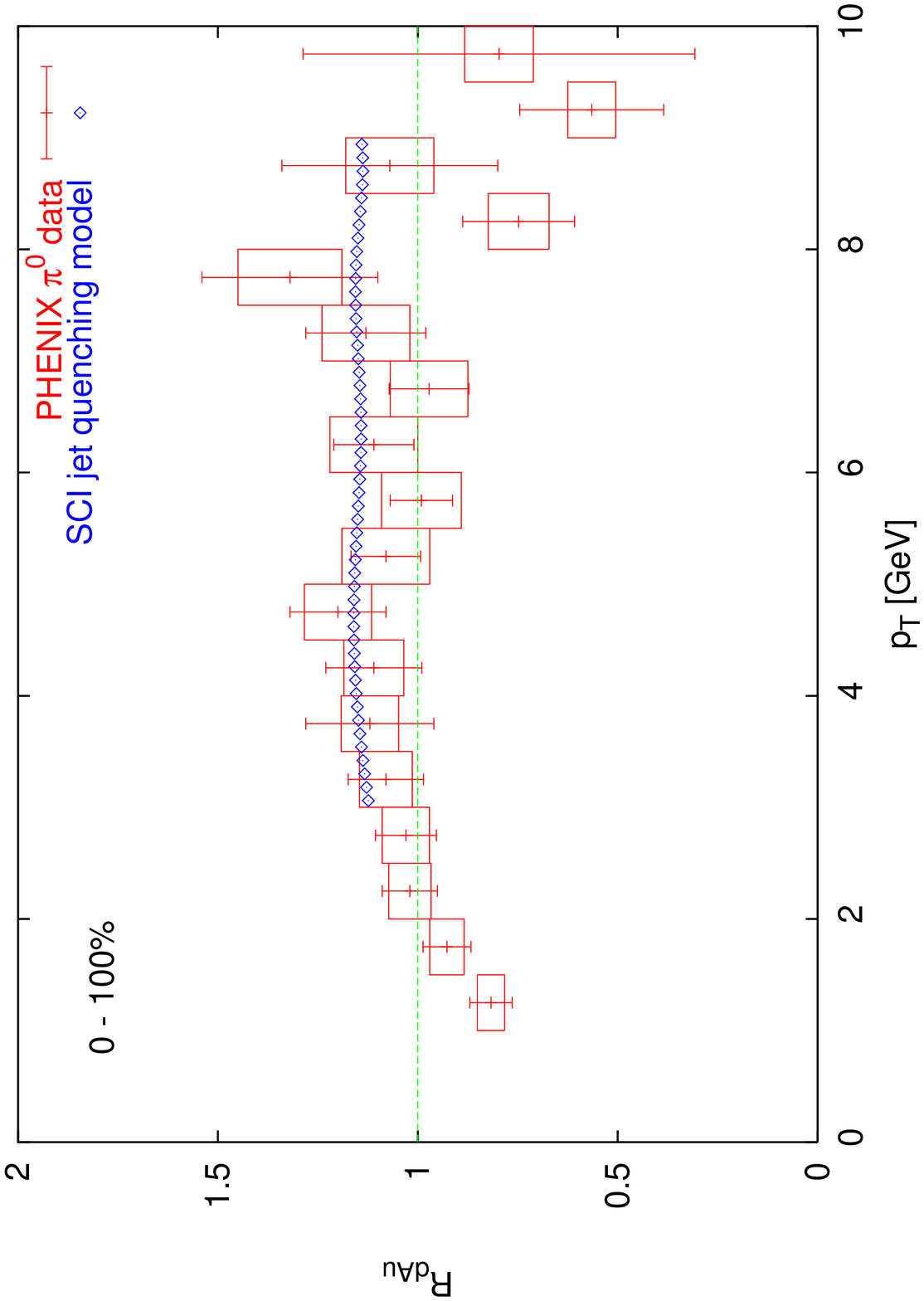}{-90}{fig_dcronin}{Model results for dAu (charged
 particles and neutral pions at $|\eta|<0.35$) and PHENIX $\pi^0$ data
 \cite{phenix_dAu} (errorbars indicate statistical and boxes systematic
 errors)}{}

The best results were obtained with $\alpha = \unit[0.25]{GeV^2}$ and are shown
in  Figure~\ref{fig_dcronin} together with the experimental data. The agreement
is reasonably good for $3 < \pt < \unit[9]{GeV}$, i.e.\  where the $\pt$ is
large enough for the model to be reliable and not too large so that the
statistical error is moderate. 

There is, however, a problem with the value of $\alpha = \unit[0.25]{GeV^2}$: 
One might want to interpret it as the square of the mean momentum transfer per
scattering because of the way the model is constructed (Ch.~\ref{sec_cronin}
and \ref{sec_des-cronin}). But the momentum transfer should  not exceed
\unit[0.2]{GeV} if the proton is supposed to interact as a whole. This number
is motivated by the uncertainty principle: The size of a proton is roughly
\unit[1]{fm} and a resolution of $\unit[1]{fm^{-1}}$ corresponds to a momentum
transfer of \unit[0.2]{GeV}. $\alpha$ being considerably larger than
$\unit[0.04]{GeV^2}$ can be seen as evidence for other effects that build up
additional transverse momentum. However, also a highly excited or dissociated proton
interacts so that a larger value of $\alpha$ may be reasonable. The model for
the Cronin effect used here is then an effective description that should not be
taken too literally. 

A variation of $\alpha$ by $\pm \unit[0.1]{GeV^2}$ leads to only small changes
with the result  still being consistent with the data. 

\medskip

Figure~\ref{fig_wcronin} shows that the broadening of the intrinsic $\kt$ also
has a small influence on the 2-particle azimuthal correlation. The fraction of
events containing trigger particles is slightly increased from $3.16\cdot
10^{-3}$ in pp to $3.75\cdot 10^{-3}$.

\turnpic[0.4]{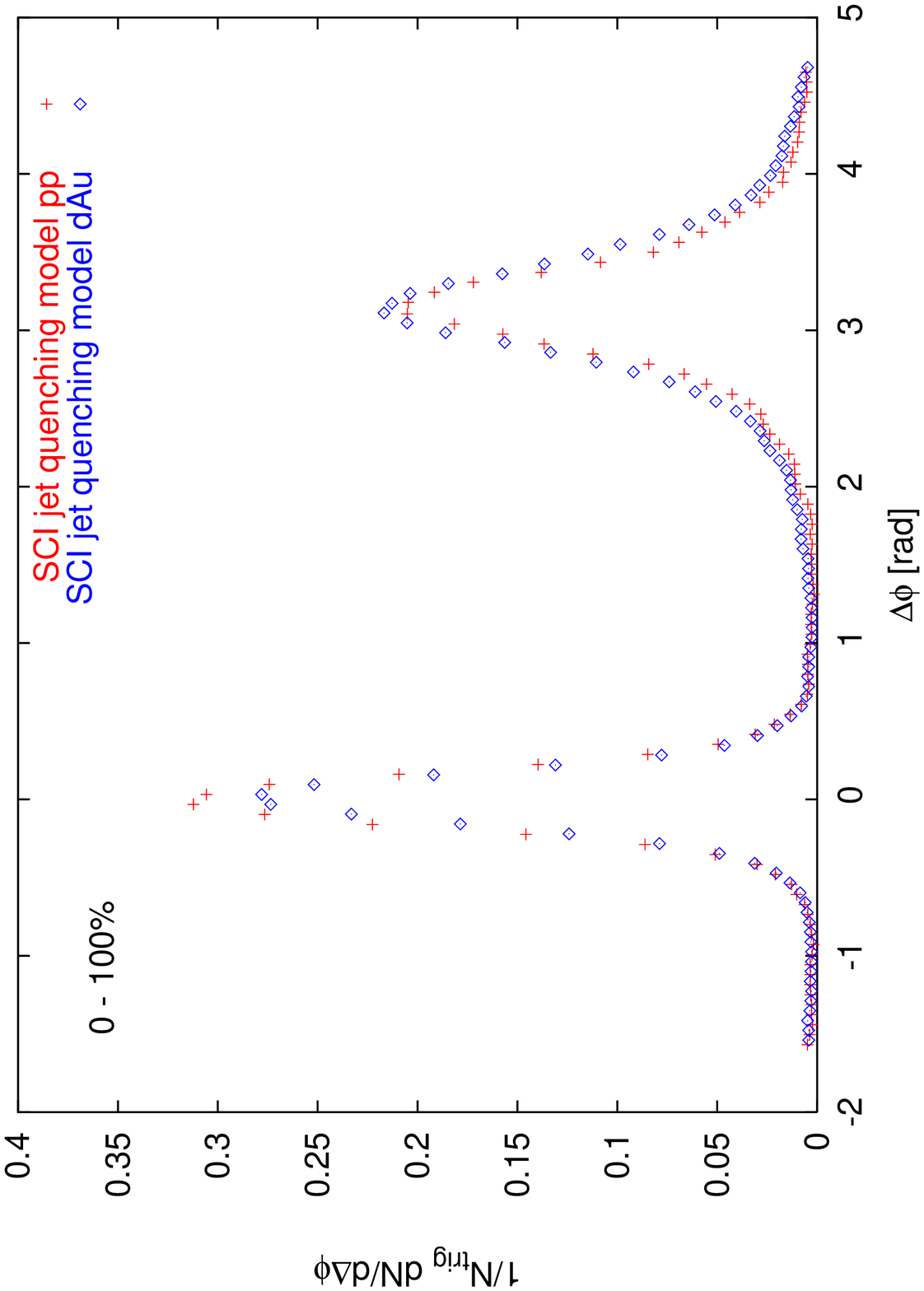}{-90}{fig_wcronin}{2-particle azimuthal
 correlation in the SCI jet quenching model for pp and dAu collisions}{}
 
\subsection{Nuclear Modification Factor} %--------------------------------------

\begin{figure}[ht]
 \centering
 \begin{turn}{-90}
  \includegraphics[scale=0.4]{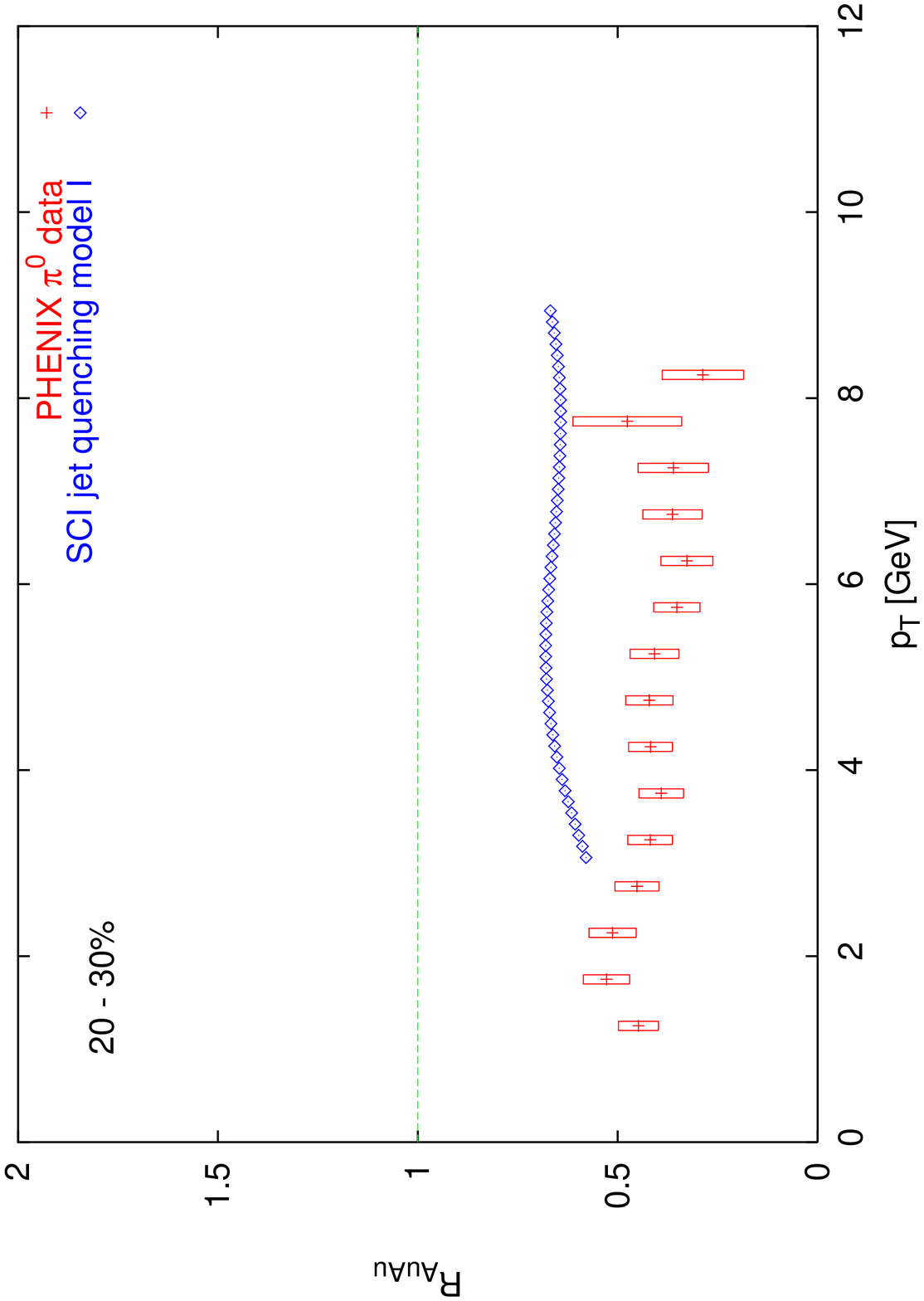} 
  \includegraphics[scale=0.4]{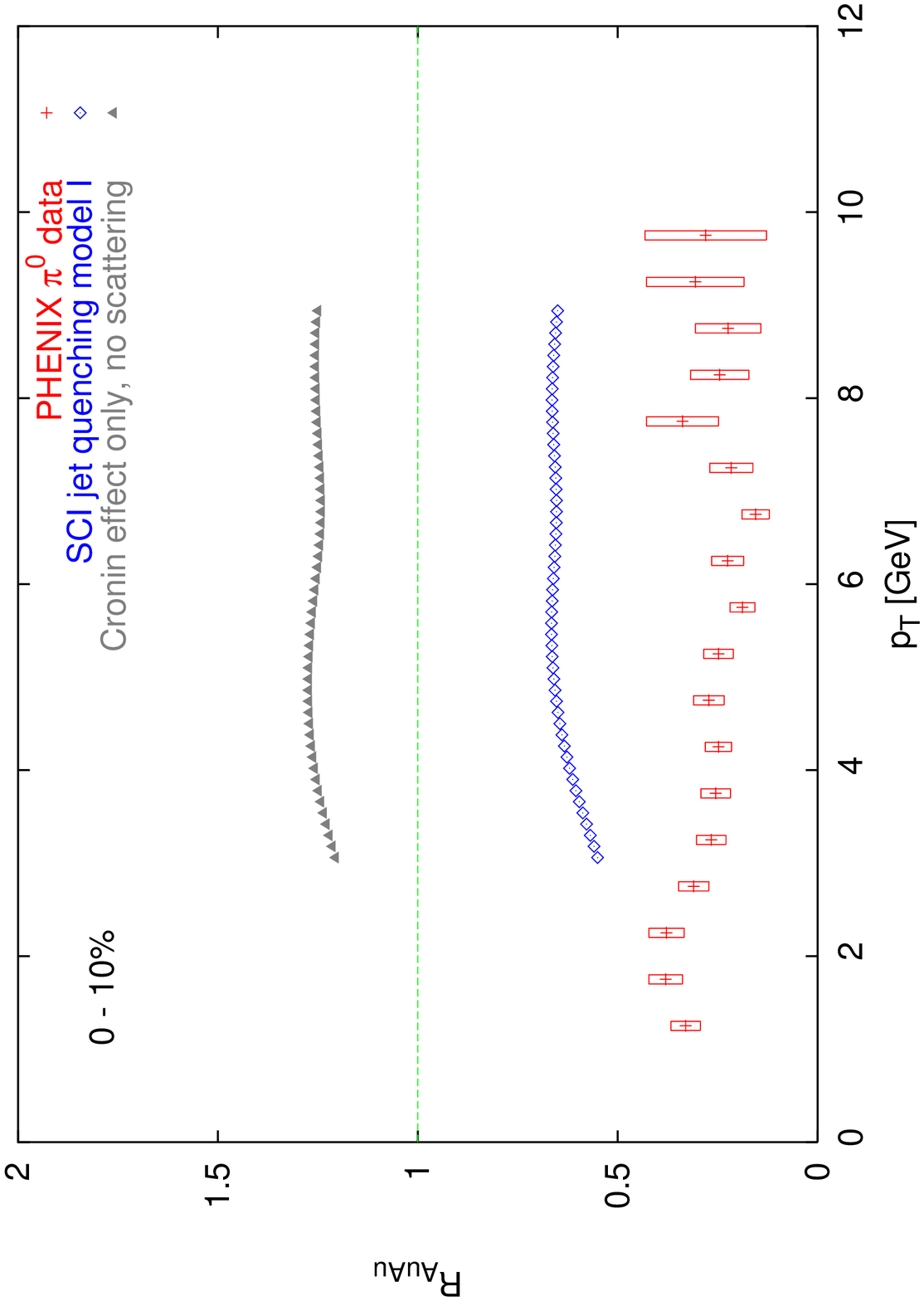} 
 \end{turn}
 \caption{Model results for the nuclear modification factor with constant energy
 density for different centrality classes and PHENIX data \cite{phenix_pi0}, for
 0-10\% and 80-92\% centrality also the effect of only $\kt$-broadening without
 partonic energy loss is shown}
 \label{fig_si41}
\end{figure}

The model results for the nuclear modification factor are shown in
Figure~\ref{fig_si41} together with the experimental data for five different
centrality classes. In the most central and most peripheral class the effect of
the $\kt$-broadening without scatterings is shown in addition. It naturally
increases with centrality because the nuclear overlap is much larger in central
collisions. Thus the Cronin effect counteracts the partonic energy loss which
is larger than it seems at first sight.

\begin{figure}[ht]
 \centering
 \begin{turn}{-90}
  \includegraphics[scale=0.4]{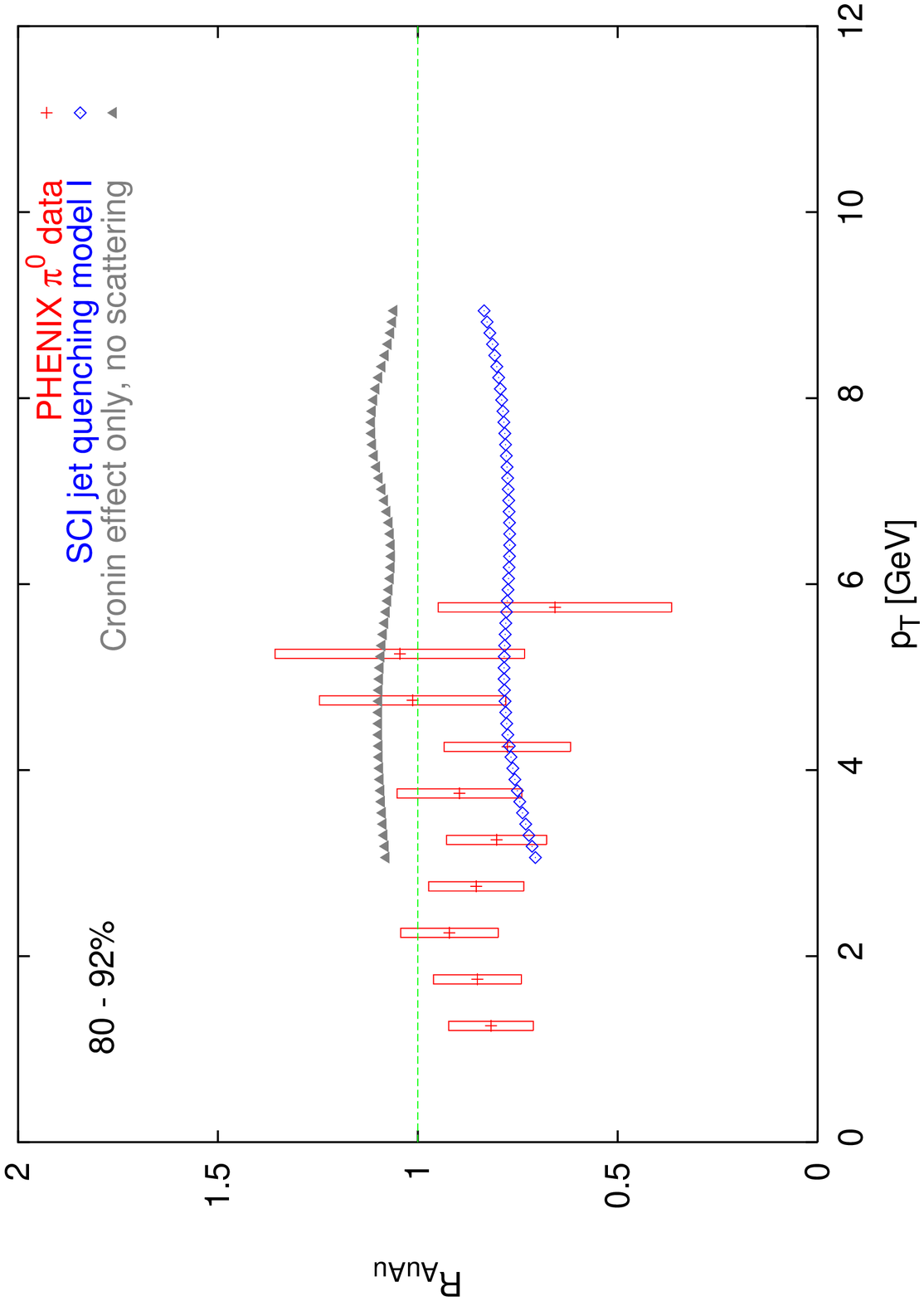} 
  \includegraphics[scale=0.4]{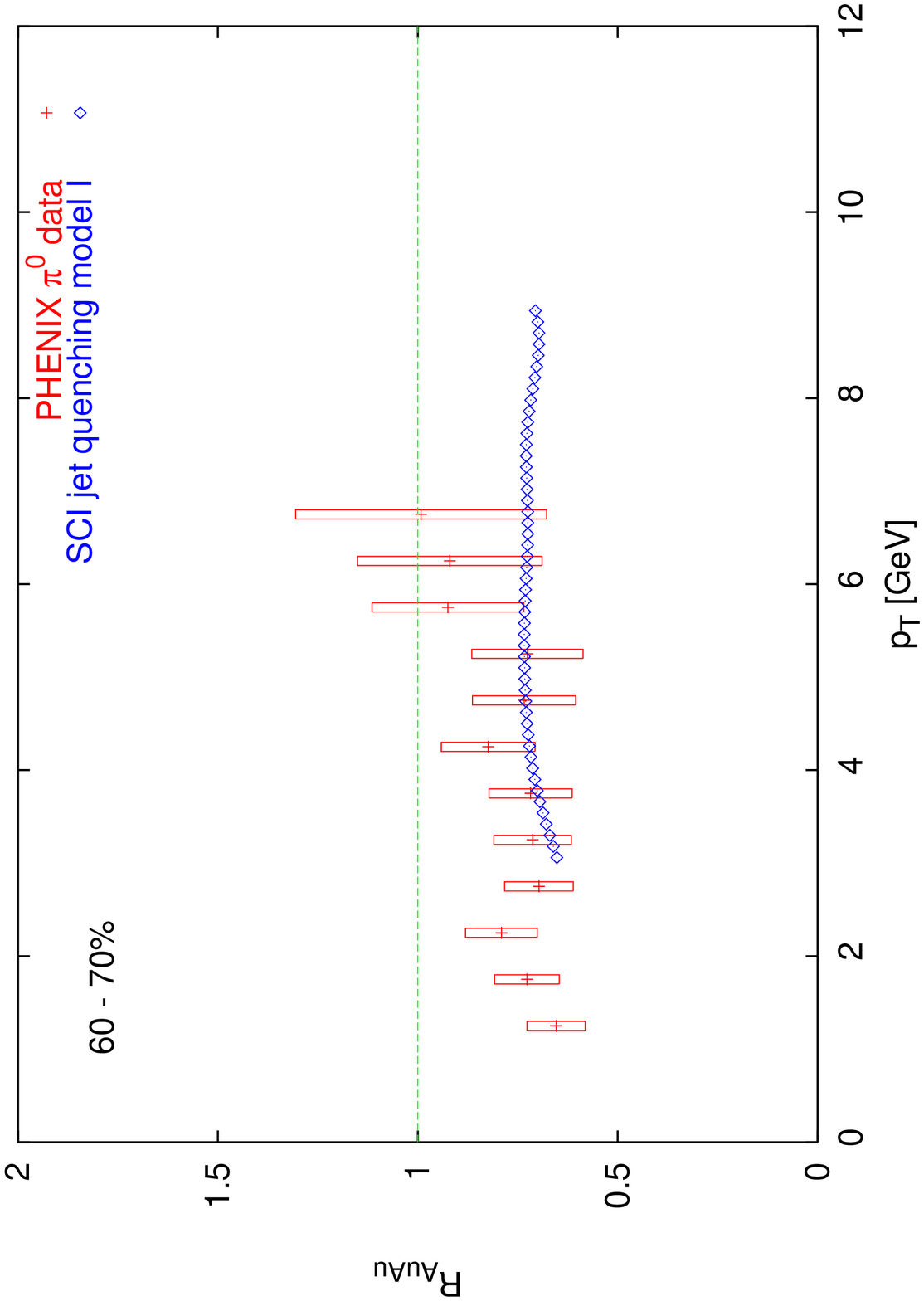} 
  \includegraphics[scale=0.4]{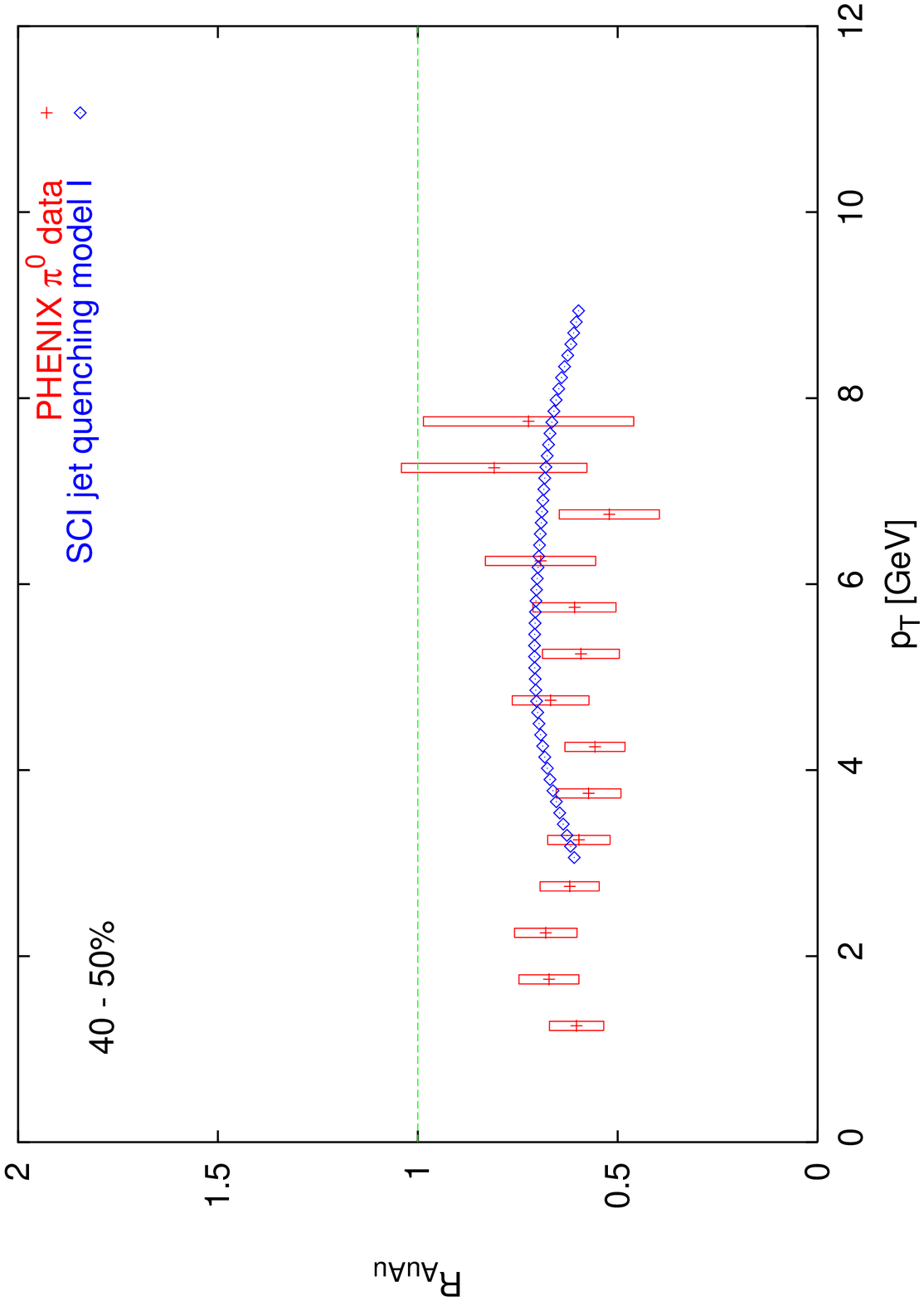} 
 \end{turn} 
\end{figure}

\clearpage

The approximately flat shape of $R_\tn{AuAu}$ for $\pt > \unit[3]{GeV}$ is well
reproduced by the model. There is a slight increase in the range between 3 and
\unit[4-5]{GeV} that is caused by the $\pt$-cut on the matrix element. Because
the intrinsic $\kt$ has to be added to the momentum transfer in the hard
scattering, there are too few particles produced also above the threshold. The
effect is more pronounced in nucleus-nucleus collisions due to the Cronin
effect and thereore it doesn't cancel completely in the ratio. As expected it
increases with centrality and also reaches out to higher $\pt$ in more central
events.

In the most peripheral class the model result is in agreement with the data but
is somewhat low. The situation is slightly better in 60-70\% but already in
40-50\% the model result falls behind the data which decrease faster. The trend
advances in the two central classes so that in the 0-10\% bin the model gives
roughly 50\% of the effect observed in the data. The centrality dependence
of the model result is rather weak so that it is practically impossible to
identify the functional dependence. 

\begin{figure}[ht]
 \centering
 \begin{turn}{-90}
  \includegraphics[scale=0.4]{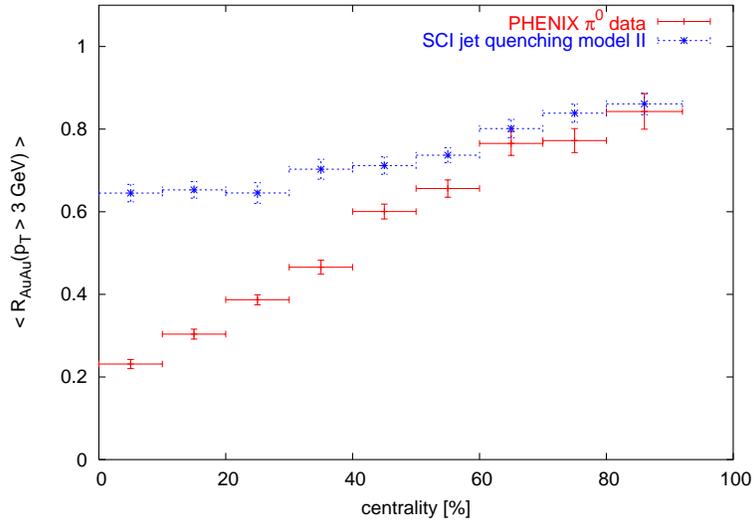}
 \end{turn}
 \caption{Comparison of the centrality dependence of the model with 
  inhomogeneous energy density profile to PHENIX data \cite{phenix_pi0}}
 \label{fig_si51-cendep}
\end{figure}

With the non-uniform energy density distribution the result for 0-10\%
centrality is nearly identical with the one obtained with a constant energy
density. But the increase when going to more peripheral collisions is stronger
because the mean energy density decreases. It thus agrees better with the data
than the other approach. This is also in reflected in
Figure~\ref{fig_si51-cendep} where it becomes clear that the model follows the
data for peripheral events, but then it levels off and stays approximately
constant. In fact, the model depends quadratically on the centrality, there is
no linear contribution. As will be discussed in Chapter~\ref{chap_genprop} this
is related to the dependence of the energy loss per interaction on the parton
energy. Figure~\ref{fig_endep} shows the energy loss in the interaction with
the  first plasma gluon (interaction probability is 1 in this case) for light
quarks as a function of the initial quark energy. Even for the highest energies
the energy loss rises slightly (the averaged values after 10 scatterings are
below those after only one interaction). On the other hand, quarks with very
small energy gain energy because the thermal energy of the environment is
higher than the quark energy.

\begin{figure}[ht]
 \centering
 \begin{turn}{-90}
  \includegraphics[scale=0.4]{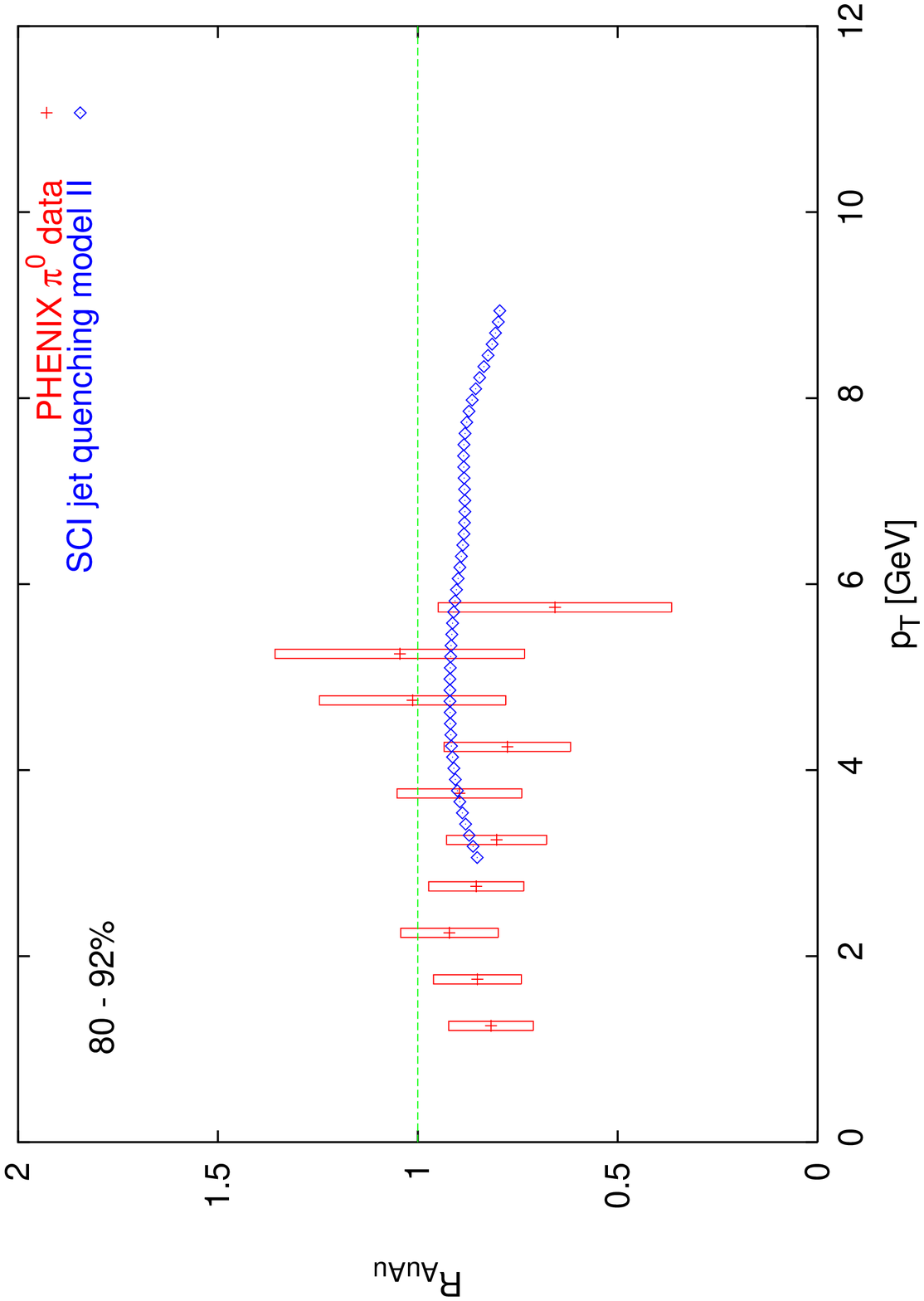}
  \includegraphics[scale=0.4]{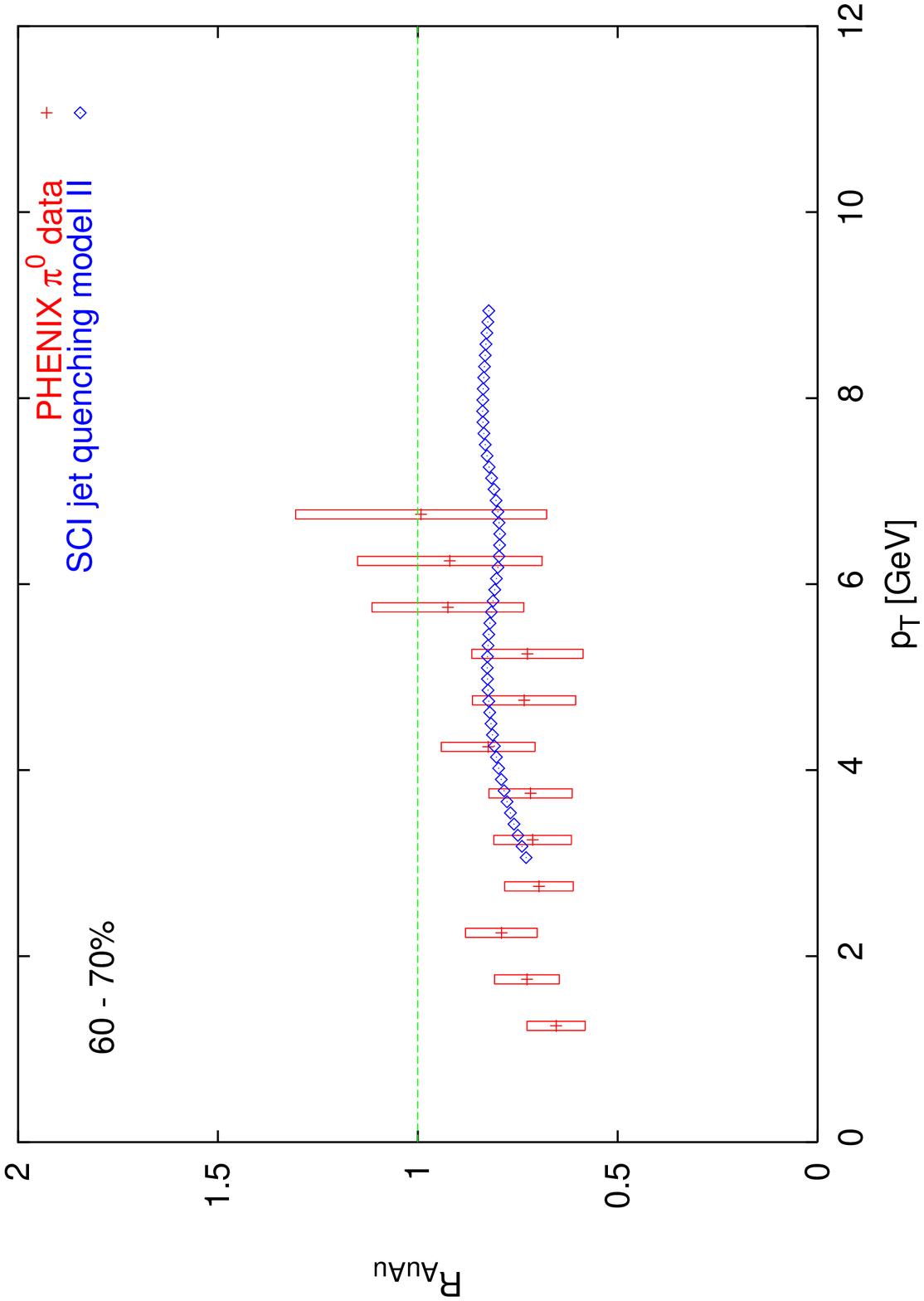}
  \includegraphics[scale=0.4]{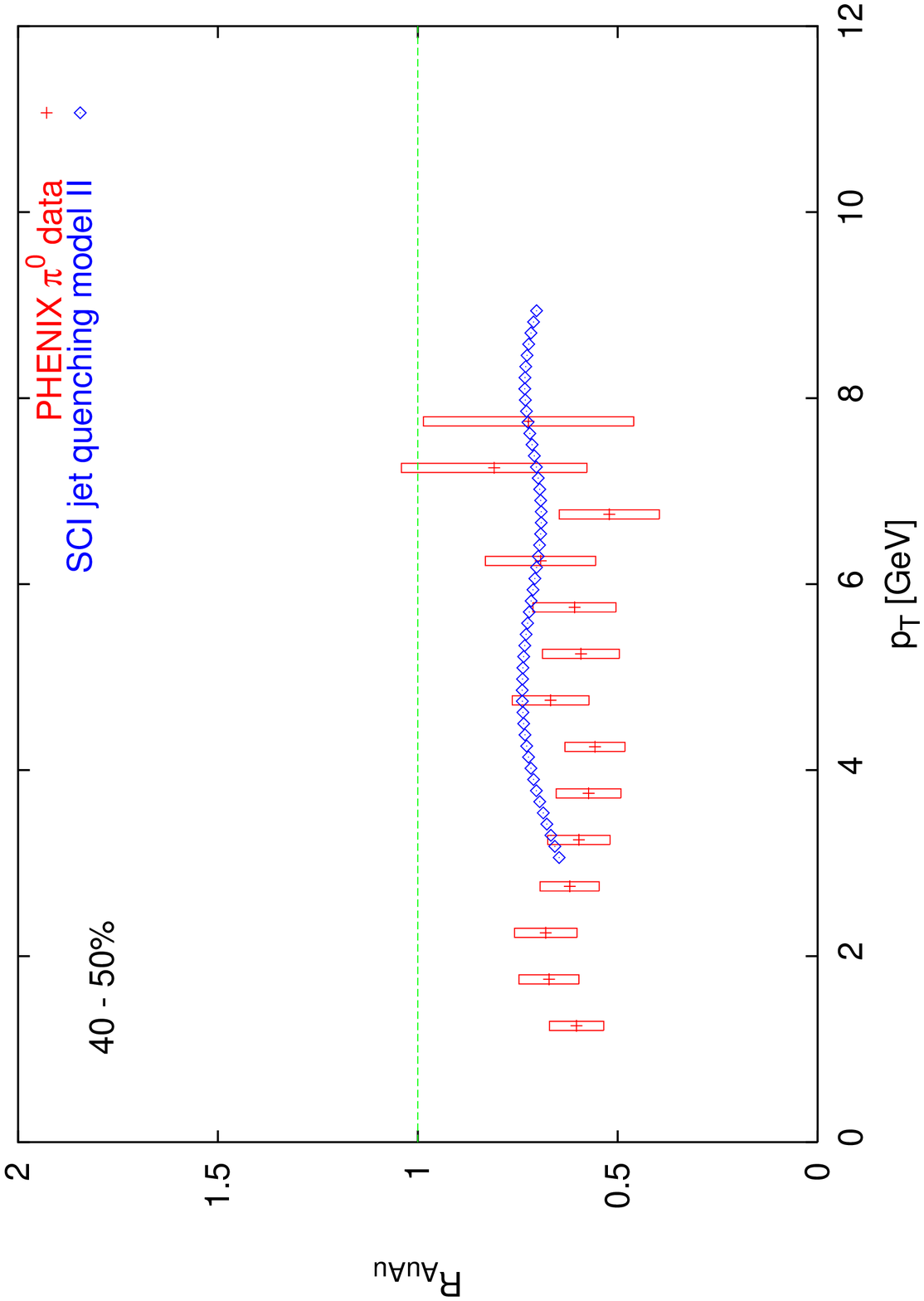}
 \end{turn}
\end{figure} 

\begin{figure}[ht]
 \centering
 \begin{turn}{-90}
  \includegraphics[scale=0.4]{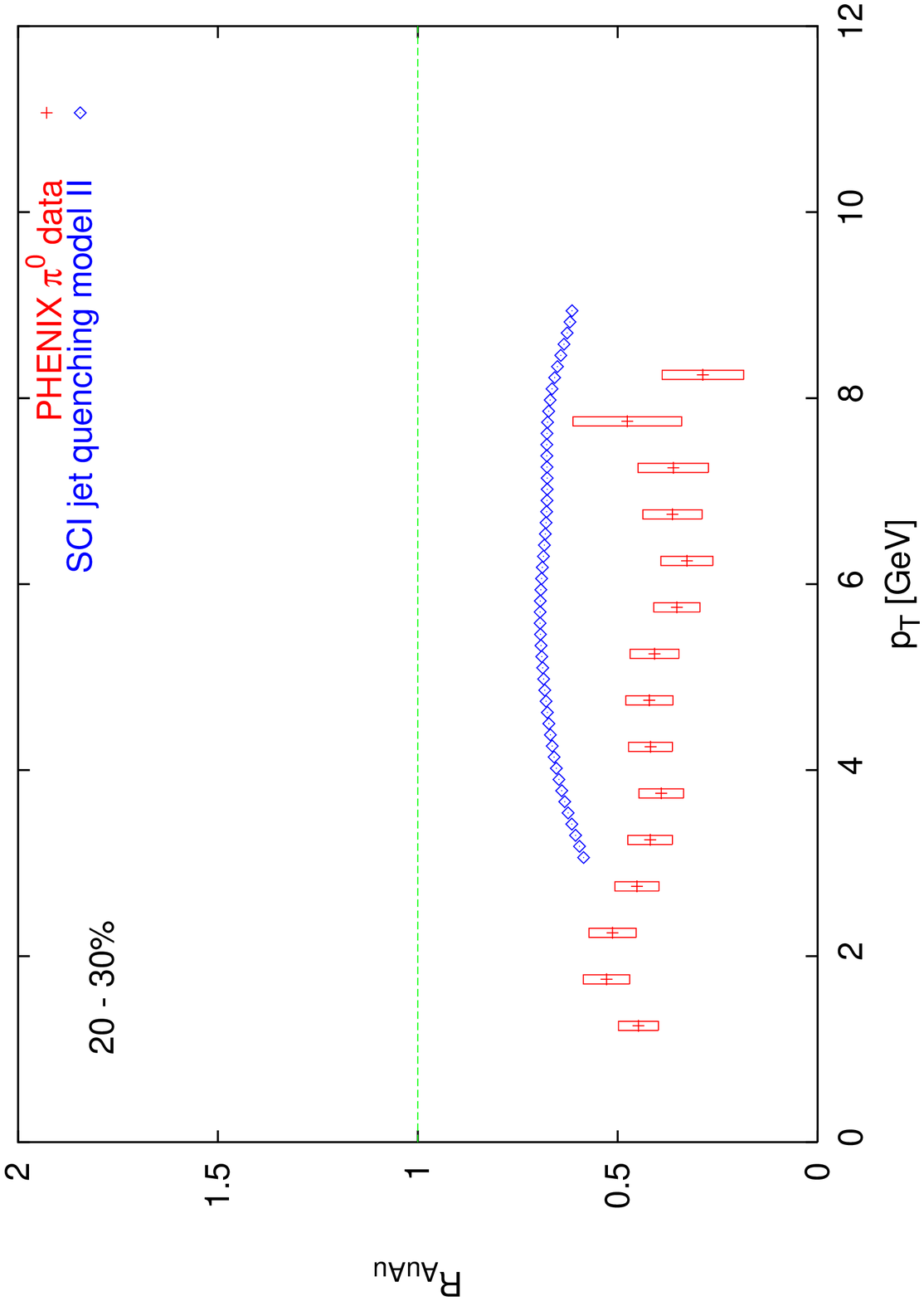}
  \includegraphics[scale=0.4]{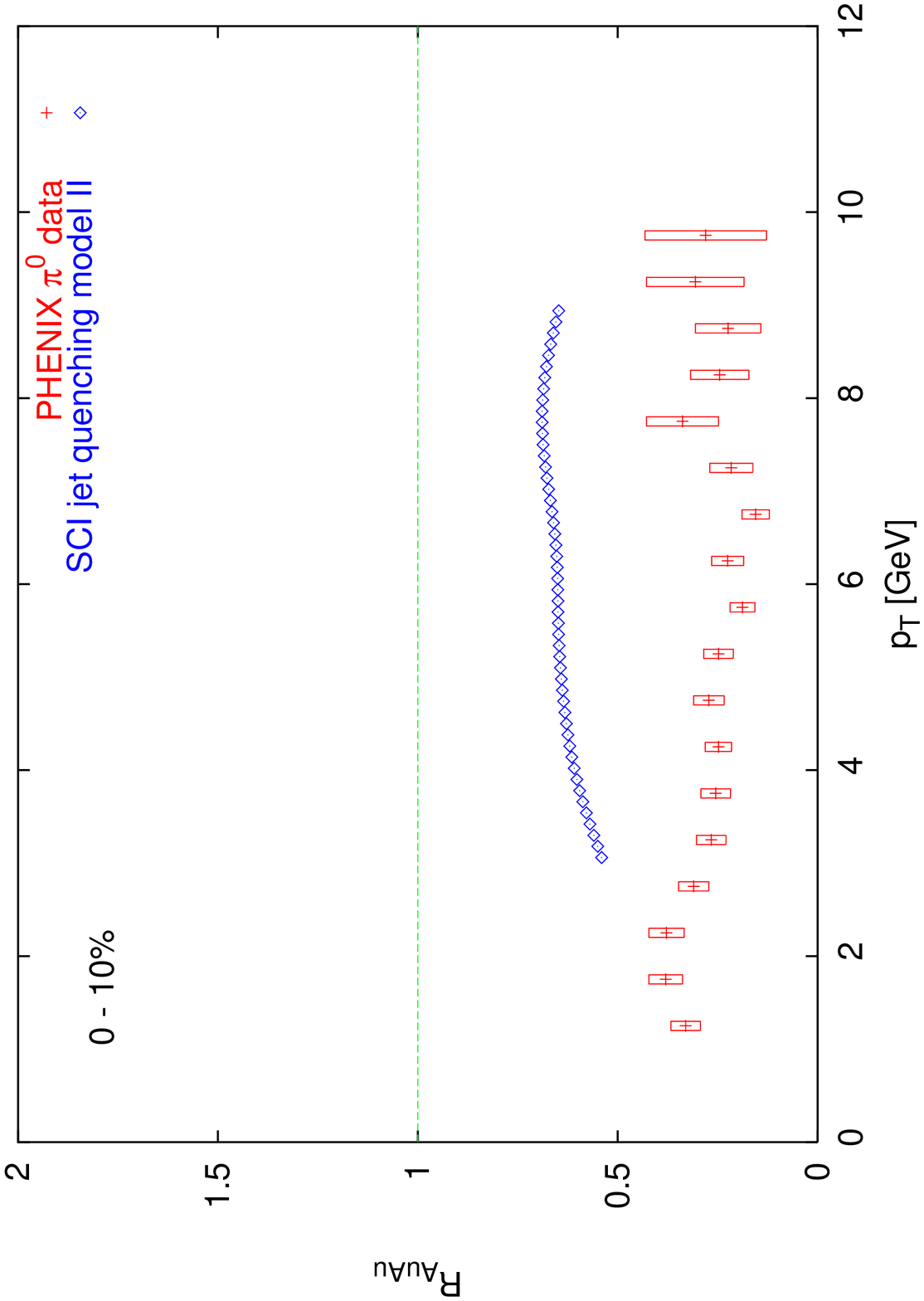}
 \end{turn}
 \caption{Model results for the nuclear modification factor with non-uniform 
 energy density for different centrality classes and PHENIX data 
 \cite{phenix_pi0}}
 \label{fig_si51}
\end{figure} 

\turnpic[0.4]{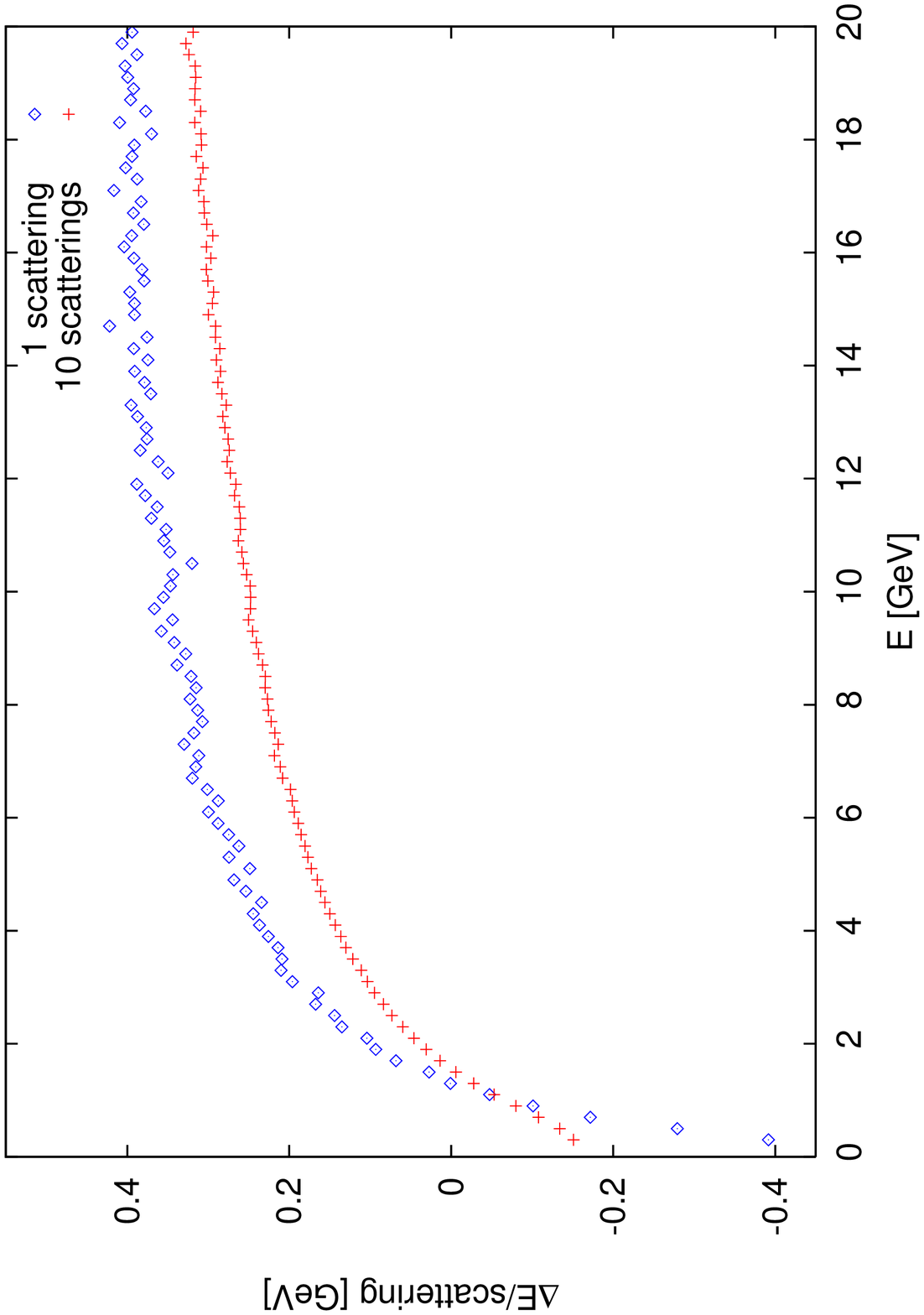}{-90}{fig_endep}{Dependence of the energy loss per
scattering on the initial energy of the quark for $u$ and $d$ quarks, 
scattering centre is the first gluon ("10 scatterings" denotes 10 interactions
with the same gluon)}{}

\FloatBarrier

\subsection{Azimuthal Correlation} %--------------------------------------------
\label{sec_scijq-azcor}

\begin{figure}[ht]
 \centering
 \begin{turn}{-90}
  \includegraphics[scale=0.4]{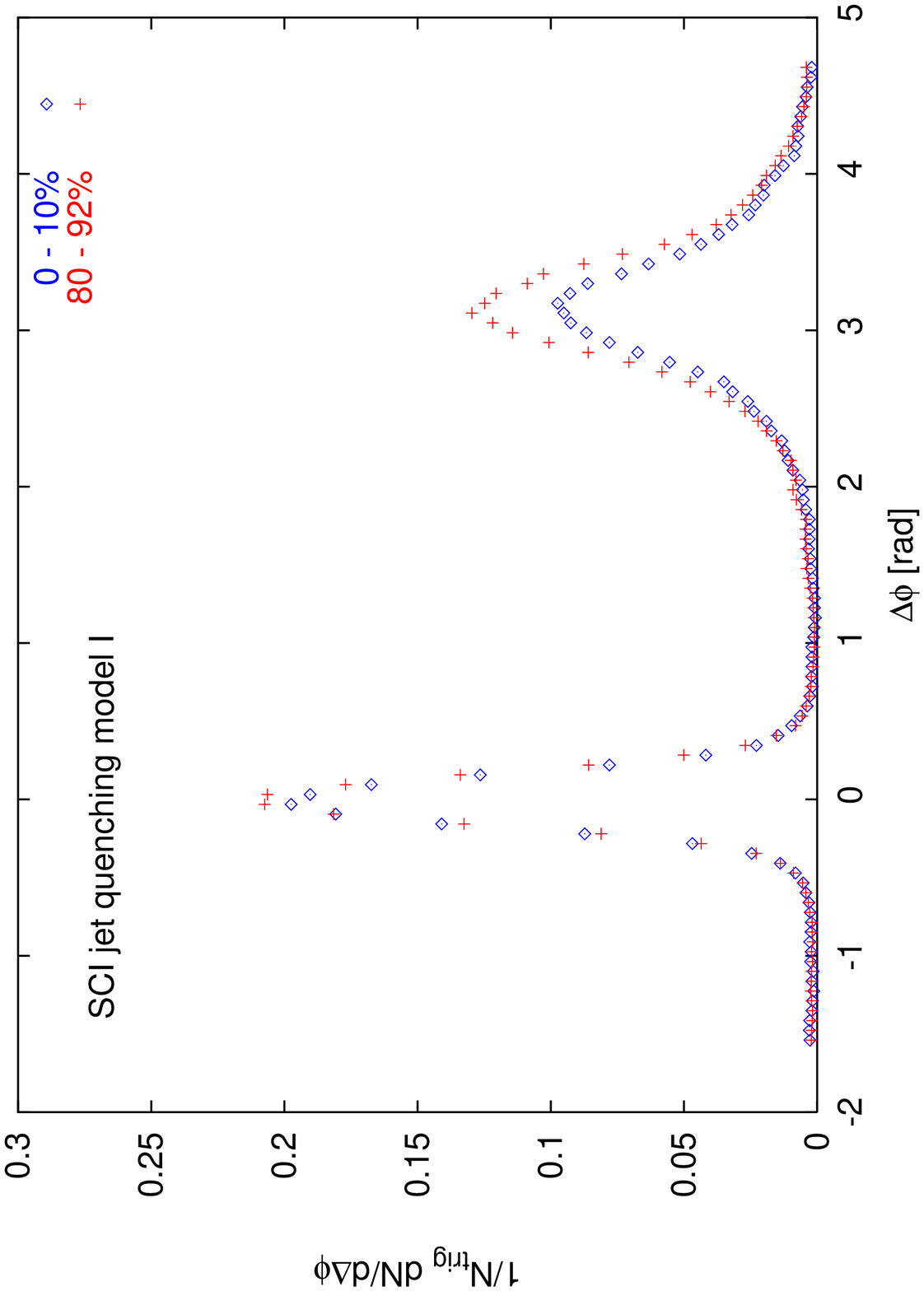}
  \includegraphics[scale=0.4]{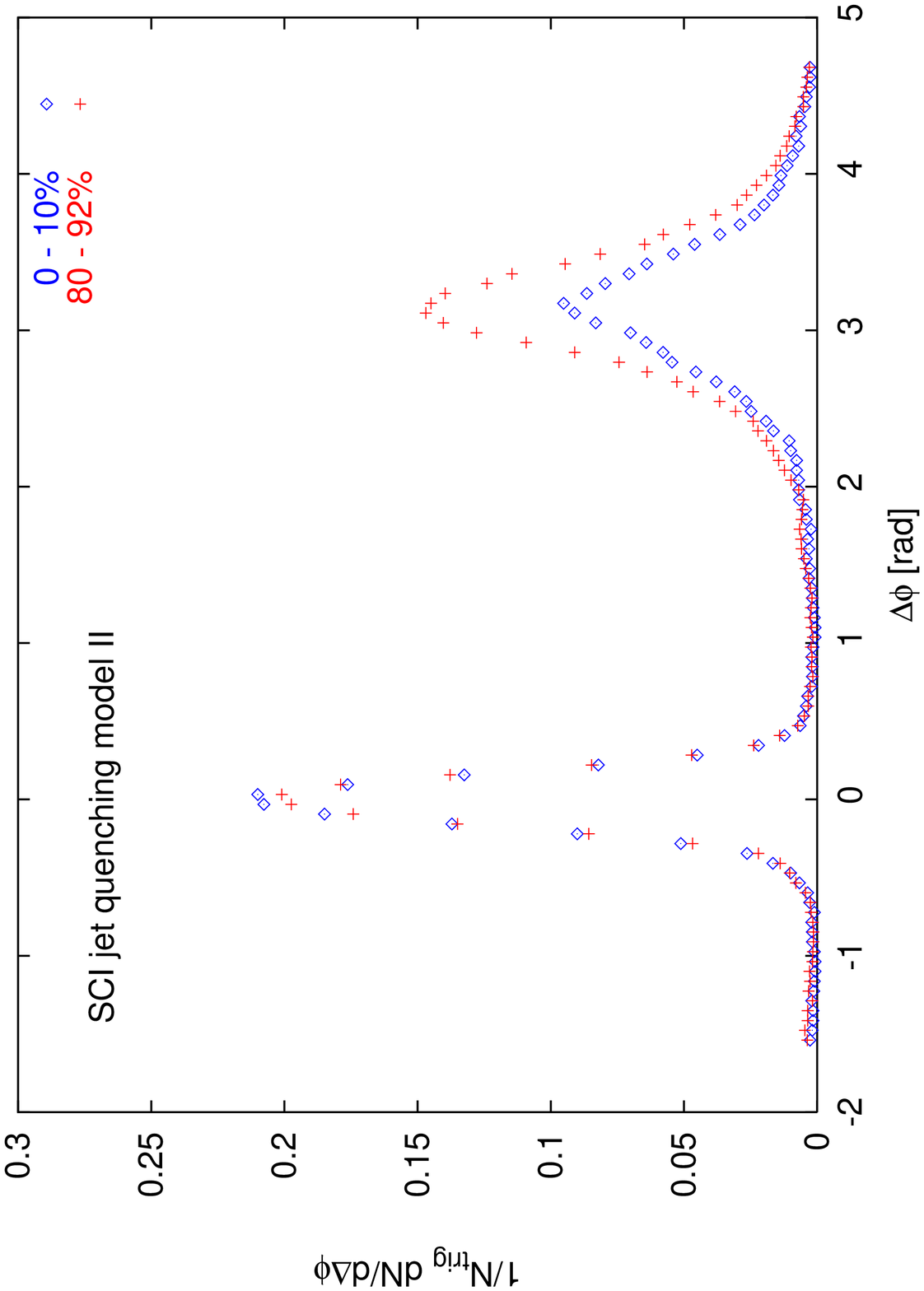}
 \end{turn}
 \caption{2-particle azimuthal correlation from the model with constant (upper 
   figure) and non-uniform energy density profile (lower figure) in peripheral
   (80-92\%) and central collisions (0-10\%)}
 \label{fig_siw}
\end{figure} 

The 2-particle azimuthal correlations for the two versions of the model are
shown in Figure~\ref{fig_siw}. There is a suppression of the away-side jet but
no disappearance as in the data. Again the ansatz with an inhomogeneous energy
density profile leads to a bigger effect. But while the Cronin effect
counteracted the energy loss in the case of the nuclear modification factor, it
now contributes to the jet quenching (Fig.~\ref{fig_wcronin2}). It leads to an
away-side peak that is lower in central than in peripheral events, although it
does not help on average (Fig.~\ref{fig_wcronin}).

\turnpic[0.4]{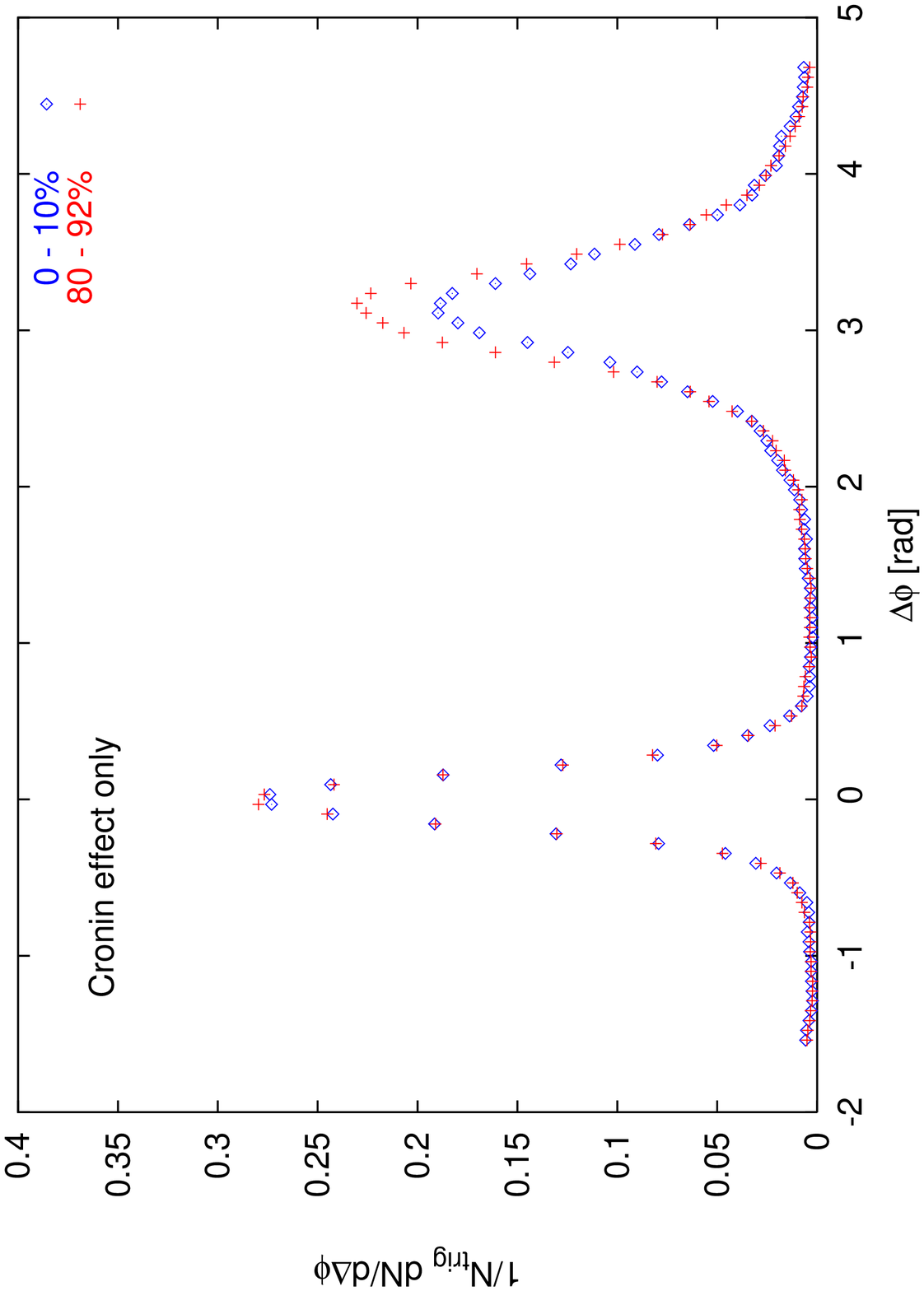}{-90}{fig_wcronin2}{Azimuthal correlation in
  AuAu collisions with only Cronin effect (no scattering) for central and
  peripheral events}{}

\smallskip

The near-side jet remains unchanged, but as discussed already in
Section~\ref{sec_2partcor} this does not necessarily mean that the respective
parton has not suffered energy loss. Since the partons are assumed to hadronise
outside the QGP and ordinary vacuum fragmentation functions are used, the shape
and the integral of the near-side peak are determined only by the momentum of
the partons. The energy loss in the QGP shifts the parton $\pt$-spectrum to
smaller momenta so that the overall yield of partons, that have enough energy to
produce a trigger particle, is smaller. But since the 2-particle azimuthal
correlation is normalized to the number of trigger particles and contains no
information about the rate at which these particles are produced, this effect is
invisible. However, there is a certain sensitivity to the shape of the parton
$\pt$-spectrum because a more energetic parton will on average produce more
associated particles. A flat spectrum will thus lead to a near-side peak with a
larger area than a steeply falling spectrum because the relative contribution
from partons with very high momenta is smaller in the latter case.

\begin{figure}[ht]
 \centering
 \begin{turn}{-90}
  \includegraphics[scale=0.4]{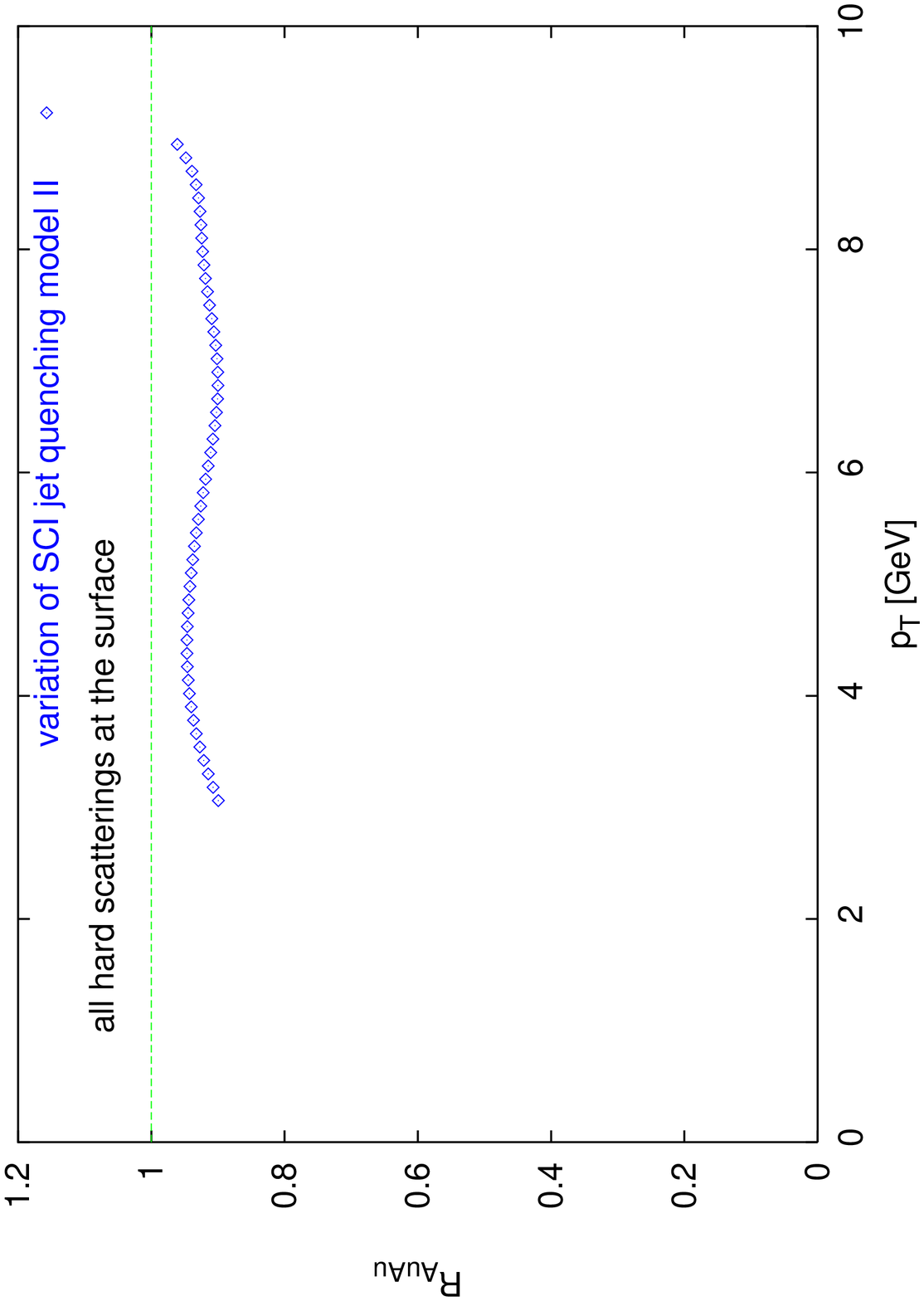}
  \includegraphics[scale=0.4]{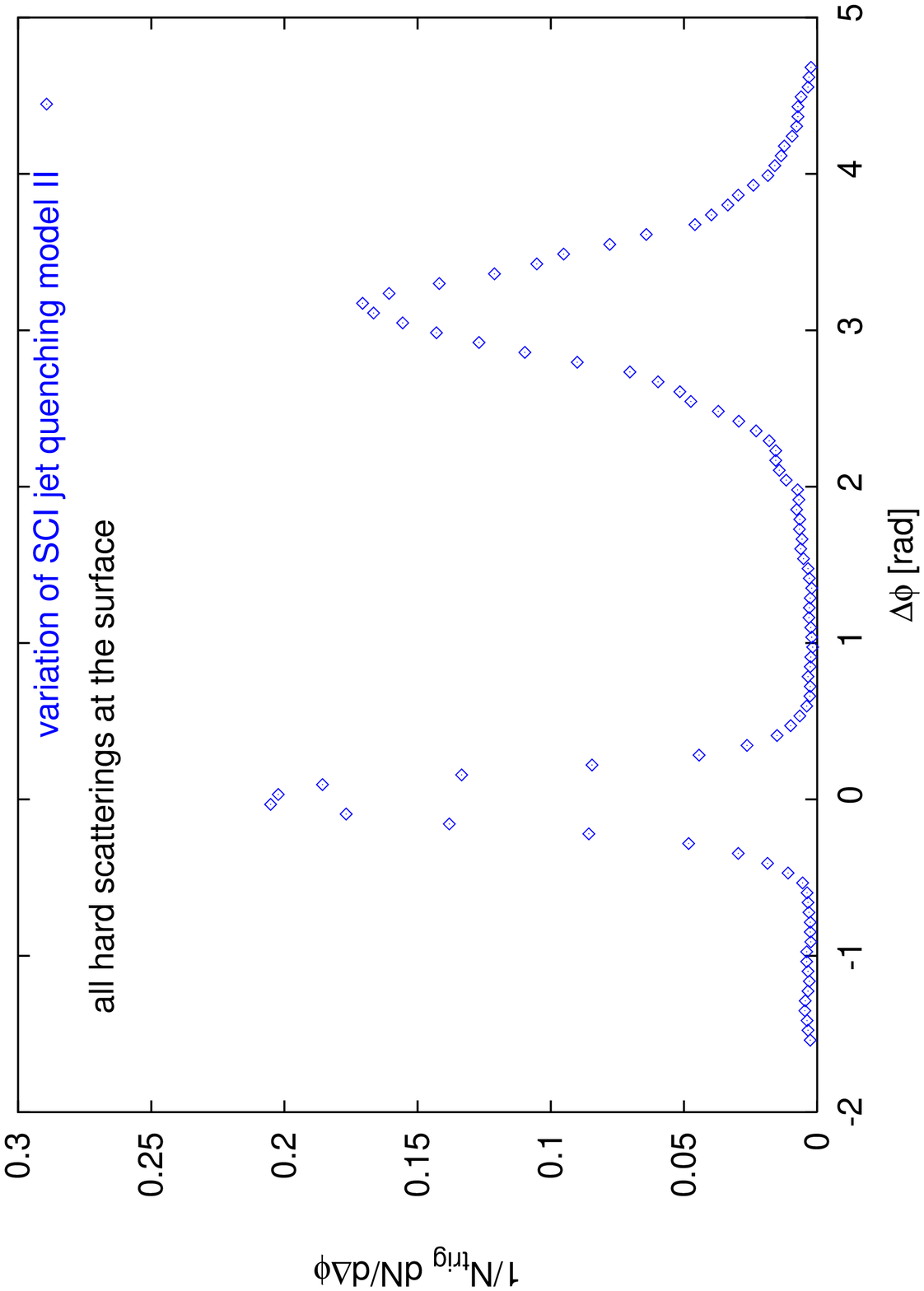}
 \end{turn}
 \caption{Results of the simulation with the SCI jet quenching model II but now
 all hard scattering points lie in a narrow shell at the edge, centrality is 0\%}
 \label{fig_rand}
\end{figure}

It is thus oversimplified to conclude that the partons producing trigger
particles have not lost a substantial amount of energy and therefore stem form
the periphery of the QGP. In fact, it can easily be shown that this is not the
case. Figure~\ref{fig_rand} shows the results from a simulation with the SCI
jet quenching model II where all hard scattering points were placed in a
\unit[0.5]{fm} thick shell at the edge of the overlap region ($b = 0$). At
first sight it is surprising that the jet quenching is very weak in this
configuration but it can be explained as a mainly geometrical effect: If a
parton is emitted from the surface with random direction in the transverse
plane (this is the case for hard scattering) only a range of $\sim \pi/2$ in
angle will lead to a sizable path through the plasma. Furthermore, the
constituent density is much smaller near the edge than in the centre. But while
a parton directed towards the centre of the QGP moves inward, the number
density drops rapidly so that the parton never reaches a region of high
density. The lifetime of the QGP is, with the parameters chosen here, $\tau_f
\simeq \unit[5]{fm/c}$ so that it hadronises even before the parton has reached
the centre. The sum of these effects leads to the very weak jet quenching
observed in Figure~\ref{fig_rand}. The jet quenching increases with the
thickness of the shell until for a thickness that equals the nuclear radius the
original result for central collisions is recovered. The result is very similar
for a constant energy density profile. This means that the observed hadrons are
the fragmentation products of partons that come from practically everywhere in
the overlap region.

\smallskip

\medskip

To get a substantial suppression of the away-side jet one parton has to lose
much more energy than the other one. Since the crucial quantity for energy loss
is the number density of constituent gluons, a situation where one parton
encounters much more gluos is needed. There are several reasons why this is
hard to achieve (and thus the effect on the azimuthal correlation is so small).
Firstly the hard scattering points are concentrated towards the centre of the
QGP and consequently the difference between the path lengths in the plasma of
the two scattered partons is typically small. What makes the situation even
worse is that the gluon density falls like $\tau^{-1}$. In fact, the energy loss
per scattering increases as the temperature falls but the decrease in number
density prevails (the consituent gluon energy is proportional to $\tau^{-1/3}$
only). This means that most of the energy loss happens at early times and small
differences in path length are further reduced because the plasma is already quite
dilute when the first parton leaves the overlap region. It is also seen in
Figure~\ref{fig_deltael}, which shows the specific energy loss of light quarks as
a function of the path length in the QGP. Quarks with low momentum gain energy
because they are below the thermal energy in the beginning. They are also more
sensitive to the plasma gluon energy than more energetic quarks, i.e.\ they
follow the evolution of the QGP more closely.

\turnpic[0.4]{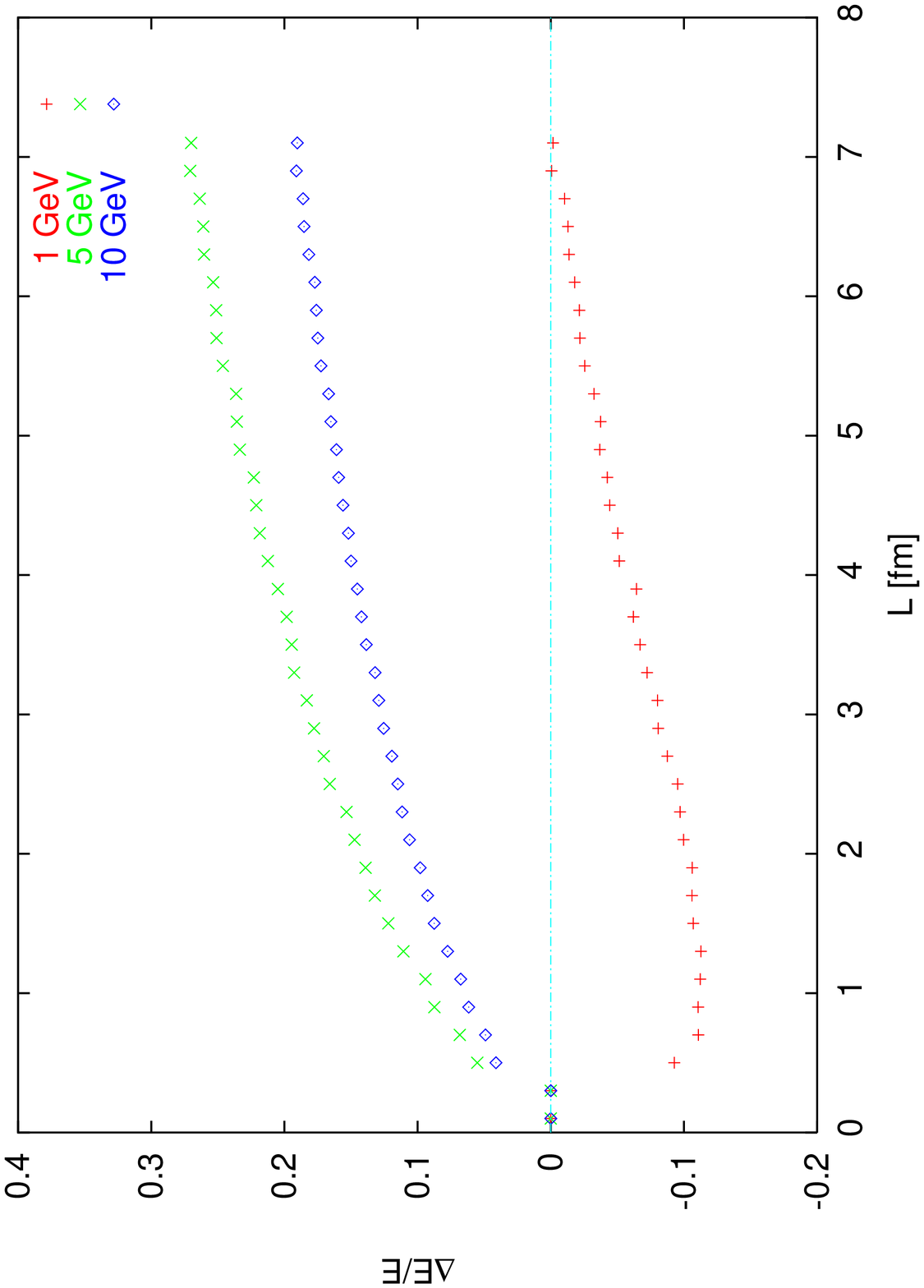}{-90}{fig_deltael}{Specific energy loss of 
monoergetic light quarks emitted from the centre of the QGP with $\theta =
\pi/4$, energy density is homogeneous}{}

Furthermore, the QGP lifetime limits the possible path length differences in
central collisions. Even for parton pairs that happen to be produced at the
surface the path length difference cannot exceed the lifetime of $\sim
\unit[5]{fm}$. The model with the non-uniform energy density profile works
better because it helps to amplify differences. If the hard scattering occured
not directly in the centre the parton with the longer path traverses the denser
central region and experiences more interactions than the other one which only
moves through the more dilute outer parts of the QGP.

\smallskip

The asymmetry between the two hard scattered partons is enhanced by the matrix
element since $\sim $50\% of all parton level processes are $q + g \to q + g$
scatterings. The gluons interact more strongly and lose more energy than the
quarks. 

\bigskip

\begin{figure}[ht]
 \centering
 \begin{turn}{-90}
  \includegraphics[scale=0.4]{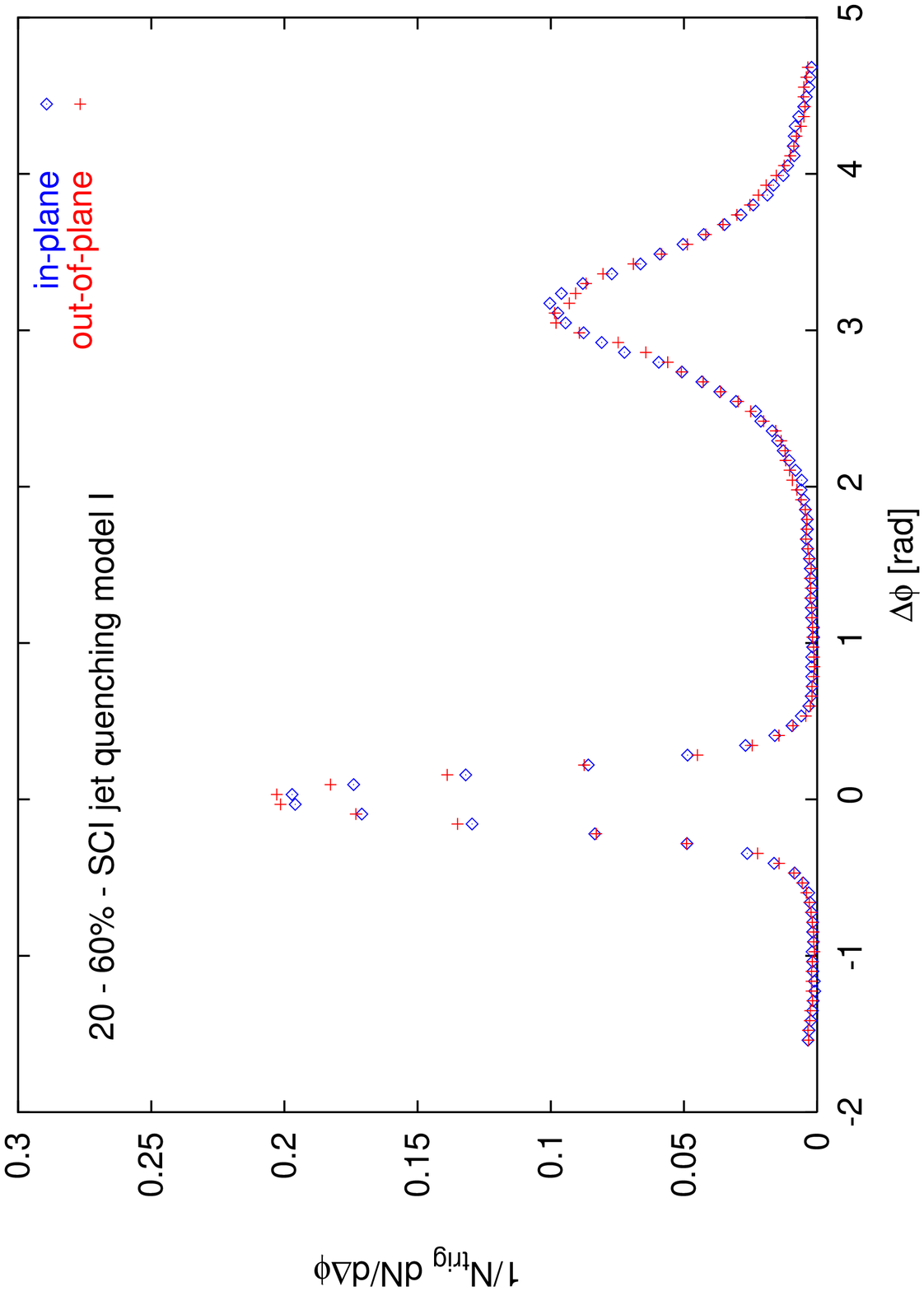}
  \includegraphics[scale=0.4]{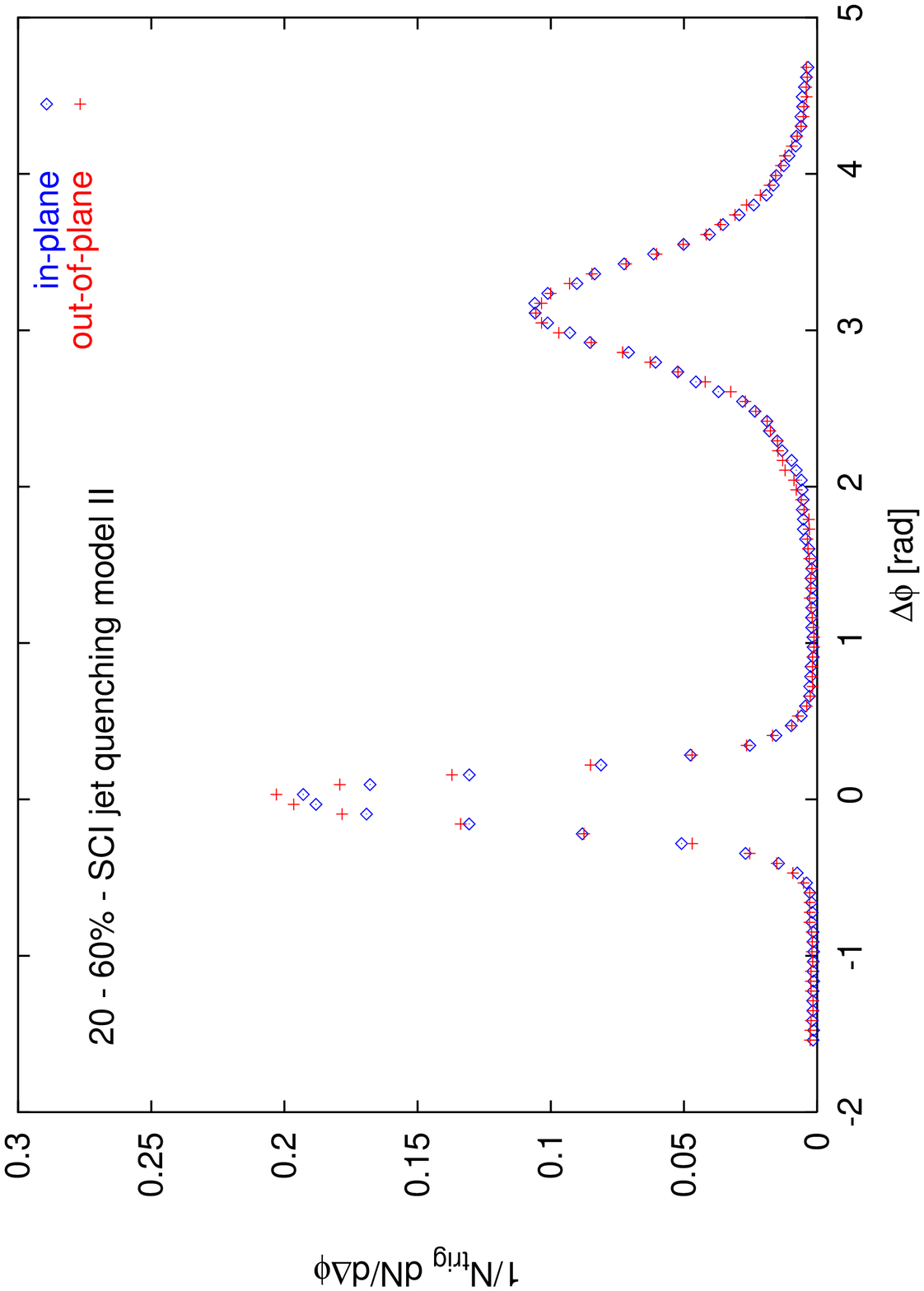}
 \end{turn}
 \caption{Correlation of the 2-particle azimuthal correlation with the reaction
   plane from the model with constant (upper figure) and non-uniform energy 
   density profile lower figure) in mid-central collisions (20-60\%)}
 \label{fig_siw-reacplane}
\end{figure} 
 
In Figure~\ref{fig_siw-reacplane} the correlation with the reaction plane is
shown for mid-central events. In both cases there is no difference between the
particles emitted in-plane and out-of-plane. The reason is that this observable
suffers even more from the problems with the similar path lengths. The mean
radius of the overlap region is \unit[3.0]{fm} in-plane and \unit[4.3]{fm}
out-of-plane for 40\% centrality. The difference is at most \unit[1.3]{fm} in
the whole range from 20 to 60\% centrality so that the differences in path
length of the two partons are small. Due to the other difficulties discussed
before, it is not surprising that there is no visible effect on the azimuthal
correlation.

\section{Discussion} %==========================================================

The SCI jet quenching model yields good results for the nuclear modification
factor. The shape is similar to the data, namely approximately flat for $\pt >
\unit[3]{GeV}$ and the magnitude can reach 50\% of the observed effect in
central collisions. The rest is presumably due to gluon bremsstrahlung which
should naturally come along with the scatterings. The centrality dependence is,
however, not linear as in the data but quadratic. It is not clear how this will
change when gluon bremsstrahlung is included.

There is also a clear suppression of the away-side jet but the effect is much
smaller than observed in data. The reasons for this are mostly understood: The
hard scattering points are concentrated towards the centre of the QGP, the gluon
density in the plasma falls rapidly and the lifetime limits the available path
length differences between hard scattered partons. These are not details of this
model but very general features.

\turnpic[0.4]{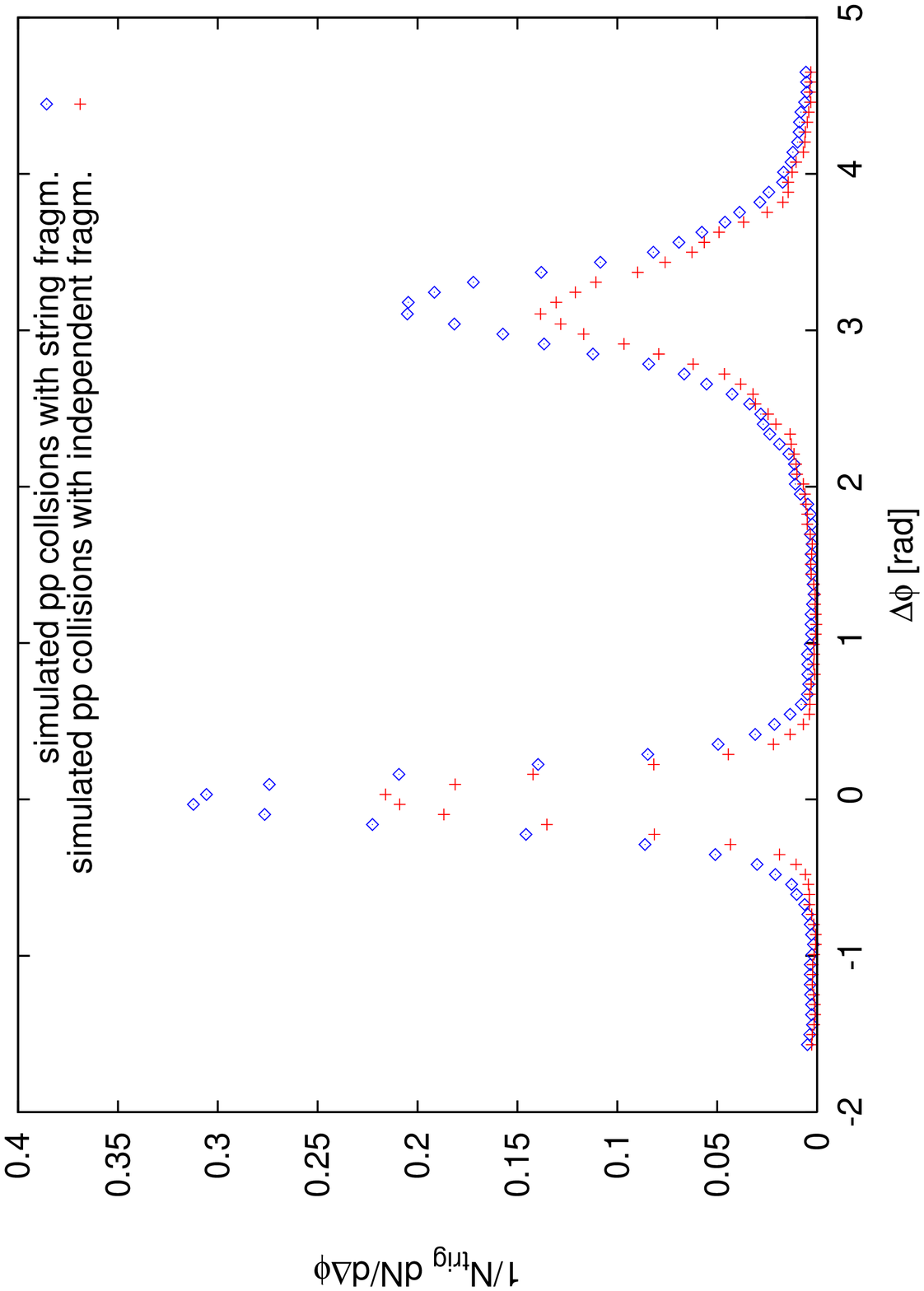}{-90}{fig_wfragm}{2-particle azimuthal
correlations in pp collisions for different fragmentation schemes}{}

Another aspect is that the azimuthal correlation depends on the details of the
fragmentation scheme. This is illustrated in Figure~\ref{fig_wfragm} which
shows a comparison of Lund string fragmentation to independent fragmentation in
pp simulations. The fragmentation function is the same in both cases (namely
the Lund symmetric function). The independent fragmentation scheme produces
somewhat less particles at midrapidity than the string fragmentation, but this
is a small effect (the ratio of the $\pt$- spectra is $\sim 0.95$). For the
2-particle azimuthal correlation, however,  this makes a big difference.
Modifications of the fragmentation scheme and function are a point that has not
been addressed in detail in this model but it is propbable that the presence of
a QGP will also influence the fragmentation. This may lead to substantial
changes in the azimuthal correlations, a softer fragmentation function for
instance will enhance the suppression of the away-side jet.

Still it can be concluded that it is indeed possible to obtain a considerable
contribution to the observed jet quenching from energy loss due to soft
scatterings. The scenario with an energy density that is higher near the centre
that at the edge of the QGP yields better results.

\medskip

The energy loss per scattering is governed by the width of the $t$ distribution
$\sigma_t$ and, to a smaller extent, the gluon mass $m_g$. A more massive gluon
leads to a smaller energy loss. These parameters affect the overall energy loss
and the centrality dependence. 

The QGP formation time, initial energy density and critical temperature
determine the number of scatterings but don't affect the centrality dependence.
A shorter formation time for example will add a certain number of encountered
gluons to each parton irrespective of the centrality. Because of the high
energy density in the early stage the overall energy loss depends quite
strongly on the formation time and moderately on the initial energy density but
hardly on the critical temperature. 

The Cronin parameter is fixed independently with the help of the dAu data but
there is still some freedom. A stronger $\kt$-broadening reduces the energy loss
such that the centrality dependence of $R_\tn{AuAu}$ becomes flatter but it
increases the suppression of the away-side jet in central collisions.

The effects of a hadronic layer around the plasma phase have not been studied
here, but their overall characteristics can be qualitatively described. It
would decrease the energy loss, but not very much as long as the hadronic layer
is thin. This may in turn alter the influence of the centrality, depending on
how the thickness changes with it. A more important effect could enter
in connection with modified fragmentation. Partons that originate from
scatterings in the hadronic layer may (depending on their direction of motion)
escape without any interactions and hadronise as in vacuum. This has a
potentially sizable impact on the azimuthal correlation.

\chapter{Exploring General Properties of Partonic Energy Loss}
\label{chap_genprop}
\section{Motivation} %==========================================================

\enlargethispage{-\baselineskip}

It has been shown in Section~\ref{sec_gluonrad} that the dependence of the
energy loss $\Delta E$ on the parton energy $E$ and the path length $L$ (in a
static medium) is sensitive to the physics of the energy loss mechanism. Matter
induced gluon radiation can for instance lead to a typical $L^2$-dependence, if
there are coherence effects. The energy loss by incoherent scatterings on the
other hand is proportional to $L$. It is thus obvious to ask if it is possible,
with the presently available data on jet quenching, to distinguish between
different scenarios.

Now it is fortunate that the SCI jet quenching model is constructed in a
modular fashion. It is thus possible to keep the model for the QGP but replace
the energy loss by something less specific, so that different dependencies can
be tested and compared to data. This can of course not replace a physical
model, but it can easily be used to investigate if the data favour a certain
type of energy loss mechanism.

\section{Energy Loss in the Toymodel} %=========================================

The only part of the model that is different from the SCI jet quenching model
is the energy loss. It is not simulated explicitely in the framework of the
Toymodel, instead a certain amount of energy is substracted for each parton and
deflection is not taken into account. This very general ansatz makes it
possible to explore different dependencies of the overall energy loss. Since
the path length is not a meaningful quantity in an expanding QGP, the number of
plasma gluons $N_g$, that a parton encounters on its way, is used as a measure
for the "amount" of plasma seen by a parton (this is sensible since also for
gluon bremsstrahlung the number of scatterings is the important quantity). The
dependence of the energy loss on $N_g$ and on the parton energy can hence be
varied according to the different cases listed in Table~\ref{tab_toymodel}. The
constant of proportionality is a free parameter that has to be adjusted. Again
gluons lose more energy than quarks, in the Toymodel this is achieved by
doubling the parameter. If the energy of a parton drops below a certain value
$\mu_\tn{th} + \sigma_\tn{th}$, it is assigned a new momentum vector with energy
picked from a Gaussian with mean $\mu_\tn{th}$ and width $\sigma_\tn{th}$, the
direction of motion is chosen randomly. This is of course a very rough way to
model the thermalisation of partons, but since they are anyway far below the
trustworthy region in $\pt$ of the model the details of the process are not
important.

\begin{tabelle}{|l|c|c|c|c|}{tab_toymodel}{Versions of energy loss in the
Toymodel, energies are in units of \unit[1]{GeV}}
\hline
Name & \multicolumn{3}{c|}{Dependencies} & Constant \\ \hline
     & path length & energy & energy density & \\ \hline
Toymodel I\,a\,$\alpha$  & $N_g$ & $E$ & uniform & $\unit[0.065]{GeV}$\\
Toymodel I\,b\,$\alpha$  & $N_g^2$ & $E$ & uniform & $\unit[0.016]{GeV}$\\
Toymodel II\,a\,$\alpha$ & $N_g$ & $E$ & non-uniform & $\unit[0.070]{GeV}$\\    
Toymodel II\,b\,$\alpha$ & $N_g^2$ & $E$ & non-uniform & $\unit[0.018]{GeV}$\\    
Toymodel II\,a\,$\beta$ & $N_g$ & $\sqrt{E}$ & non-uniform & $\unit[0.300]{GeV}$\\
Toymodel II\,b\,$\beta$ & $N_g^2$ & $\sqrt{E}$ & non-uniform & $\unit[0.020]{GeV}$\\
Toymodel II\,a\,$\gamma$ & $N_g$ & const. & non-uniform & $\unit[0.425]{GeV}$\\
Toymodel II\,b\,$\gamma$ & $N_g^2$ & const. & non-uniform & $\unit[0.200]{GeV}$\\
\hline    
\end{tabelle}     

\medskip

The other parameters that describe the QGP are the same as in the SCI jet
quenching model. A complete list with the default values is given in
Table~\ref{tab_toyparam}. The fragmentation scheme is again independent
fragmentation.

\begin{tabelle}{|p{1.2cm}|p{4cm}|l|l|l|}{tab_toyparam}{Parameters of the 
Toymodel, their default values and constraints}
\hline
& \multicolumn{2}{c|}{Parameter} & default value & sensible range \\ \hline
QGP & size of hadronic layer (only with homogeneous energy density) & $d$ 
			& \unit[0]{fm} & \unit[0\dots3]{fm} (?) \\
& QGP formation time & $\tau_i$ & \unit[0.6]{fm/c} & \unit[0.2\dots 1]{fm/c}
							\cite{huovinen} \\
& initial energy density  ($\tau_0 = \unit[1]{fm/c}$) & $\epsilon_0$ &
		$\unit[5.5]{GeV\,fm^{-3}}$ & 	\unit[4\dots 6.5]{GeV\,fm$^{-3}$}
		\cite{huovinen} \\
& critical temperature & $T_c$ & \unit[0.17]{GeV} & \unit[0.17\dots
		0.18]{GeV} \\		
& radius of parton cylinder & $R_\tn{cyl}$ & \unit[0.3]{fm} & $\sim$
			\unit[0.3]{fm} \\ 
 \hline
Cronin effect& increase of $\sigma_{\kt}^2$ per scattering & $\alpha$ & 
								$\unit[0.25]{GeV^2}$ & ? \\
\hline
energy & mean thermal energy & $\mu_\tn{th}$ & \unit[0.6]{GeV} & \\
loss   & width of th. energy distr. & $\sigma_\tn{th}$ & \unit[0.1]{GeV} &
		\\
		& constant of proportionality & & & \\
\hline
\end{tabelle}

The constant is adjusted in one centrality bin (either the most central or the
most peripheral) and then the other classes are generated with the same value.

\enlargethispage{-\baselineskip}

\section{Results} %=============================================================

\enlargethispage{-\baselineskip}

In this section only a fraction of the results is shown, the complete set can
be found in Appendix~\ref{app_toymodel}. The shape of the nuclear modification
factor is very similar in all versions: It is nearly flat with a tendency to
decrease slightly towards high $\pt$, only when $\Delta E$ is independent of
$E$ it rises. The centrality dependence and the azimuthal correlation are more
interesting, because they are sensitive to the differences in the model assumed
for the energy loss.

\smallskip

\turnpic[0.4]{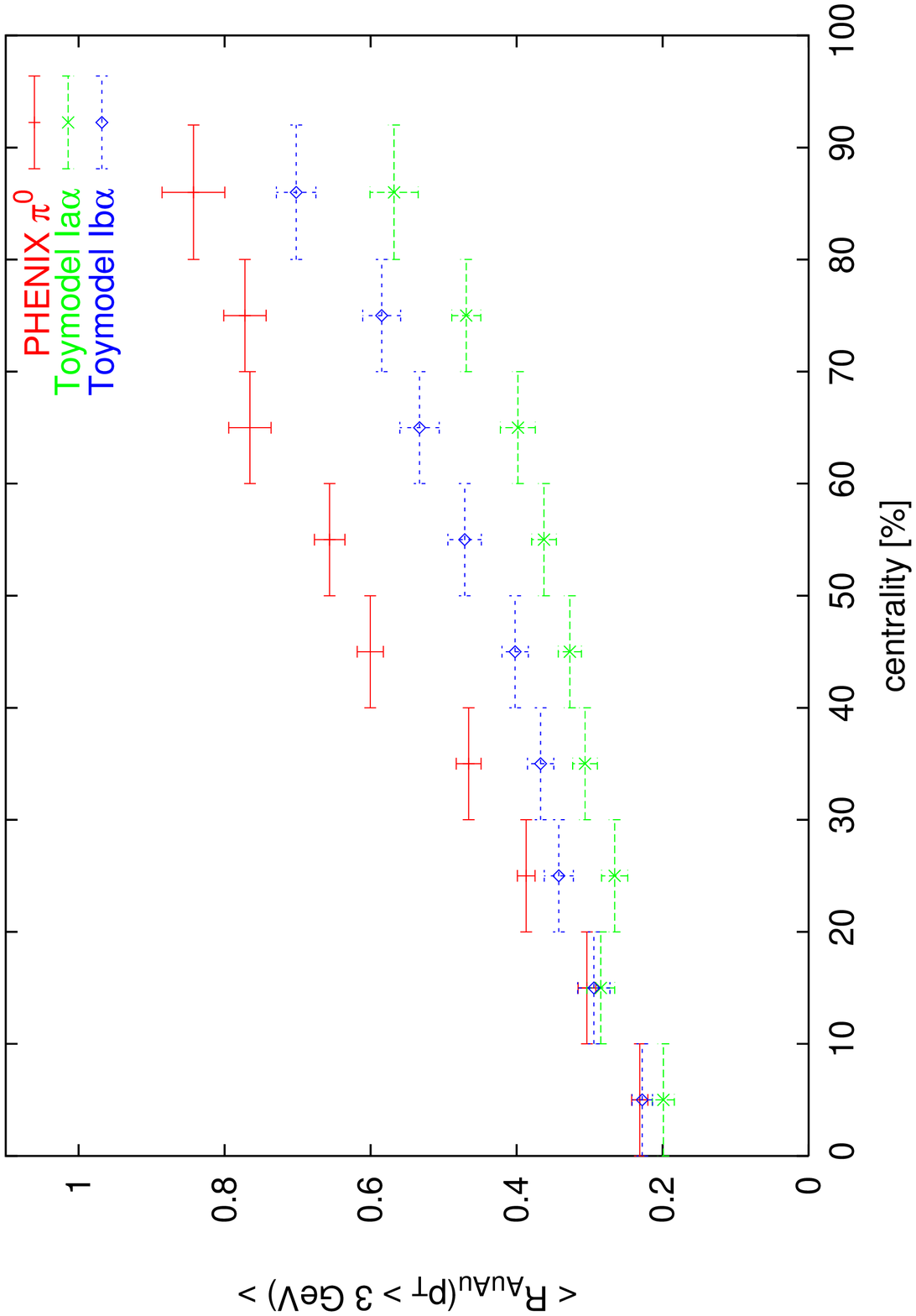}{-90}{fig_toy51-c}{Centrality dependence
of $R_\tn{AuAu}$ for the Toymodel I\,a\,$\alpha$ with $\Delta E \propto N_g E$
and Toymodel I\,b\,$\alpha$ with $\Delta E \propto N_g^2 E$, both with uniform
energy density profile}{}

\turnpic[0.4]{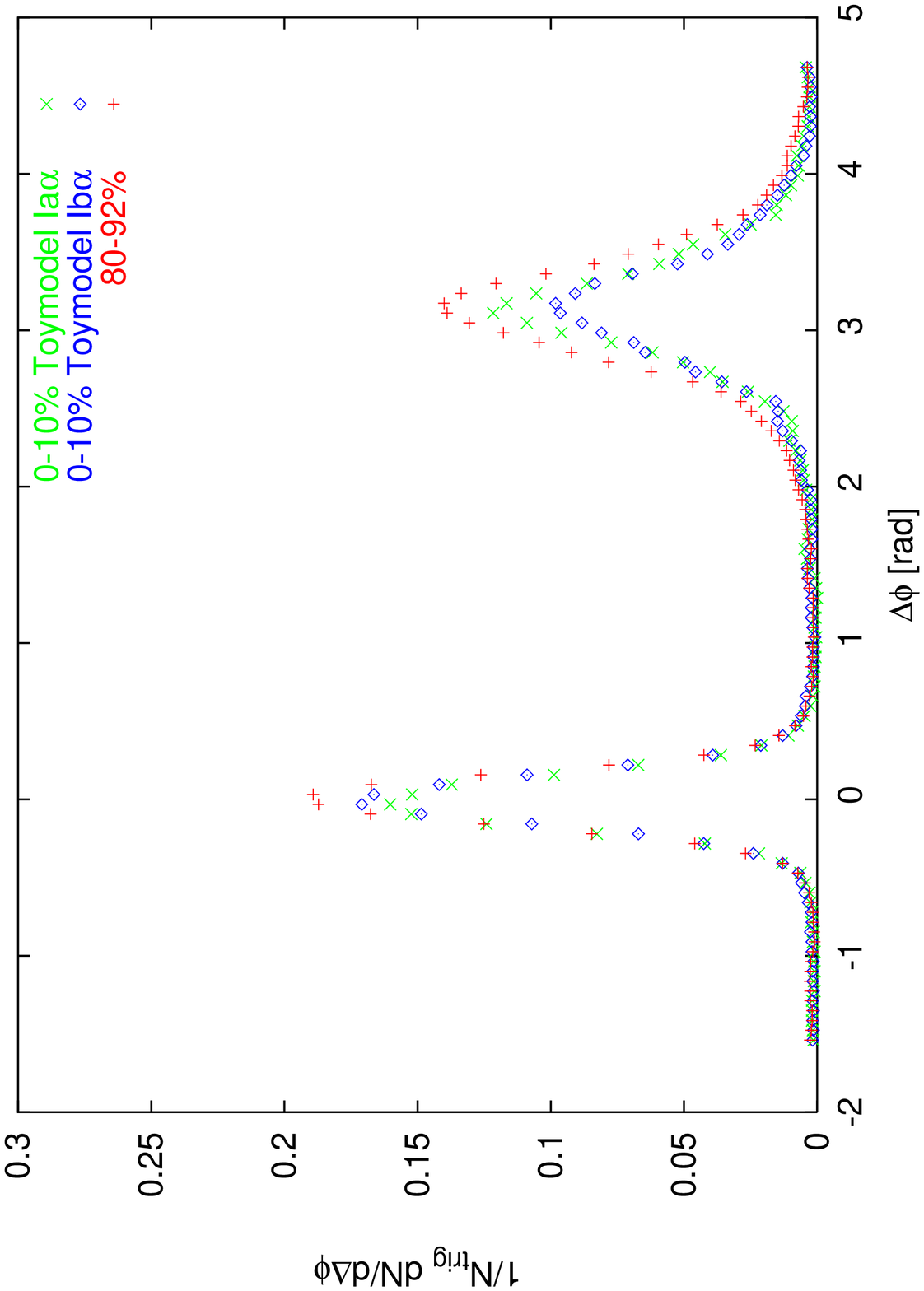}{-90}{fig_toy51-w}{2-particle azimuthal
correlation  for the Toymodel I\,a\,$\alpha$ with $\Delta E \propto N_g E$ and
Toymodel I\,b\,$\alpha$ with $\Delta E \propto N_g^2 E$, both with uniform
energy density profile, the 80-92\% data are the same in both cases}{}

Figure~\ref{fig_toy51-c} shows the centrality dependence of $R_\tn{AuAu}$ for
the configurations with a homogeneous energy density profile Toymodel
I\,a\,$\alpha$ and I\,b\,$\alpha$. The shape is in both cases similar, although
the linear version (I\,a\,$\alpha$) seems to be somewhat flatter. The course of
I\,a\,$\alpha$ is mildly curved while I\,b\,$\alpha$ is more straight. Both
model results are significantly flatter than the data. The absolut magnitude of
the nuclear modification factor is not meaningful, since it is governed by a free
parameter. 

The effect on the 2-particle correlation is in both cases small, but stronger in
I\,b\,$\alpha$ (Fig.~\ref{fig_toy51-w}). So altogether the configuration with
$\Delta E \propto N_g^2$ appears to be slightly closer to the experimental data.

\smallskip

\turnpic[0.4]{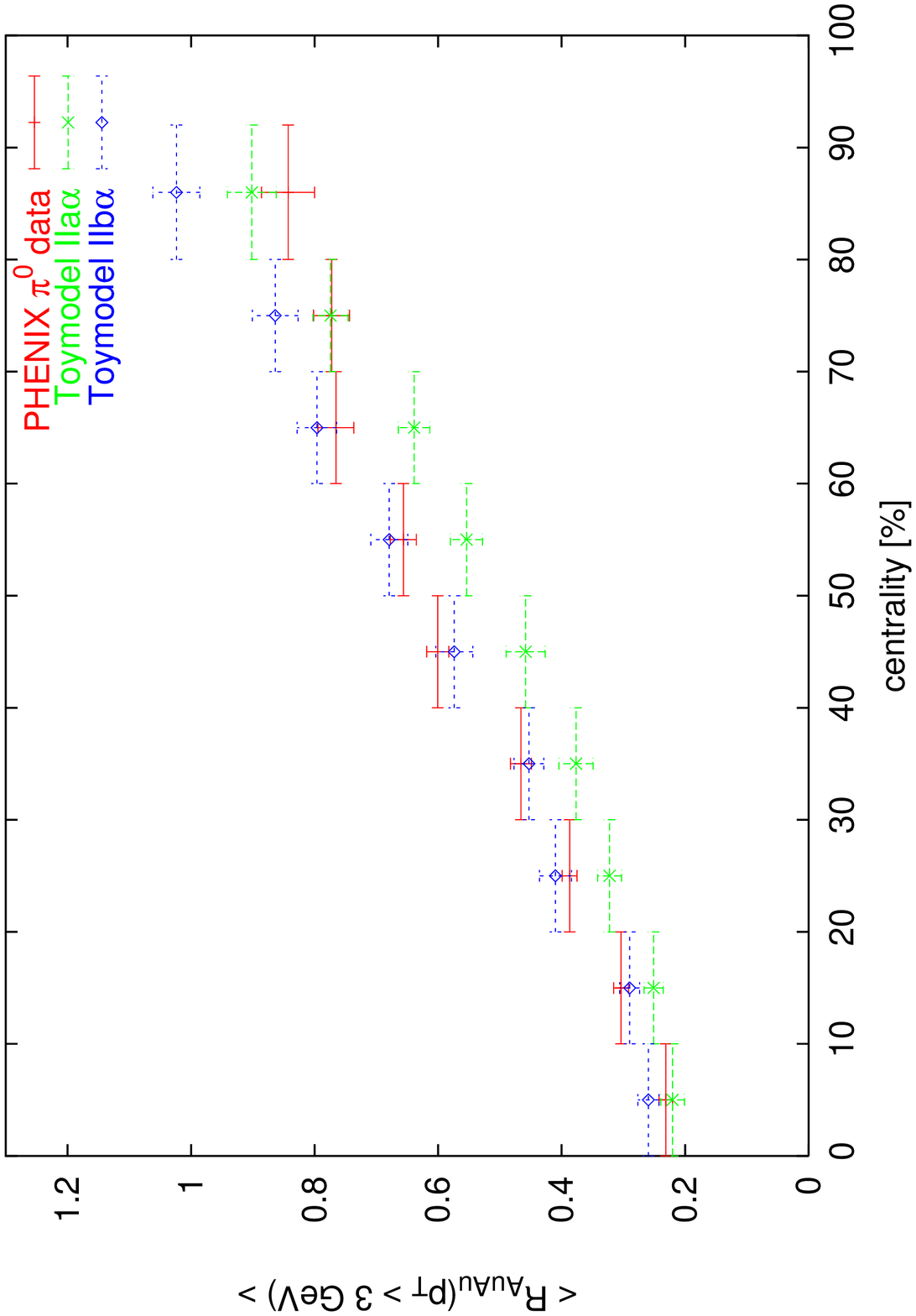}{-90}{fig_toy61-c}{Centrality dependence
of $R_\tn{AuAu}$ for the Toymodel II\,a\,$\alpha$ with $\Delta E \propto N_g E$
and Toymodel II\,b\,$\alpha$ with $\Delta E \propto N_g^2 E$, both with
non-uniform energy density profile}{}

\turnpic[0.4]{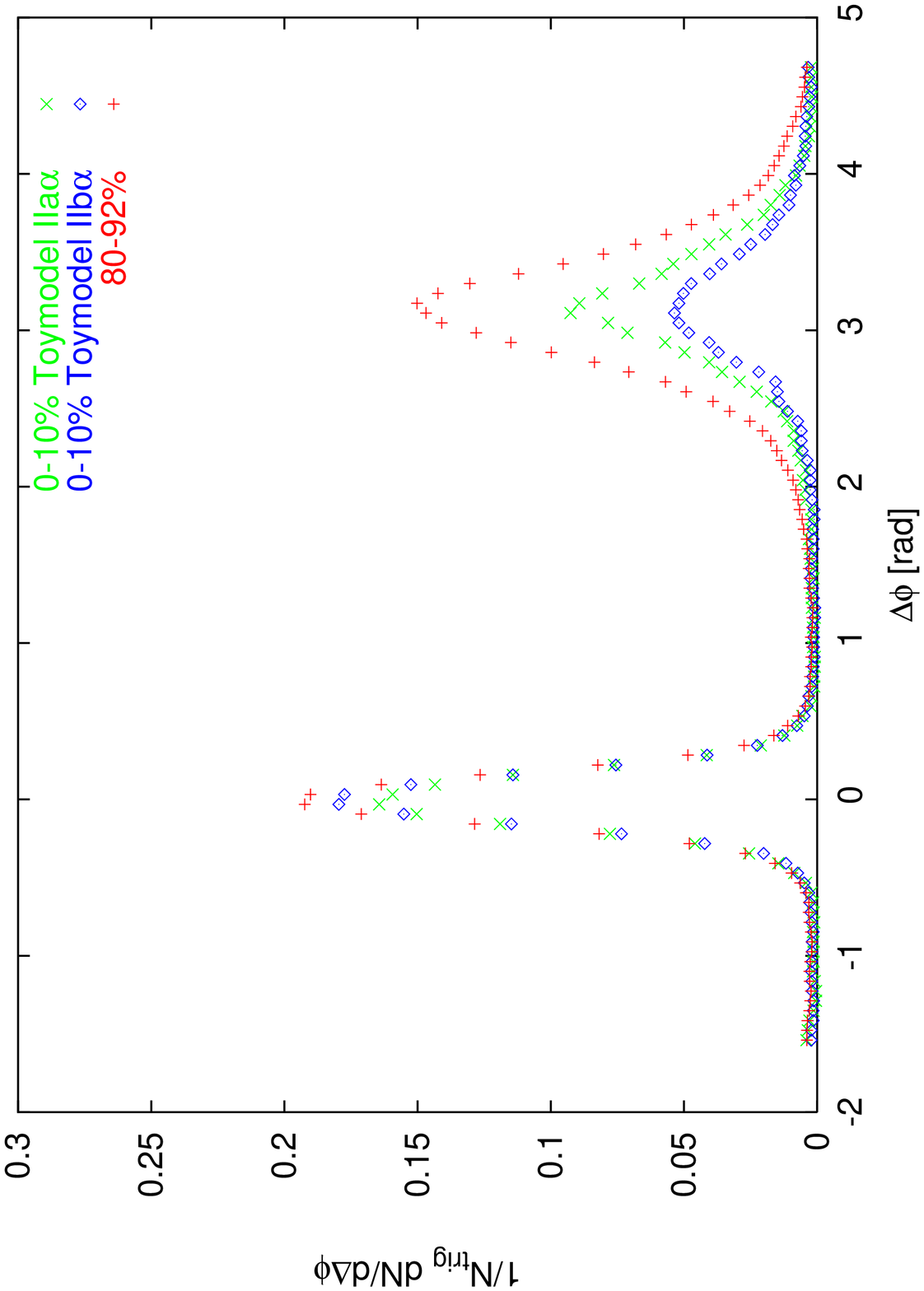}{-90}{fig_toy61-w}{2-particle azimuthal
correlation  for the Toymodel II\,a\,$\alpha$ with $\Delta E \propto N_g E$ and
Toymodel II\,b\,$\alpha$ with $\Delta E \propto N_g^2 E$, both with non-uniform
energy density profile, the 80-92\% data are the same in both cases}{}

In the configuration with a non-uniform energy density the mean energy density
increases with centrality. Thus the expectation is that the shape of the
centrality dependence of $R_\tn{AuAu}$ is steeper than with a constant energy
density. This is in fact observed (Fig.~\ref{fig_toy61-c}). Although both
calculation results are slightly curved, they now give a reasonably good
description of the data. The bending is again more distinct in the linear
ansatz. 

The suppression of the away-side jet has also improved in both cases as
compared to Toymodel~I (Fig.~\ref{fig_toy61-w}). This matches with the
observation made with the SCI jet-quenching model where also the inhomogeneous
energy density distribution yields better results.

\smallskip

\turnpic[0.4]{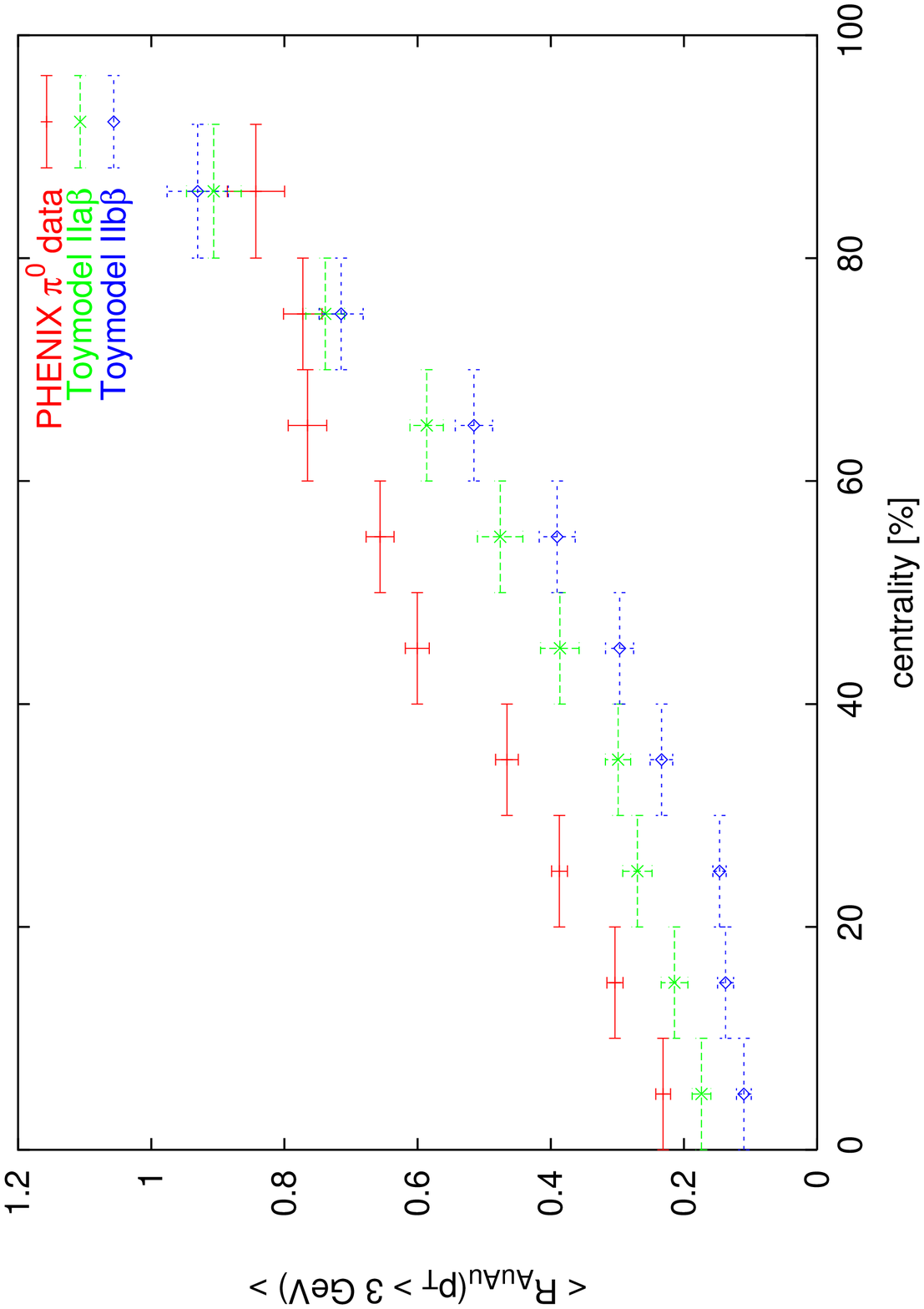}{-90}{fig_toy63-c}{Centrality dependence
of $R_\tn{AuAu}$ for the Toymodel II\,a\,$\beta$ with $\Delta E \propto N_g
\sqrt{E}$ and Toymodel II\,b\,$\beta$ with $\Delta E \propto N_g^2 \sqrt{E}$,
both with non-uniform energy density profile}{}

\turnpic[0.4]{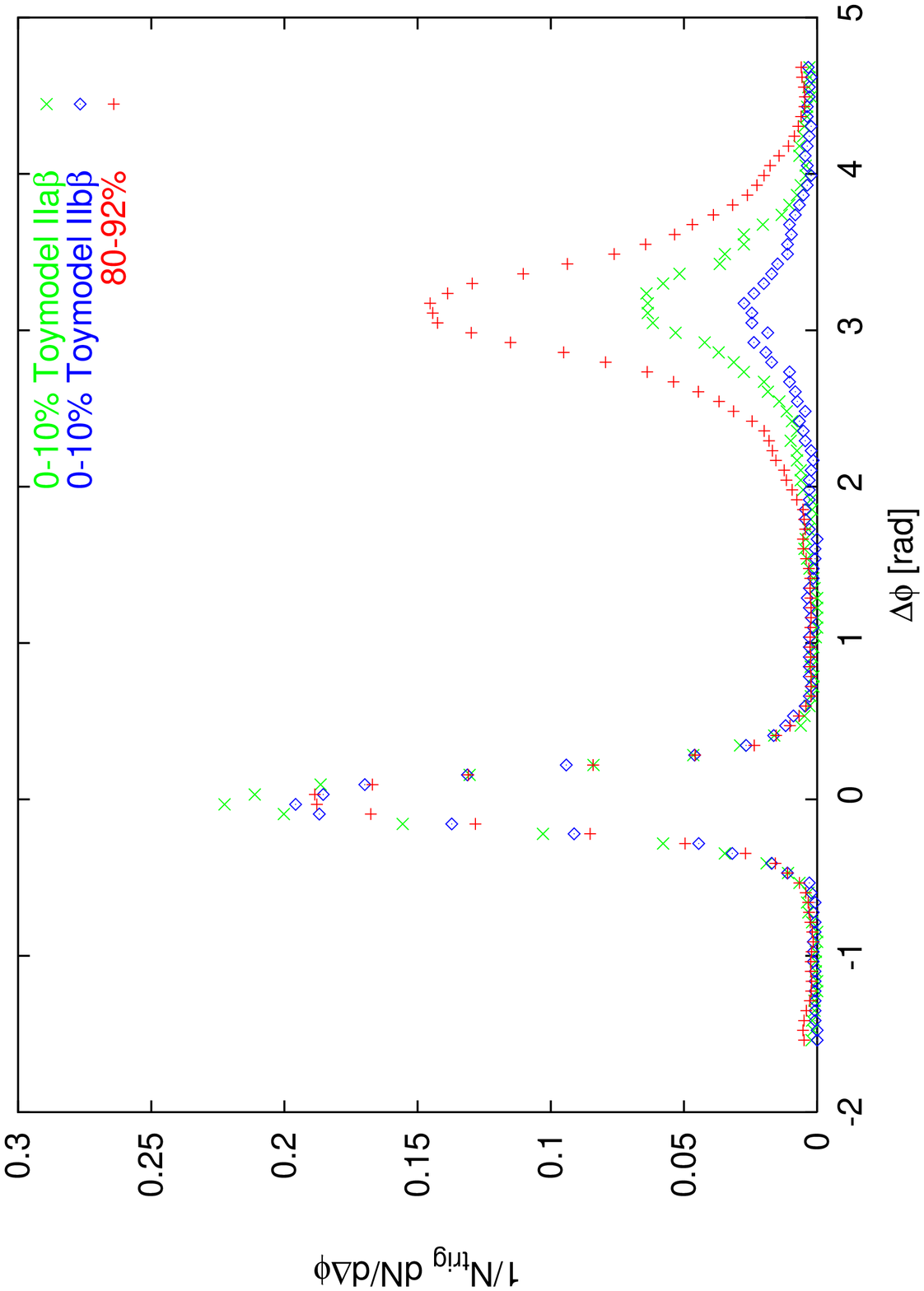}{-90}{fig_toy63-w}{2-particle azimuthal
correlation  for the Toymodel II\,a\,$\beta$ with $\Delta E \propto N_g
\sqrt{E}$ and Toymodel II\,b\,$\beta$ with $\Delta E \propto N_g^2 \sqrt{E}$,
both with non-uniform energy density profile, the 80-92\% data are the same in
both cases}{}

When $\Delta E \propto \sqrt{E}$ the centrality dependence is parabolic (Fig.~\ref{fig_toy63-c}) without a linear component and hence qualitatively different
from the data. The two scenarios Toymodel~II\,a\,$\beta$ and II\,b\,$\beta$ are
again very similar, but this time the curvature is somewhat stronger in the
quadratic setup II\,b\,$\beta$.  

\enlargethispage{-\baselineskip}

In the 2-particle correlation on the other hand the situation is as before: The
suppression obtained with II\,b\,$\beta$ is clearly stronger than with
II\,a\,$\beta$. The effect is even bigger than before and has reached a
reduction by more than 50\% (Fig.~\ref{fig_toy63-w}).

\smallskip

\turnpic[0.4]{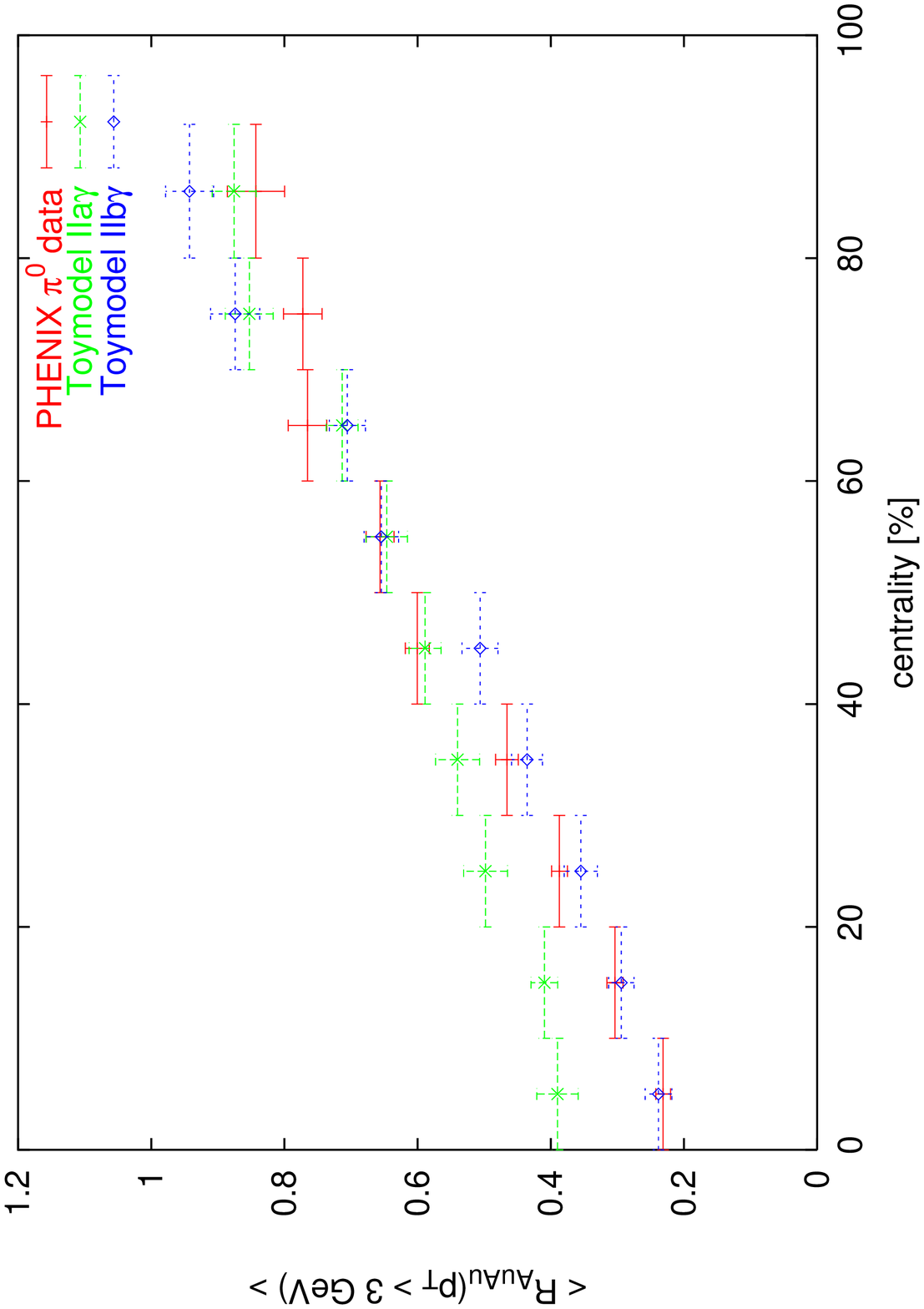}{-90}{fig_toy64-c}{Centrality dependence
of $R_\tn{AuAu}$ for the Toymodel II\,a\,$\gamma$ with $\Delta E \propto N_g$
and Toymodel II\,b\,$\gamma$ with $\Delta E \propto N_g^2$, both with
non-uniform energy density profile}{}

\turnpic[0.4]{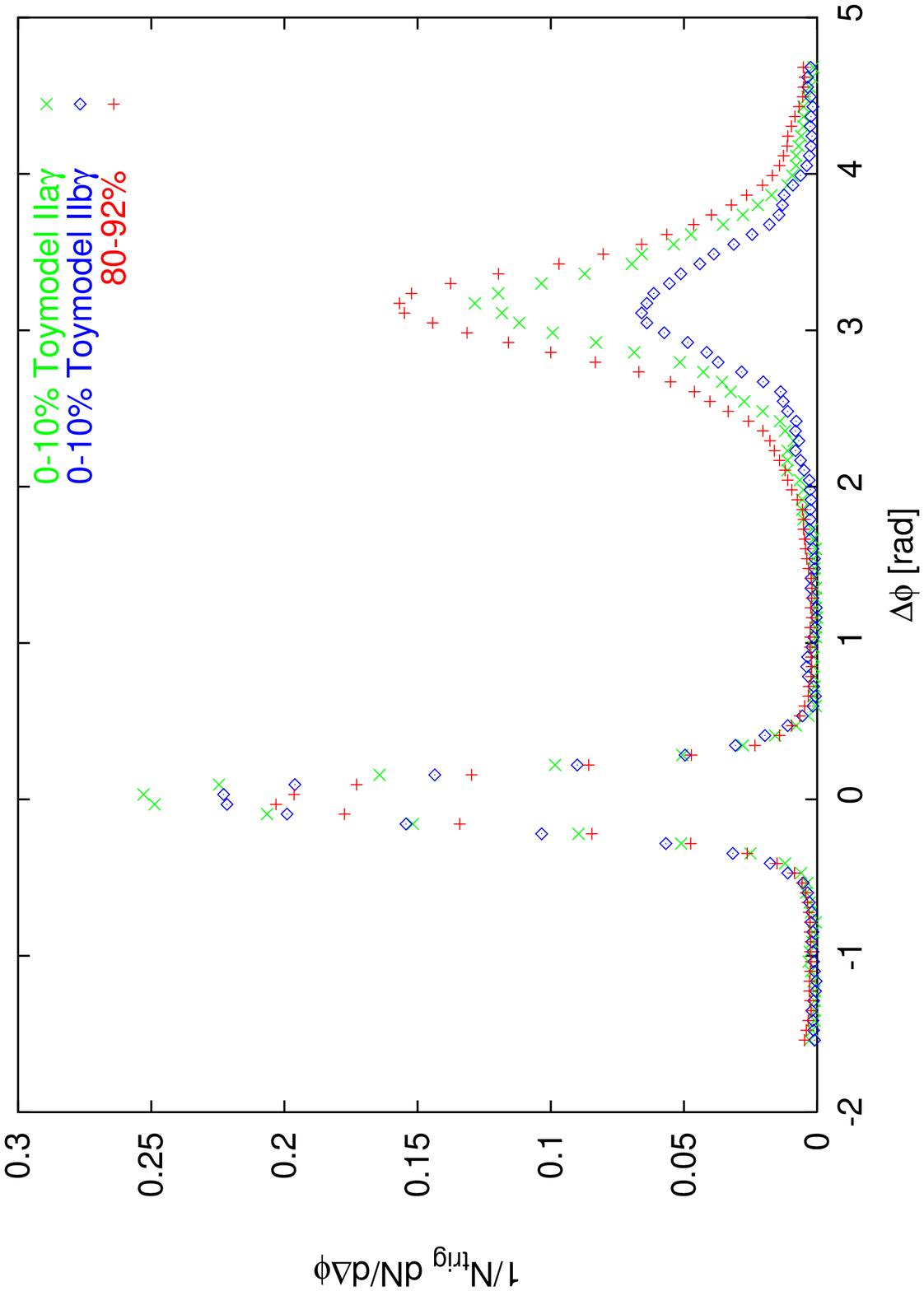}{-90}{fig_toy64-w}{2-particle azimuthal
correlation  for the Toymodel II\,a\,$\gamma$ with $\Delta E \propto N_g$ and
Toymodel II\,b\,$\gamma$ with $\Delta E \propto N_g^2$, both with non-uniform
energy density profile, the 80-92\% data are the same in both cases}{}

Surprisingly the results of the ansatz with $\Delta E$ independent of $E$ are
very similar to those obtained with $\Delta E \propto E$. The centrality
dependence is now linear in both cases and again the II\,a\,$\gamma$ is slightly
flatter than II\,b\,$\gamma$. While the centrality dependence of $R_\tn{AuAu}$
agrees quite well with the data the azimuthal correlation is not as convincing.
There is a suppression which is again stronger for the quadratic version
II\,b\,$\gamma$, but this is still far from a disappearance. It is also slightly
worse than the results obtained with $\Delta E \propto E$.

\enlargethispage{\baselineskip}

\section{Summary and Discussion} %==============================================

In summary, the differences in $R_\tn{AuAu}$ and its centrality dependence
between the configurations with $\Delta E \propto N_g$ and $\Delta E \propto
N_g^2$ are small. $\Delta E \propto N_g^2$ has the tendency to look slightly
better and also the suppression of the away-side jet is stronger. However,
these differences are not big enough to come to a clear cut conclusion. The
$N_g^2$ dependence seems to be favoured, but the $N_g$ dependence is not ruled
out.

\smallskip

The situation becomes partly clearer when it comes to the energy dependence.
Here the $\Delta E \propto \sqrt{E}$ scenario leads to a centrality dependence
of $R_\tn{AuAu}$ that has a clear parabolic shape and is thus qualitatively
different from the linear behaviour observed in the data. The suppression of
the away-side jet is stronger than in the other versions, but since this
observable is not very sensitive to the energy dependence this is not a strong
argument. Besides, the azimuthal correlation is obviously quite sensitive to
details of the fragmentation which is not well under control in this model. So
it is justified to conclude that $\Delta E \propto \sqrt{E}$ is clearly
disfavoured. 

But then it becomes more difficult since it is practically impossible to
distinguish between $\Delta E \propto E$ and $\Delta E$ independent of $E$.
Perhaps the centrality dependence is slightly better in the latter case, but
for the azimuthal correlation it is just the other way around. There are
minimal differences in the shape of $R_\tn{AuAu}$: It has a tendency to fall
mildly when $\Delta E \propto E$ and rise when $\Delta E$ does not depend on
$E$, but the data don't favour one or the other alternative
(Fig.~\ref{fig_toy61vs64-r}). With the presently available data it is therefore
not possible to differentiate between the two options and also with better
statistics it is not trivial since the differences in $R_\tn{AuAu}$ are also at
higher $\pt$ not big. 

There is a difference between the two scenarios when going to higher $\pt$ with
the 2-particle azimuthal correlation. This is shown with an example for the
$\Delta E \propto N_g^2$ configurations in Figure~\ref{fig_toy61vs64-1} and
\ref{fig_toy61vs64-2}. The suppression of the away-side jet is clearly stronger
in Toymodel~II\,b\,$\alpha$ than in II\,b\,$\gamma$. It is also observed that
the height of the away-side peak relative to the near-side peak increases when
going to higher $\pt$, although the nuclear modification factor stays more or
less flat. The increase is thus not due to a "punch-through", i.e.\ particles
that have so much energy that they lose only a very small fraction and emerge
nearly undisturbed from the plasma. In fact, it is connected to the observation
that the jet quenching becomes inefficient when the hard scattering occurs near
the surface of the QGP made in Section~\ref{sec_scijq-azcor}. When the trigger
condition gets more restrictive only jets from partons that suffered hardly any
energy loss are selected. This means that the hard scattering points of the
selected events move more and more towards the surface and the quenching
becomes more inefficient. Consequently the away-side jet reappears even though
the nuclear modification factor stays constant.

\turnpic[0.4]{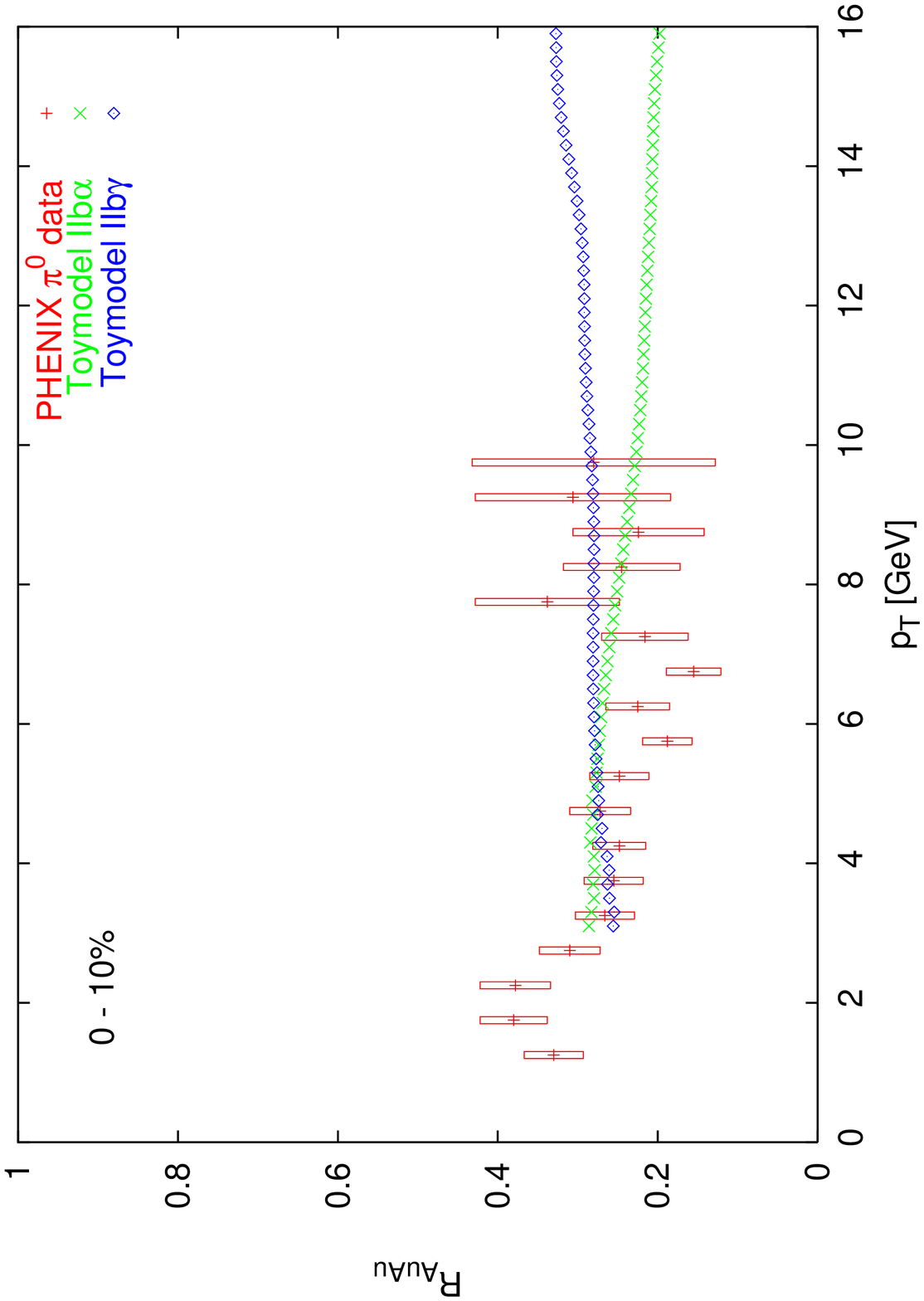}{-90}{fig_toy61vs64-r}{$R_\tn{AuAu}$ in
  Toymodel~II\,b\,$\alpha$ with $\Delta E \propto N_g^2E$ and
  Toymodel~II\,b\,$\gamma$ with $\Delta E \propto N_g^2$}{}
\turnpic[0.4]{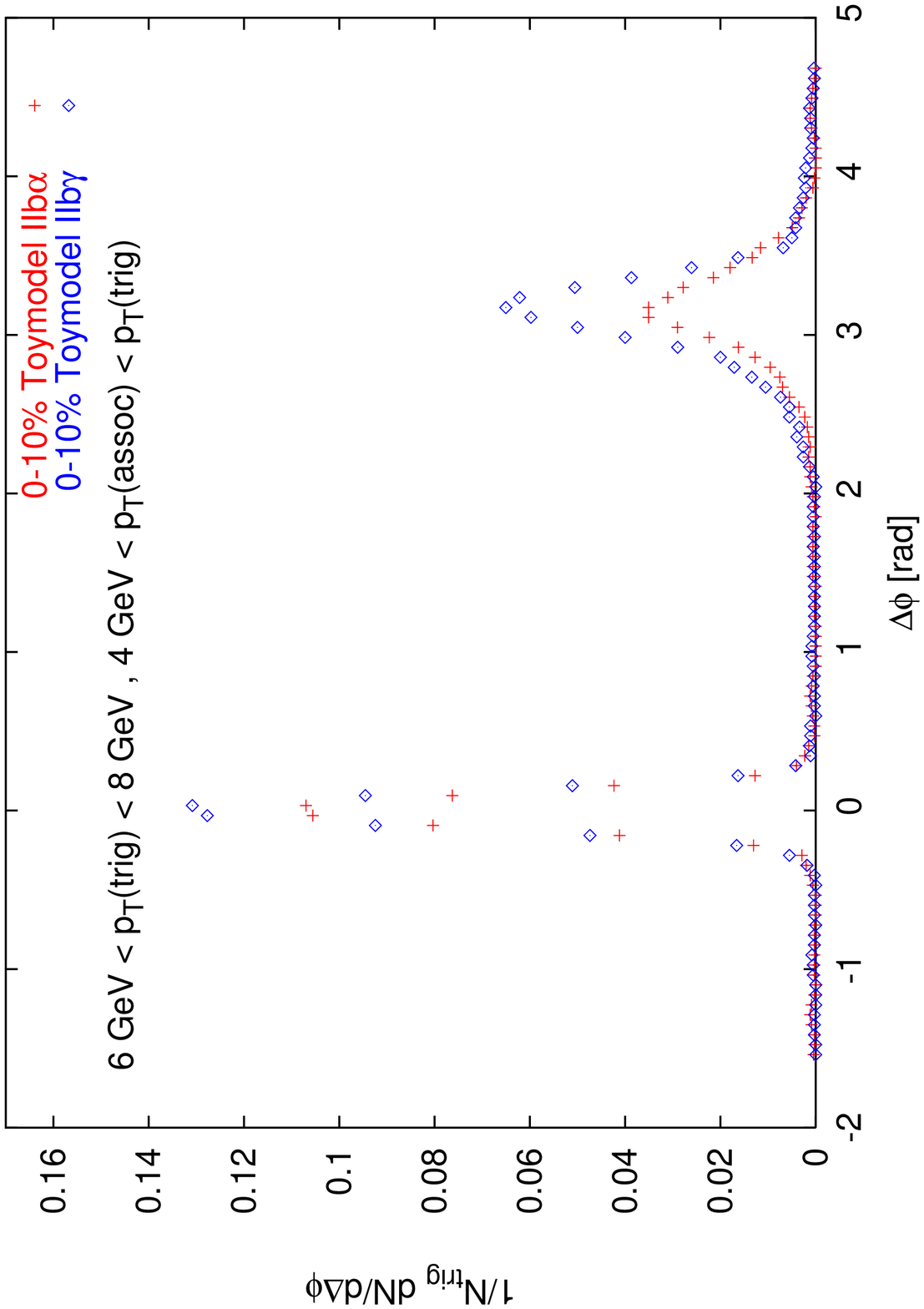}{-90}{fig_toy61vs64-1}{2-particle azimuthal
correlation  for the Toymodel II\,b\,$\alpha$ with $\Delta E \propto N_g^2E$ and
Toymodel II\,b\,$\gamma$ with $\Delta E \propto N_g^2$ for 0-10\% centrality,
trigger particles satisfy $\unit[6]{GeV} < \pt(\tn{trig}) < \unit[8]{GeV}$ and
associated particles have $\unit[4]{GeV} < \pt(\tn{assoc}) < \pt(\tn{trig})$}{}
\turnpic[0.4]{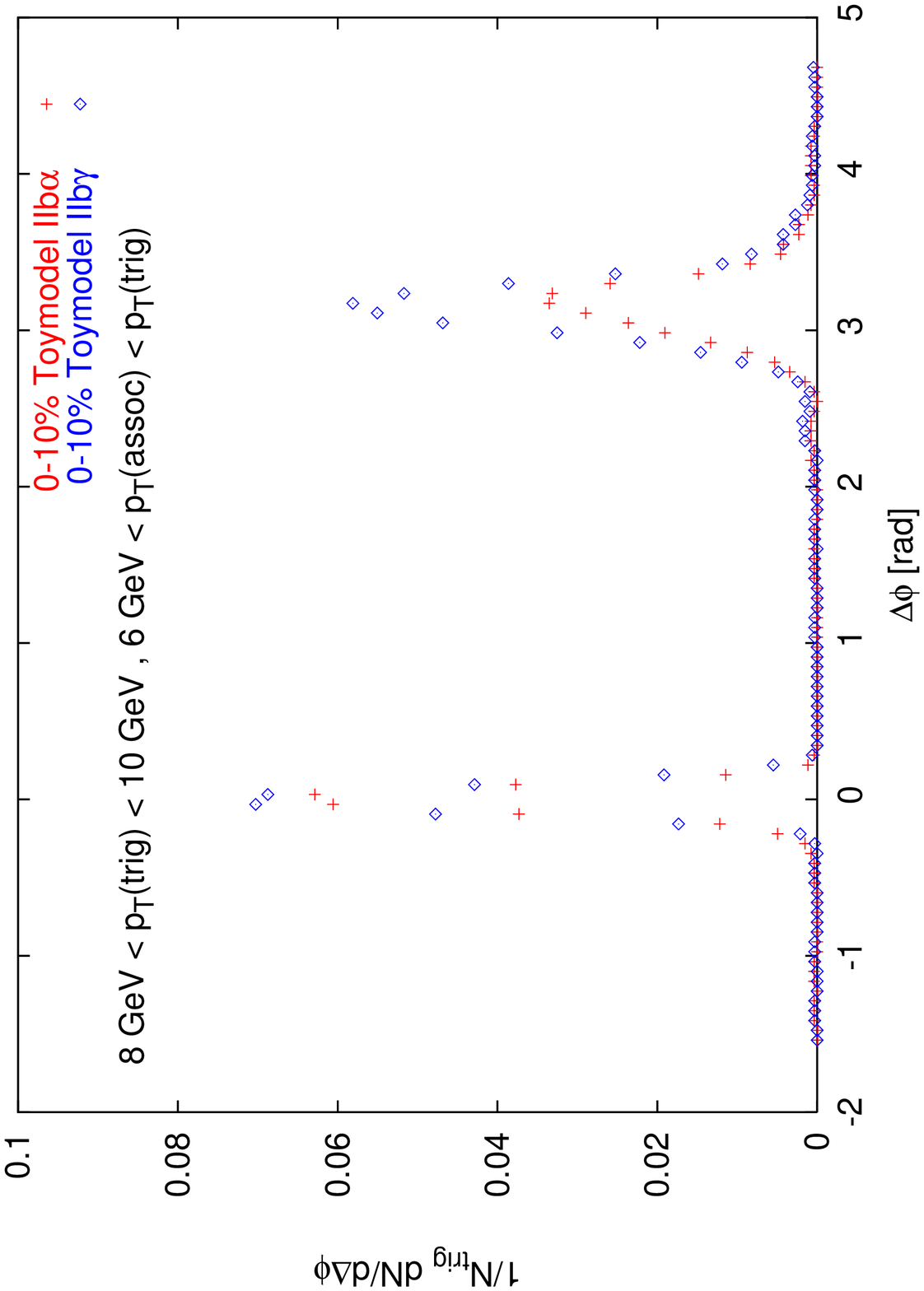}{-90}{fig_toy61vs64-2}{2-particle azimuthal
correlation  for the Toymodel II\,b\,$\alpha$ with $\Delta E \propto N_g^2E$ and
Toymodel II\,b\,$\gamma$ with $\Delta E \propto N_g^2$ for 0-10\% centrality,
trigger particles satisfy $\unit[8]{GeV} < \pt(\tn{trig}) < \unit[10]{GeV}$ and
associated particles have $\unit[6]{GeV} < \pt(\tn{assoc}) < \pt(\tn{trig})$}{}

\smallskip

\enlargethispage{\baselineskip}

The results of this section show that the gluon bremsstrahlung scenario in the
high energy limit (Eq.~\ref{eq_gluonrad1}) is within the bounds of possibility
while the result obtained for $L > L_\tn{cr}$ (Eq.~\ref{eq_gluonrad2}) seems to
be ruled out, at least when it is not combined with some other mechanism. 

The SCI jet quenching model also has an energy dependence
(Fig.~\ref{fig_endep}) which is not consistent with the data. But here it is
clear that there is an additional effect missing since it can account only for
50\% of the observed jet quenching. And it cannot be expected that the SCI jet
quenching model describes the whole effect. So there is still freedom to
resolve the present difficulties when the model is combined with induced gluon
radiation (or possibly even more effects) to get a full description of the
energy loss.

\smallskip

The inhomogeneous energy density distribution works better than the
homogeneous. This agrees with the findings of the SCI jet quenching model and
there is no reason why the energy density should be homogeneous. Concerning a
possible hadronic layer the same considerations as for the SCI jet quenching
model apply, namely that it can have an impact on the influence of the
centrality depending on how the thickness of the layer changes with centrality.

\smallskip

The fact that the centrality dependence does not level off and become flat at a
certain level illustrates that the available path lengths are limited
predominantly by the geometry and not by the QGP lifetime. Otherwise, at some
point an increase of the size of the overlap region (i.e.\ going to more central
events) would not result in longer path lengths and the nuclear modification
factor would be unchanged.

\smallskip

Since the Toymodel makes use of the same model for the geometry (e.g.\
distribution of hard scatterings in the overlap region), the QGP etc.\ as the
SCI jet quenching model, it is expected to suffer from the same problems with
the 2-particle azimuthal correlations. However, because they are mainly due to
very general properties of the QGP one would think that the energy loss
mechanism has to compensate the difficulties. The fact that even those versions
of the Toymodel, which yield a good description of the nuclear modification
factor, fail to describe the disappearance of the away-side jet strengthens the
case for the hypothesis that the azimuthal correlation is fairly sensitive to
the fragmentation scheme and function. However, it seems that it is also
influenced by the path length dependence of the energy loss mechanism.

The nuclear modification factor on the other hand was found to depend only very
weakly on the exact energy and path length correlation with the energy loss and
is also insensitive to the energy density distribution. The centrality
dependence of $R_\tn{AuAu}$ probes mainly the energy dependence.

Now one can also try to answer the question of how the linear centrality
dependence comes about. Strangely enough, it seems that a lot of different
effects conspire to yield a very simple form in the end. It has been shown that
apart from the geometry (and the QGP lifetime) also the Cronin effect, the
energy density distribution, the energy dependence of the energy loss and to
some extend the impact of the path length on the energy loss influence the
behaviour with the centrality.

\chapter{Conclusions and Outlook}
Given the wealth of different observables that point towards QGP formation at
RHIC, it seems that indeed the long-sought quark-gluon plasma has been created.
Among them, the jet quenching plays an important role. For SPS the question
whether a QGP is formed or not is not so easy to answer. But as has been shown
in this thesis, the shapes of the $\pt$-spectra are not very different from
those measured at RHIC. Since this applies also for other observables, it is
very well conceivable that the phase boundary has already been crossed in
central collisions at SPS. However, from SPS data it is very difficult to
investigate the properties of the new state of matter. This is to some extend
possible at RHIC, but still a very difficult task that requires a profound
understanding in particular of the high-$\pt$ physics and the jet quenching.

\medskip

It was shown that soft rescattering of a hard scattered parton in a QGP can
contribute significantly to the overall energy loss. With the model introduced
in the first part of this thesis, roughly 50\% of the observed effect in
central collisions was reproduced. The magnitude depends on the momentum
transfer per scattering which is a parameter, but it is clear that losses
induced by scattering may not simply be neglected. 

It is not a problem that the model cannot account for the whole effect, since
it cannot be considered complete in the sense that all processes contributing
to the energy loss are included. The most obvious example is medium induced
gluon radiation. It seems natural that scattering of the parton should induce
gluon bremsstrahlung resulting in further energy loss. Additional processes
will of course also change the centrality dependence so that at this stage it
should not be overemphasised that the centrality dependence found in the SCI
jet quenching model does not match the data.

Nevertheless, the model is in two respects not very well motivated. First, it 
could be argued that the assumption of independent scatterings is not
justified, at least at the early times when the density in the QGP is high,
since soft interactions happen on relatively long time scales. Then
interferences might become important and it is not clear how they should be
treated in a Monte Carlo generator. The second point is that since the momentum
transfer is small it is impossible to resolve individual gluons. This would
mean that the plasma has to be treated as a colour charged gluonic field with
which the partons would nevertheless interact. But the parton-field interaction
may not be very different from the parton-gluon interaction. Since there is
poor theoretical guidance for soft interactions, the parton-gluon scattering
can be used as a model.

It should also be noted that the (single) gluon exchange model should not be
taken too literally. The Gaussian momentum transfer should rather be
interpreted as due to a superposition of many-gluon exchanges. The exact
details of this process are not relevant for this study.

The main conclusion is unaffected by these problems that are unescapable when
dealing with soft physics: Soft scattering can result in a sizable energy loss.

\bigskip

The lesson to learn from the investigations of the overall features of partonic
energy loss is that it is difficult to distinguish between different scenarios
given the presently available data. $\Delta E \propto \sqrt{E}$ seems
unprobable because it yields the wrong centrality dependence of $R_\tn{AuAu}$.
But $\Delta E \propto E$ and $\Delta E \propto \tn{const}$ are practically
undistinguishable, this will maybe become possible with better data at higher
$\pt$. $\Delta E \propto N_g$ and $\Delta E \propto N_g^2$ are also similar,
$\Delta E \propto N_g^2$ looks slightly better. This may be interpreted as a
hint for coherent gluon radiation but it is not a strong indication. In
summary, it has not been possible to get to a clear cut conclusion for or
against coherent gluon bremsstrahlung.

\smallskip

The 2-particle azimuthal correlation seems to depend on details of the
fragmentation procedure, an issue that has not been addressed in great detail
in this study. This makes it very difficult to judge the results of the models
but also to understand the data when attention is paid to the details. Beyond
that, the azimuthal correlation shows a certain sensitivity to the path length
dependence of the energy loss. In contrast the nuclear modification factor is
insensitive to many details and the centrality dependence of $R_\tn{AuAu}$
is ruled mainly by the energy dependence.

\bigskip

For the future it is desirable to gain a better understanding of the interplay
of scattering and gluon radiation. Unfortunately, including radiation in the SCI
jet quenching model is not free of difficulties. While incoherent radiation is
rather straightforward the coherence effects are extremely difficult to handle
in a Monte Carlo framework.

Another unsolved problem are the medium modifications to hadronisation. This is
also highly non-trivial, mainly due to lack of information. Moreover, it is
difficult to disentangle the effect of modified fragmentation from other medium
modifications. Hopefully, one can get some guidance from the observation of
modified fragmentation in cold nuclear matter (e.g.\ \cite{hermes}).

If one wants to fully exploit jet tomography, i.e.\ use the jet quenching to
determine the properties of the QGP, one first has to come to a consistent
picture and get the partonic energy loss theoretically better under control.
This means that further effort has to be made in order to gain a better
understanding of the underlying physical principles.

\chapter*{Acknowledgements}
\pagestyle{plain}

I want to thank my supervisors Johanna Stachel and Gunnar Ingelman for the
support I got from them, the freedom I had and the confidence they had in me.
Many thanks also to the rest of the theory group in Uppsala for a wonderful
time I spend there and many interesting discussions, especially to Johan
Rathsman for help with physics and my code. And last but not least I want to
thank Alessio Mangiarotti and Klaus Schneider for proof-reading my thesis and
Marc Stockmeier and the FOPI groug for an enjoyable time in Heidelberg.

\cleardoublepage

\pagestyle{headings}

\begin{appendix}

\chapter{Fit Parameters for Identified Particle Spectra}
\label{app_spectra}
\begin{figure}[ht]
\centering
\begin{turn}{-90}
\includegraphics[scale=0.39]{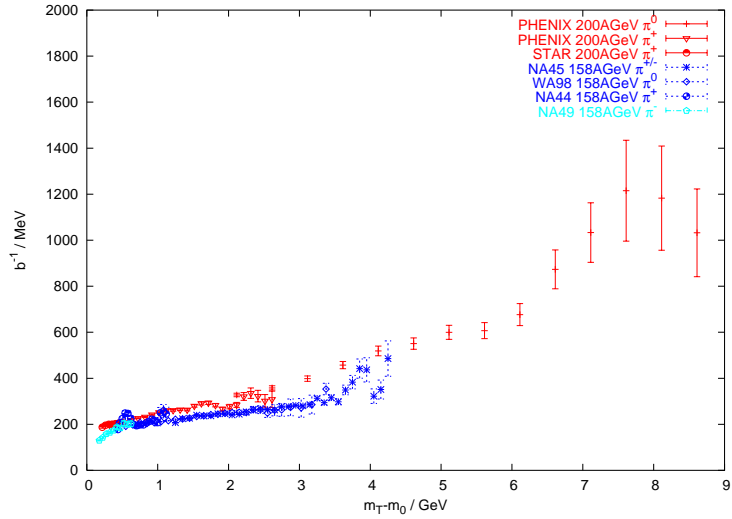}
\end{turn}
\caption{Fit parameters for the pion spectra}
\label{fig_pion}
\end{figure}

\begin{figure}[ht]
\centering
\begin{turn}{-90}
\includegraphics[scale=0.39]{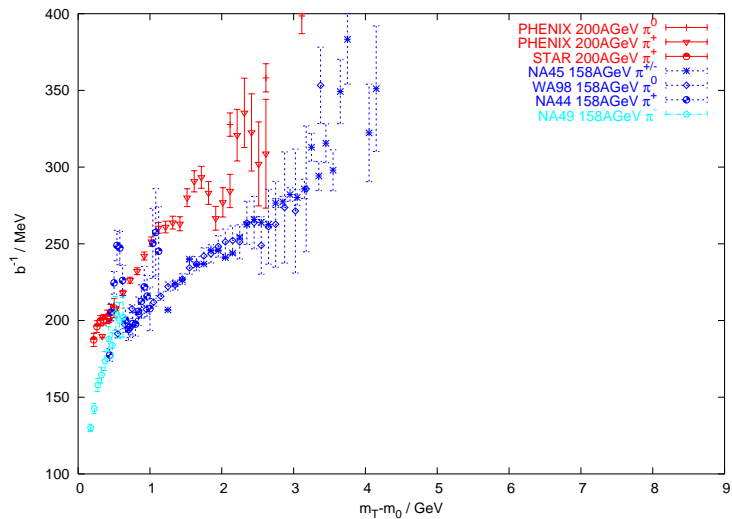}
\end{turn}
\caption{Enlarged low $\pt$ region of the above pion results}
\label{fig_pion2}
\end{figure}

\begin{figure}[ht]
\centering
\begin{turn}{-90}
\includegraphics[scale=0.39]{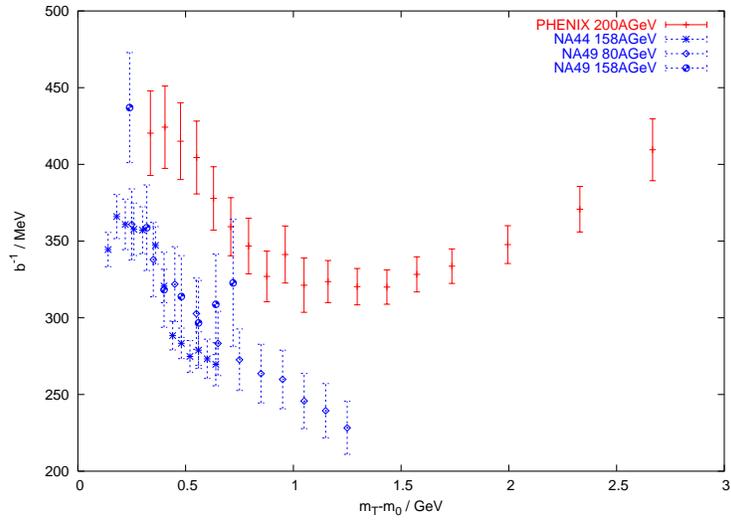}
\end{turn}
\caption{Fit parameters for the feed-down corrected proton spectra}
\label{fig_proton}
\end{figure}

\begin{figure}[ht]
\centering
\begin{turn}{-90}
\includegraphics[scale=0.39]{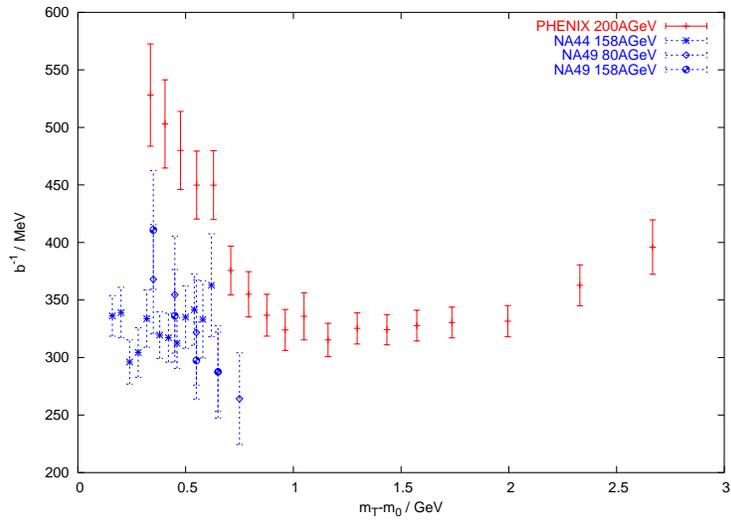}
\end{turn}
\caption{Fit parameters for the feed-down corrected antiproton spectra}
\label{fig_pbar}
\end{figure}

\begin{figure}[ht]
\centering
\begin{turn}{-90}
\includegraphics[scale=0.39]{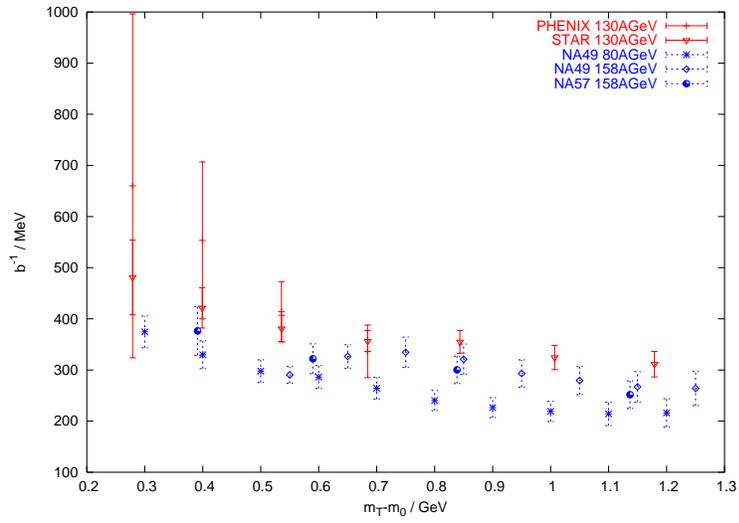}
\end{turn}
\caption{Fit parameters for the $\Lambda$ spectra}
\label{fig_lambda}
\end{figure}

\begin{figure}[ht]
\centering
\begin{turn}{-90}
\includegraphics[scale=0.39]{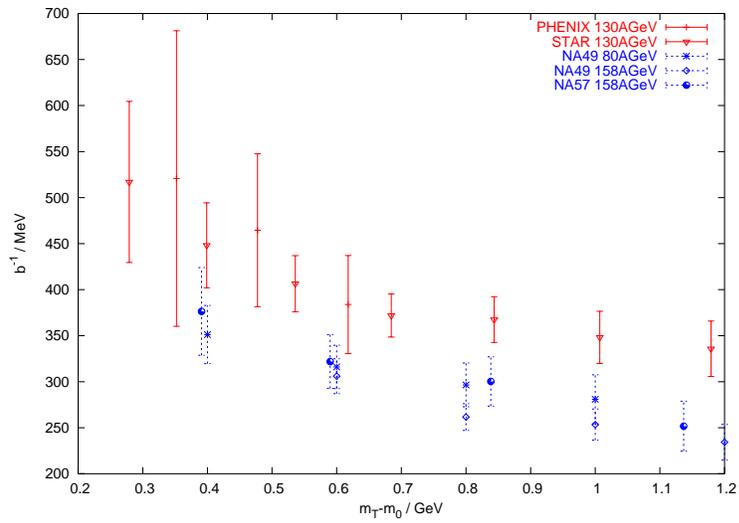}
\end{turn}
\caption{Fit parameters for the $\bar \Lambda$ spectra}
\label{fig_lambdabar}
\end{figure}

\chapter{Complete Results from the Toymodel} 
\label{app_toymodel}
\begin{figure}[ht]
 \centering
 \begin{minipage}{.45\textwidth}
  \centering
  \includegraphics[scale=0.325]{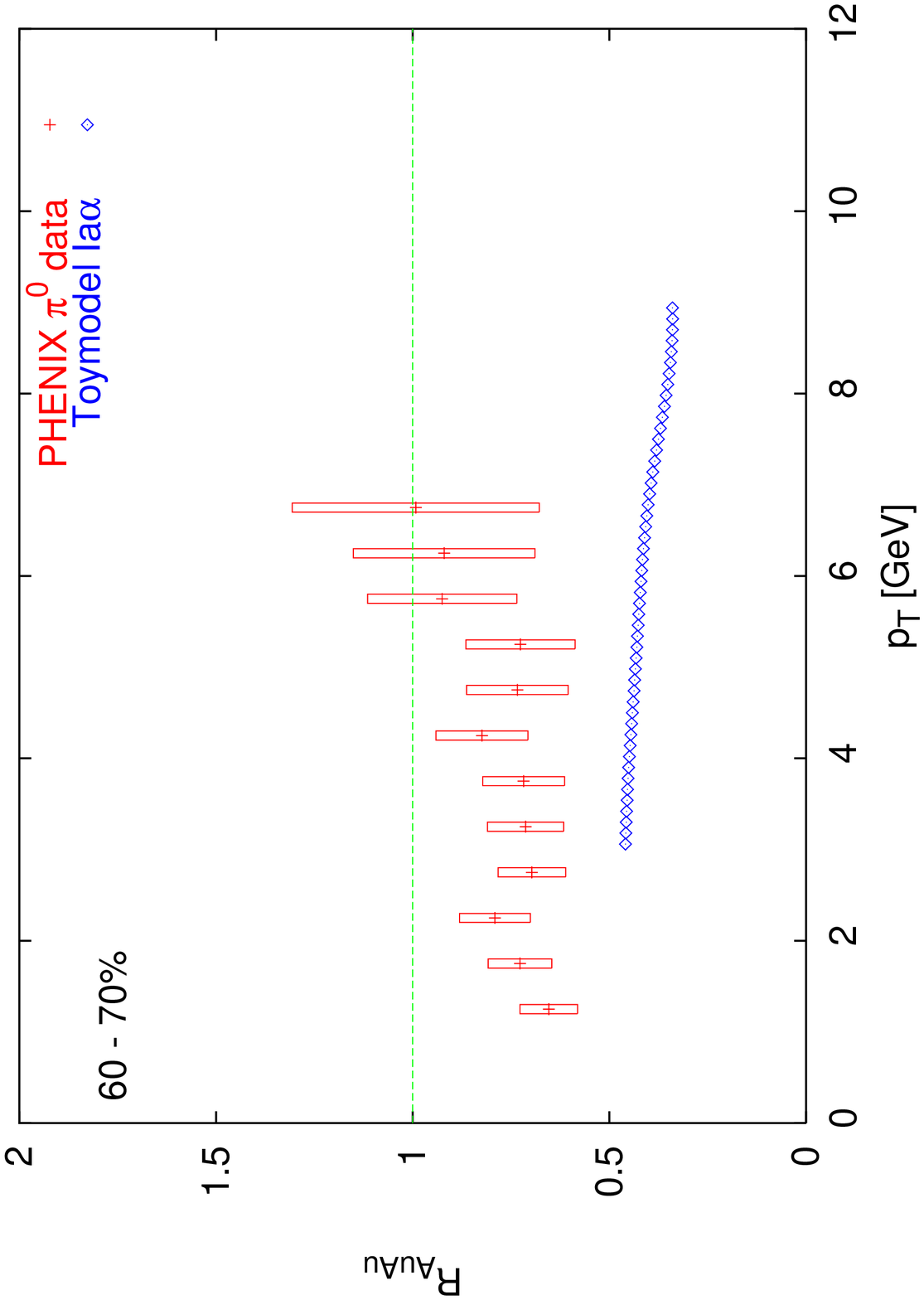}
 \end{minipage} 
 \begin{minipage}{.45\textwidth}
  \centering
  \includegraphics[scale=0.325]{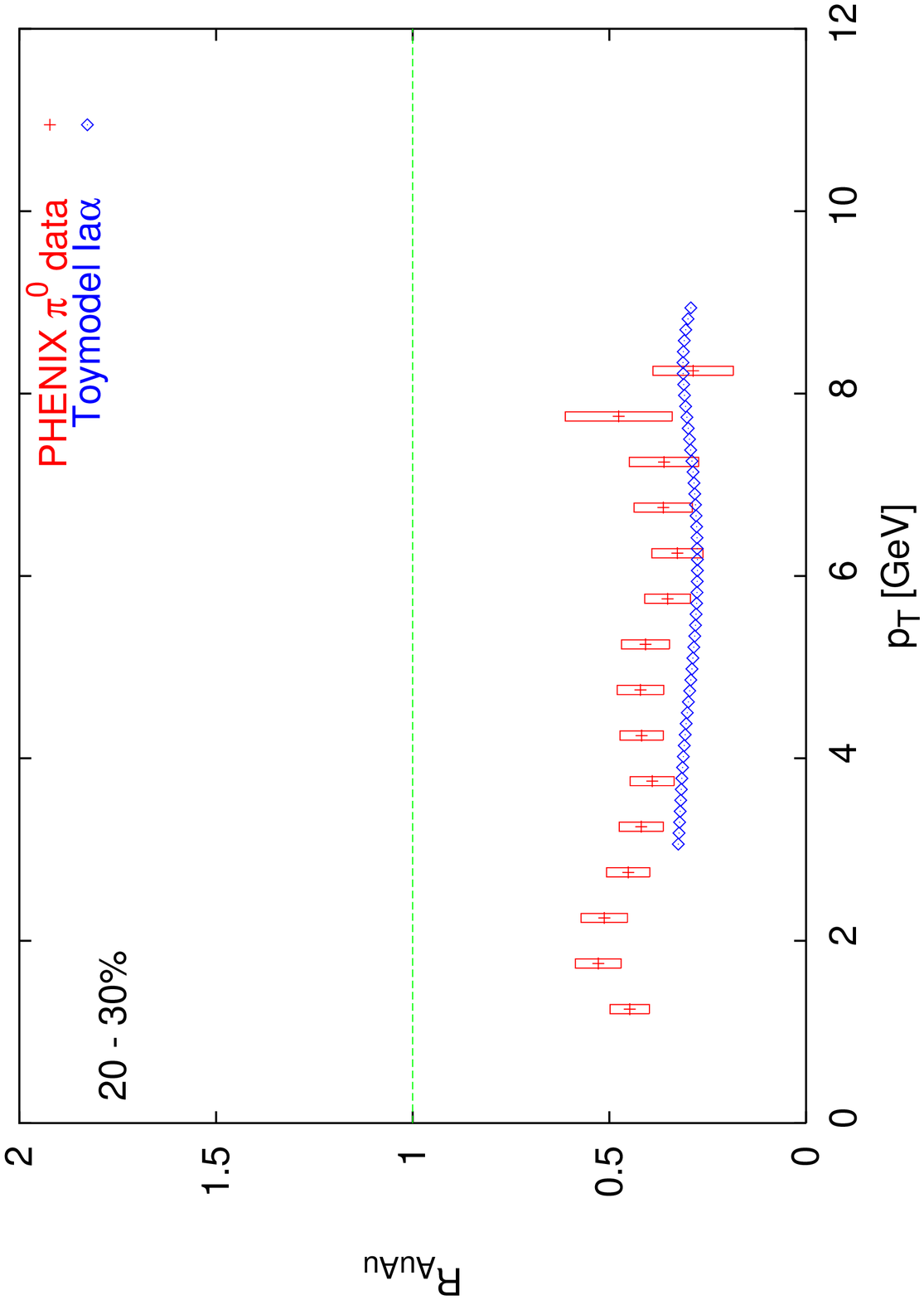}
 \end{minipage} 
 \begin{minipage}{.45\textwidth}
  \centering
  \includegraphics[scale=0.325]{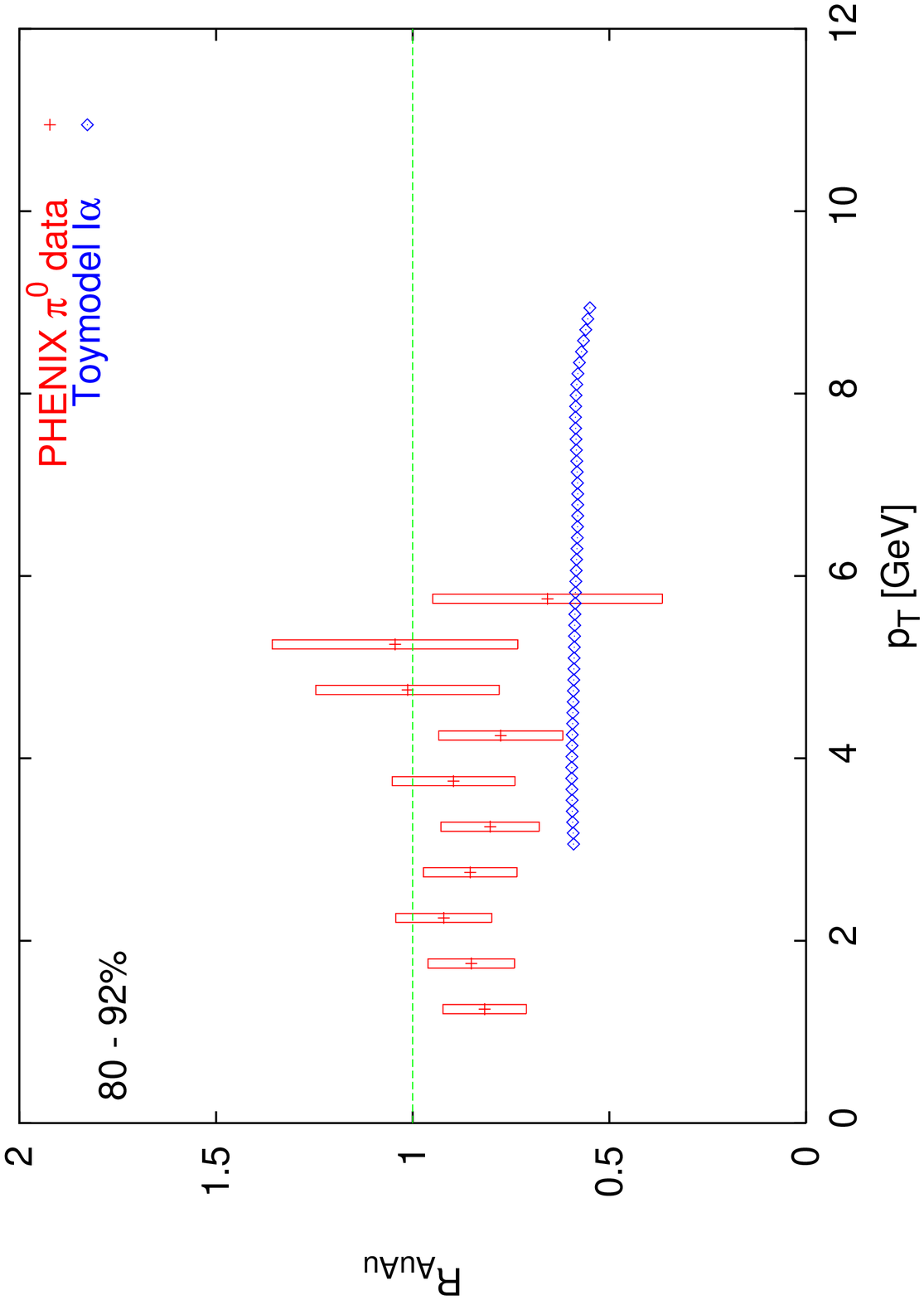}
 \end{minipage} 
 \begin{minipage}{.45\textwidth}
  \centering
  \includegraphics[scale=0.325]{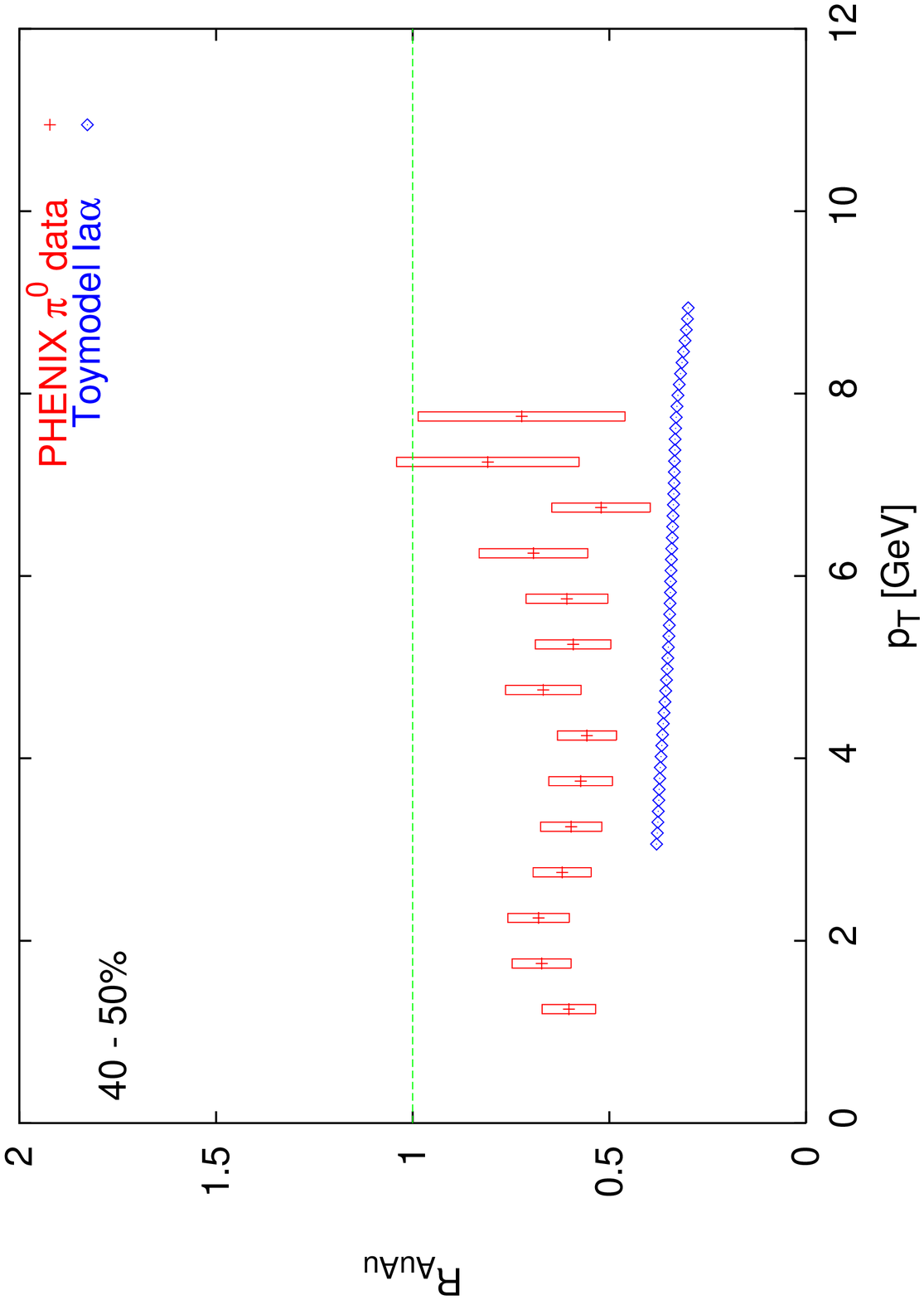}
 \end{minipage} 
\end{figure}

\begin{figure}[ht]
 \centering
 \begin{minipage}{.45\textwidth}
  \centering
  \includegraphics[scale=0.325]{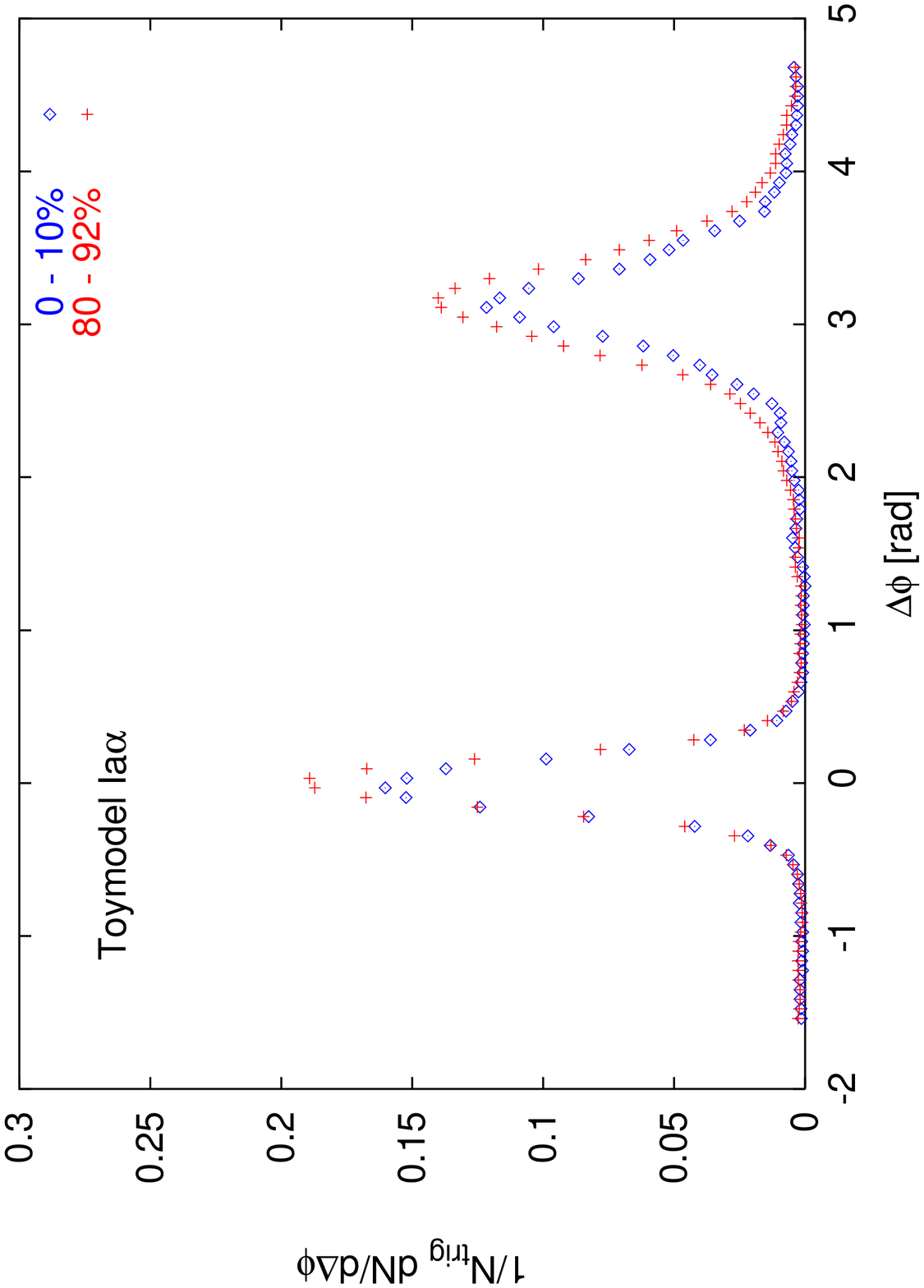}
 \end{minipage} 
 \begin{minipage}{.45\textwidth}
  \centering
  \hspace*{.4\textwidth}
 \end{minipage} 
 \begin{minipage}{.45\textwidth}
  \centering
  \includegraphics[scale=0.325]{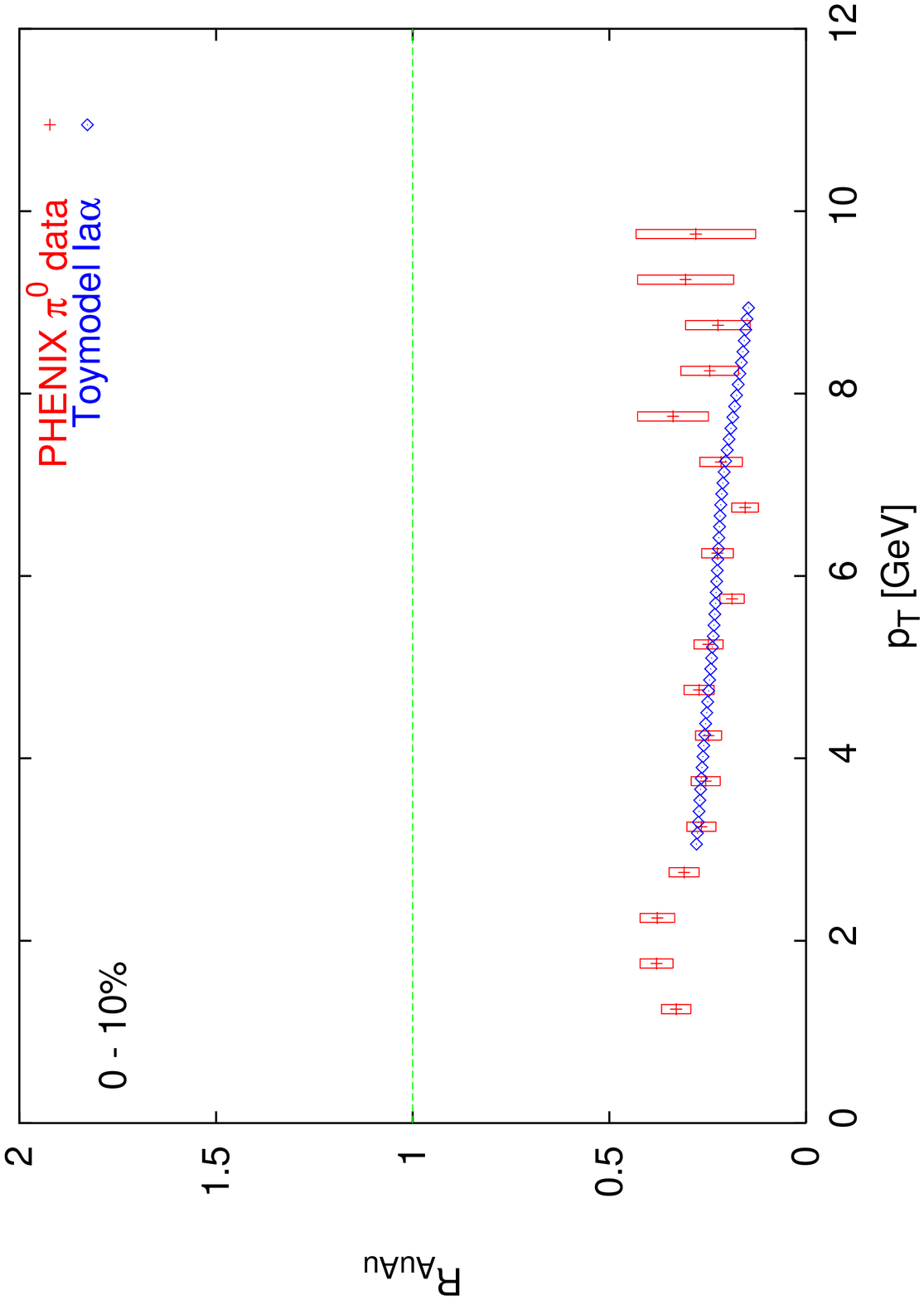}
 \end{minipage} 
 \begin{minipage}{.45\textwidth}
  \centering
  \includegraphics[scale=0.325]{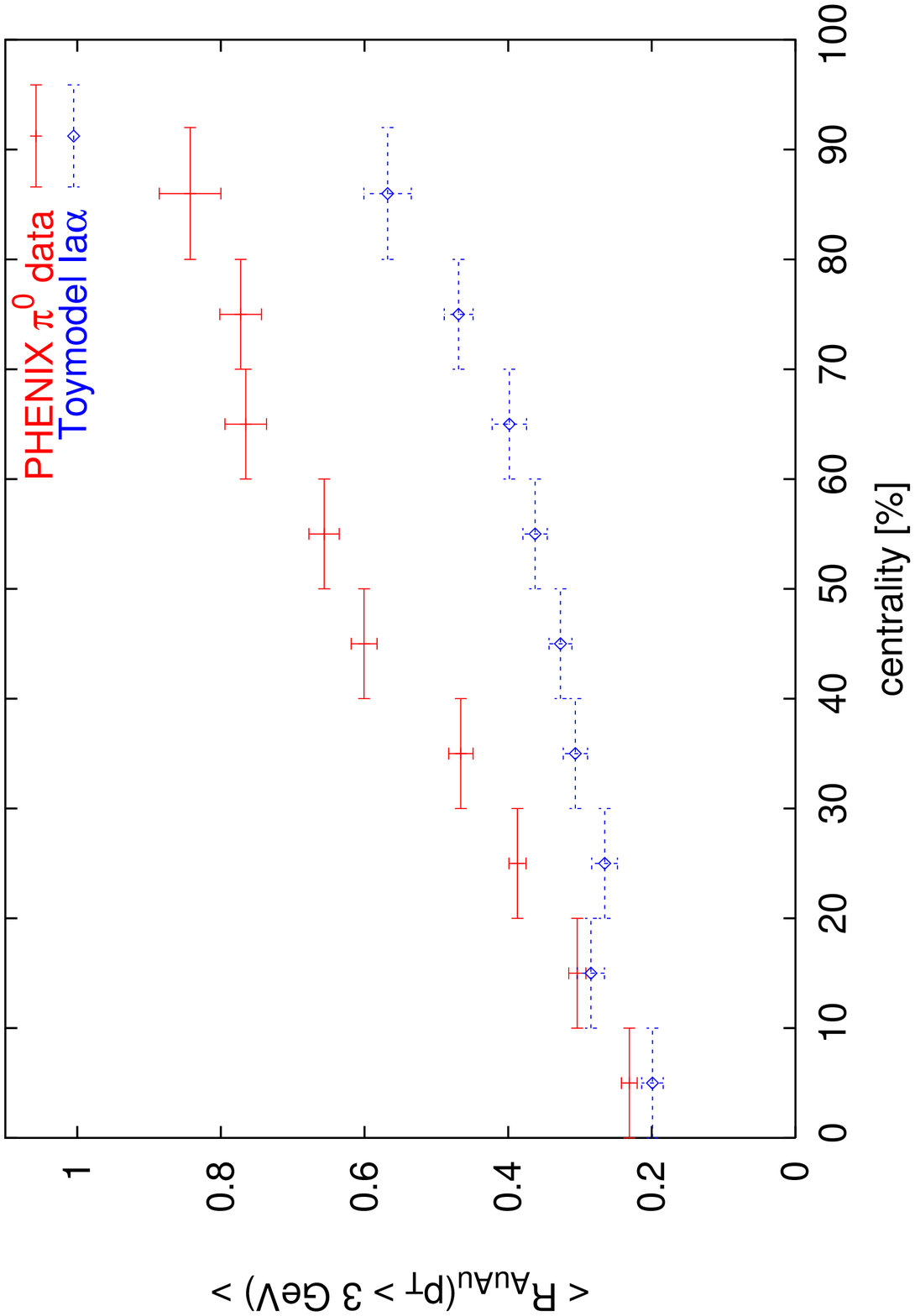}
 \end{minipage} 
 \caption{Results for the Toymodel I\,a\,$\alpha$ with $\Delta E \propto N_g E$
 and homogeneous energy density distribution}
\end{figure}

\begin{figure}[ht]
 \centering
 \begin{minipage}{.45\textwidth}
  \centering
  \includegraphics[scale=0.325]{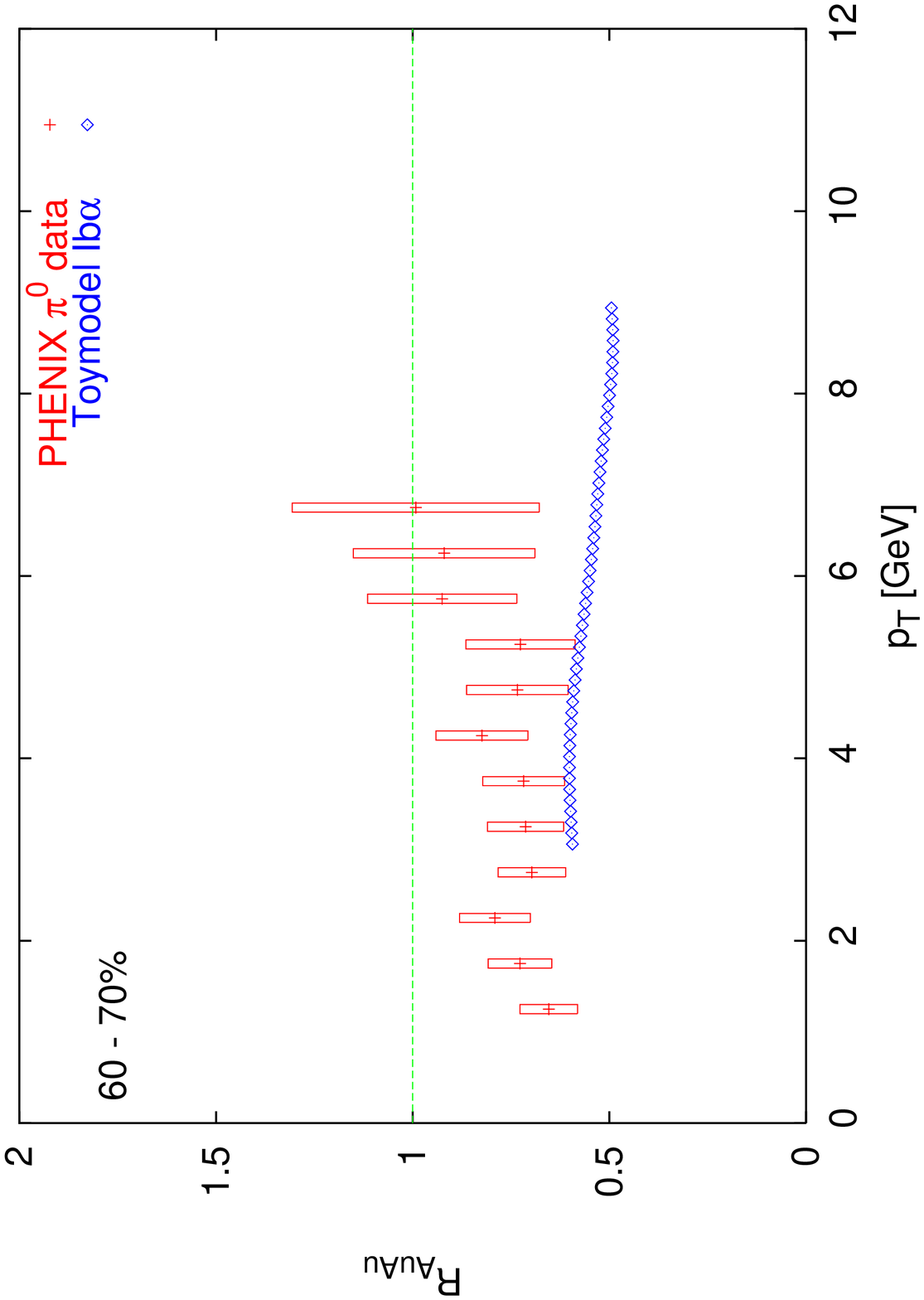}
 \end{minipage} 
 \begin{minipage}{.45\textwidth}
  \centering
  \includegraphics[scale=0.325]{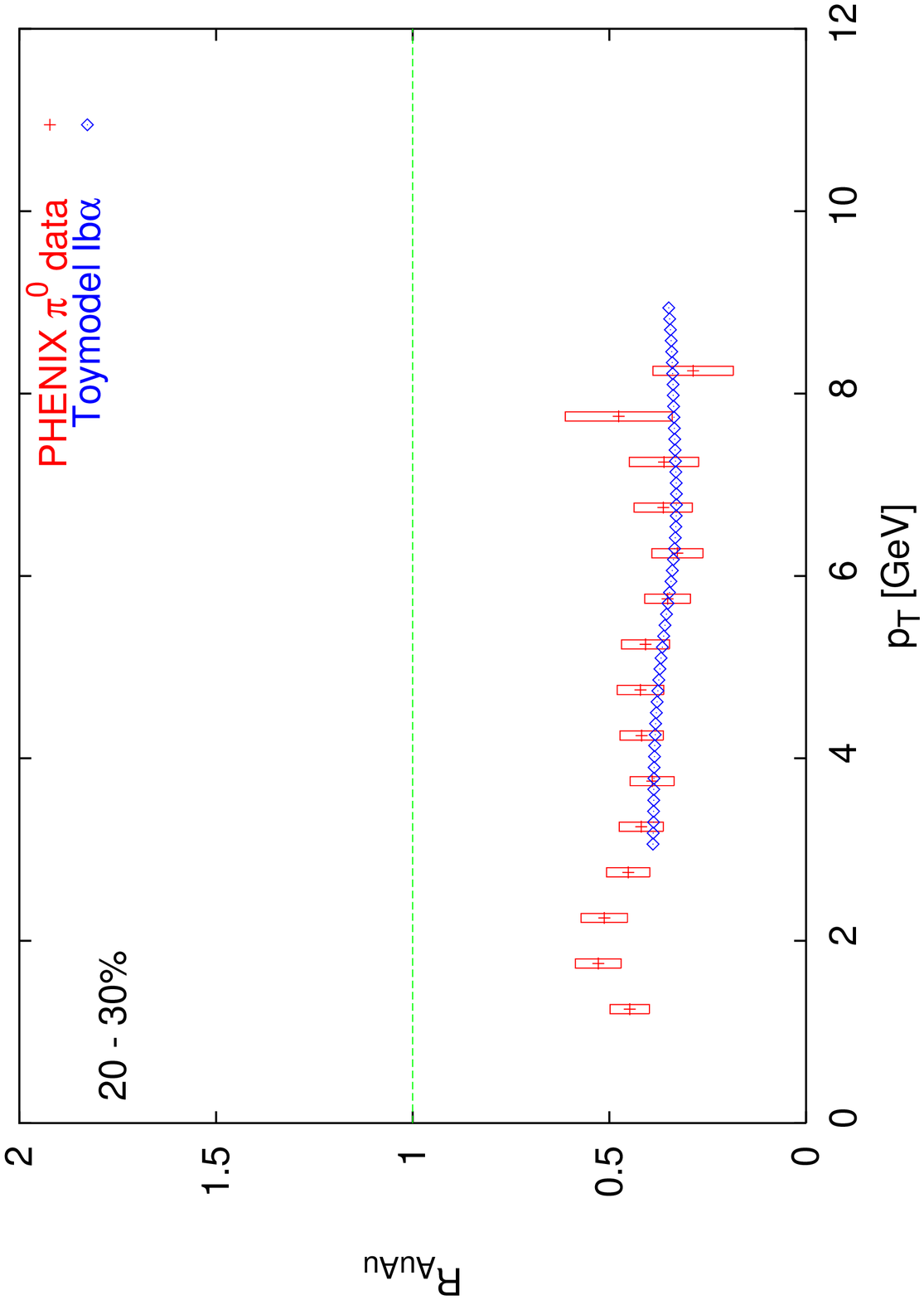}
 \end{minipage} 
 \begin{minipage}{.45\textwidth}
  \centering
  \includegraphics[scale=0.325]{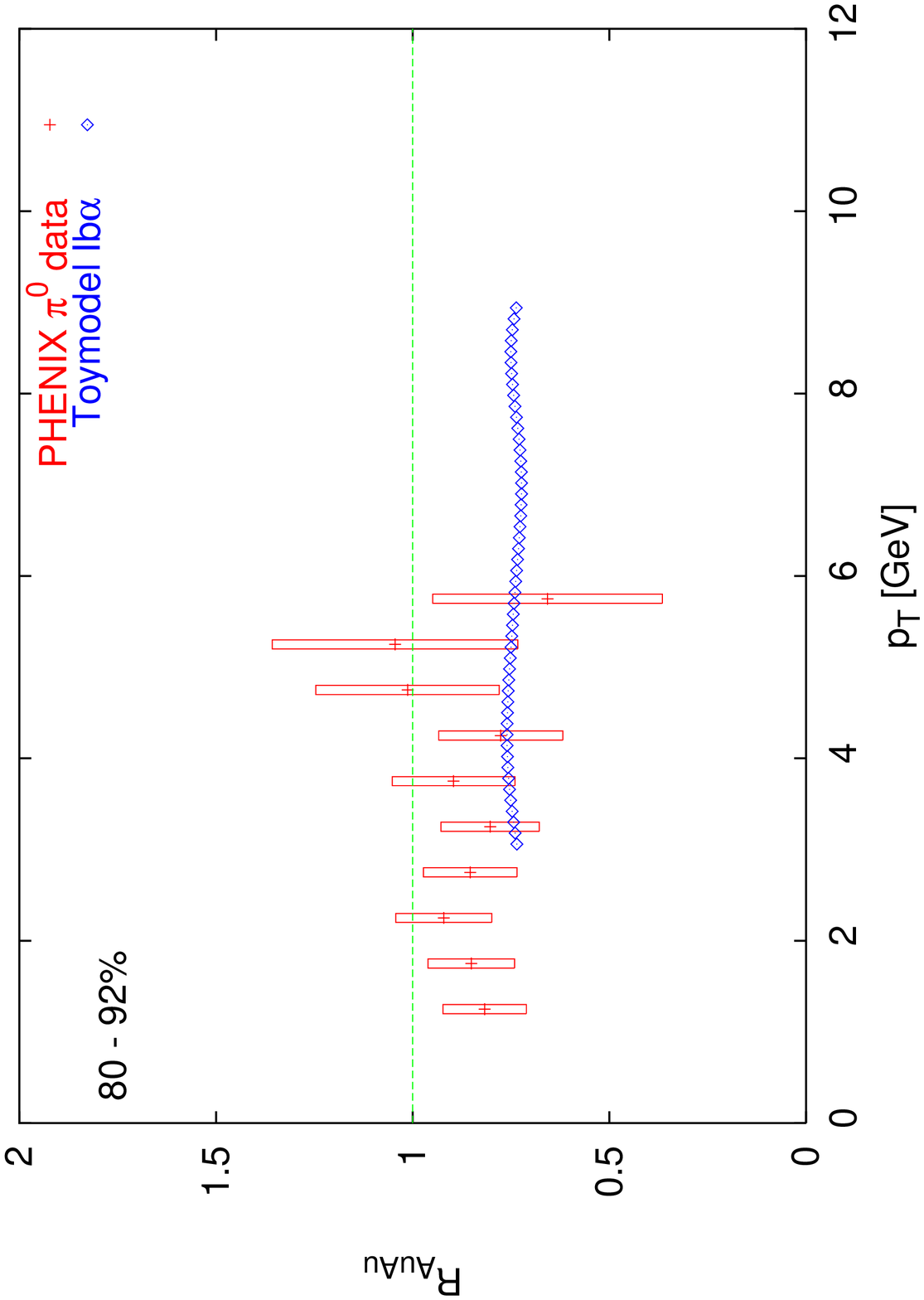}
 \end{minipage} 
 \begin{minipage}{.45\textwidth}
  \centering
  \includegraphics[scale=0.325]{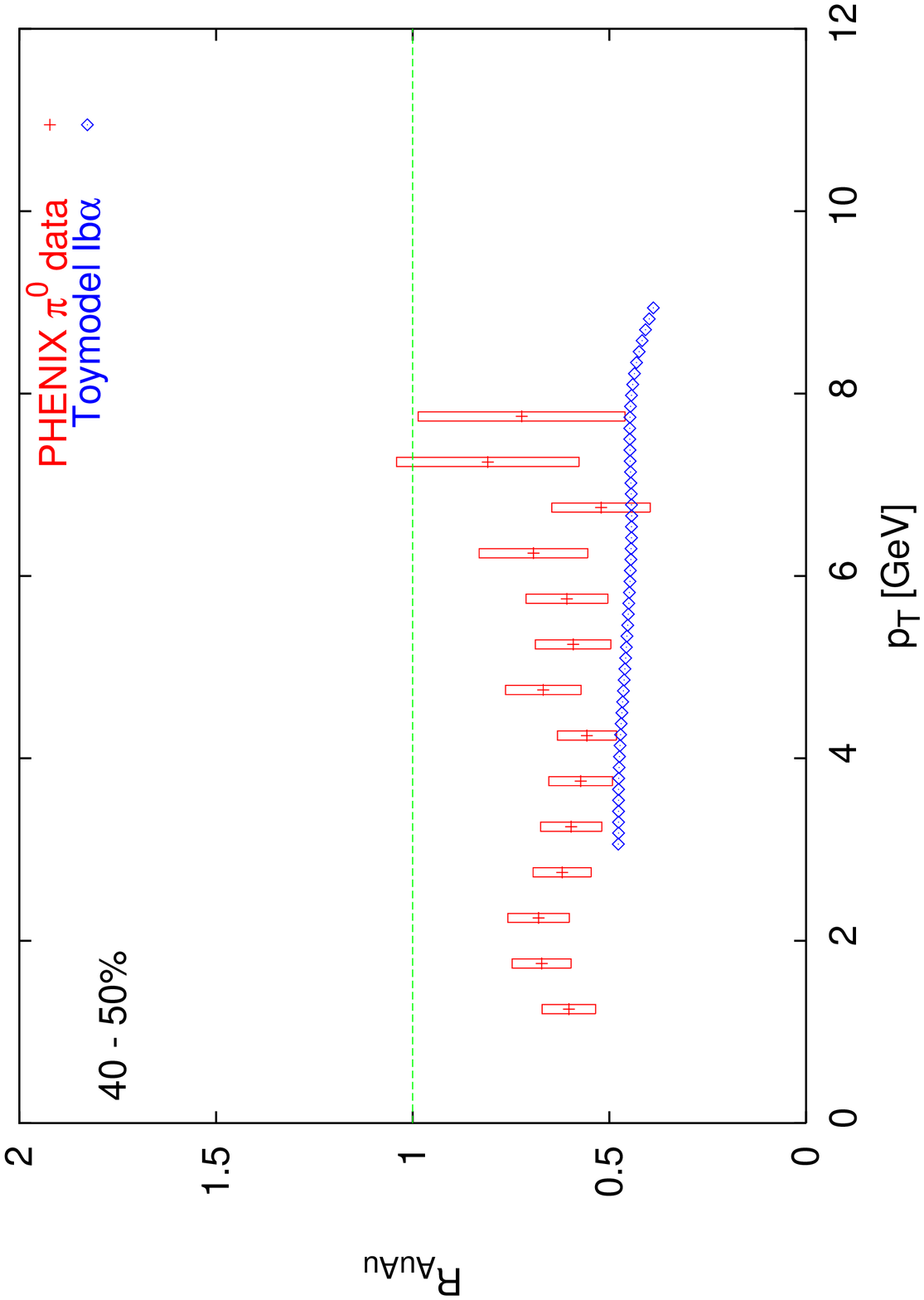}
 \end{minipage} 
\end{figure}

\begin{figure}[ht]
 \centering
 \begin{minipage}{.45\textwidth}
  \centering
  \includegraphics[scale=0.325]{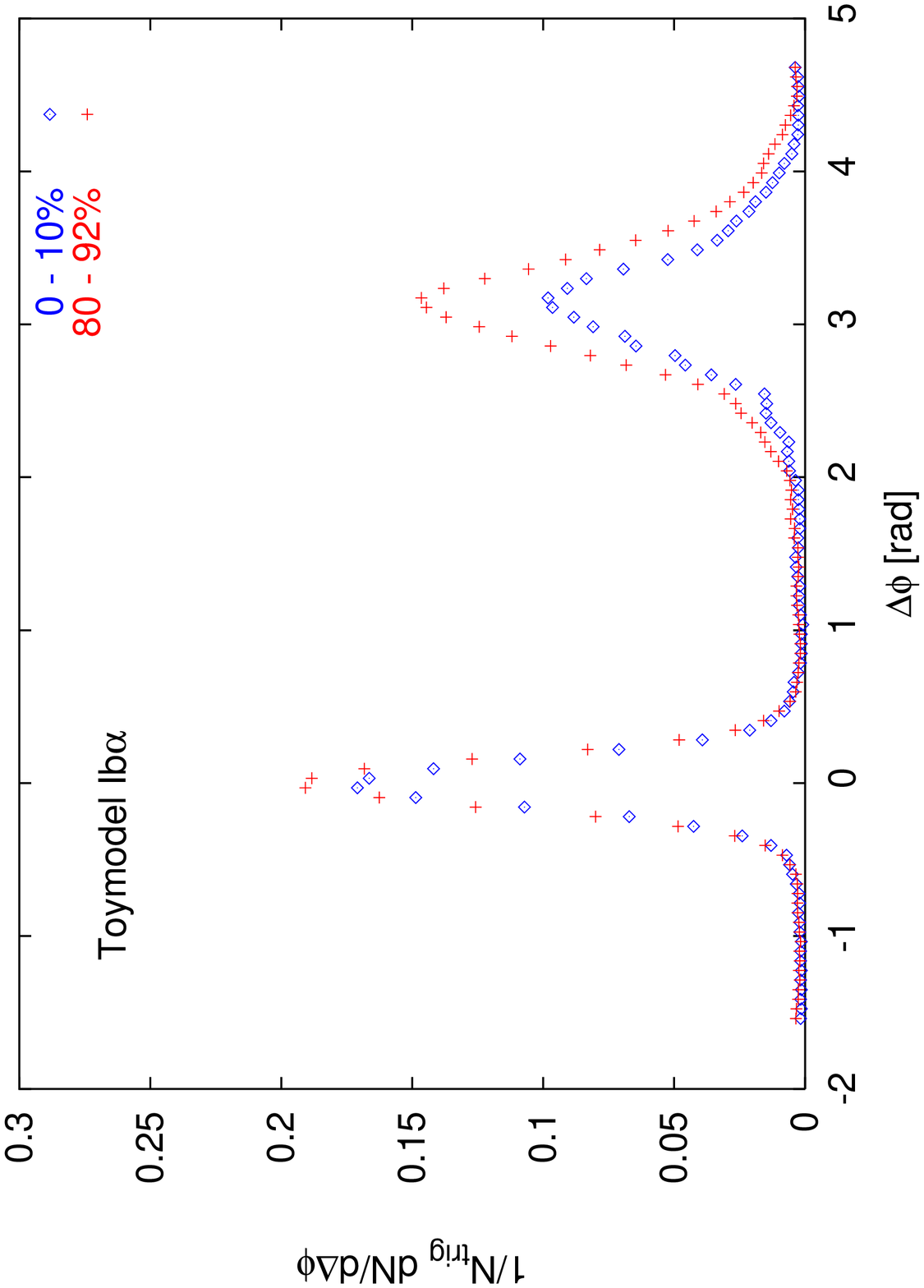}
 \end{minipage} 
 \begin{minipage}{.45\textwidth}
  \centering
  \hspace*{.4\textwidth}
 \end{minipage} 
 \begin{minipage}{.45\textwidth}
  \centering
  \includegraphics[scale=0.325]{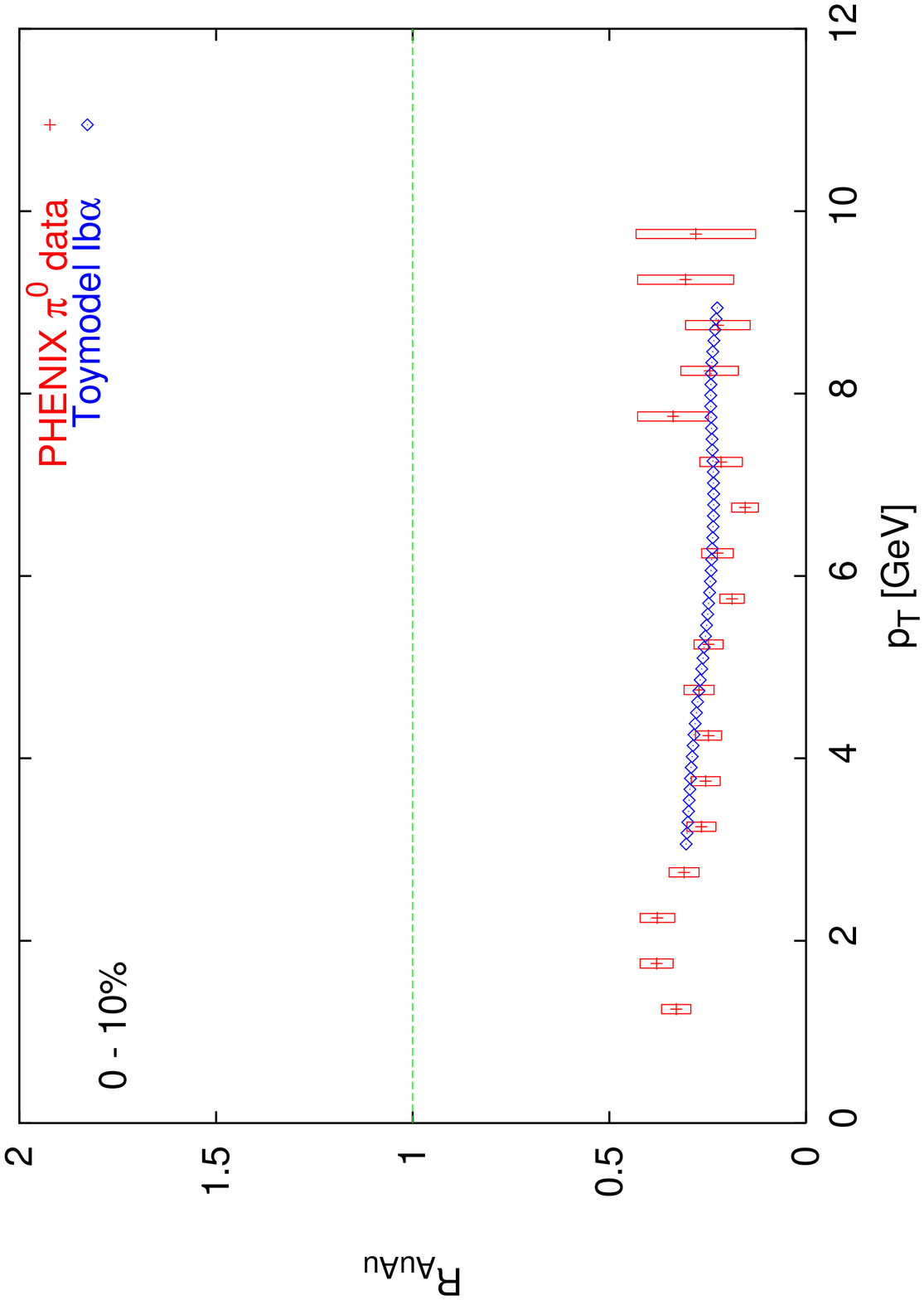}
 \end{minipage} 
 \begin{minipage}{.45\textwidth}
  \centering
  \includegraphics[scale=0.325]{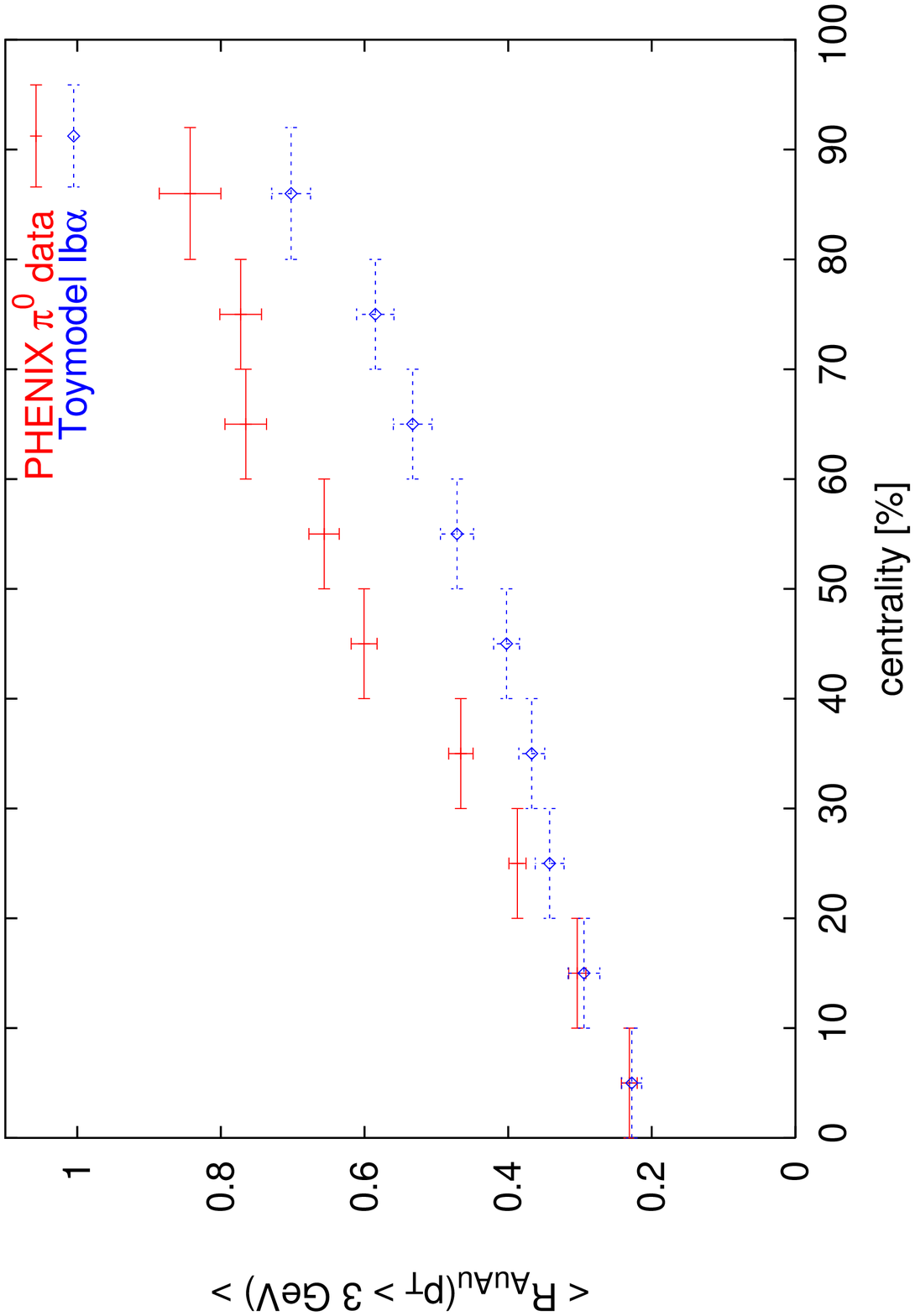}
 \end{minipage} 
 \caption{Results for the Toymodel I\,b\,$\alpha$ with $\Delta E \propto N_g^2 E$
 and homogeneous energy density distribution}
\end{figure}

\begin{figure}[ht]
 \centering
 \begin{minipage}{.45\textwidth}
  \centering
  \includegraphics[scale=0.325]{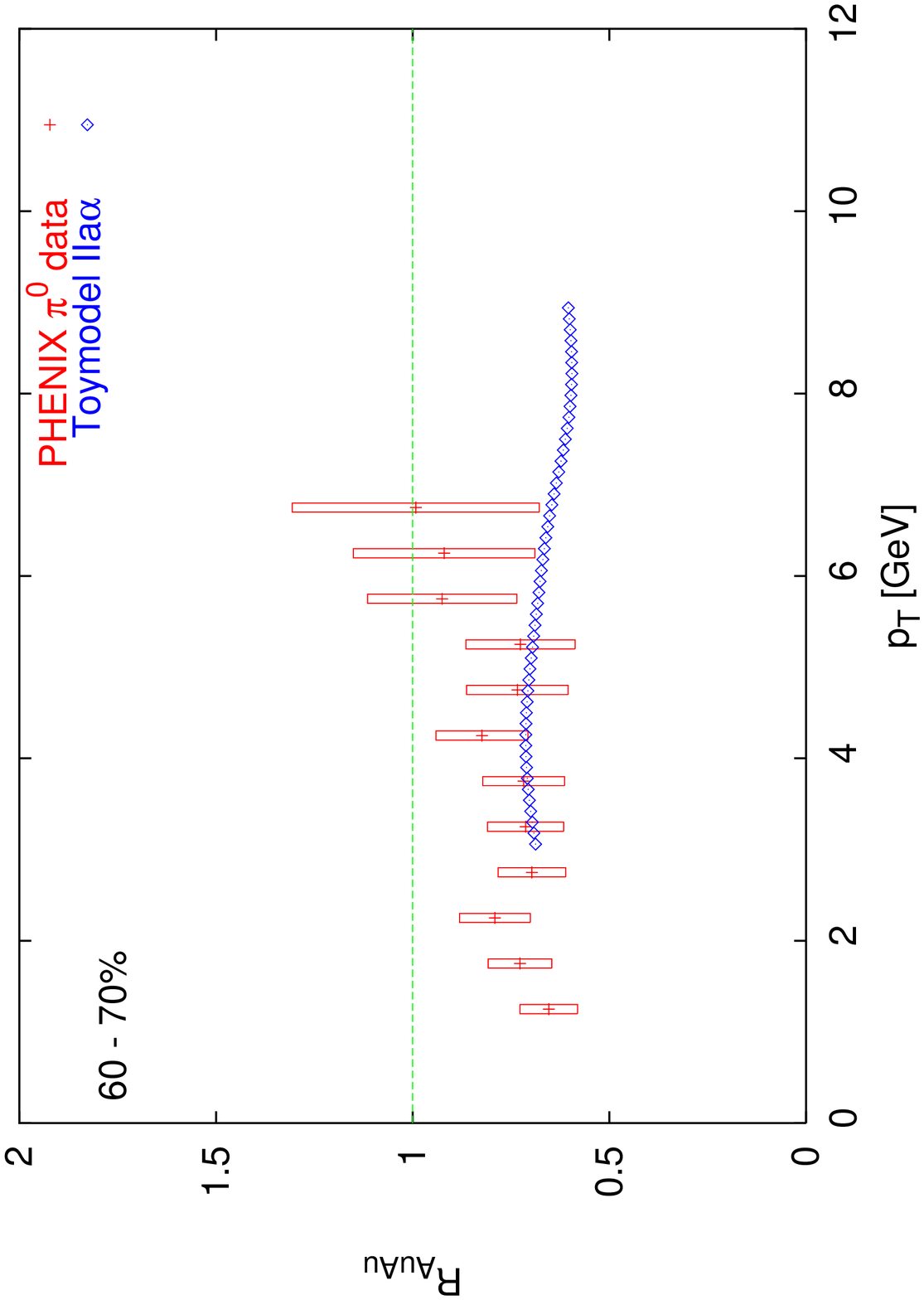}
 \end{minipage} 
 \begin{minipage}{.45\textwidth}
  \centering
  \includegraphics[scale=0.325]{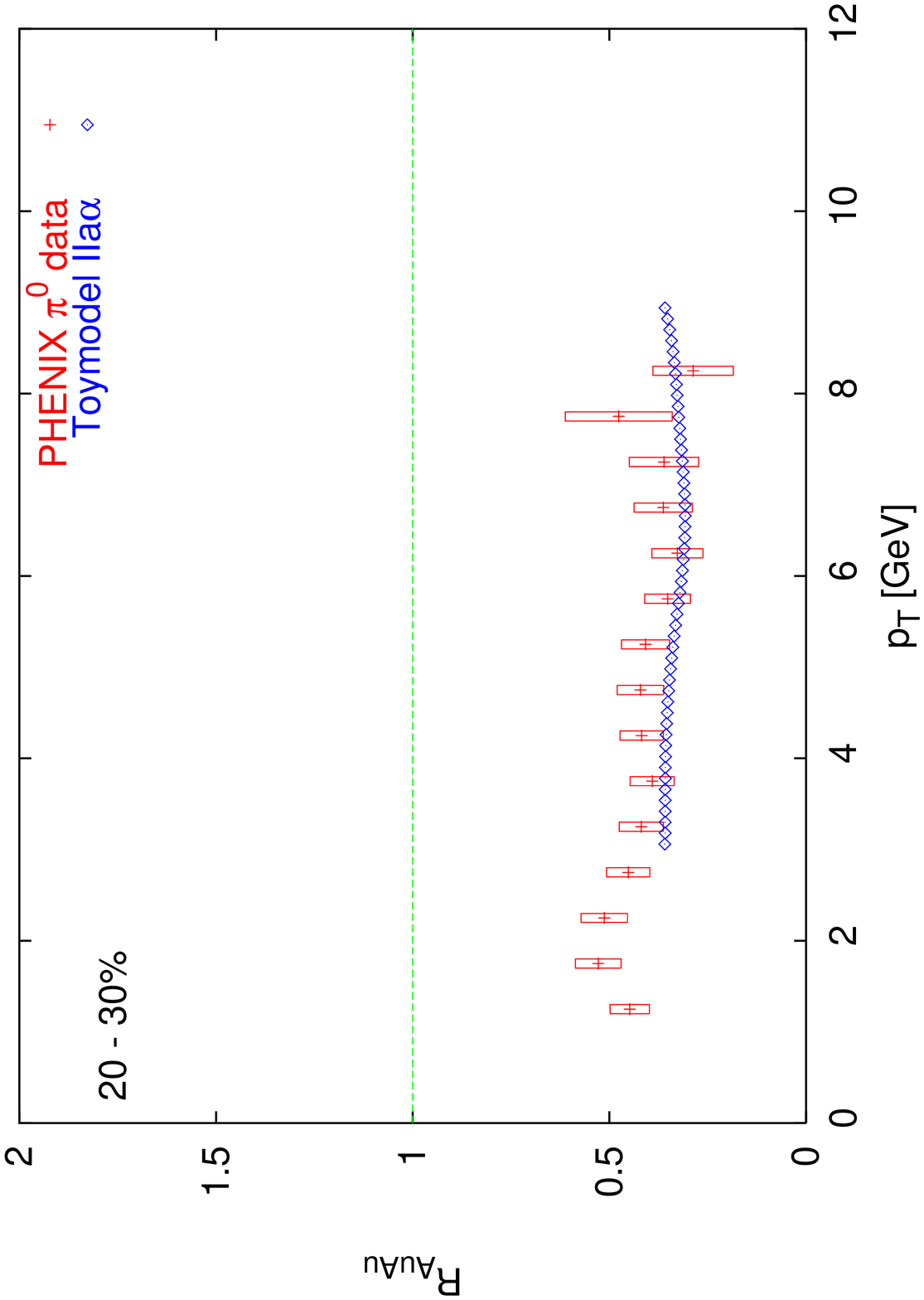}
 \end{minipage} 
 \begin{minipage}{.45\textwidth}
  \centering
  \includegraphics[scale=0.325]{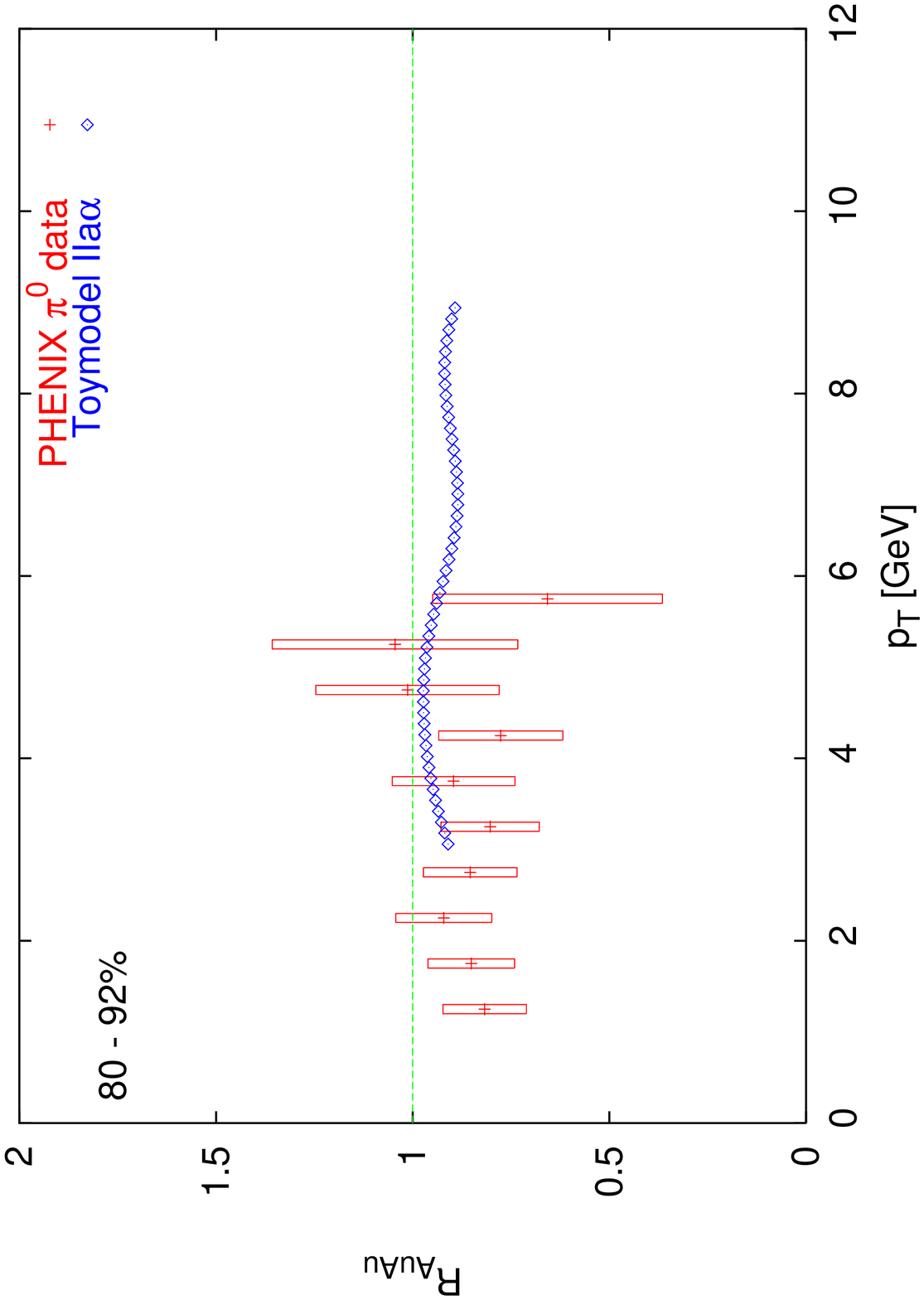}
 \end{minipage} 
 \begin{minipage}{.45\textwidth}
  \centering
  \includegraphics[scale=0.325]{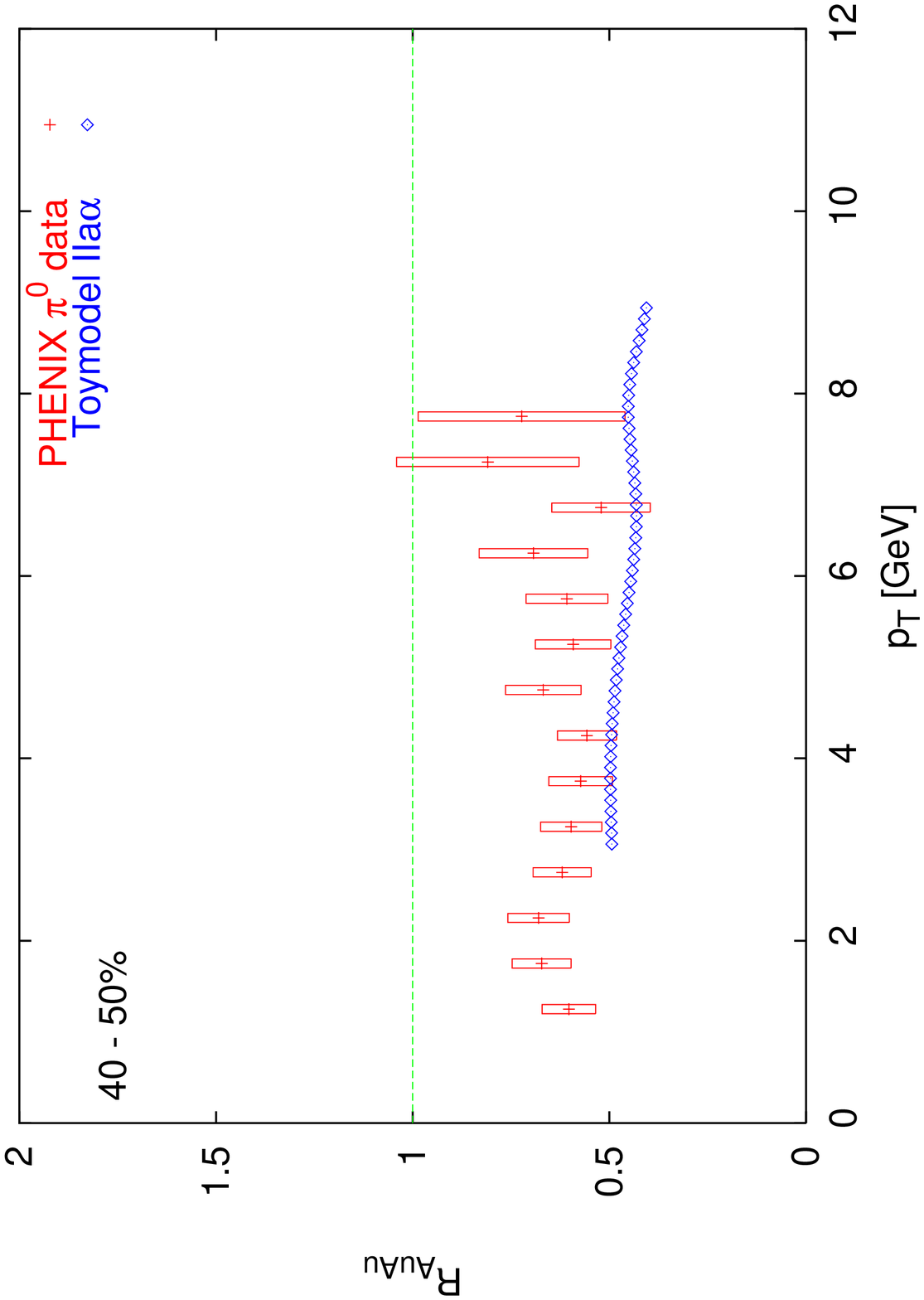}
 \end{minipage} 
\end{figure}

\begin{figure}[ht]
 \centering
 \begin{minipage}{.45\textwidth}
  \centering
  \includegraphics[scale=0.325]{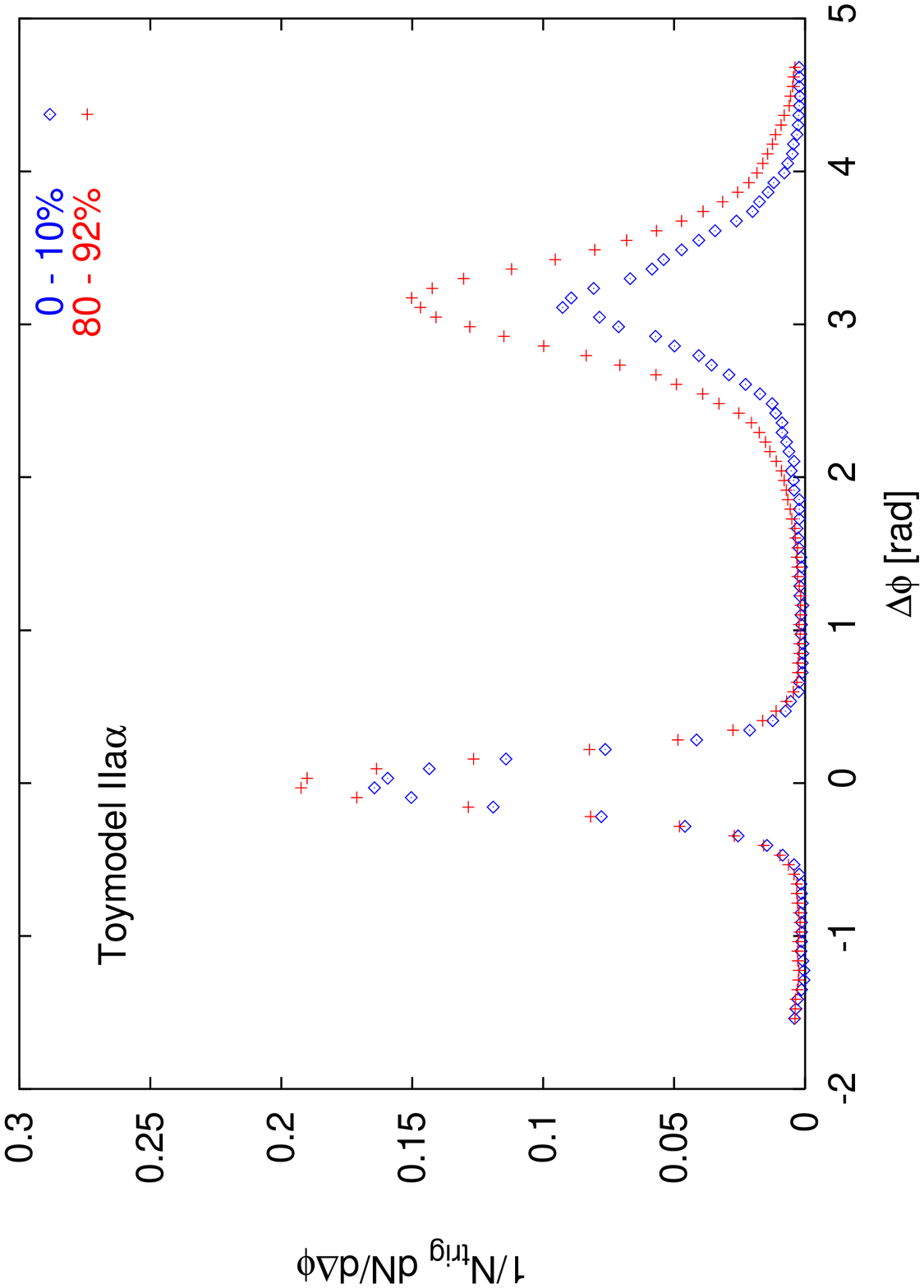}
 \end{minipage} 
 \begin{minipage}{.45\textwidth}
  \centering
  \hspace*{.4\textwidth}
 \end{minipage} 
 \begin{minipage}{.45\textwidth}
  \centering
  \includegraphics[scale=0.325]{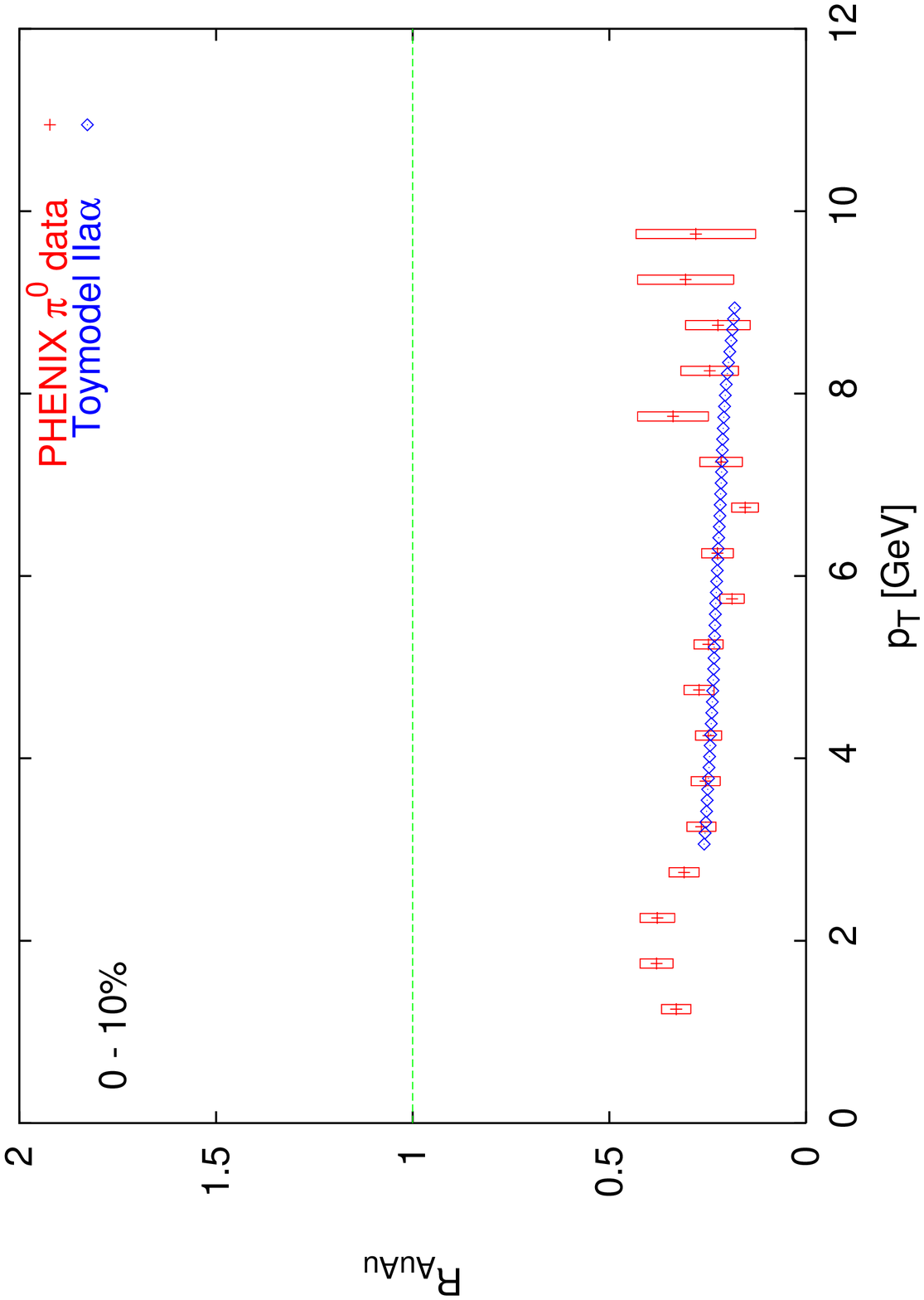}
 \end{minipage} 
 \begin{minipage}{.45\textwidth}
  \centering
  \includegraphics[scale=0.325]{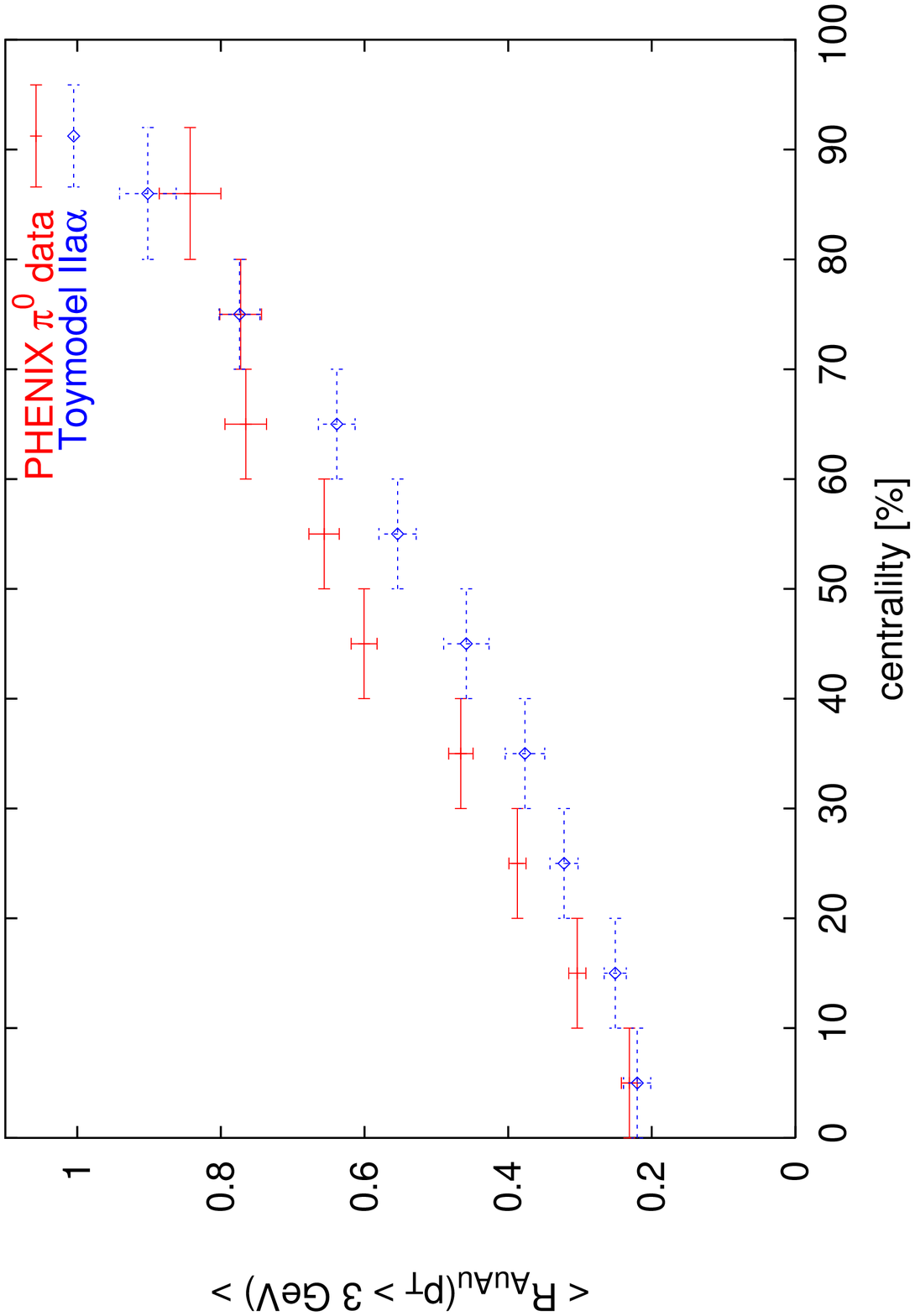}
 \end{minipage} 
 \caption{Results for the Toymodel II\,a\,$\alpha$ with $\Delta E \propto N_g E$
 and inhomogeneous energy density distribution}
\end{figure}

\begin{figure}[ht]
 \centering
 \begin{minipage}{.45\textwidth}
  \centering
  \includegraphics[scale=0.325]{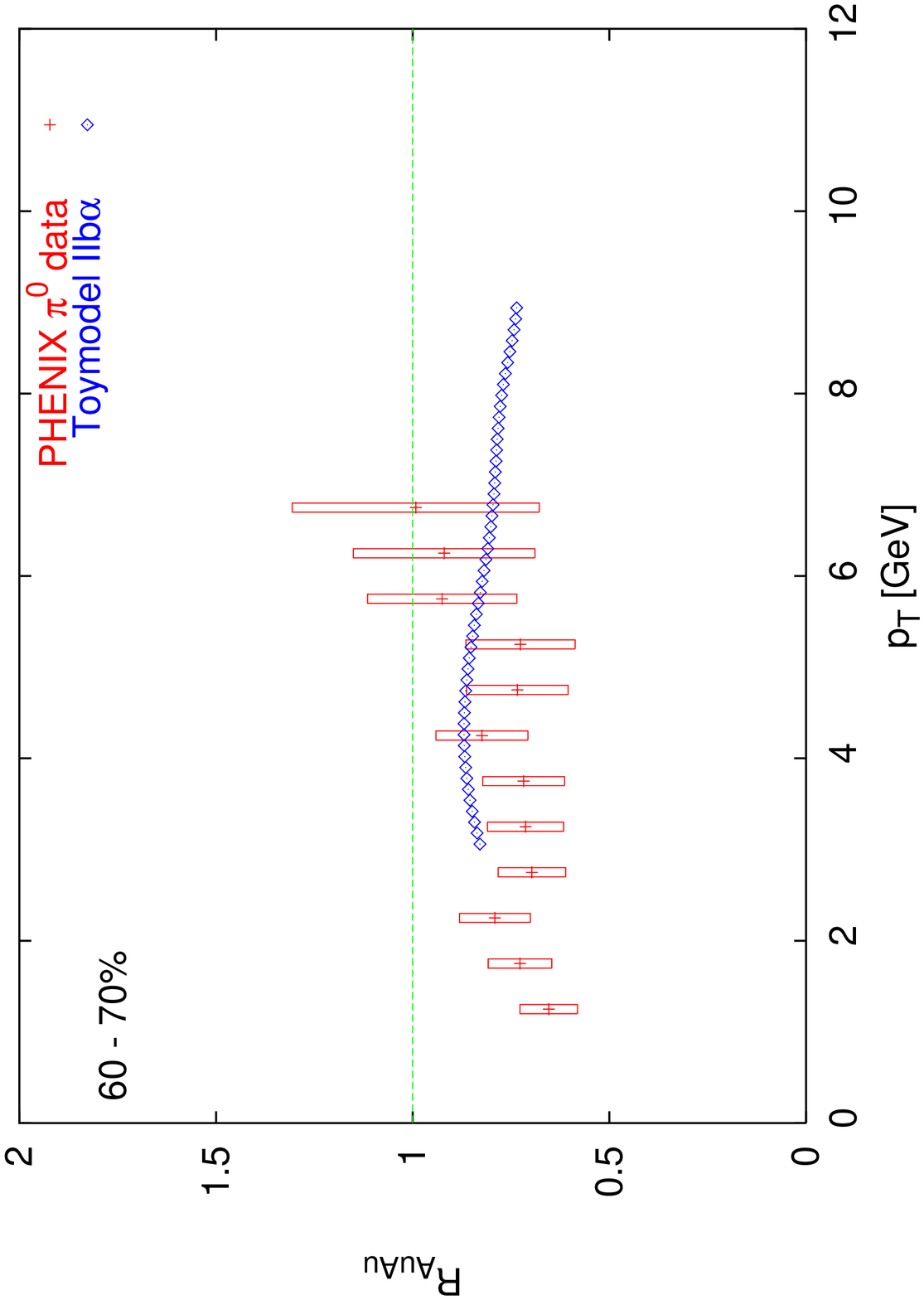}
 \end{minipage} 
 \begin{minipage}{.45\textwidth}
  \centering
  \includegraphics[scale=0.325]{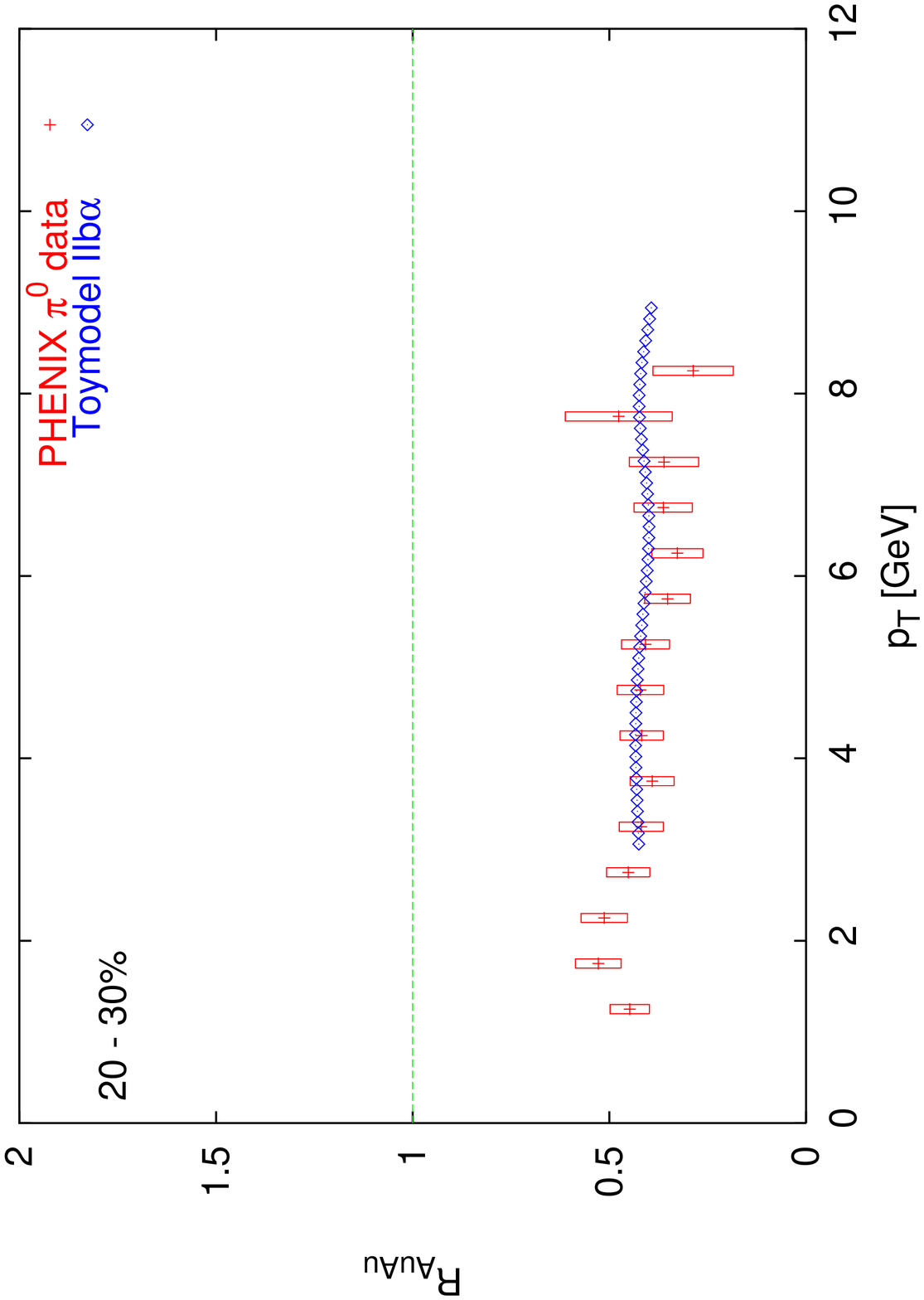}
 \end{minipage} 
 \begin{minipage}{.45\textwidth}
  \centering
  \includegraphics[scale=0.325]{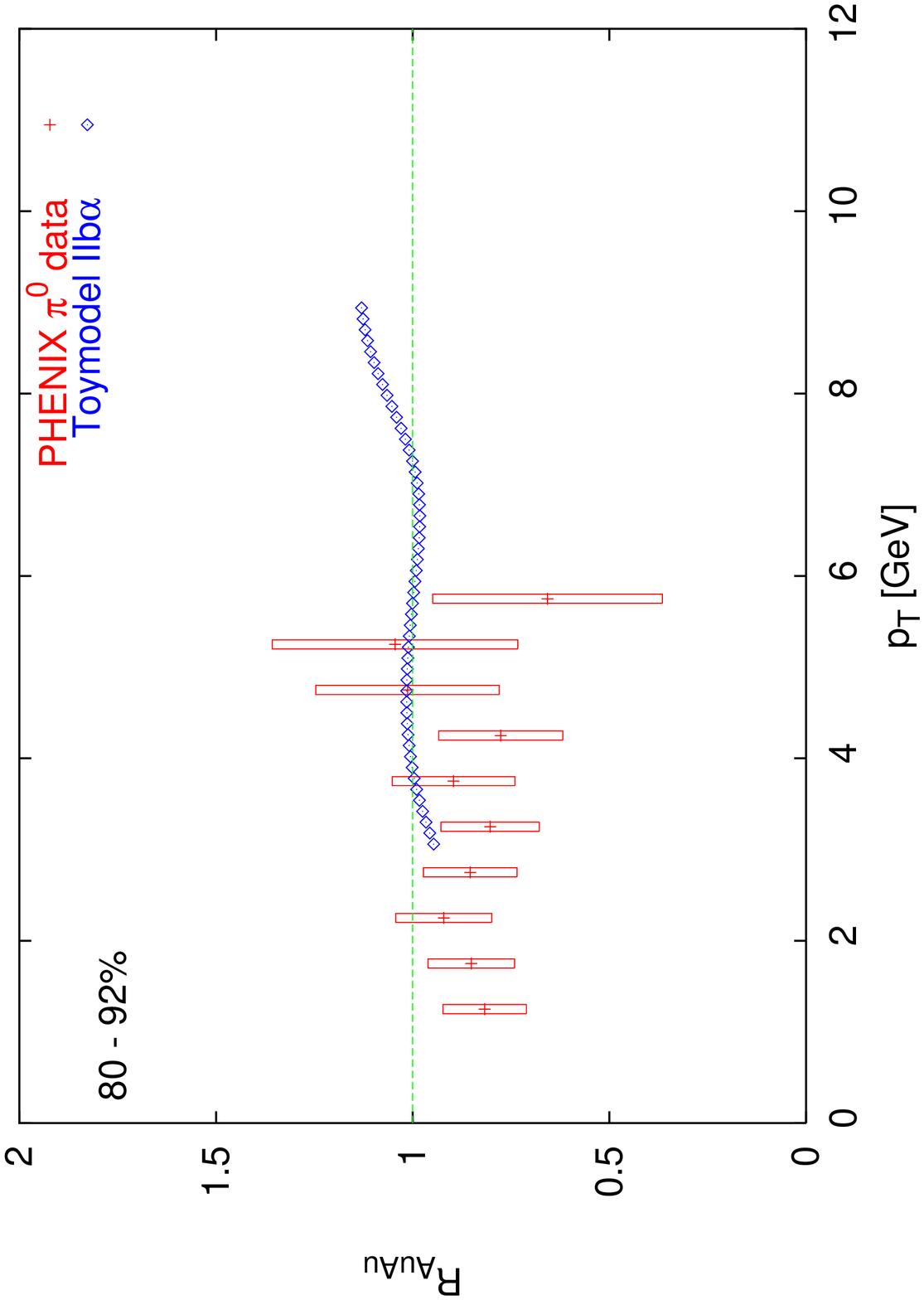}
 \end{minipage} 
 \begin{minipage}{.45\textwidth}
  \centering
  \includegraphics[scale=0.325]{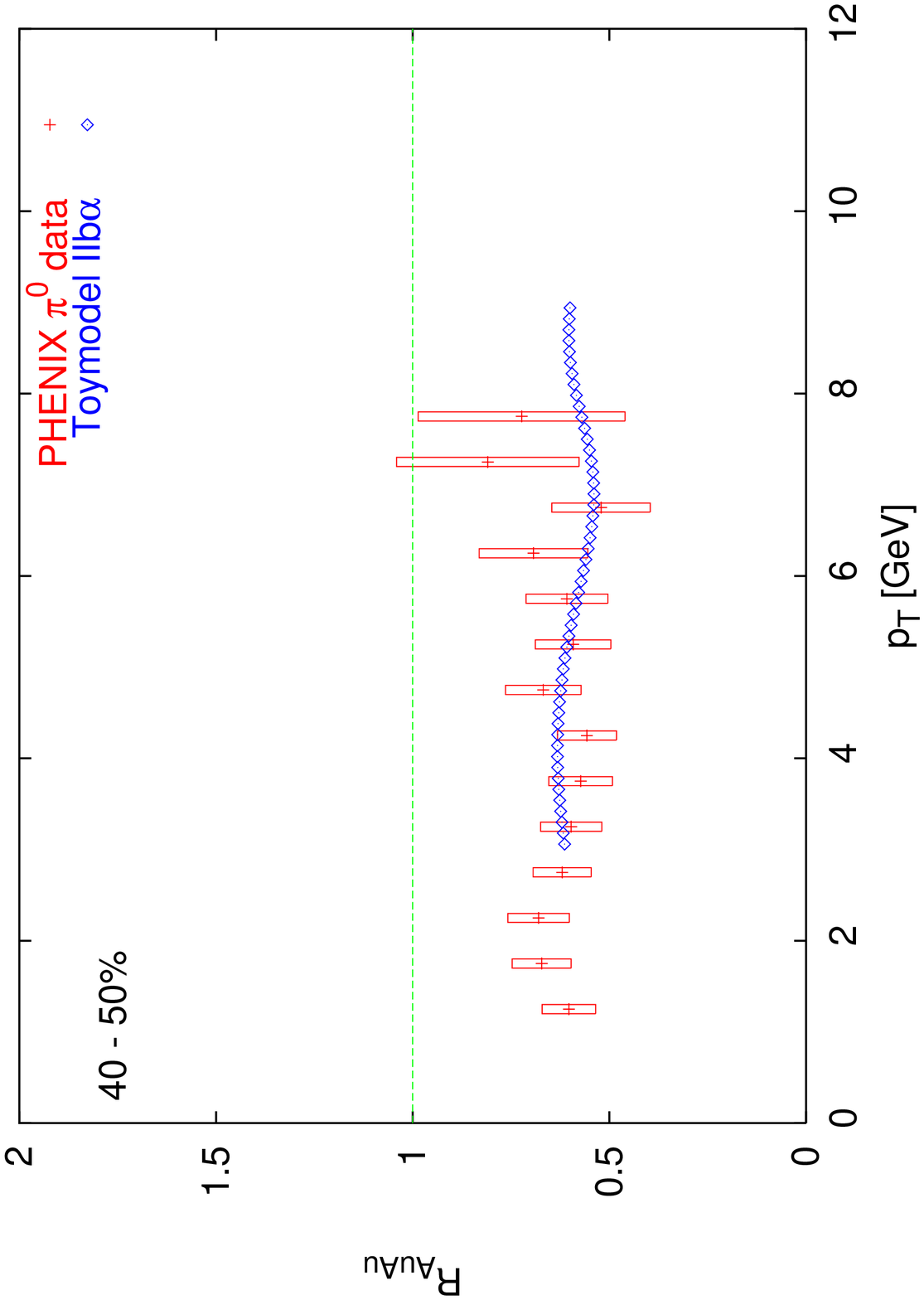}
 \end{minipage} 
\end{figure}

\begin{figure}[ht]
 \centering
 \begin{minipage}{.45\textwidth}
  \centering
  \includegraphics[scale=0.325]{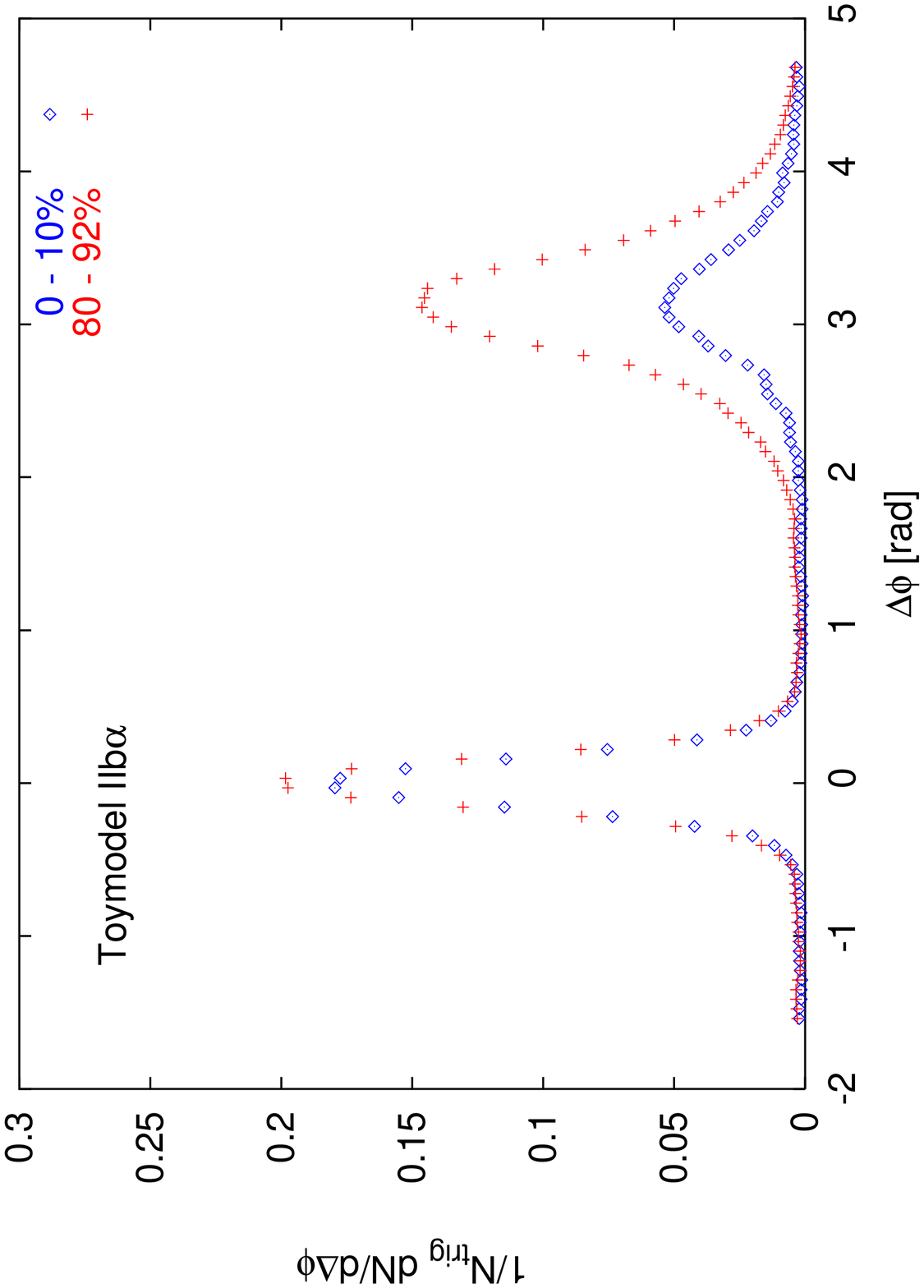}
 \end{minipage} 
 \begin{minipage}{.45\textwidth}
  \centering
  \hspace*{.4\textwidth}
 \end{minipage} 
 \begin{minipage}{.45\textwidth}
  \centering
  \includegraphics[scale=0.325]{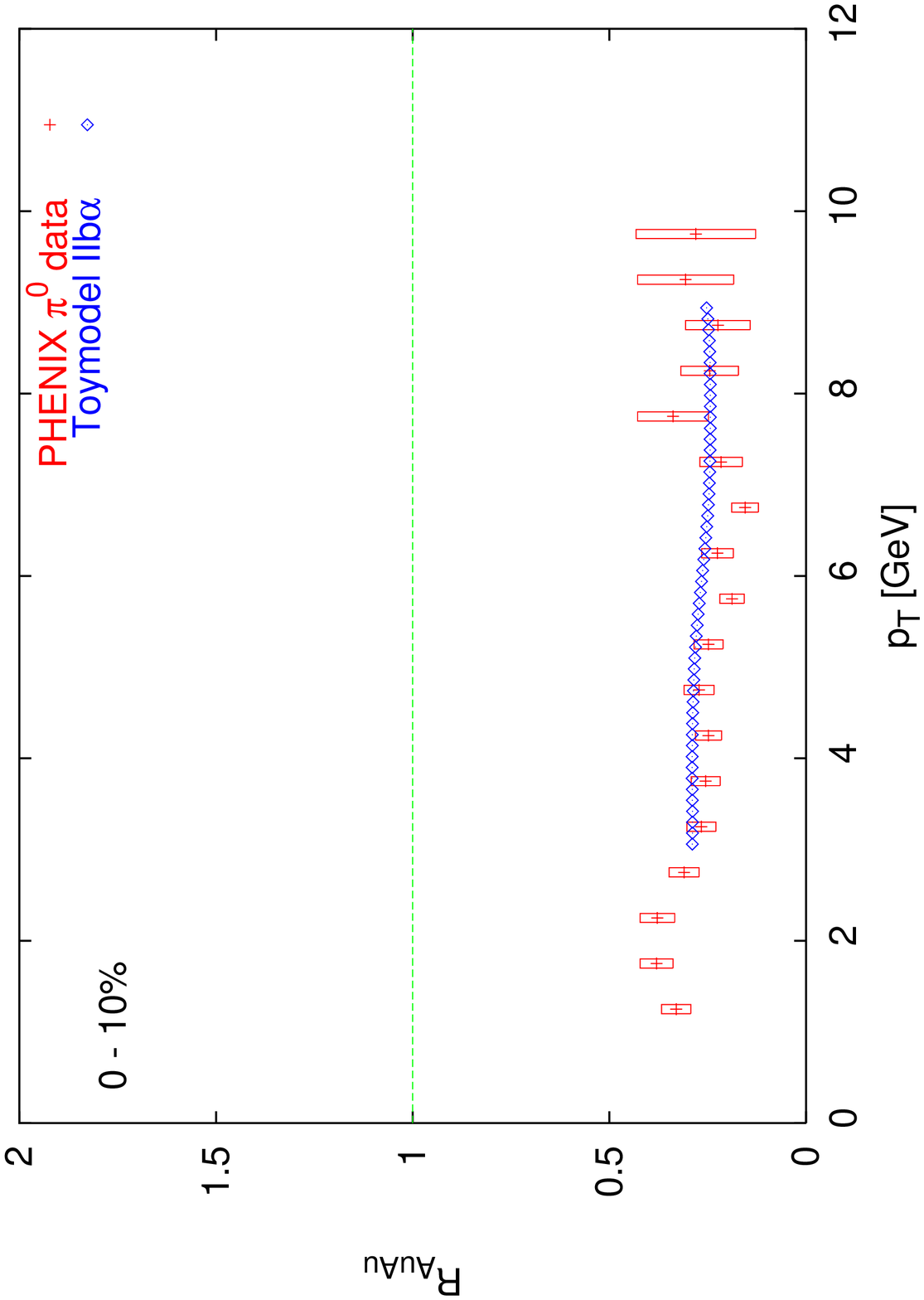}
 \end{minipage} 
 \begin{minipage}{.45\textwidth}
  \centering
  \includegraphics[scale=0.325]{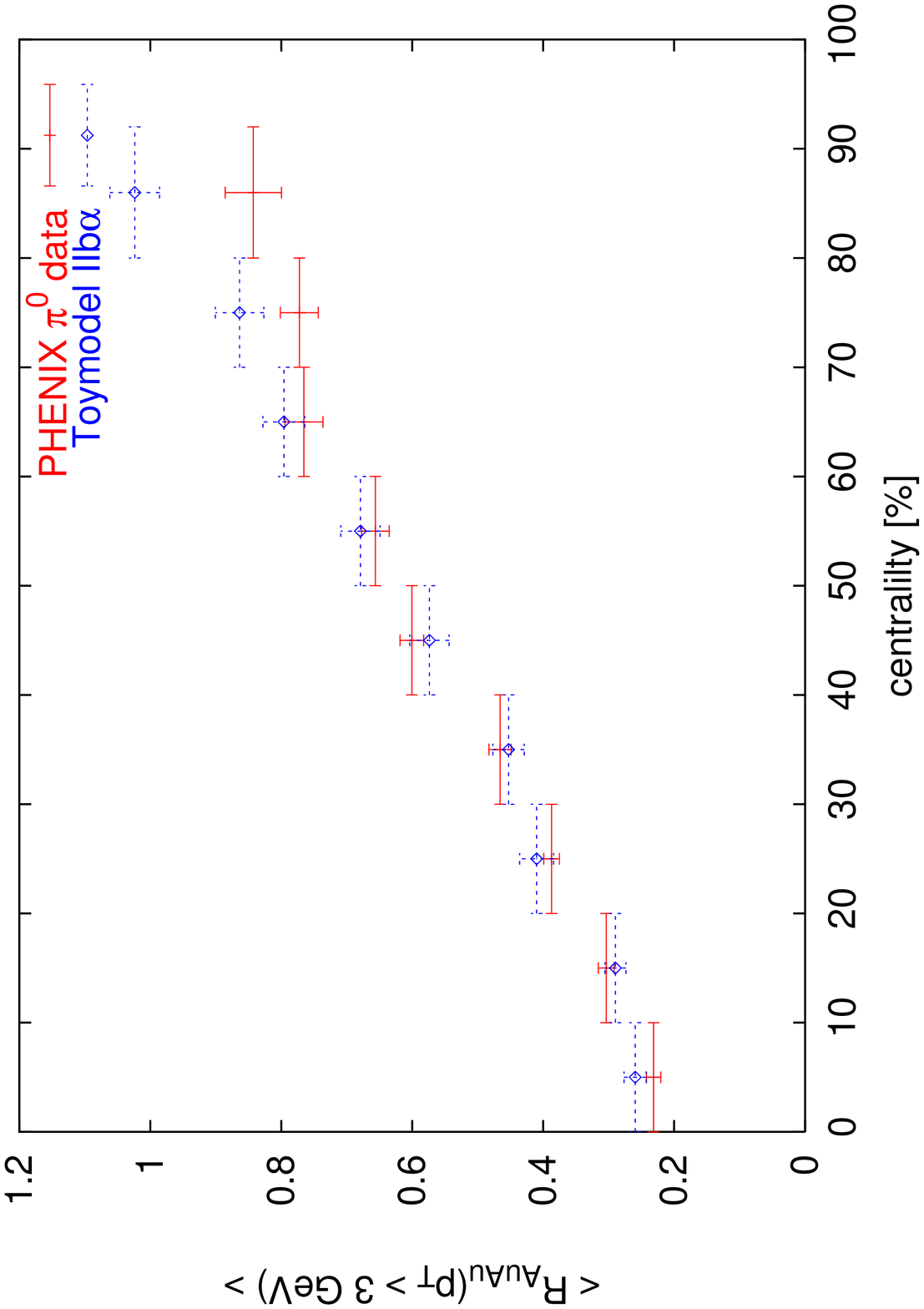}
 \end{minipage} 
 \caption{Results for the Toymodel II\,b\,$\alpha$ with $\Delta E \propto N_g^2 E$
 and inhomogeneous energy density distribution}
\end{figure}

\begin{figure}[ht]
 \centering
 \begin{minipage}{.45\textwidth}
  \centering
  \includegraphics[scale=0.325]{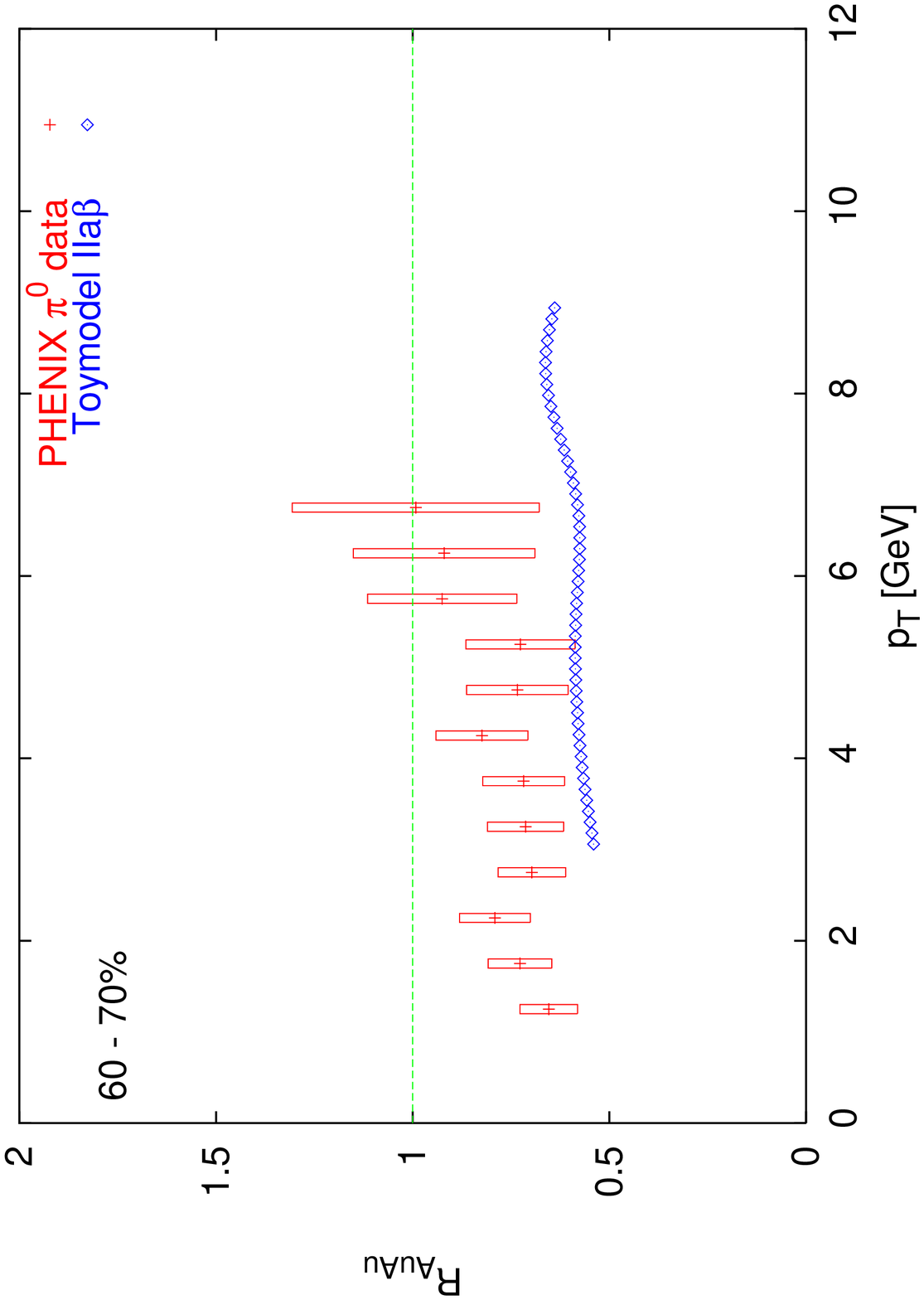}
 \end{minipage} 
 \begin{minipage}{.45\textwidth}
  \centering
  \includegraphics[scale=0.325]{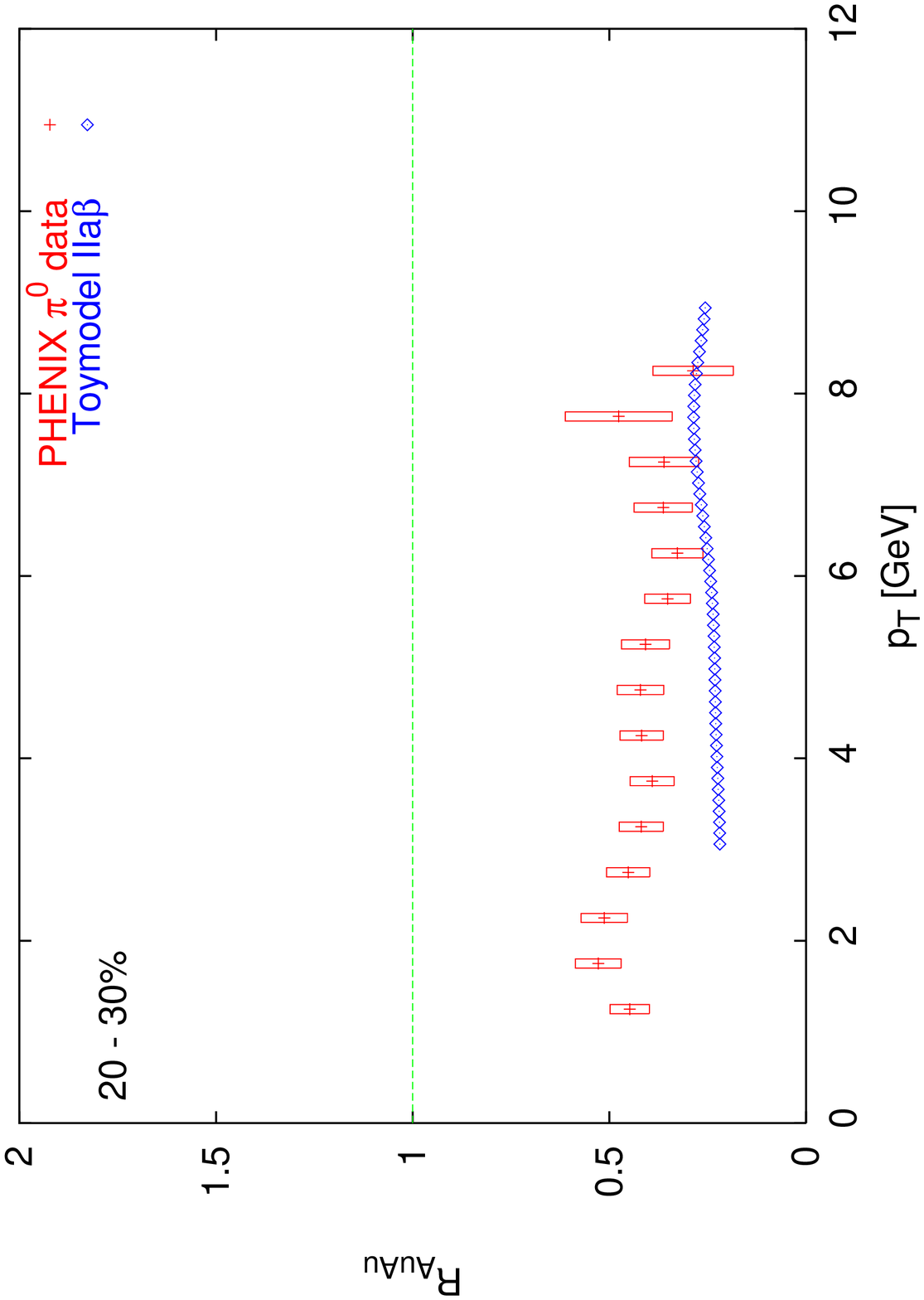}
 \end{minipage} 
 \begin{minipage}{.45\textwidth}
  \centering
  \includegraphics[scale=0.325]{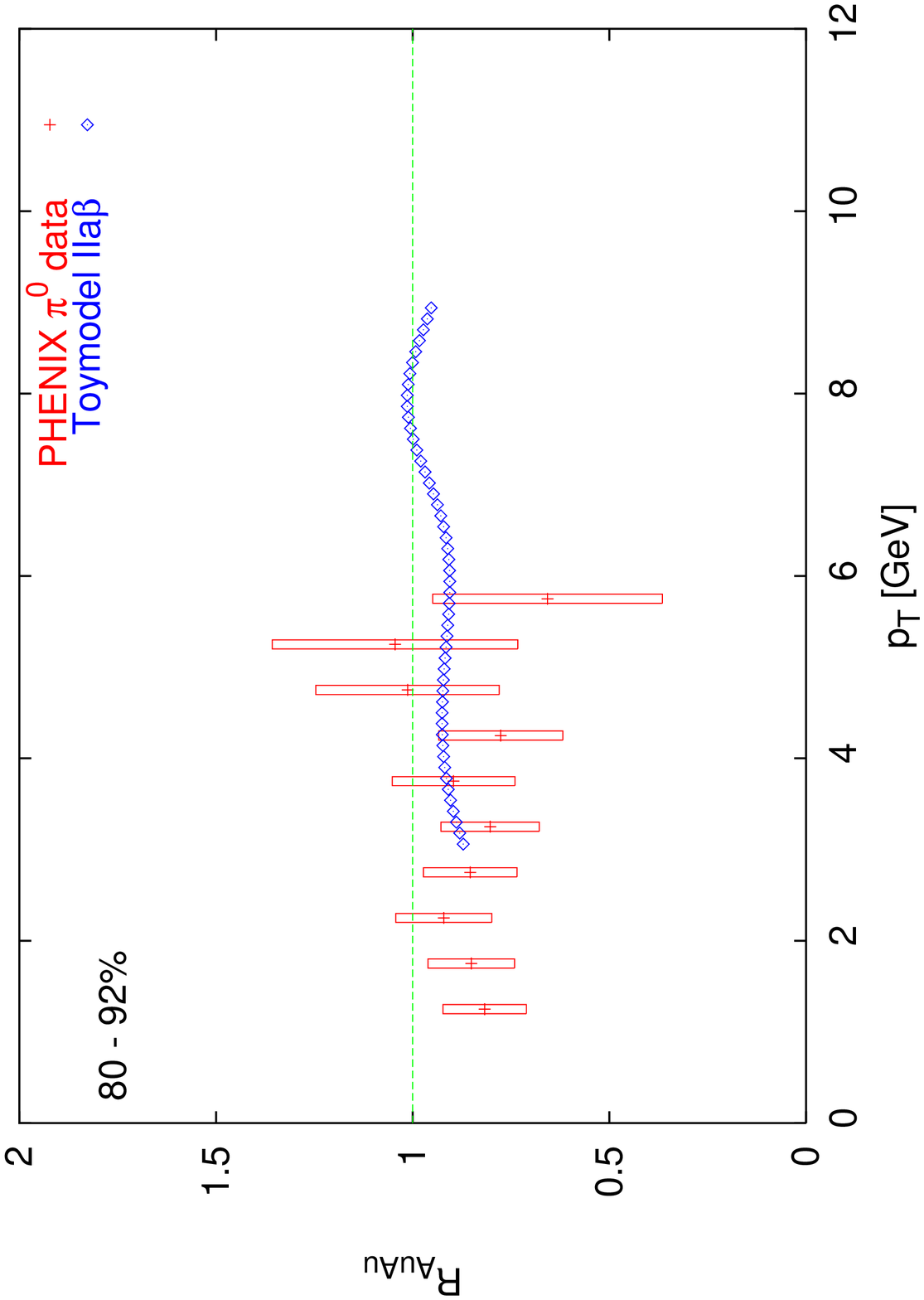}
 \end{minipage} 
 \begin{minipage}{.45\textwidth}
  \centering
  \includegraphics[scale=0.325]{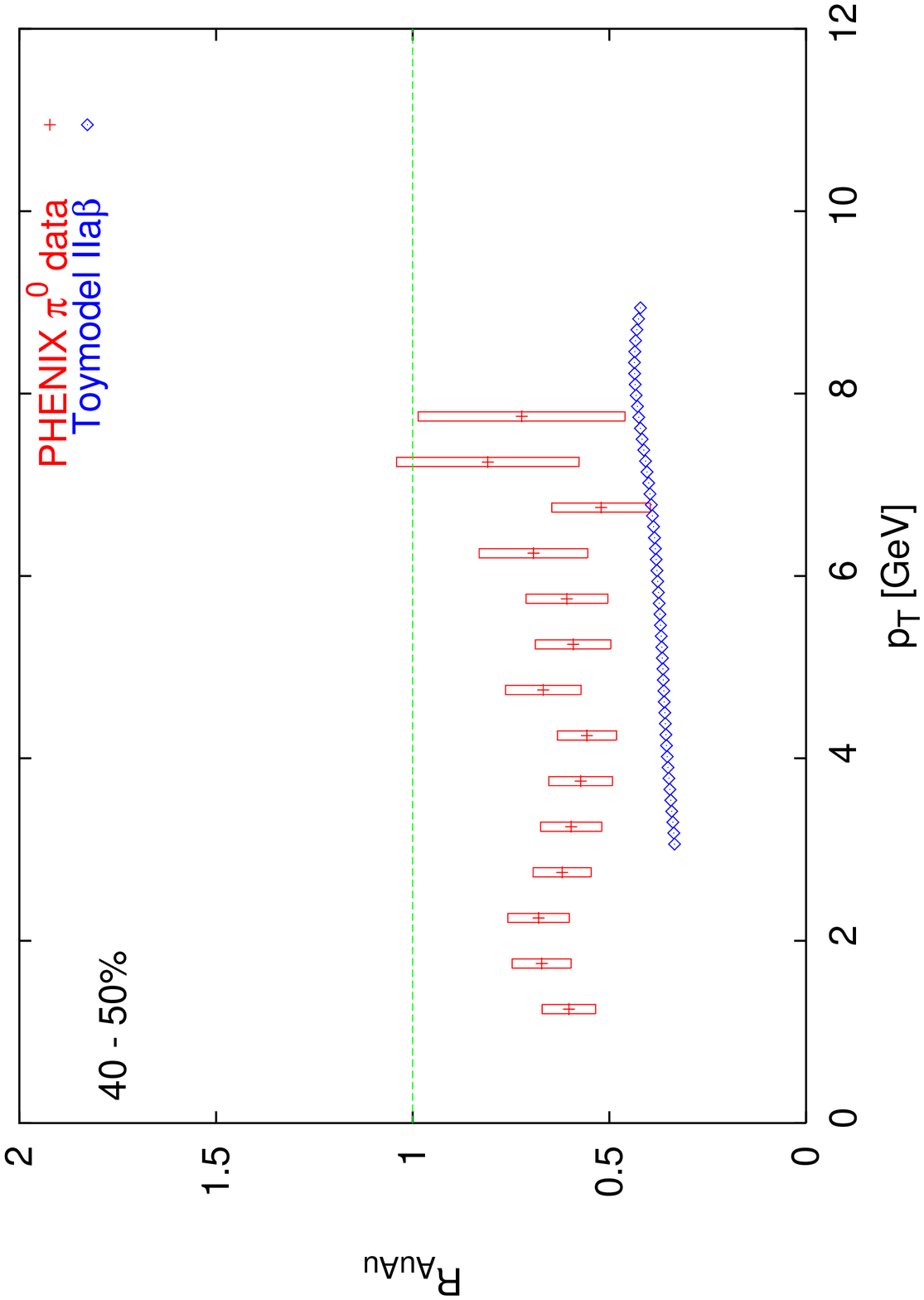}
 \end{minipage} 
\end{figure}

\begin{figure}[ht]
 \centering
 \begin{minipage}{.45\textwidth}
  \centering
  \includegraphics[scale=0.325]{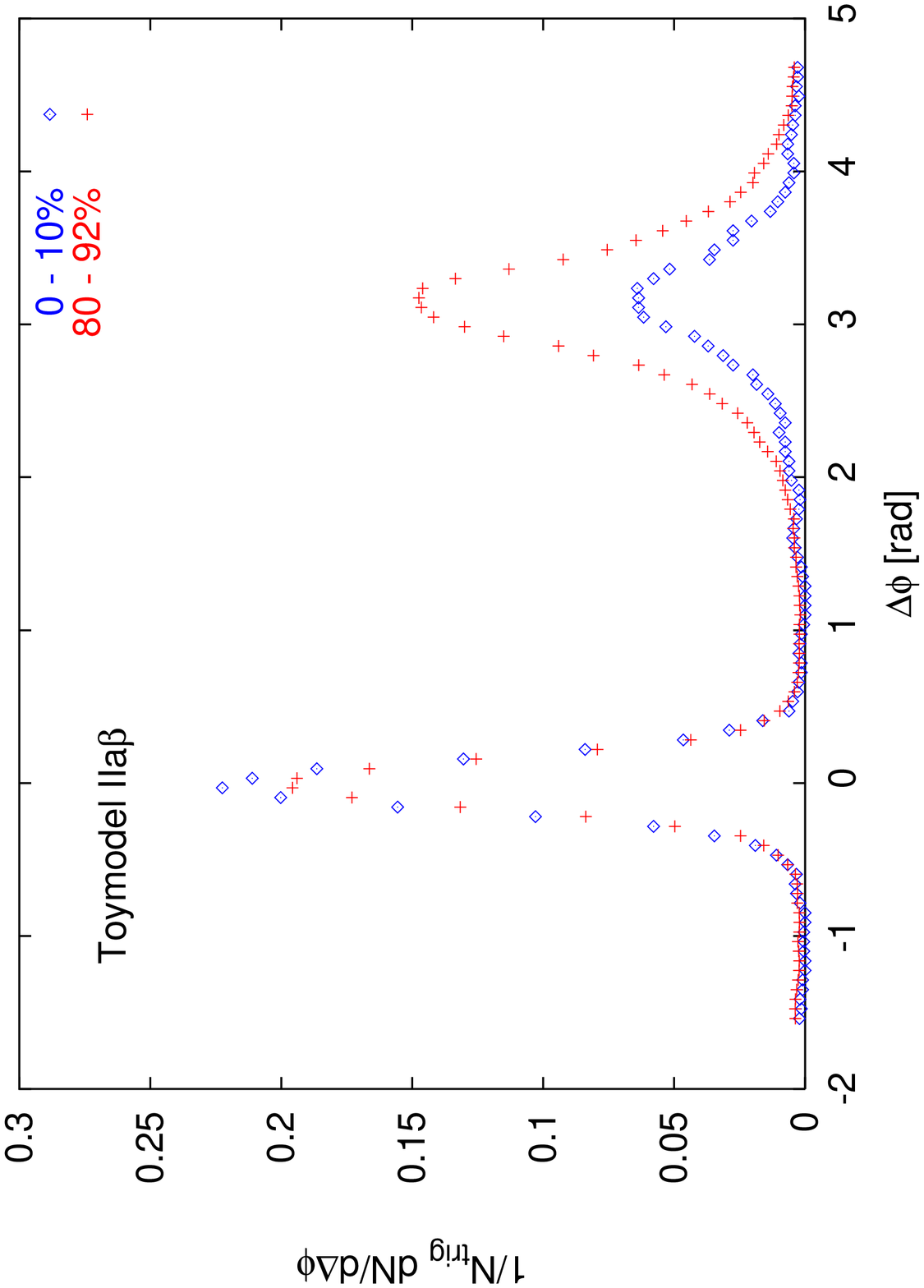}
 \end{minipage} 
 \begin{minipage}{.45\textwidth}
  \centering
  \hspace*{.4\textwidth}
 \end{minipage} 
 \begin{minipage}{.45\textwidth}
  \centering
  \includegraphics[scale=0.325]{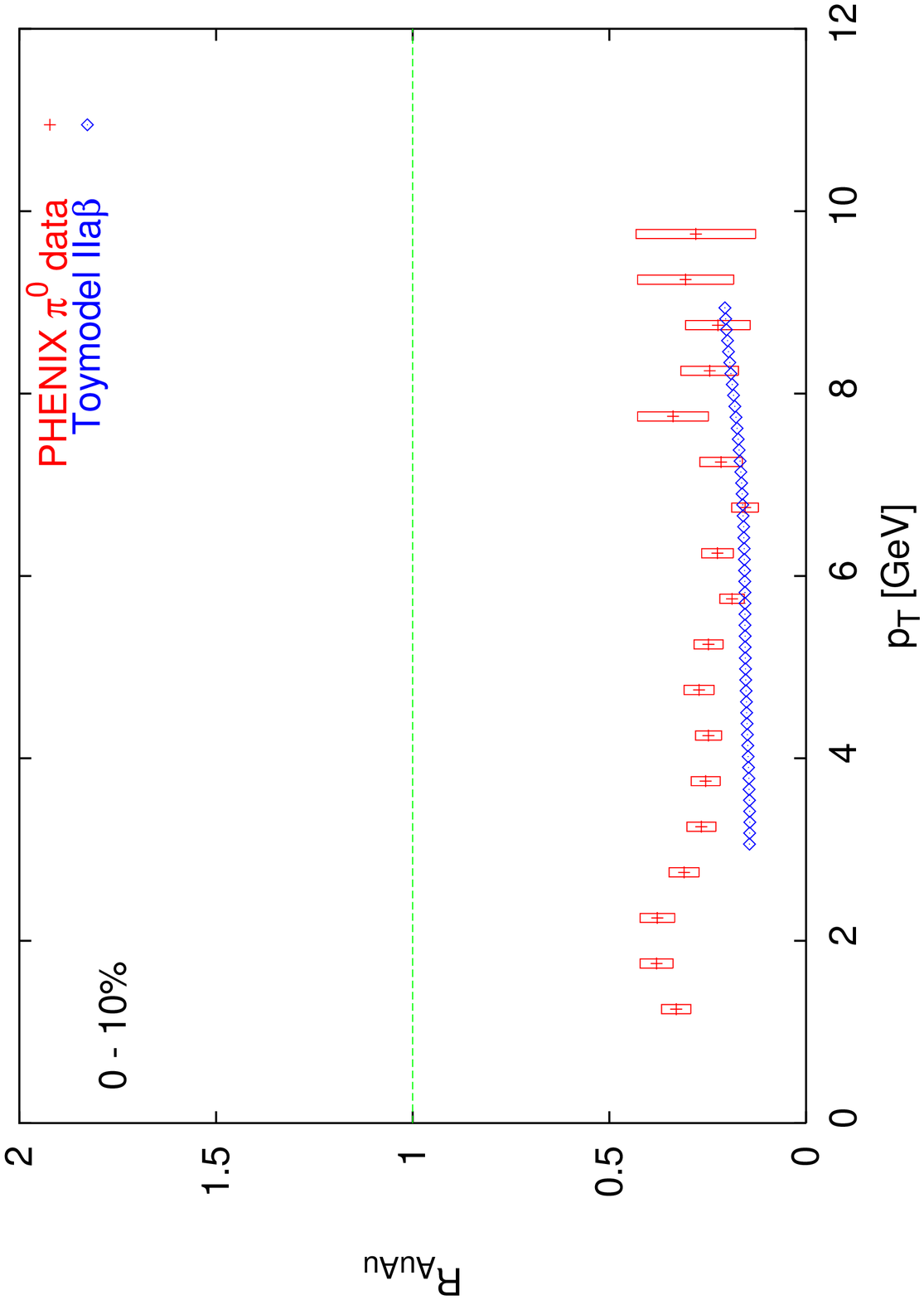}
 \end{minipage} 
 \begin{minipage}{.45\textwidth}
  \centering
  \includegraphics[scale=0.325]{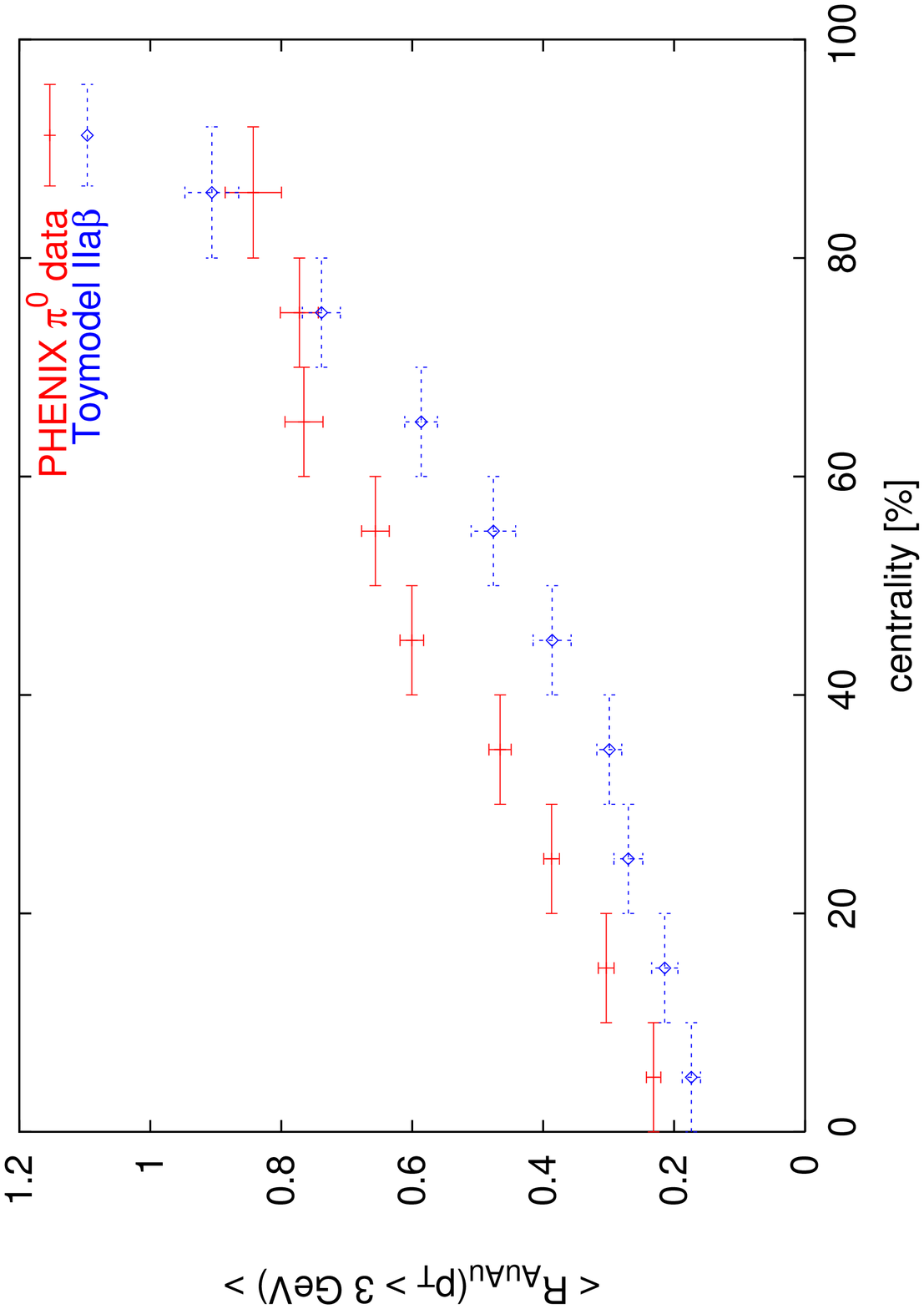}
 \end{minipage} 
 \caption{Results for the Toymodel II\,a\,$\beta$ with $\Delta E \propto N_g
 \sqrt{E}$ and inhomogeneous energy density distribution}
\end{figure}

\begin{figure}[ht]
 \centering
 \begin{minipage}{.45\textwidth}
  \centering
  \includegraphics[scale=0.325]{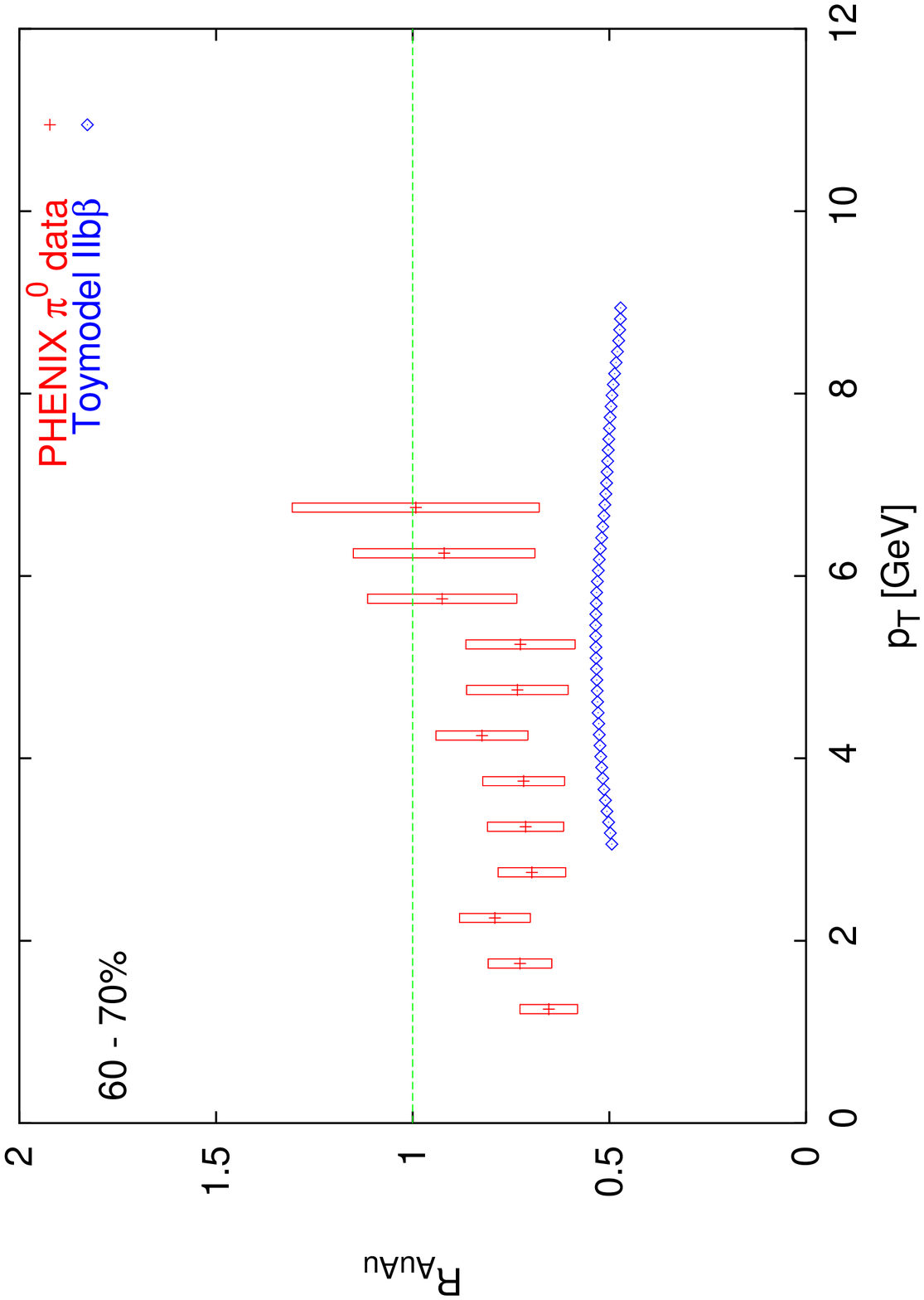}
 \end{minipage} 
 \begin{minipage}{.45\textwidth}
  \centering
  \includegraphics[scale=0.325]{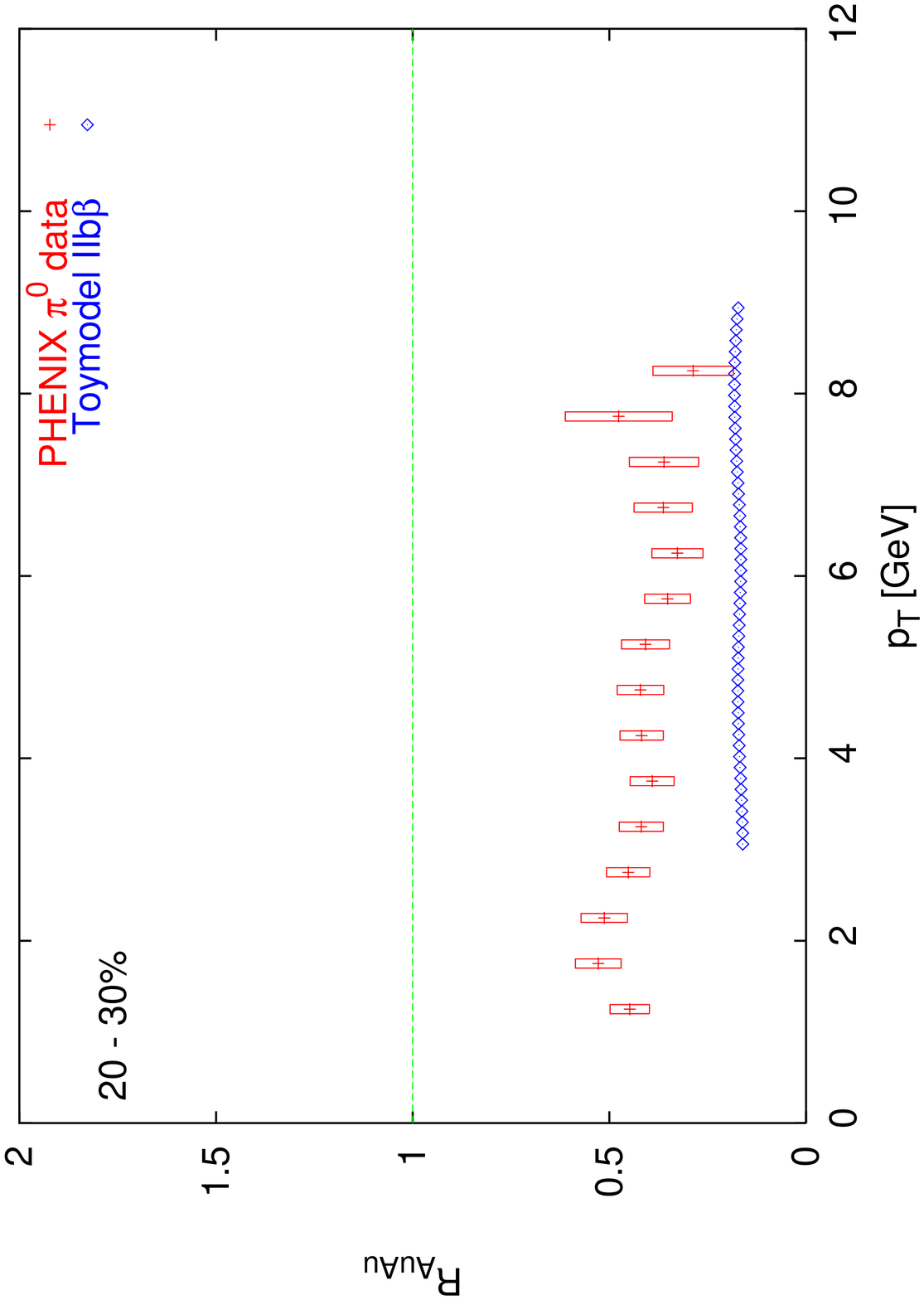}
 \end{minipage} 
 \begin{minipage}{.45\textwidth}
  \centering
  \includegraphics[scale=0.325]{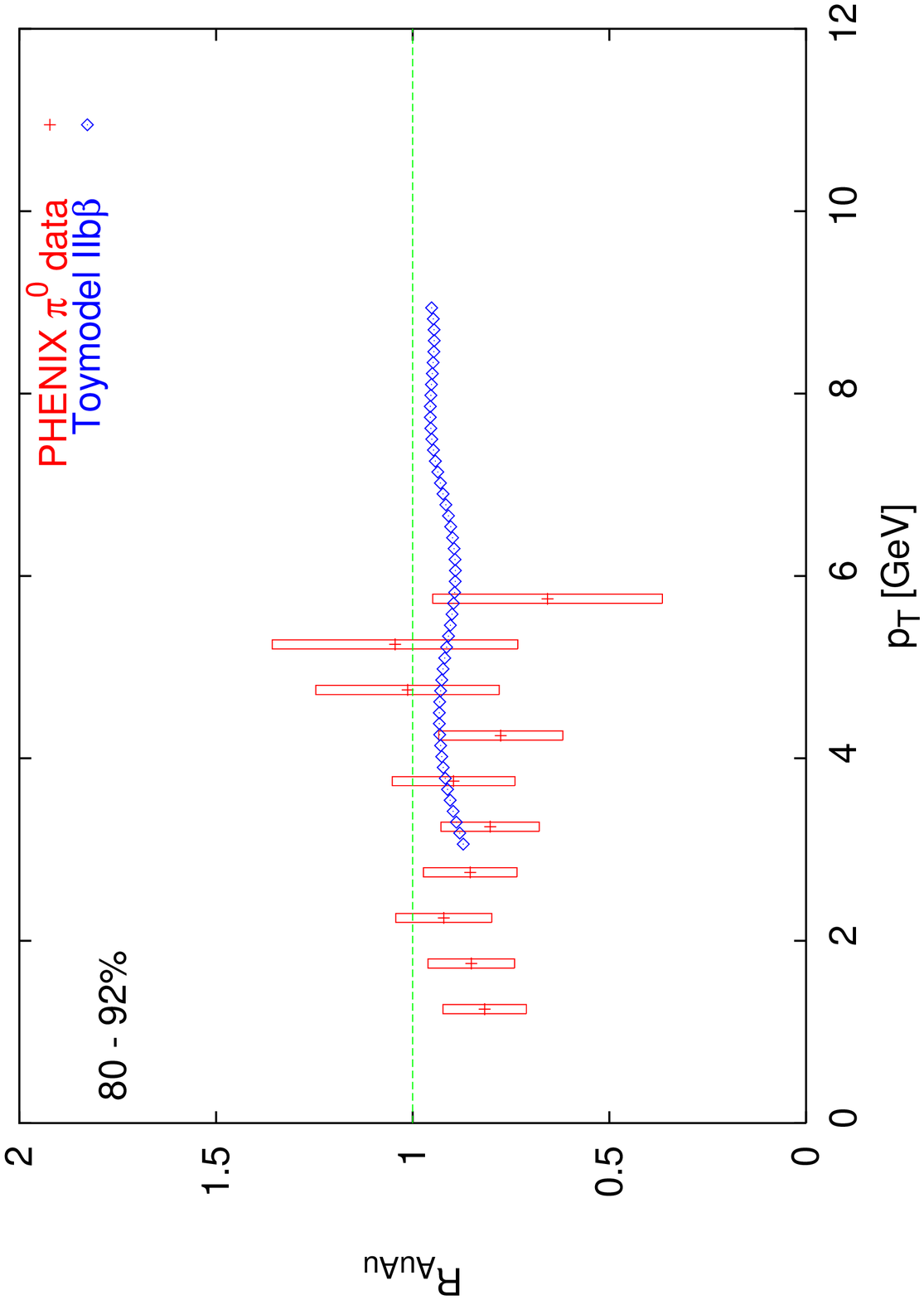}
 \end{minipage} 
 \begin{minipage}{.45\textwidth}
  \centering
  \includegraphics[scale=0.325]{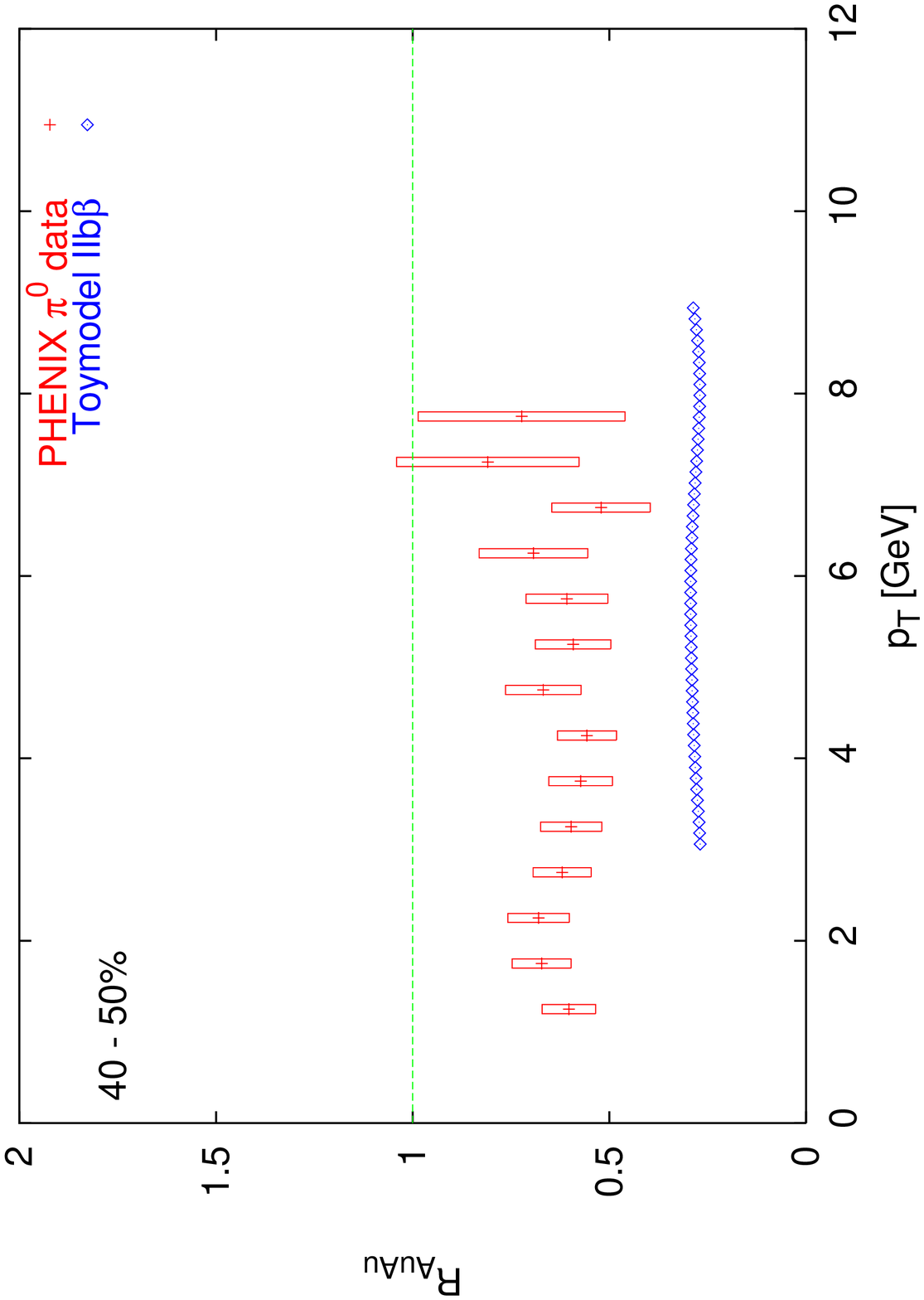}
 \end{minipage} 
\end{figure}

\begin{figure}[ht]
 \centering
 \begin{minipage}{.45\textwidth}
  \centering
  \includegraphics[scale=0.325]{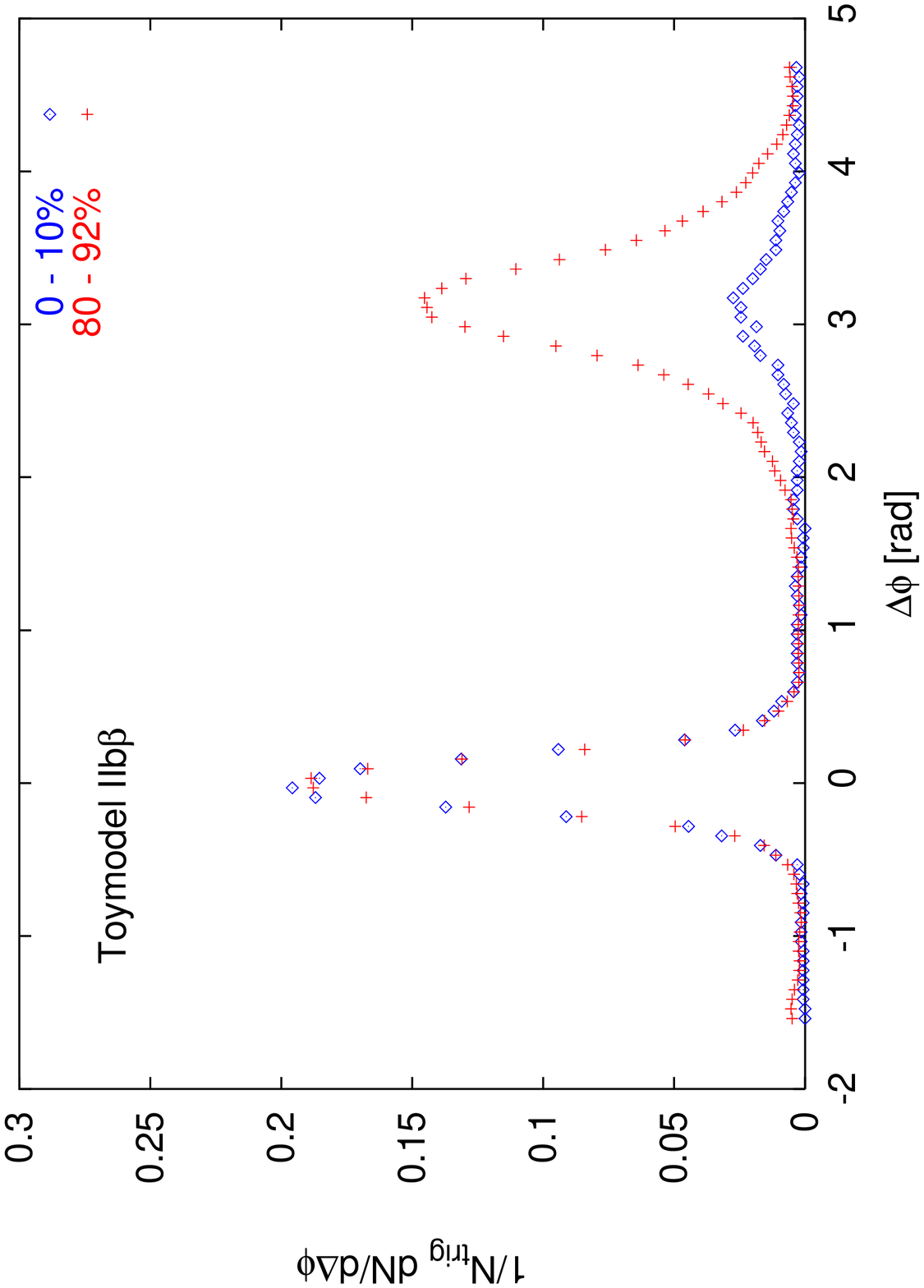}
 \end{minipage} 
 \begin{minipage}{.45\textwidth}
  \centering
  \hspace*{.4\textwidth}
 \end{minipage} 
 \begin{minipage}{.45\textwidth}
  \centering
  \includegraphics[scale=0.325]{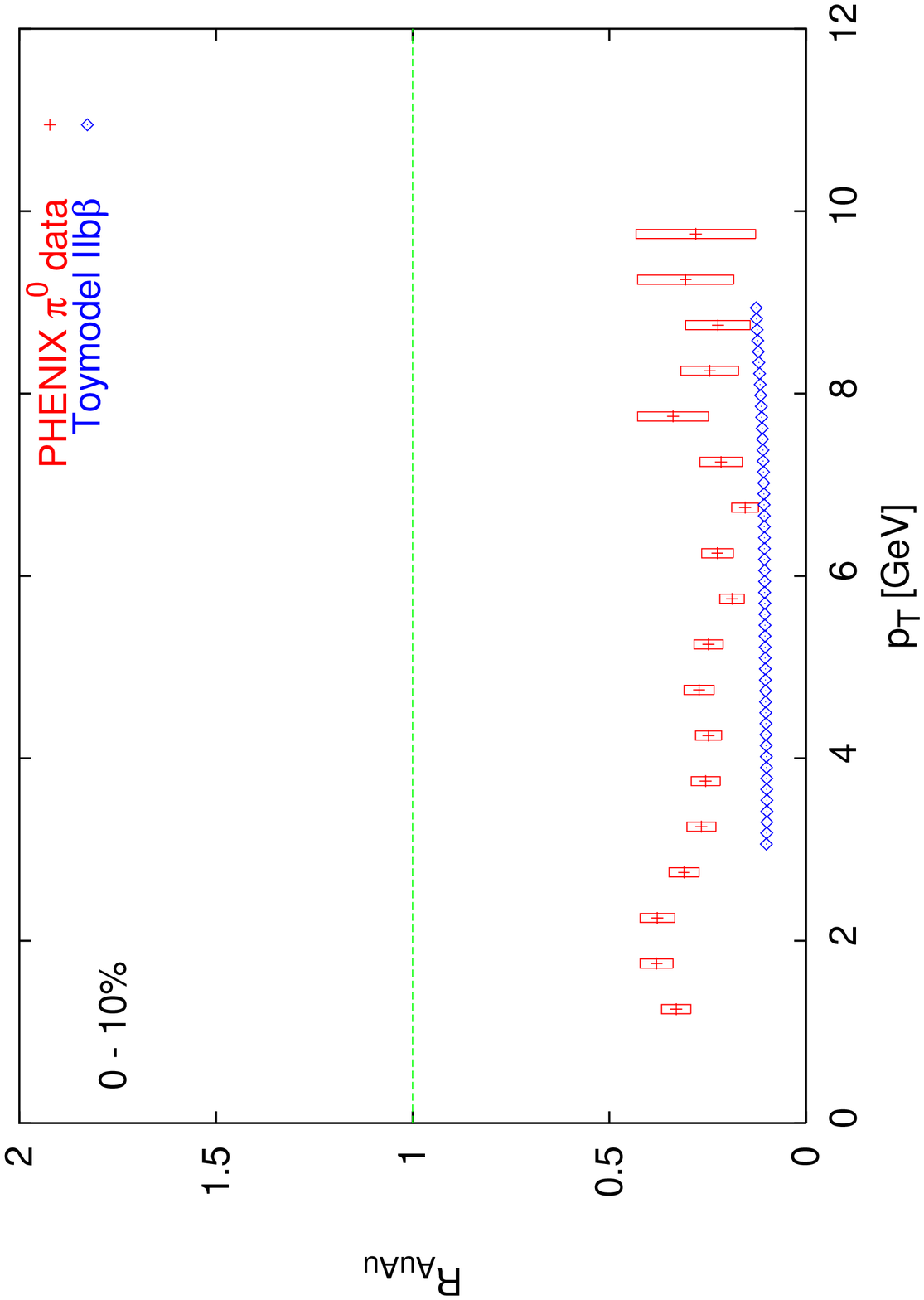}
 \end{minipage} 
 \begin{minipage}{.45\textwidth}
  \centering
  \includegraphics[scale=0.325]{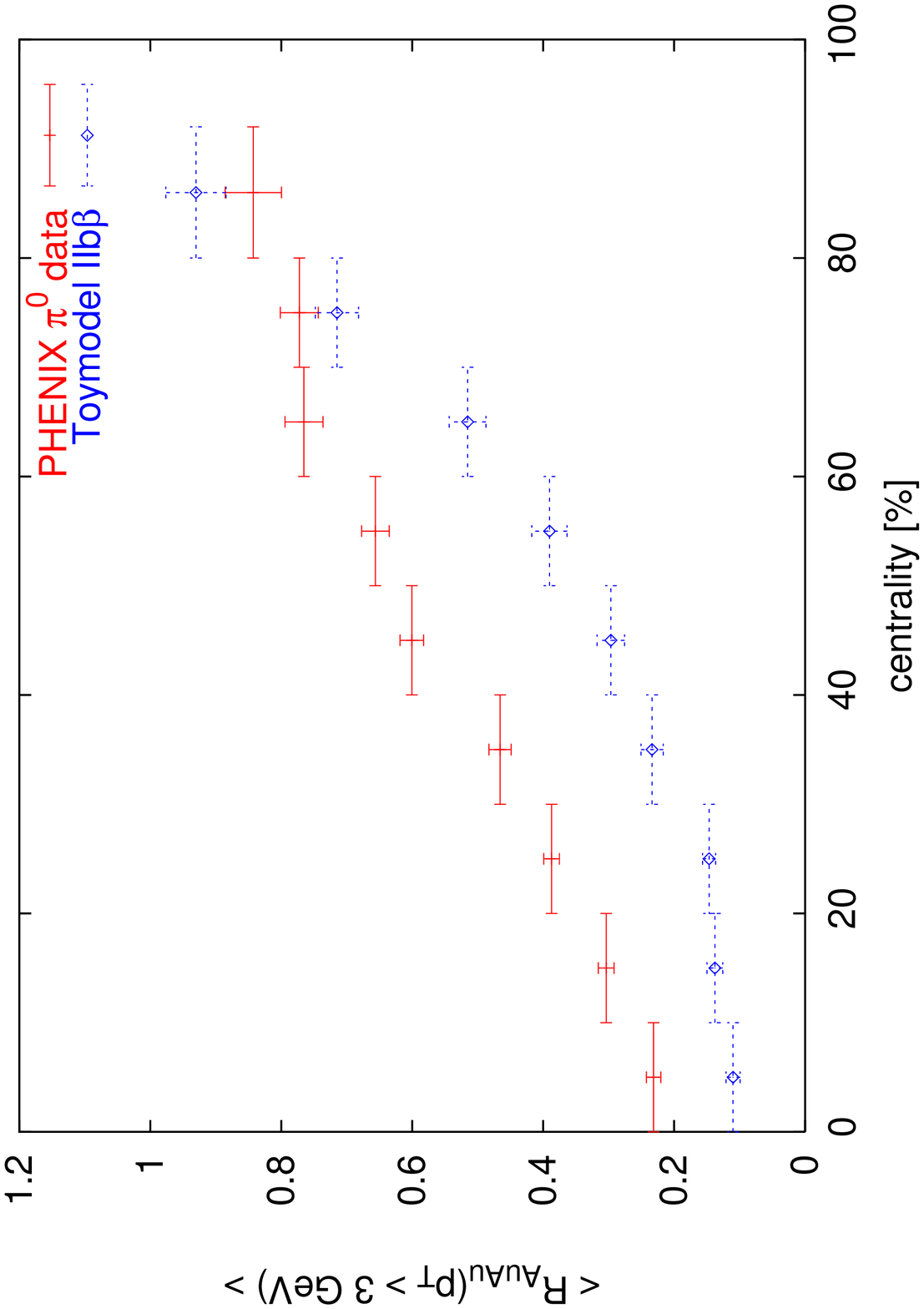}
 \end{minipage} 
 \caption{Results for the Toymodel II\,b\,$\beta$ with $\Delta E \propto N_g^2
 \sqrt{E}$ and inhomogeneous energy density distribution}
\end{figure}

\begin{figure}[ht]
 \centering
 \begin{minipage}{.45\textwidth}
  \centering
  \includegraphics[scale=0.325]{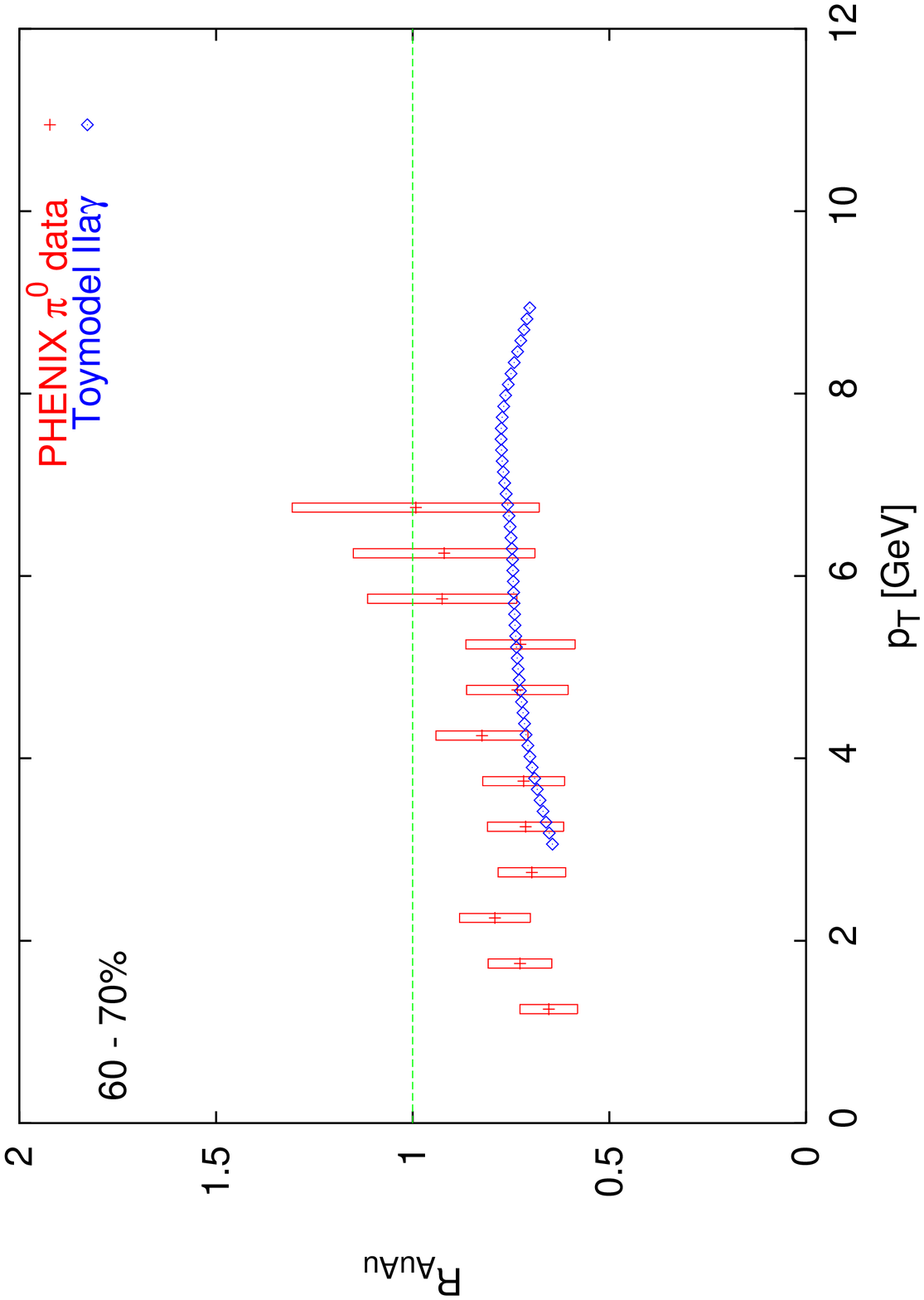}
 \end{minipage} 
 \begin{minipage}{.45\textwidth}
  \centering
  \includegraphics[scale=0.325]{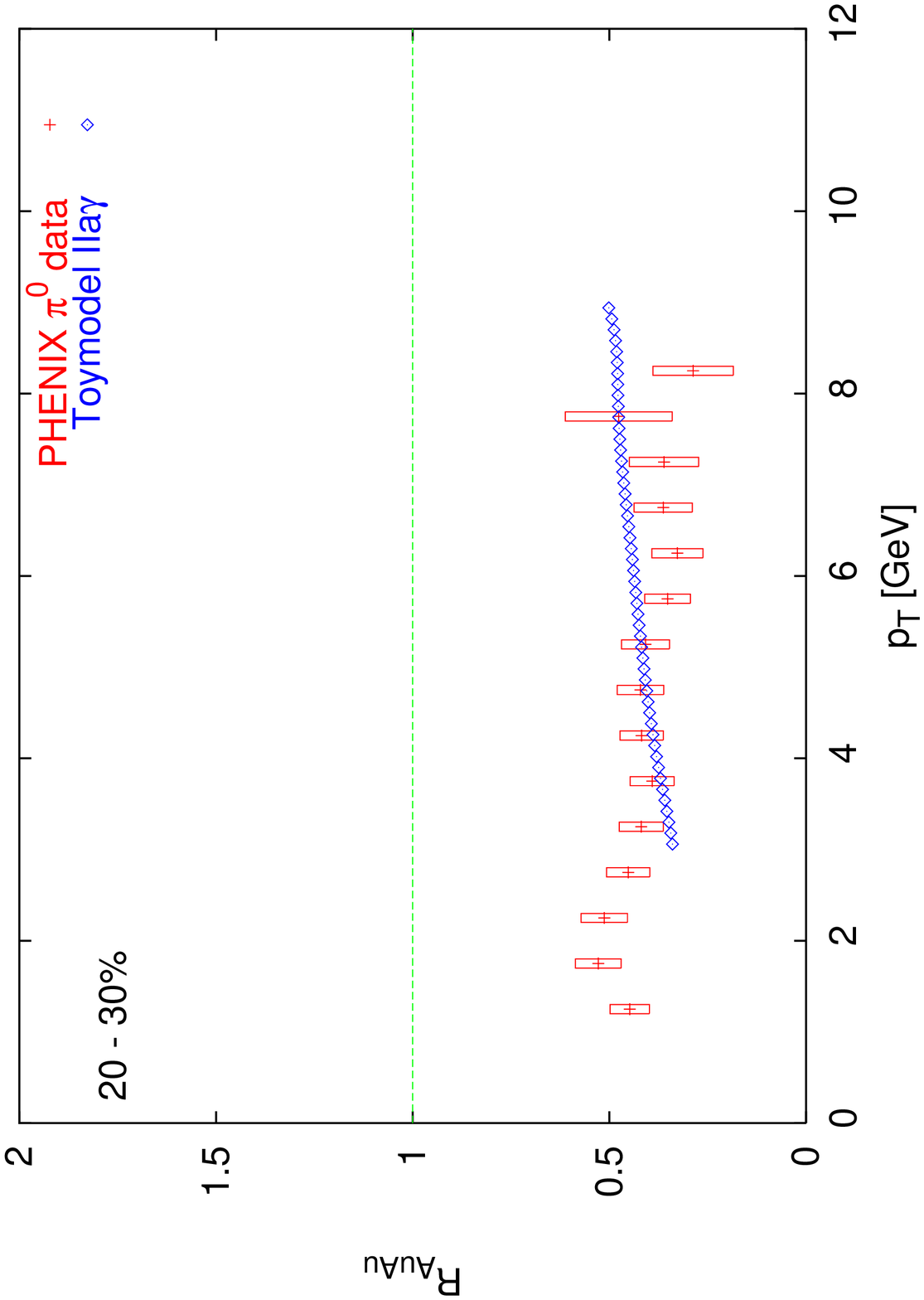}
 \end{minipage} 
 \begin{minipage}{.45\textwidth}
  \centering
  \includegraphics[scale=0.325]{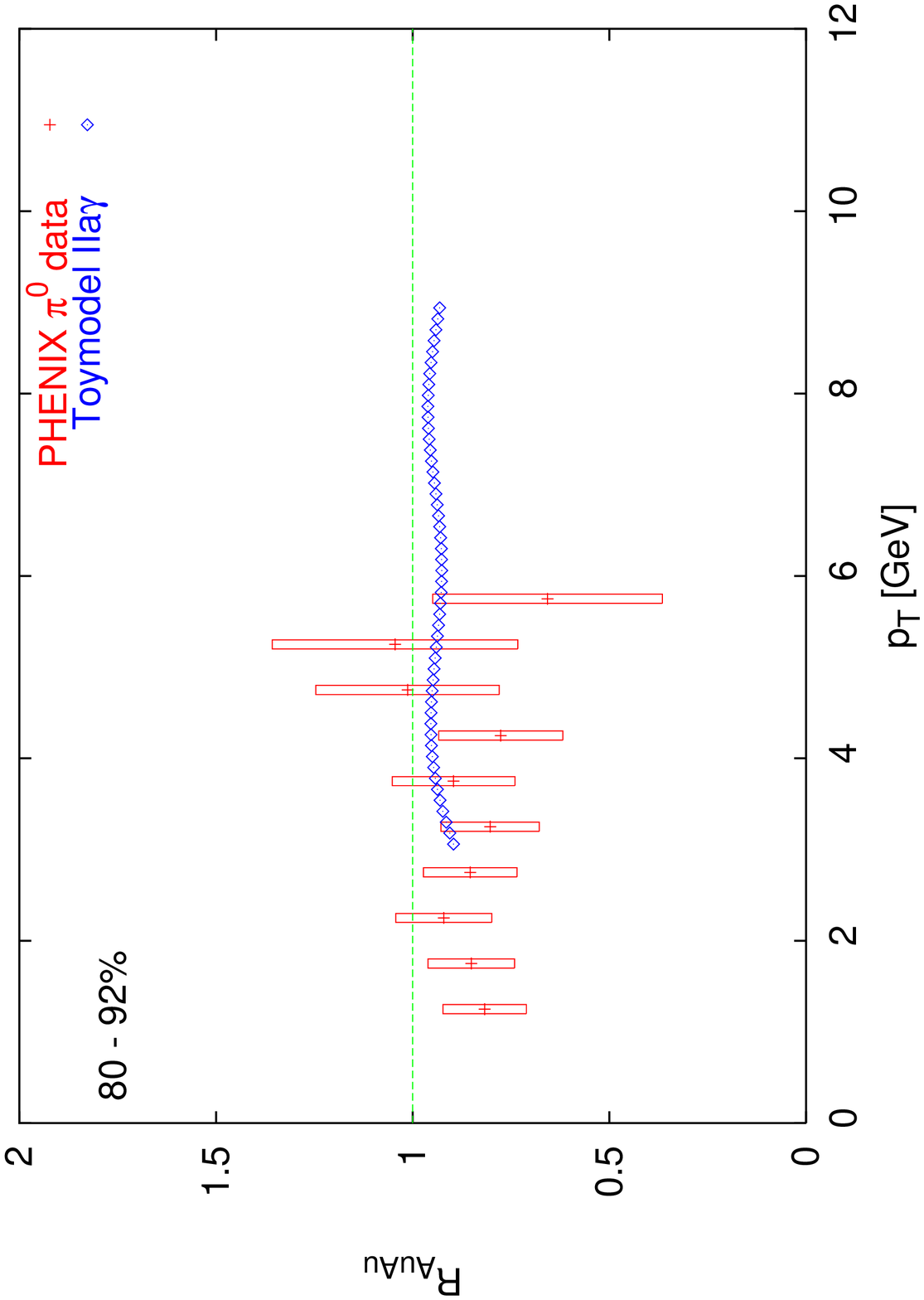}
 \end{minipage} 
 \begin{minipage}{.45\textwidth}
  \centering
  \includegraphics[scale=0.325]{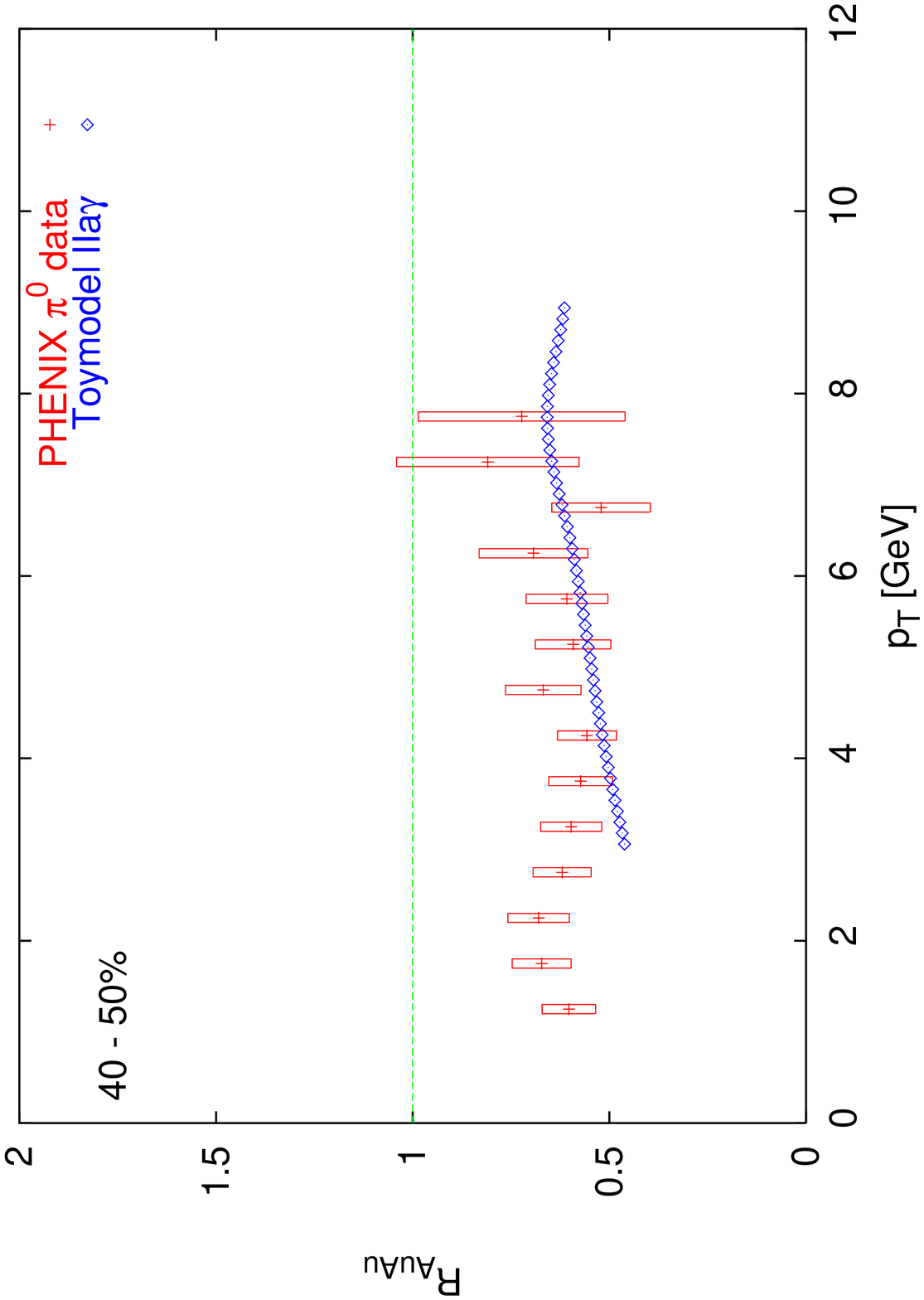}
 \end{minipage} 
\end{figure}

\begin{figure}[ht]
 \centering
 \begin{minipage}{.45\textwidth}
  \centering
  \includegraphics[scale=0.325]{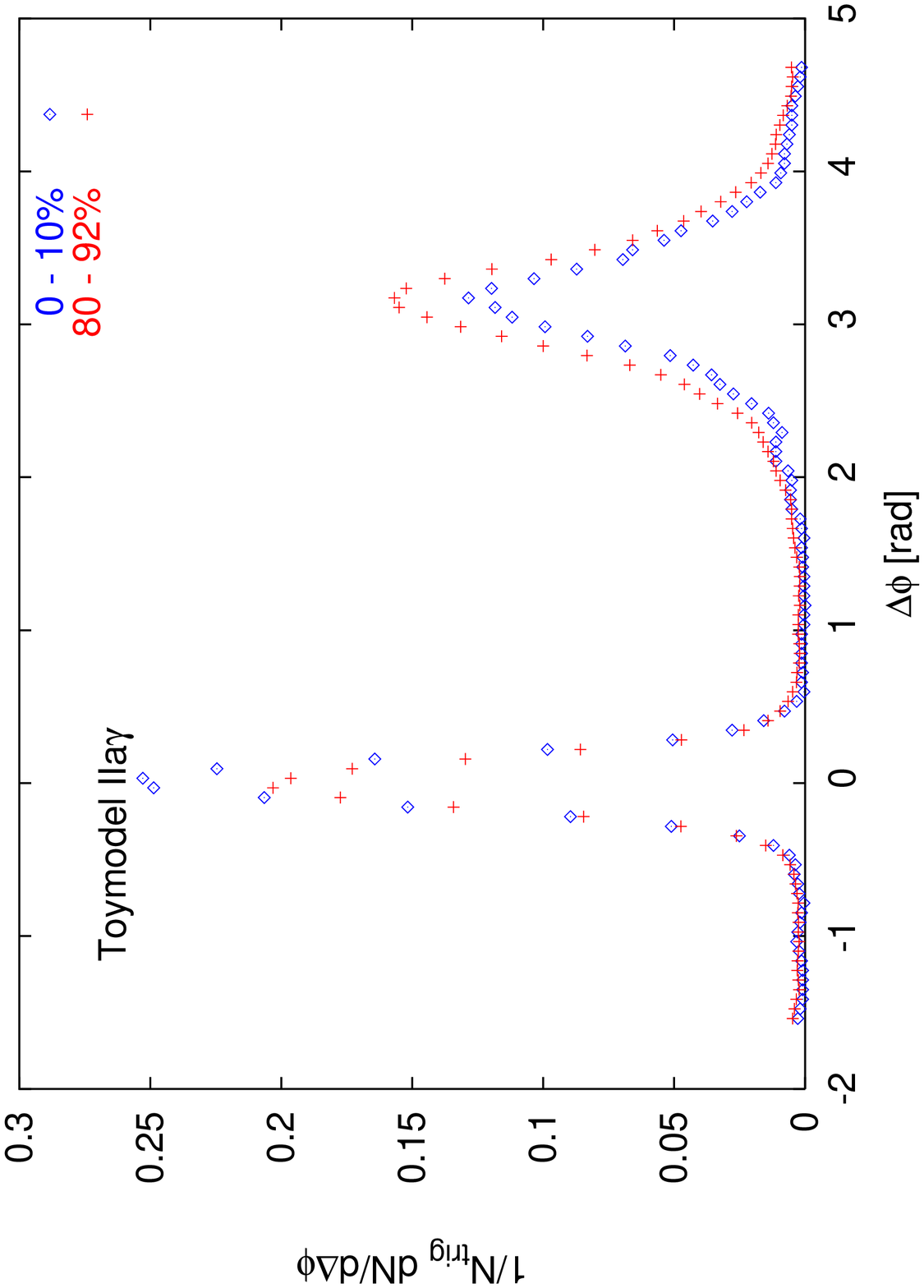}
 \end{minipage} 
 \begin{minipage}{.45\textwidth}
  \centering
  \hspace*{.4\textwidth}
 \end{minipage} 
 \begin{minipage}{.45\textwidth}
  \centering
  \includegraphics[scale=0.325]{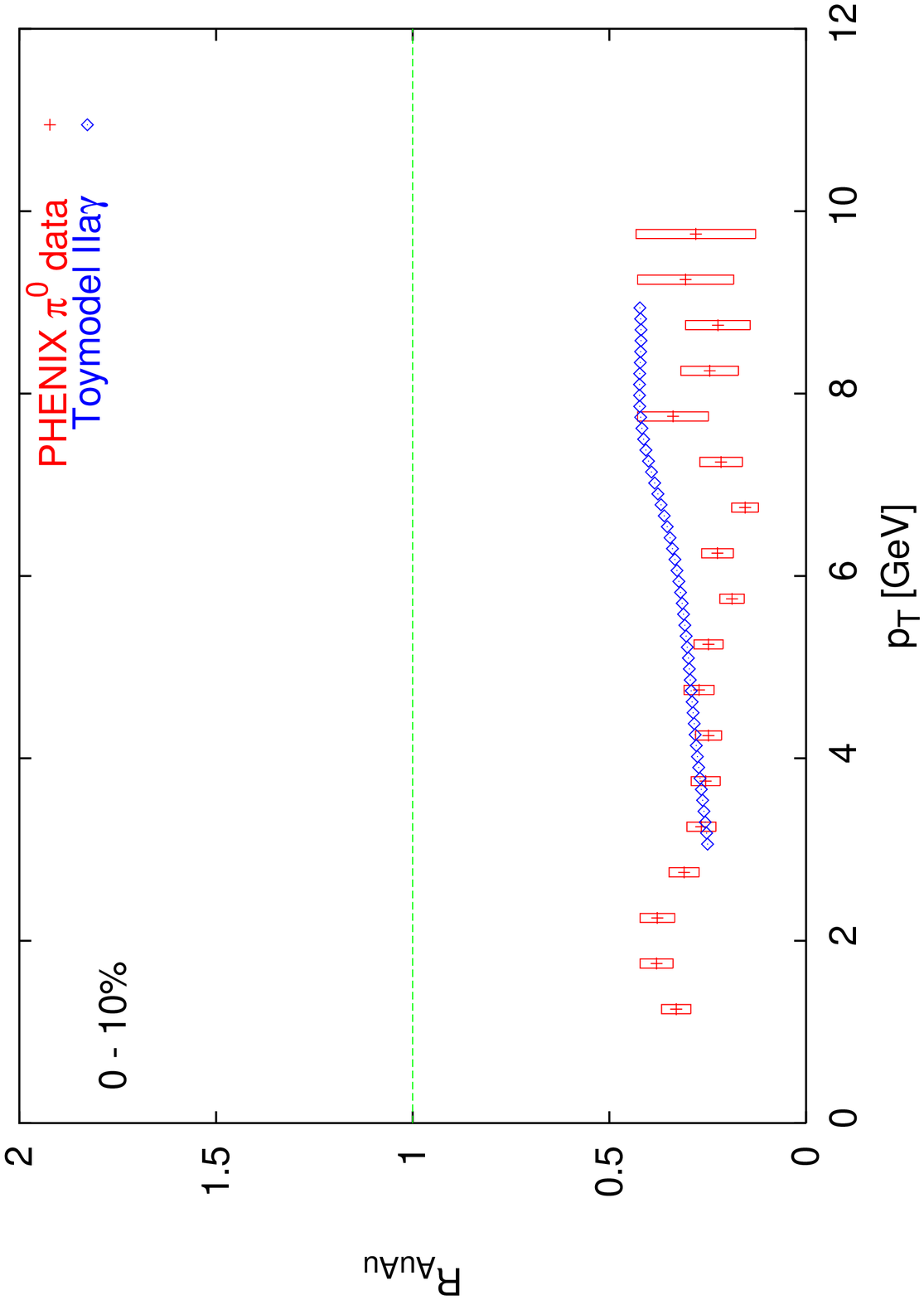}
 \end{minipage} 
 \begin{minipage}{.45\textwidth}
  \centering
  \includegraphics[scale=0.325]{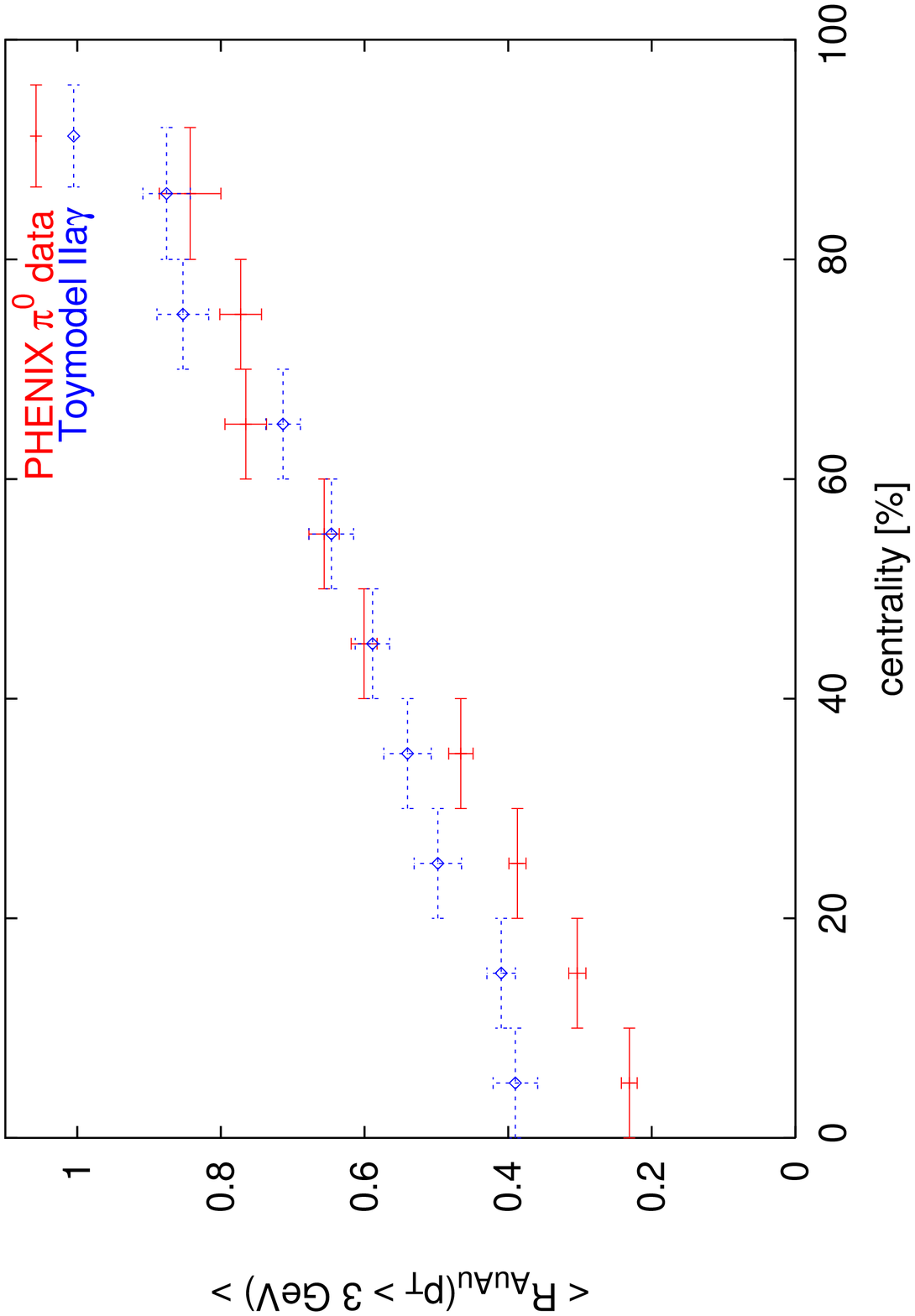}
 \end{minipage} 
 \caption{Results for the Toymodel II\,a\,$\gamma$ with $\Delta E \propto N_g$ 
 and inhomogeneous energy density distribution}
\end{figure}

\begin{figure}[ht]
 \centering
 \begin{minipage}{.45\textwidth}
  \centering
  \includegraphics[scale=0.325]{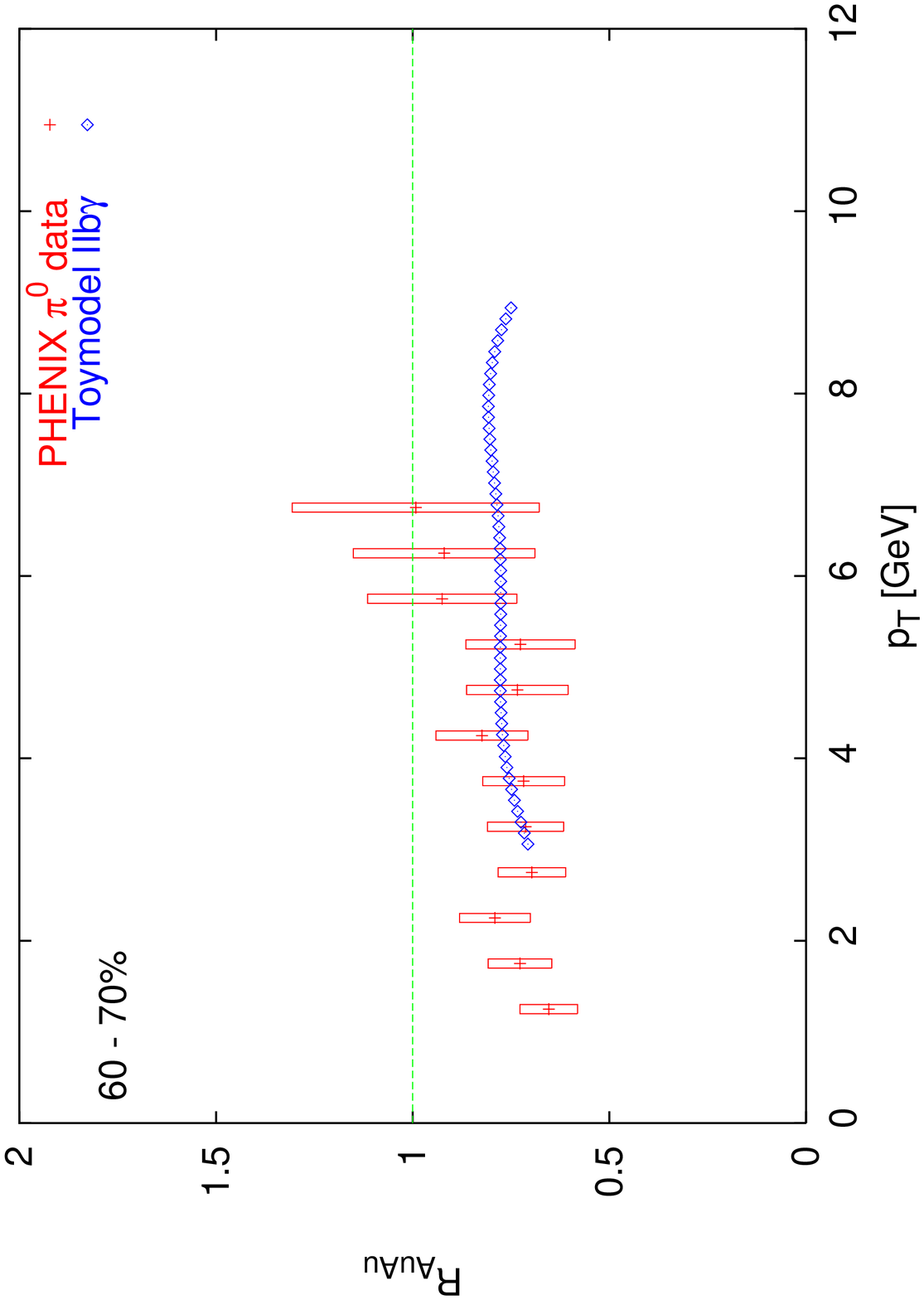}
 \end{minipage} 
 \begin{minipage}{.45\textwidth}
  \centering
  \includegraphics[scale=0.325]{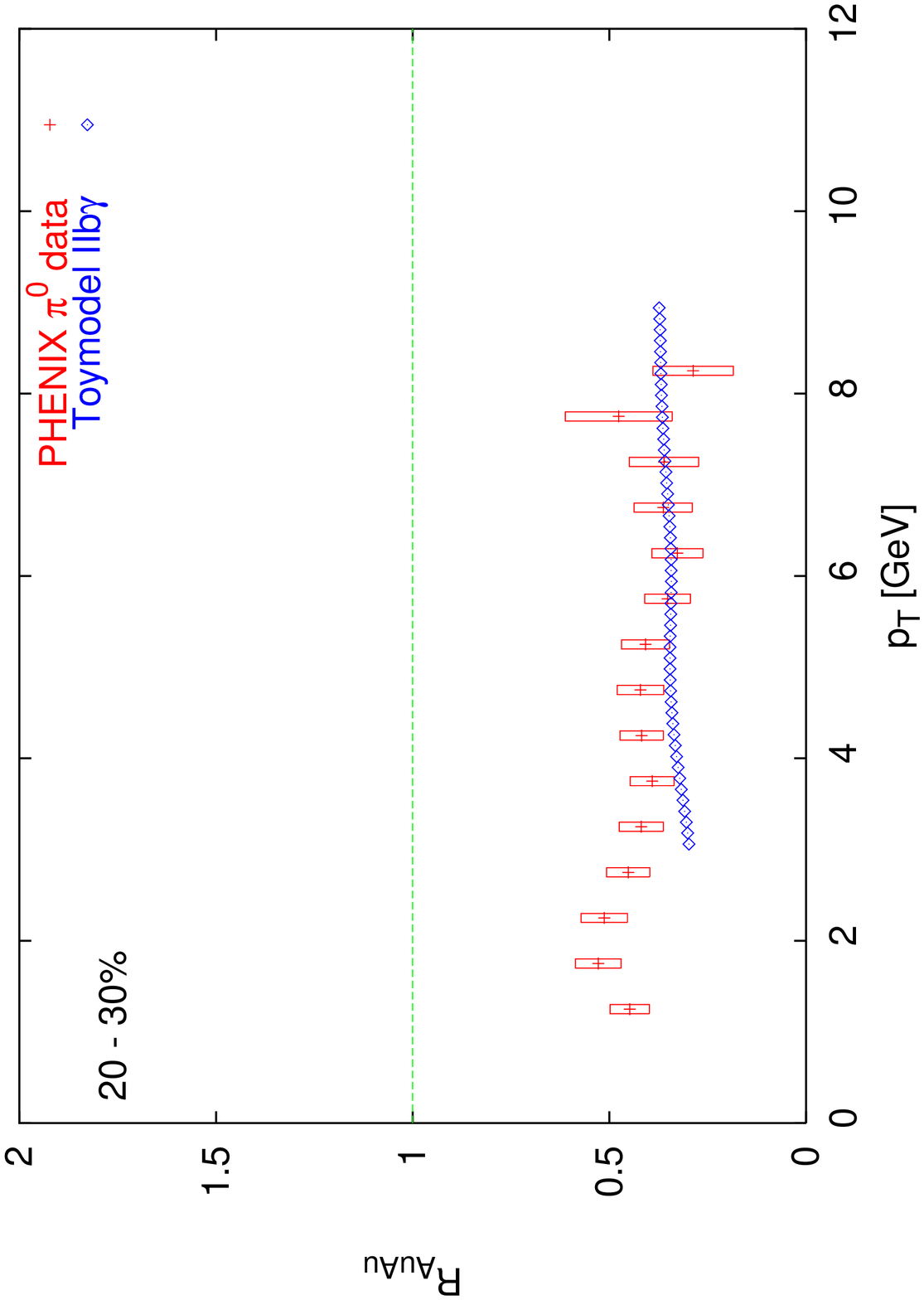}
 \end{minipage} 
 \begin{minipage}{.45\textwidth}
  \centering
  \includegraphics[scale=0.325]{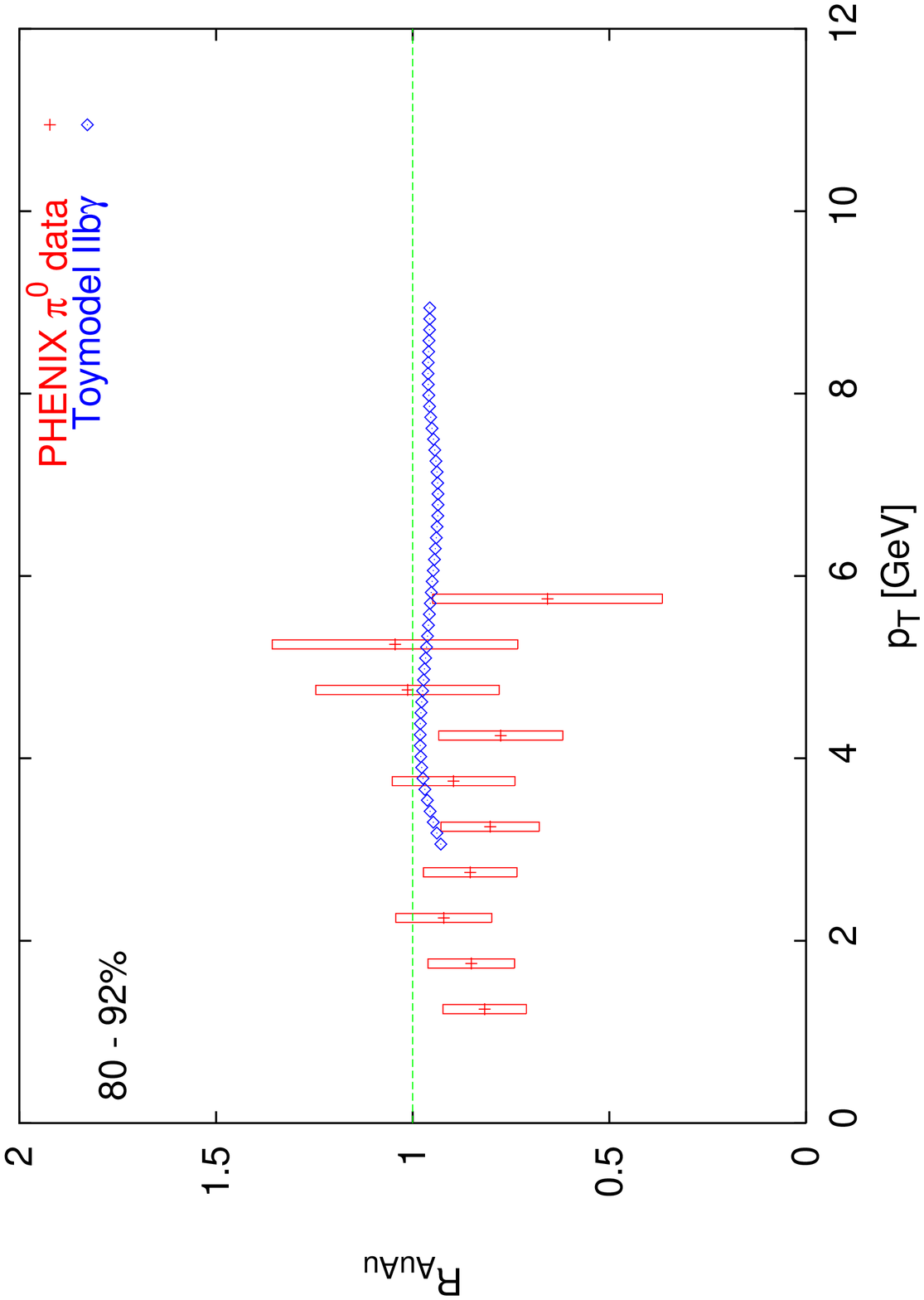}
 \end{minipage} 
 \begin{minipage}{.45\textwidth}
  \centering
  \includegraphics[scale=0.325]{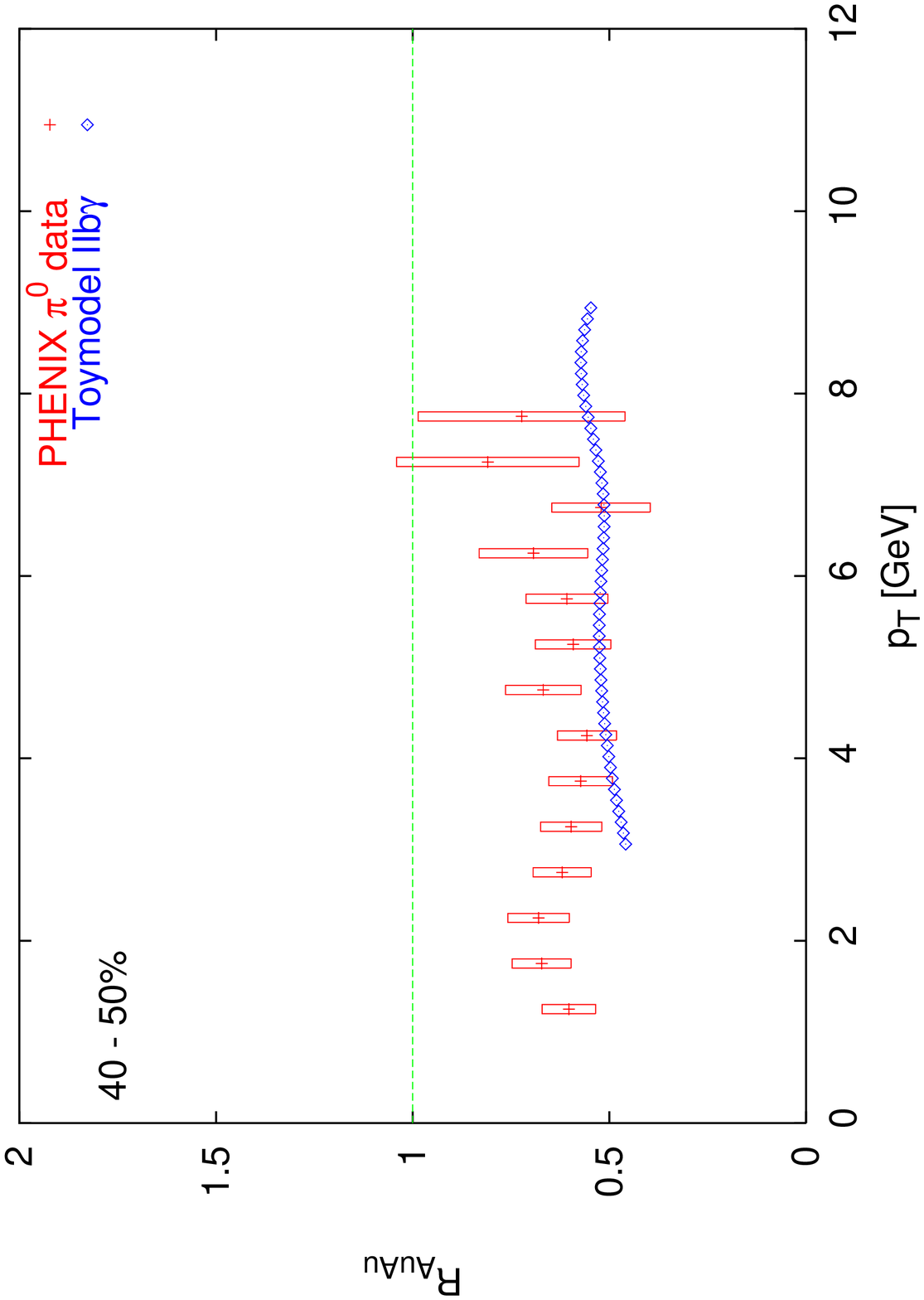}
 \end{minipage} 
\end{figure}

\begin{figure}[ht]
 \centering
 \begin{minipage}{.45\textwidth}
  \centering
  \includegraphics[scale=0.325]{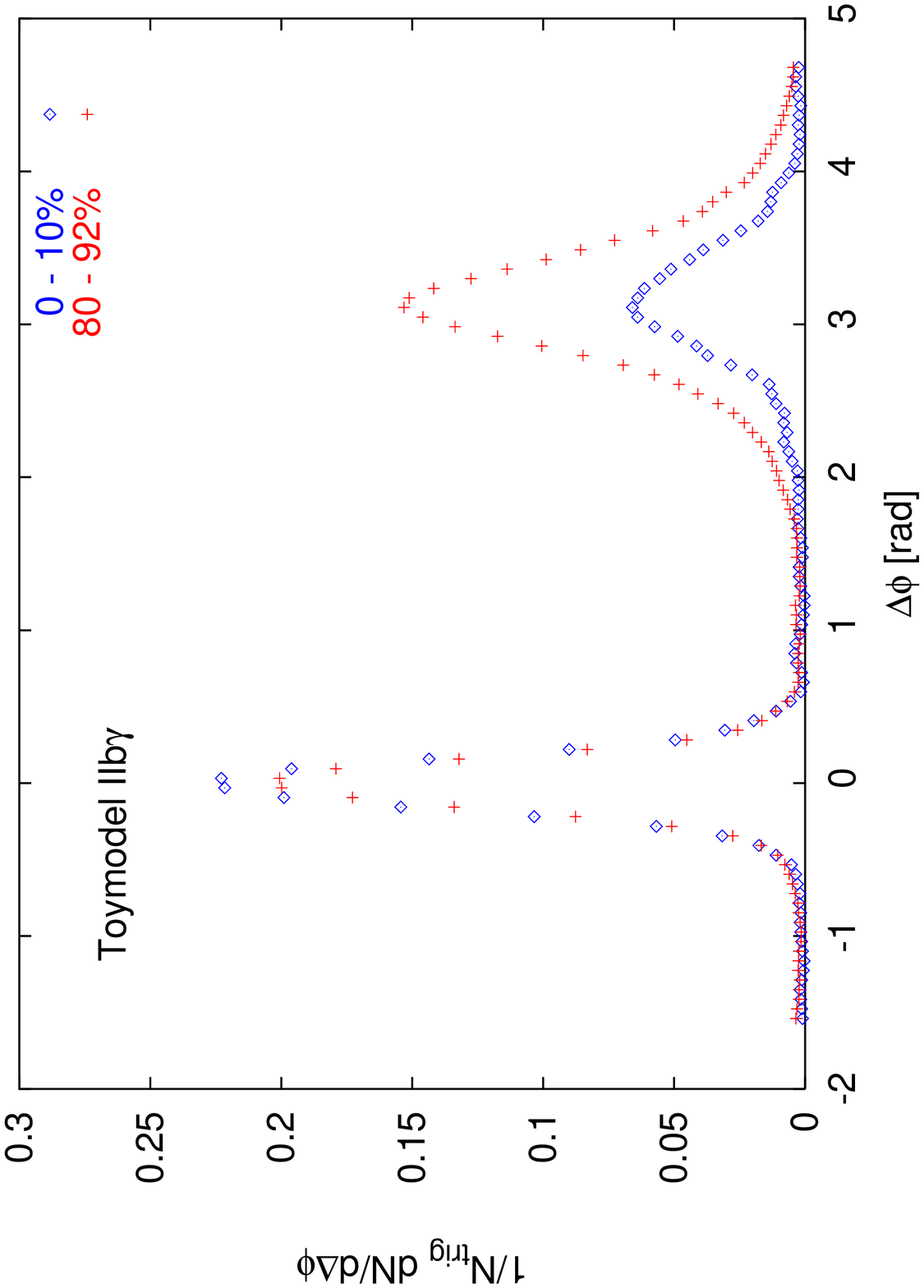}
 \end{minipage} 
 \begin{minipage}{.45\textwidth}
  \centering
  \hspace*{.4\textwidth}
 \end{minipage} 
 \begin{minipage}{.45\textwidth}
  \centering
  \includegraphics[scale=0.325]{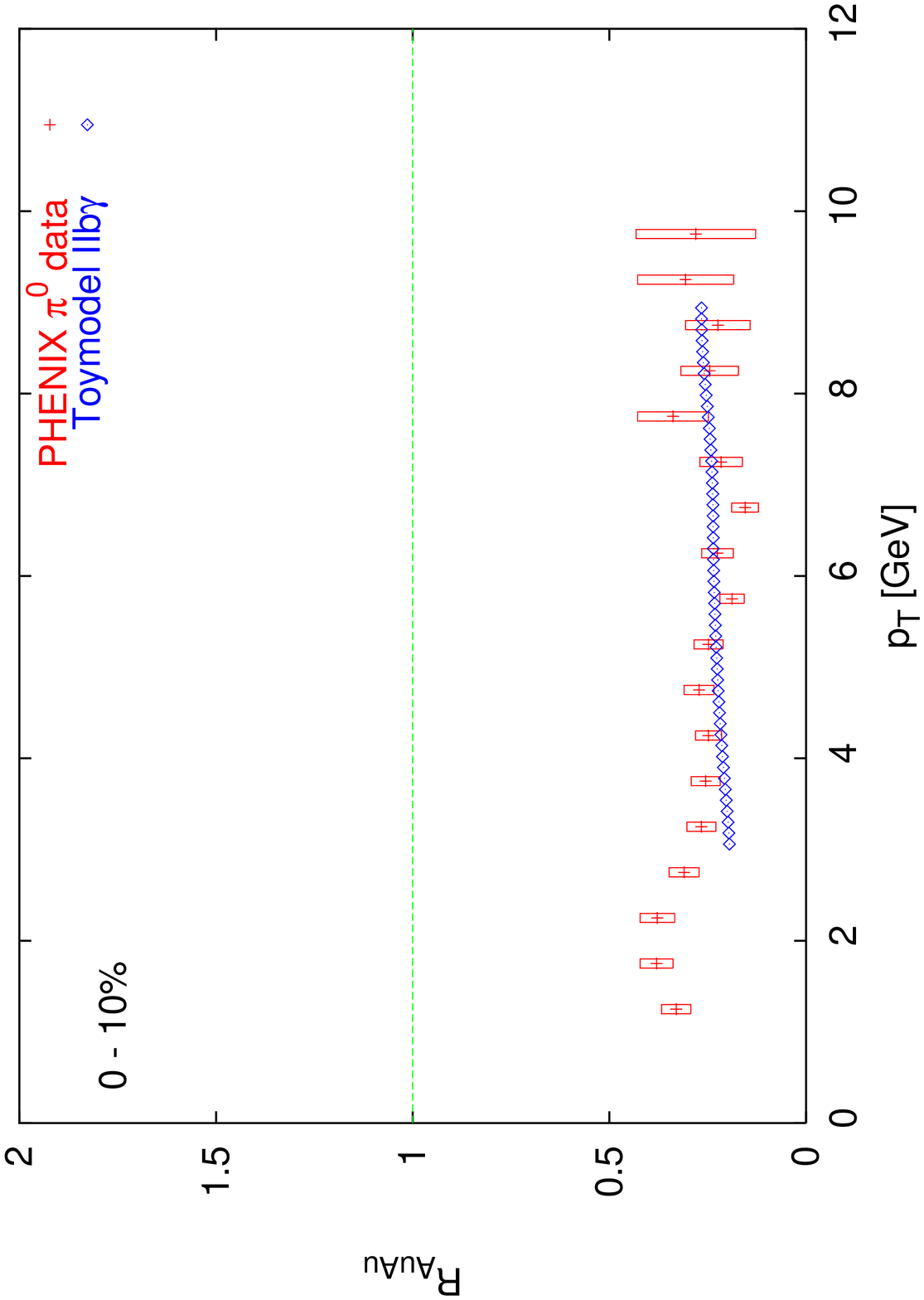}
 \end{minipage} 
 \begin{minipage}{.45\textwidth}
  \centering
  \includegraphics[scale=0.325]{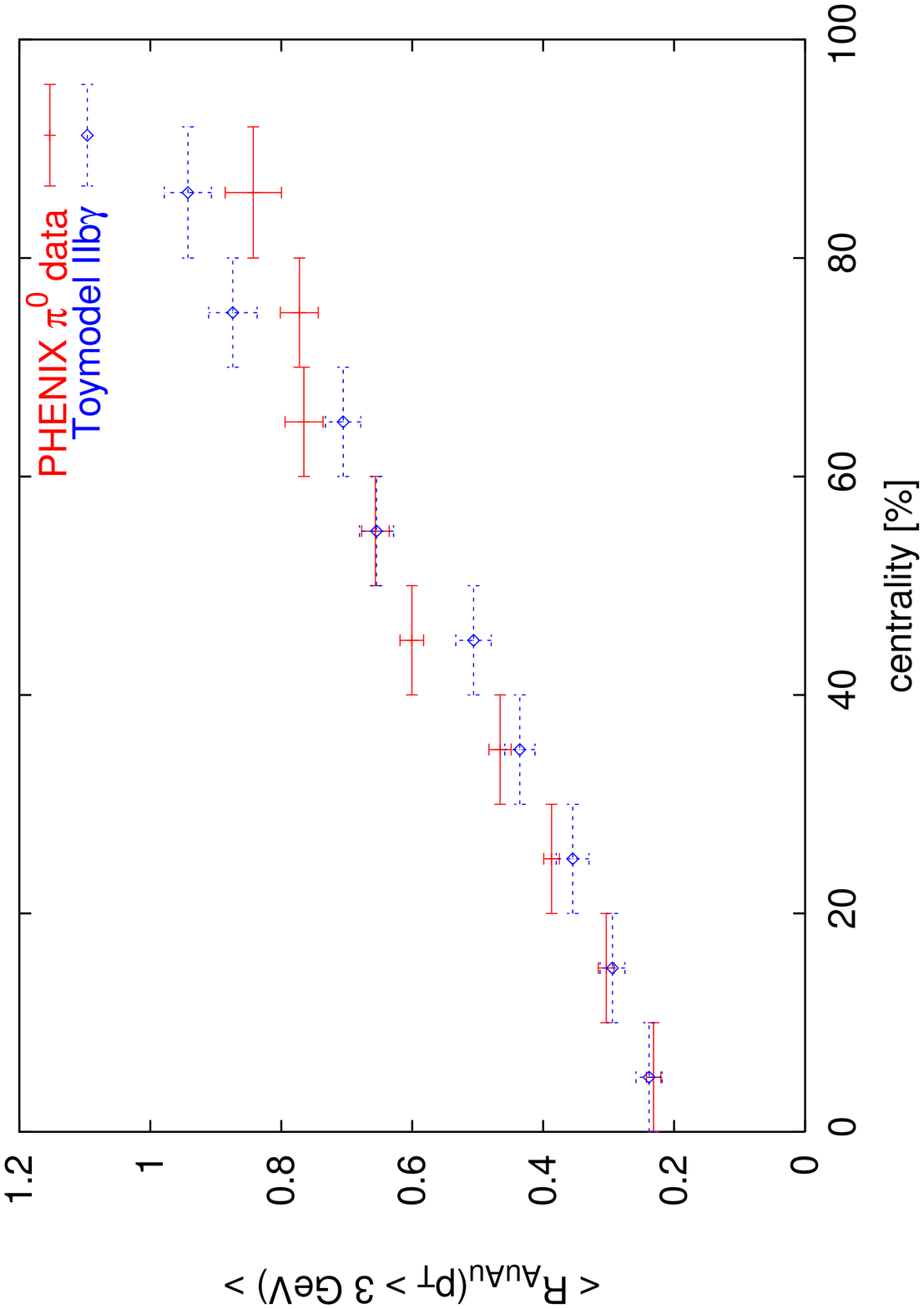}
 \end{minipage} 
 \caption{Results for the Toymodel II\,b\,$\gamma$ with $\Delta E \propto N_g^2$ 
 and inhomogeneous energy density distribution}
\end{figure}

\end{appendix}

\end{document}